**Naoto Shiraishi and Shinji Takesue**

# Complete ergodicity in one-dimensional reversible cellular automata

**Abstract** Exactly ergodicity in boundary-driven semi-infinite cellular automata (CA) are investigated. We establish all the ergodic rules in CA with 3, 4, and 5 states. We analytically prove the ergodicity for 12 rules in 3-state CA and 118320 rules in 5-state CA with any ergodic and periodic boundary condition, and numerically confirm all the other rules non-ergodic with some boundary condition. We classify ergodic rules into several patterns, which exhibit a variety of ergodic structure.

## 1 Introduction

Ergodicity is an important property expected to chaotic systems. Here, the dynamics is called ergodic if there is only one orbit on a sector in a finite phase space restricted by possible conserved quantities [12]. The notion of ergodicity was first introduced by Boltzmann [1] in order to justify the use of ensembles in statistical mechanics. This idea triggered a fruitful mathematical research field named ergodic theory [2, 3].

However, the existence of ergodic orbits is rarely proved except for several hard ball systems [5], and generic Hamiltonian systems are shown to be non-ergodic [7]. For example, nearly integrable Hamiltonian systems are not ergodic due to the existence of invariant tori, which result from the Kolmogorov-Arnold-Moser theorem [6]. In spite of much effort for discovering ergodic systems, including Rannou's reversible integer mapping obtained by coarse-graining a two-dimensional area-preserving mapping [4], few ergodic systems have been established.

In conventional cellular automata (CA) with finite volume, no periodic orbit covers the entire phase space because uniform states are mapped only to uniform states, i.e., there is no path from uniform states to nonuniform states. However, if a CA has conserved quantities [9] and we consider a sector restricted by these conserved quantities, an ergodic orbit may reside on some subset specified by the conserved quantities. This is possible because some conservation laws in CA break the uniformity (translation invariance) of states. One example is rule 12R, which is a reversible variant of Wolfram's elementary cellular automata [8, 11]. Rule 12R, whose state at time $t$ and site $n$ is given by a pair of binary values $(\hat{x}_n^t, x_n^t) \in \{(0,0),(0,1),(1,0),(1,1)\}$, is defined by the following equation

$$\hat{x}_n^{t+1} = x_n^t, \quad x_n^{t+1} = (1 - x_{n-1}^t)x_n^t - \hat{x}_n^t \mod 2. \tag{1}$$

This rule has the following conservation law: If $(\hat{x}_n^0, x_n^0) = (0,0)$ at the initial time, then they stay as $(\hat{x}_n^t, x_n^t) = (0,0)$ for any time $t$. On the contrary, if $(\hat{x}_n^0, x_n^0) \neq (0,0)$, then this site never takes $(\hat{x}_n^t, x_n^t) =$

Naoto Shiraishi
Faculty of arts and sciences, University of Tokyo, 3-8-1 Komaba, Meguro-ku, Tokyo, Japan
E-mail: shiraishi@phys.c.u-tokyo.ac.jp
Shinji Takesue
Department of Physics, Kyoto University, 606-8502, Japan
E-mail: takesue.shinji.4s@kyoto-u.jp

$(0,0)$ for any time $t$. Suppose that sites $x=0$ and $x=N+1$ have the value $(\hat{x}_0, x_0) = (0,0)$ and sites 1 through $N$ do not, which induces the corresponding sector with $3^N$ states; $\{(0,1),(1,0),(1,1)\}^{\otimes N}$. We numerically find and analytically prove that such a sector contains only one periodic orbit. That is, rule 12R has ergodic orbits.

Observe that in rule 12R the dynamics of a cell at site $n$ depends only on itself ($x=n$) and its left cell ($x=n-1$) and does not depend on its right cell ($x=n+1$). Namely, this CA is driven from left to right, and this driving induces ergodic dynamics persisting any length $N$. This ergodic behavior of rule 12R motivates us to investigate a semi-infinite reversible cellular automaton where site 1 periodically evolves and the time-evolution of site $n>1$ is given by a permutation determined by the state of site $n-1$. In rule 12R, site 1 evolves with period 3 like $(\hat{x}_1, x_1) = (1,1) \to (1,0) \to (0,1) \to (1,1)$, because $x_0^t = 0$ for any $t$. The time evolution of site $n$ is given by

$$(\hat{x}_n^{t+1}, x_n^{t+1}) = \begin{cases} (1,1) & \text{if } (\hat{x}_n^t, x_n^t) = (0,1) \\ (0,1) & \text{if } (\hat{x}_n^t, x_n^t) = (1,0) \\ (1,0) & \text{if } (\hat{x}_n^t, x_n^t) = (1,1) \end{cases} \tag{2}$$

if $x_{n-1}^t = 0$, and

$$(\hat{x}_n^{t+1}, x_n^{t+1}) = (x_n^t, \hat{x}_n^t) \tag{3}$$

if $x_{n-1}^t = 1$. They are permutations. Moreover, because the rule does not depend on the right nearest neighbor, site $N+1$ plays the role only restricting the system size.

In this paper, we investigate this kind of semi-infinite CA and clarify the condition for the existence of ergodic orbits. We thoroughly examine semi-infinite CA with 3, 4, and 5 states. We numerically find and analytically prove that there are 12, 0, and 118320 ergodic rules in CA with 3, 4, and 5 states, respectively. The 12 ergodic rules in CA with 3 states are classified into 2 types, both of which are proven to be ergodic. The ergodicity of rule 12R is also proved in this context. The 118320 rules in CA with 5 states are classified into 72 types with 206 subtypes, all of which can be proven to be ergodic. Other rules are numerically shown to be non-ergodic. In other words, we succeed in determining all ergodic rules in CA with 3, 4, and 5 states.

A rule of CA where the dynamic of a cell depends only on itself and its left cells is known as a one-way CA. One-way CAs are studied with different boundary conditions from ours [10]. However, as far as we know, our study is the first exhaustive classification of semi-infinite one-way CAs with periodic driving of the leftmost cell.

This paper is organized as follows. In Section. 2, we introduce definitions and notation in terms of symmetric groups. In Section. 3, we prove all ergodic rules in 3-state CA in detail, which plays as a good introduction to the proof of ergodicity in 5-state CA. In Section. 4, we treat 4-state CA, which has no ergodic rules. In Section. 5, we summarize the results on 5-state CA, and present the proof strategy and symbols employed in the proof. In Section. 6, we prove the ergodicity of rules in 5-state CA. We classify rules into 5 patterns with some sub-categories, and present proofs for several representative rules. In Section. 7, we consider generalizations of petterns to CA with more than 5 states. In Section. 8, we list several open problems.

We list all the subtypes of ergodic rules and their transition maps in 5-state in Sec. 5.1. The structure of these ergodic rules, which play a crucial role in proving their ergodicity, is presented in Appendix. A .

## 2 Semi-infinite reversible cellular automata

### 2.1 Symmetric group

For completeness, we first describe some basic facts of symmetric groups. The symmetric group $S_k$ is composed of all bijections from $\{0,1,\ldots,k-1\}$ to itself. The elements of $S_k$ are called the permutations. Each permutation $\pi \in S_k$ is specified in the two-line form $\begin{pmatrix} 0 & 1 & \cdots & k-1 \\ \pi(0) & \pi(1) & \cdots & \pi(k-1) \end{pmatrix}$. A permutation $\pi$ is called a *p-cycle* or a cycle of length $p$ if $\pi^p(x) = x$ and $\pi^q(x) \neq x$ for $0 \leq q < p$. In the *cycle notation*, a $p$-cycle $\pi$ is expressed by the enumeration of states as $(x\ \pi(x)\ \pi^2(x)\dots\pi^{p-1}(x))$. Every permutation is represented by a product of non-overlapping cycles. An example of the cycle notation is

$$\begin{pmatrix} 0\ 1\ 2\ 3\ 4\ 5 \\ 1\ 3\ 4\ 0\ 2\ 5 \end{pmatrix} = (0\,1\,3)(2\,4). \tag{4}$$

In this paper, we omit 1-cycles from the cycle notation for brevity. Let $n_k$ be the numbers of $k$-cycles in $\pi$, and using this we introduce the *cycle type* of $\pi$ as $1^{n_1}\ 2^{n_2}\cdots k^{n_k}$, where we can omit the factors with $n_k = 0$. In this expression, the superscript of $m$ represents the number of $m$-cycles in this permutation. For example, the cycle type of the permutation in Eq. (4) is $2^1 3^1$. We call permutations $\pi$ and $\sigma$ *conjugate* if they satisfy $\pi = \tau\sigma\tau^{-1}$ for some $\tau$. Two permutations are conjugate if and only if they have the same cycle type. The permutations with the same cycle type compose a conjugacy class.

We encode each permutation $\pi \in S_k$ by a natural number as

$$n(\pi) = \sum_{i=0}^{k-2} r(\pi, i)(k-1-i)!, \tag{5}$$

where $r(\pi, i) = |\{i \le j \mid \pi(j) \le \pi(i)\}| - 1$, and $|A|$ means the number of elements in set $A$. Noting $0 \le r(\pi, i) \le k - i - 1$, we confirm that this encoding provides a one-to-one correspondence between $k!$ possible permutations and integers $0 \le n \le k! - 1$. In this paper, we specify permutations by using integers with this encoding.

In the case of $k = 3$, we have the following 6 permutations

$$\sigma_0 = \begin{pmatrix} 0\ 1\ 2 \\ 0\ 1\ 2 \end{pmatrix} = \text{id.}, \qquad \sigma_1 = \begin{pmatrix} 0\ 1\ 2 \\ 0\ 2\ 1 \end{pmatrix} = (1\,2), \qquad \sigma_2 = \begin{pmatrix} 0\ 1\ 2 \\ 1\ 0\ 2 \end{pmatrix} = (0\,1), \tag{6}$$

$$\sigma_3 = \begin{pmatrix} 0\ 1\ 2 \\ 1\ 2\ 0 \end{pmatrix} = (0\,1\,2), \qquad \sigma_4 = \begin{pmatrix} 0\ 1\ 2 \\ 2\ 0\ 1 \end{pmatrix} = (0\,2\,1), \qquad \sigma_5 = \begin{pmatrix} 0\ 1\ 2 \\ 2\ 1\ 0 \end{pmatrix} = (0\,2). \tag{7}$$

As seen from this, $S_3$ is decomposed into the sum of conjugacy classes; $(1^3) = \{\sigma_0\}$, $(1^1 2^1) = \{\sigma_1, \sigma_2, \sigma_5\}$, and $(3^1) = \{\sigma_3, \sigma_4\}$.

In the case of $k = 5$, we have 120 permutations listed in Table. 1:

## 2.2 Semi-infinite reversible CA

Throughout this paper, we consider the semi-infinite reversible cellular automata with $k$ states defined by the following time-evolution rules:

$$x_1^{t+1} = \pi_{\mathrm{B}}(x_1^t) \tag{8}$$

$$x_n^{t+1} = \pi_{x_{n-1}^t}(x_n^t), \quad (n \ge 2) \tag{9}$$

where $x_n^t$ takes values in the set $\{0, 1, \dots, k-1\}$, $\pi_{\mathrm{B}}$ is a fixed cycle of length $k$, and $\pi_x$, $(x \in \{0, 1, \dots, k-1\})$ is a permutation. The fixed permutation $\pi_{\mathrm{B}}$ in the first line is the boundary driving on $n = 1$. The second line means that the permutation acting on site $n(\ge 2)$ at time $t$ is determined by the state on site $n - 1$ at the same time $t$.

The reversibility of the dynamics is evident because $x_1^t = \pi_{\mathrm{B}}^{-1}(x_1^{t+1})$, and $x_n^t$ is inductively obtained from $(x_1^{t+1}, \dots, x_n^{t+1})$ by using the relation $x_n^t = \pi^{-1}_{x_{n-1}^t}(x_n^{t+1})$. The rule of CA is specified by $\pi_{\mathrm{B}}$ and a set of $k$ permutations $(\pi_0, \dots, \pi_{k-1})$. If $\pi_s$ $(0 \le s \le k-1)$ is encoded by $i_s$, we also refer to the rule as $(i_0, \dots, i_{k-1})$ (with boundary dynamics $\pi_{\mathrm{B}}$). Thus, there are $(k!)^k$ rules and $(k-1)!$ boundary dynamics in the $k$-state semi-infinite reversible cellular automata.

The phase space of site 1 through $n$ consists of $k^n$ states. Since the sites $m > n$ do not affect the dynamics of those $n$ sites, the sites $1 \le i \le n$ can be considered as a closed dynamical system. We say that site $n$ is ergodic if it evolves with period $k^n$ and has no shorter period, and that the rule is ergodic if every site is ergodic. Since $\pi_{\mathrm{B}}$ is a $k$ cycle, site 1 is ergodic for any rule by construction. However, the ergodicity of sites $n \ge 2$ depends on the rule.

**Table 1** Permutations $\sigma_i \in S_5$. Each row shows $i$, $\sigma_i(0)\sigma_i(1)\sigma_i(2)\sigma_i(3)\sigma_i(4)$ and its cycle notation.

| | | | | | | | | |
|---|---|---|---|---|---|---|---|---|
| 0 | 01234 | id | 40 | 13402 | (013)(24) | 80 | 31204 | (03) |
| 1 | 01243 | (34) | 41 | 13420 | (01324) | 81 | 31240 | (034) |
| 2 | 01324 | (23) | 42 | 14023 | (01432) | 82 | 31402 | (03)(24) |
| 3 | 01234 | (234) | 43 | 14032 | (0142) | 83 | 31420 | (0324) |
| 4 | 01423 | (243) | 44 | 14203 | (0143) | 84 | 32014 | (0312) |
| 5 | 01432 | (24) | 45 | 14230 | (014) | 85 | 32041 | (03412) |
| 6 | 02134 | (12) | 46 | 14302 | (01423) | 86 | 32104 | (03)(12) |
| 7 | 02143 | (12)(34) | 47 | 14320 | (014)(23) | 87 | 32140 | (034)(12) |
| 8 | 02314 | (123) | 48 | 20134 | (021) | 88 | 32401 | (03)(124) |
| 9 | 02341 | (1234) | 49 | 20143 | (021)(34) | 89 | 32410 | (03124) |
| 10 | 02413 | (1243) | 50 | 20314 | (0231) | 90 | 34012 | (03142) |
| 11 | 02431 | (124) | 51 | 20341 | (02341) | 91 | 34021 | (032)(14) |
| 12 | 03124 | (132) | 52 | 20413 | (02431) | 92 | 34102 | (03)(142) |
| 13 | 03142 | (1342) | 53 | 20431 | (0241) | 93 | 34120 | (03214) |
| 14 | 03214 | (13) | 54 | 21034 | (02) | 94 | 34201 | (03)(14) |
| 15 | 03241 | (134) | 55 | 21043 | (02)(34) | 95 | 34210 | (0314) |
| 16 | 03412 | (13)(24) | 56 | 21304 | (023) | 96 | 40123 | (04321) |
| 17 | 03421 | (1324) | 57 | 21340 | (0234) | 97 | 40132 | (0421) |
| 18 | 04123 | (1432) | 58 | 21403 | (0243) | 98 | 40213 | (0431) |
| 19 | 04132 | (142) | 59 | 21430 | (024) | 99 | 40231 | (041) |
| 20 | 04213 | (143) | 60 | 23014 | (02)(13) | 100 | 40312 | (04231) |
| 21 | 04231 | (14) | 61 | 23041 | (02)(134) | 101 | 40321 | (041)(23) |
| 22 | 04312 | (1423) | 62 | 23104 | (0213) | 102 | 41023 | (0432) |
| 23 | 04321 | (14)(23) | 63 | 23140 | (02134) | 103 | 41032 | (042) |
| 24 | 10234 | (01) | 64 | 23401 | (02413) | 104 | 41203 | (043) |
| 25 | 10243 | (01)(34) | 65 | 23410 | (024)(13) | 105 | 41230 | (04) |
| 26 | 10324 | (01)(23) | 66 | 24013 | (02)(143) | 106 | 41302 | (0423) |
| 27 | 10342 | (01)(234) | 67 | 24031 | (02)(14) | 107 | 41320 | (04)(23) |
| 28 | 10423 | (01)(243) | 68 | 24103 | (02143) | 108 | 42013 | (04312) |
| 29 | 10432 | (01)(24) | 69 | 24130 | (0214) | 109 | 42031 | (0412) |
| 30 | 12034 | (012) | 70 | 24301 | (023)(14) | 110 | 42103 | (043)(12) |
| 31 | 12043 | (012)(34) | 71 | 24310 | (02314) | 111 | 42130 | (04)(12) |
| 32 | 12304 | (0123) | 72 | 30124 | (0321) | 112 | 42301 | (04123) |
| 33 | 12340 | (01234) | 73 | 30142 | (03421) | 113 | 42310 | (04)(123) |
| 34 | 12403 | (01243) | 74 | 30214 | (031) | 114 | 43012 | (042)(13) |
| 35 | 12430 | (0124) | 75 | 30241 | (0341) | 115 | 43021 | (04132) |
| 36 | 13024 | (0132) | 76 | 30412 | (031)(24) | 116 | 43102 | (04213) |
| 37 | 13042 | (01342) | 77 | 30421 | (03241) | 117 | 43120 | (04)(132) |
| 38 | 13204 | (013) | 78 | 31024 | (032) | 118 | 43201 | (0413) |
| 39 | 13240 | (0134) | 79 | 31042 | (0342) | 119 | 43210 | (04)(13) |

Given a temporal sequence on site $n$ between $t_0 \le t \le t_1$ as $x_n^{t_0} x_n^{t_0+1} \cdots x_n^{t_1-1}$, the transformation on site $n+1$ from $t = t_0$ to $t = t_1$ denoted by $T_{n+1}^{t_0 \to t_1} : x_{n+1}^{t_0} \mapsto x_{n+1}^{t_1}$ is determined by the product of permutations as

$$T_{n+1}^{t_0 \to t_1} = \pi_{x_n^{t_1-1}} \pi_{x_n^{t_1-2}} \cdots \pi_{x_{n+1}^{t_0+1}} \pi_{x_n^{t_0}} . \tag{10}$$

Thus, $T_{n+1}^{t_0 \to t_1}$ itself is a permutation. If site $n$ is ergodic, $T_{n+1}^{0 \to k^n}$ is a $k$-cycle. Conversely, if the period of a site $n$ is $k^n$ and $T_{n+1}^{0 \to k^n}$ is a $k$-cycle, the period of site $n+1$ is $k^{n+1}$.

## 3 CA with 3 states

For $k = 3$, there are $6^3 = 216$ rules and two boundary dynamics. We first numerically examine the ergodicity of these rules with boundary dynamics (012).

As a result of numerical simulation, we find that the following two types of rules can be ergodic. The rule in type 1 is illustrated by the state transition diagram depicted in Fig. 1. In the state transition diagram, the arrow from $x$ to $y$ with index $z$ represents the rule $\pi_z(x) = y$. An asterisk stands for any values (i.e., $* = \{a, b', c'\}$ in this case). We set the symbols $(a, b, c)$ as $(0, 1, 2)$ or its permutation, and set $(b', c') = (b, c)$ or $(c, b)$. In the case of $(a, b, c) = (1, 0, 2)$ and $(b', c') = (c, b) = (2, 0)$, as an example, the permutations are $\pi_1 = \pi_2 = (02)$ and $\pi_0 = (102)$. We have 12 rules in this type. If we identify $a = (1,1)$, $b = (1,0)$, $c = (0,1)$, and $(b', c') = (c, b)$, the rule corresponds to rule 12R on the sector $\{(0,1),(1,0),(1,1)\}^{\mathbb{N}}$. The rule in type 2 is illustrated in Fig. 2. Because the exchange of $b$ and $c$ does not change the diagram, there are 6 rules in this type.

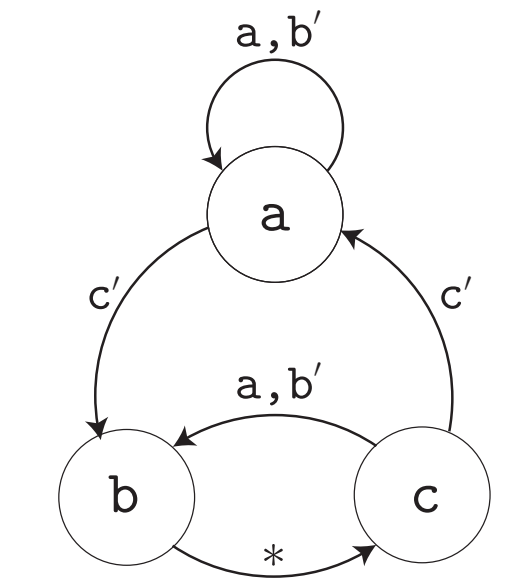

| | | | (b′, c′) | |
|---|---|---|---|---|
| a | b | c | (b, c) | (c, b) |
| 0 | 1 | 2 | (1,1,3) | (1,3,1) |
| 0 | 2 | 1 | (1,4,1) | (1,1,4) |
| 1 | 0 | 2 | (5,5,4) | (4,5,5) |
| 1 | 2 | 0 | (3,5,5) | (5,5,3) |
| 2 | 0 | 1 | (2,3,2) | (3,2,2) |
| 2 | 1 | 0 | (4,2,2) | (2,4,2) |

**Fig. 1** Type 1 ergodic rules for $k = 3$. The state transition diagram and the rules that belong to this type.

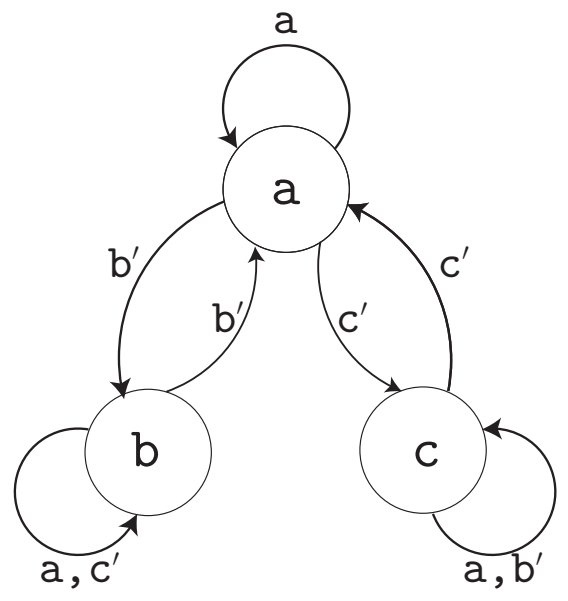

| | | | (b′, c′) | |
|---|---|---|---|---|
| a | b | c | (b, c) | (c, b) |
| 0 | 1 | 2 | (0,2,5) | (0,5,2) |
| 1 | 0 | 2 | (2,0,1) | (1,0,2) |
| 2 | 0 | 1 | (5,1,0) | (1,5,0) |

**Fig. 2** Type 2 ergodic rules for $k = 3$.

Now we shall prove the ergodicity of these two types with boundary dynamics $\pi_{\mathrm{B}} = (012)$. Note that both types' rules keep ergodic when using boundary dynamics (021).

Proof of ergodicity of Type 1: Type 1 ergodic CA consists of the permutations $\pi_a = \pi_b = (bc)$ and $\pi_c = (abc)$ for $(b', c') = (b, c)$, or $\pi_a = \pi_c = (bc)$ and $\pi_b = (abc)$ for $(b', c') = (c, b)$. For $n \geq 2$, we can prove by induction the following two properties [1] (see an example of sequences in the first, second, and third cells shown in Fig. 4):

1. Site $n$ has period $3^n$.
2. The sequence of site $n$, $\{x_n^t \mid 0 \leq t \leq 3^n - 1\}$ is a concatenation of one $a$ and units of $bc$ and $aa$.

Unit $bc$ at site $n \geq 2$ can be understood from the facts that $b$ is necessarily mapped to $c$ and that the preimage of state $c$ is $b$ only. Thus, only the behavior of $a$'s is nontrivial.

We now start mathematical induction. We can directly confirm that site 2 satisfies these two properties for both boundary rules. Now we assume that site $n$ satisfies these two properties. Then, property 1 suggests that there must be the same number of $a$s, $b$s, and $c$s in the sequence $\{x_n^t \mid 0 \leq t \leq 3^n - 1\}$. Accordingly, the sequence in site $n$ consists of a single $a$, $(3^{n-1} - 1)/2$ $aa$'s, and $3^{n-1}$ $bc$'s. Then, in the transformation from $x_{n+1}^0$ to $x_{n+1}^{3^n}$ (i.e., $T_{n+1}^{0 \to 3^n}$), we apply $\pi_a$ once, $\pi_a^2$ by $(3^{n-1} - 1)/2$ times, and $\pi_c \pi_b$ by $3^{n-1}$ times. Using the relations $\pi_a^2 = \mathrm{id}$ and $\pi_c \pi_b$ is a transposition, we find

$$T_{n+1}^{0 \to 3^n} = \begin{cases} \pi_c \pi_b \pi_a & \text{or} \\ \pi_a \pi_c \pi_b. \end{cases} \tag{11}$$

[1] More precisely, we can show these properties for $n \geq 1$ when $\pi_{\mathrm{B}} = (012)$, and for $n \geq 2$ when $\pi_{\mathrm{B}} = (021)$.

In either case, it is a cycle of length 3. This means that site $n+1$ has period $3^{n+1}$, which is property 1.

Next, we prove that site $n+1$ satisfies property 2. To this end, in the following, we treat the case of $(b', c') = (b, c)$ for convenience of explanation. A similar argument holds for $(b', c') = (c, b)$. Consider a temporal region of site $n$ consisting of units $bc$ and $aa$. Then, the corresponding temporal region of site $n+1$ also consists of units $bc$ and $aa$, because $\pi_a\pi_a$ induces transitions

$$a \xrightarrow{a} a \xrightarrow{a} a,$$
$$b \xrightarrow{a} c \xrightarrow{a} b,$$
$$c \xrightarrow{a} b \xrightarrow{a} c,$$

and $\pi_c\pi_b$ induces transitions

$$a \xrightarrow{b} a \xrightarrow{c} b,$$
$$b \xrightarrow{b} c \xrightarrow{c} a,$$
$$c \xrightarrow{b} b \xrightarrow{c} c.$$

As clearly seen from these transitions, if the initial state of site $n+1$ is $c$, then both $\pi_c\pi_b$ and $\pi_a\pi_a$ induce repeated $bc$ as $c \to b \to c \to b \to \cdots$. On the other hand, if the initial state of site $n+1$ is $a$ or $b$, then $\pi_a\pi_a$ keeps the unit $aa$ and $bc$ in site $n+1$, while $\pi_c\pi_b$ switches these two units, $aa \leftrightarrow bc$, in site $n+1$ (See also Fig. 3). In addition, single $\pi_a$ is transposition $(bc)$, which plays the role of switching the initial state between $b$ and $c$ with keeping the structure of units. Note that single $\pi_a$ acts on state $a$ exactly once in a single period with length $3^{n+1}$, and when single $\pi_a$ applies to state $a$, this plays the role of adding a single $a$ in site $n+1$, which results in a single $a$ besides unit $aa$ in the sequence in site $n+1$ with length $3^{n+1}$. The overall argument confirms property 2.

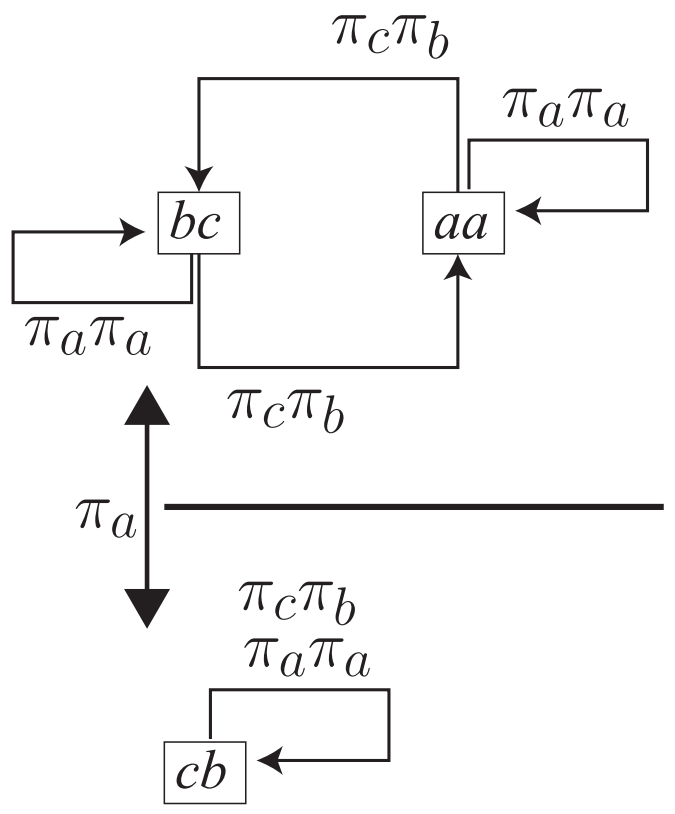

**Fig. 3** The flow chart of the dynamics of Type 1 CA with $(b', c') = (b, c)$. If the state is $a$ or $b$, the application of $\pi_c\pi_b$ and $\pi_a\pi_a$ provides units $bc$ and $aa$. In contrast, if the state is $c$, the application of $\pi_c\pi_b$ and $\pi_a\pi_a$ always provides $c \to b \to c \to b \to c \to \cdots$. A single $\pi_a$ switches these two phases if the state is $b$ or $c$.

Proof of ergodicity of Type 2: Type 2 ergodic CA consists of $\pi_a = \mathrm{id}$, $\pi_{b'} = (ab)$, and $\pi_{c'} = (ac)$. We fix the initial value $x_n^0 = a$ without loss of generality (If we prove the ergodicity in this case, it holds for any initial conditions). We prove the following two properties by mathematical induction (see an example of sequences in the first, second, and third cells shown in Fig. 5):

1. Site $n$ has period $3^n$.
2. The sequence $\{x_n^t \mid 0 \leq t \leq 3^n - 1\}$ is divided into two intervals. One consists of $a$ and $b$ only, and the other consists of $a$ and $c$ only.

| | | | |
|---|---|---|---|
| 1st cell | 0 1 2 0 1 2 0 1 2 | 0 1 2 0 1 2 0 1 2 | 0 1 2 0 1 2 0 1 2 |
| 2nd cell | 0 0 0 1 2 1 2 1 2 | 0 0 0 1 2 1 2 1 2 | 0 0 0 1 2 1 2 1 2 |
| 3rd cell | 0 0 0 0 0 1 2 0 0 | 1 2 1 2 1 2 1 2 1 | 2 1 2 1 2 0 0 1 2 |

**Fig. 4** The dynamics of a Type-1 rule with $(a, b, c) = (0, 1, 2)$ and $(b', c') = (b, c) = (1, 2)$. The first, second, and third lines have period 3, $9 = 3^2$, and $27 = 3^3$.

Site 1 clearly satisfies both properties 1 and 2. Now we assume that site $n$ satisfies both properties 1 and 2. Combining property 2 and the fact that $\pi_a$ is the identity, we find

$$T_n^{3^n} = \begin{cases} \pi_{b'}^{3^{n-1}} \pi_{c'}^{3^{n-1}} = \pi_{b'}\pi_{c'} & \text{or} \\ \pi_{c'}^{3^{n-1}} \pi_{b'}^{3^{n-1}} = \pi_{c'}\pi_{b'}. \end{cases} \tag{12}$$

In either case, it is a 3-cycle, and accordingly, site $n+1$ has period $3^{n+1}$, which implies property 1.

To prove property 2, we look into the dynamics more in detail. In the interval of site $n$ with $a$ or $b'$ only, site $n+1$ either takes "$a$ or $b$ only" or stays at $c$, because $\pi_a$ is the identity and $\pi_{b'}$ is a transposition between $a$ and $b$. For a similar reason, in the interval of site $n$ with $a$ or $c'$ only, site $n+1$ either takes "$a$ or $c$ only" or stays at $b$. These facts assure that property 2 holds for site $n+1$, which is exhibited in Fig. 6.

| | | | |
|---|---|---|---|
| 1st cell | 0 1 2 0 1 2 0 1 2 | 0 1 2 0 1 2 0 1 2 | 0 1 2 0 1 2 0 1 2 |
| 2nd cell | 0 0 1 1 1 0 2 2 2 | 0 0 1 1 1 0 2 2 2 | 0 0 1 1 1 0 2 2 2 |
| 3rd cell | 0 0 0 1 0 1 1 1 1 | 1 1 1 0 1 0 0 2 0 | 2 2 2 2 2 2 2 0 2 |

**Fig. 5** The dynamics of a Type-2 rule with $(a, b, c) = (0, 1, 2)$ and $(b', c') = (b, c) = (1, 2)$. The first, second, and third lines have period 3, $9 = 3^2$, and $27 = 3^3$.

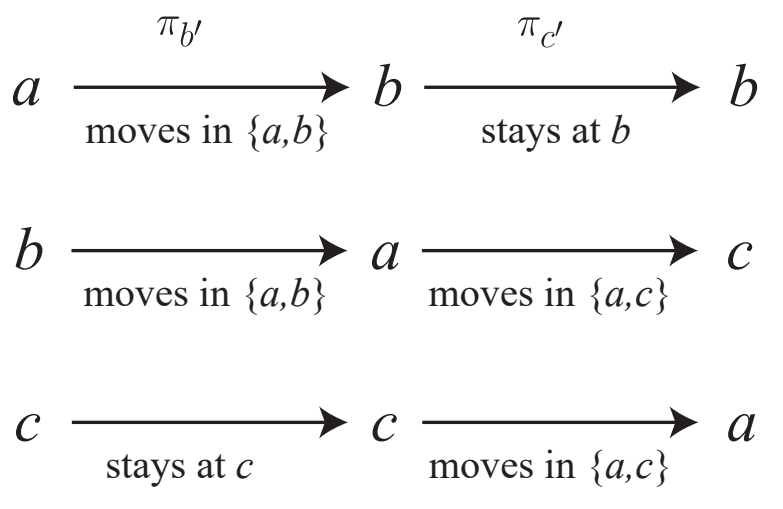

**Fig. 6** We draw transitions in three periods in the case of $T_n^{3^n} = \pi_{c'}\pi_{b'}$. It directly shows that if property 2 is satisfied in site $n$, it is also satisfied in site $n+1$.

## 4 CA with 4 states

It is numerically confirmed that there are no ergodic rules in the case $k = 4$. The period of site 2 is shorter than $4^2$ in any rule and any boundary dynamics.

## 5 CA with 5 states: summary of results

### 5.1 List of ergodic CA

In the case $k = 5$, there are 120 permutations listed in Table 1. The semi-infinite reversible CA of $k = 5$ has $(5!)^5 = 24883200000$ rules and $4! = 24$ boundary rules. In order to moderate the number of ergodic rules, we restrict ourselves to the rules which are ergodic for any boundary rules. In the following, we use the term ergodic in this sense.

We numerically searched candidates of ergodic rules with increasing system size $N$. Namely, if a rule generates an orbit of period $5^N$ for every boundary dynamics, it is a candidate. The numbers of the candidates for $N$ are exhibited in Table 5.1. Though it looks convergent at $N = 11$, true convergence is reached at $N = 15$, where we have 118320 rules. Thanks to the boundary condition, if a rule is ergodic, its conjugate rules are all ergodic. Thus, we can classify the ergodic rules with the numbers of rules in conjugacy classes into 72 types as Table 3. Further, we classify the rules in subtypes by their state transition diagrams. The list of ergodic rules classified into subtypes is presented in the form of state transition diagrams in Fig. 7 with the parameter choice shown in Table 4. Notice that although only representative rules are shown, their conjugate rules are also ergodic. The number of ergodic rules includes those conjugate rules.

The remainder of this paper is devoted to the proof of ergodicity of these rules.

**Table 2** The numbers of candidates for ergodic rules for system size $N$. It is analytically proven that the number of candidates 118320 at $N = 15$ is true convergence.

| $N$ | Number of candidates |
|---|---|
| 2 | 13972800 |
| 3 | 1626840 |
| 4 | 486000 |
| 5 | 226920 |
| 6 | 164040 |
| 7 | 134280 |
| 8 | 125880 |
| 9 | 120240 |
| 10 | 119040 |
| 11 | 118560 |
| 12 | 118560 |
| 13 | 118560 |
| 14 | 118560 |
| 15 | 118320 |

**Table 3** Types of the $k = 5$ ergodic rules classified by the conjugacy classes and the number of ergodic rules in each type.

| | Conjugacy classes | | | | | | | |
|---|---|---|---|---|---|---|---|---|
| Type | $1^5$ | $1^3 2^1$ | $1^2 3^1$ | $1^1 2^2$ | $1^1 4^1$ | $2^1 3^1$ | $5^1$ | Number of rules |
| 01 | 3 | 1 | 0 | 0 | 1 | 0 | 0 | 720 |
| 02 | 3 | 1 | 0 | 0 | 0 | 1 | 0 | 1200 |
| 03 | 3 | 0 | 2 | 0 | 0 | 0 | 0 | 480 |
| 04 | 3 | 0 | 1 | 1 | 0 | 0 | 0 | 1200 |
| 05 | 3 | 0 | 0 | 2 | 0 | 0 | 0 | 720 |
| 06 | 2 | 2 | 1 | 0 | 0 | 0 | 0 | 3840 |
| 07 | 2 | 2 | 0 | 1 | 0 | 0 | 0 | 5160 |
| 08 | 2 | 1 | 1 | 0 | 0 | 1 | 0 | 2400 |
| 09 | 2 | 1 | 0 | 1 | 1 | 0 | 0 | 5040 |
| 10 | 2 | 1 | 0 | 1 | 0 | 1 | 0 | 1440 |
| 11 | 2 | 0 | 3 | 0 | 0 | 0 | 0 | 480 |
| 12 | 2 | 0 | 2 | 1 | 0 | 0 | 0 | 720 |
| 13 | 2 | 0 | 1 | 2 | 0 | 0 | 0 | 720 |
| 14 | 2 | 0 | 1 | 0 | 1 | 1 | 0 | 1440 |
| 15 | 2 | 0 | 1 | 0 | 0 | 2 | 0 | 480 |
| 16 | 2 | 0 | 0 | 3 | 0 | 0 | 0 | 720 |
| 17 | 2 | 0 | 0 | 1 | 2 | 0 | 0 | 720 |
| 18 | 1 | 4 | 0 | 0 | 0 | 0 | 0 | 1920 |
| 19 | 1 | 3 | 0 | 0 | 1 | 0 | 0 | 5040 |
| 20 | 1 | 3 | 0 | 0 | 0 | 1 | 0 | 1320 |

| Type | $1^5$ | $1^3 2^1$ | $1^2 3^1$ | $1^1 2^2$ | $1^1 4^1$ | $2^1 3^1$ | $5^1$ | Number of rules |
|---|---|---|---|---|---|---|---|---|
| 21 | 1 | 2 | 2 | 0 | 0 | 0 | 0 | 480 |
| 22 | 1 | 2 | 1 | 1 | 0 | 0 | 0 | 4320 |
| 23 | 1 | 2 | 0 | 2 | 0 | 0 | 0 | 8160 |
| 24 | 1 | 2 | 0 | 0 | 0 | 2 | 0 | 840 |
| 25 | 1 | 1 | 2 | 0 | 1 | 0 | 0 | 720 |
| 26 | 1 | 1 | 2 | 0 | 0 | 1 | 0 | 1080 |
| 27 | 1 | 1 | 0 | 2 | 1 | 0 | 0 | 6000 |
| 28 | 1 | 1 | 0 | 2 | 0 | 1 | 0 | 600 |
| 29 | 1 | 1 | 0 | 0 | 3 | 0 | 0 | 480 |
| 30 | 1 | 1 | 0 | 0 | 1 | 2 | 0 | 720 |
| 31 | 1 | 1 | 0 | 0 | 0 | 3 | 0 | 360 |
| 32 | 1 | 0 | 4 | 0 | 0 | 0 | 0 | 120 |
| 33 | 1 | 0 | 3 | 1 | 0 | 0 | 0 | 720 |
| 34 | 1 | 0 | 2 | 0 | 1 | 1 | 0 | 720 |
| 35 | 1 | 0 | 1 | 3 | 0 | 0 | 0 | 240 |
| 36 | 1 | 0 | 0 | 4 | 0 | 0 | 0 | 960 |
| 37 | 1 | 0 | 0 | 2 | 2 | 0 | 0 | 2160 |
| 38 | 1 | 0 | 0 | 0 | 1 | 3 | 0 | 240 |
| 39 | 0 | 4 | 1 | 0 | 0 | 0 | 0 | 240 |
| 40 | 0 | 4 | 0 | 1 | 0 | 0 | 0 | 2400 |
| 41 | 0 | 3 | 1 | 0 | 0 | 1 | 0 | 480 |
| 42 | 0 | 3 | 0 | 1 | 1 | 0 | 0 | 4800 |
| 43 | 0 | 3 | 0 | 1 | 0 | 1 | 0 | 960 |
| 44 | 0 | 2 | 2 | 1 | 0 | 0 | 0 | 720 |
| 45 | 0 | 2 | 1 | 2 | 0 | 0 | 0 | 1680 |
| 46 | 0 | 2 | 1 | 0 | 0 | 2 | 0 | 240 |
| 47 | 0 | 2 | 0 | 3 | 0 | 0 | 0 | 2760 |
| 48 | 0 | 2 | 0 | 2 | 0 | 0 | 1 | 960 |
| 49 | 0 | 2 | 0 | 1 | 2 | 0 | 0 | 1680 |
| 50 | 0 | 2 | 0 | 1 | 1 | 1 | 0 | 2160 |
| 51 | 0 | 2 | 0 | 1 | 0 | 2 | 0 | 720 |
| 52 | 0 | 2 | 0 | 0 | 0 | 2 | 1 | 480 |
| 53 | 0 | 1 | 1 | 2 | 0 | 1 | 0 | 480 |
| 54 | 0 | 1 | 0 | 3 | 1 | 0 | 0 | 3600 |
| 55 | 0 | 1 | 0 | 3 | 0 | 1 | 0 | 480 |
| 56 | 0 | 1 | 0 | 1 | 3 | 0 | 0 | 480 |
| 57 | 0 | 0 | 3 | 0 | 1 | 1 | 0 | 360 |
| 58 | 0 | 0 | 2 | 3 | 0 | 0 | 0 | 720 |
| 59 | 0 | 0 | 2 | 0 | 0 | 2 | 1 | 240 |
| 60 | 0 | 0 | 1 | 4 | 0 | 0 | 0 | 720 |
| 61 | 0 | 0 | 1 | 0 | 1 | 3 | 0 | 120 |
| 62 | 0 | 0 | 0 | 5 | 0 | 0 | 0 | 240 |
| 63 | 0 | 0 | 0 | 4 | 0 | 0 | 1 | 1920 |
| 64 | 0 | 0 | 0 | 3 | 2 | 0 | 0 | 480 |
| 65 | 0 | 0 | 0 | 3 | 1 | 1 | 0 | 7200 |
| 66 | 0 | 0 | 0 | 3 | 0 | 2 | 0 | 1680 |
| 67 | 0 | 0 | 0 | 3 | 0 | 0 | 2 | 960 |
| 68 | 0 | 0 | 0 | 2 | 2 | 0 | 1 | 7680 |
| 69 | 0 | 0 | 0 | 2 | 0 | 0 | 3 | 480 |
| 70 | 0 | 0 | 0 | 1 | 3 | 1 | 0 | 3840 |
| 71 | 0 | 0 | 0 | 0 | 4 | 0 | 1 | 960 |
| 72 | 0 | 0 | 0 | 0 | 0 | 4 | 1 | 720 |

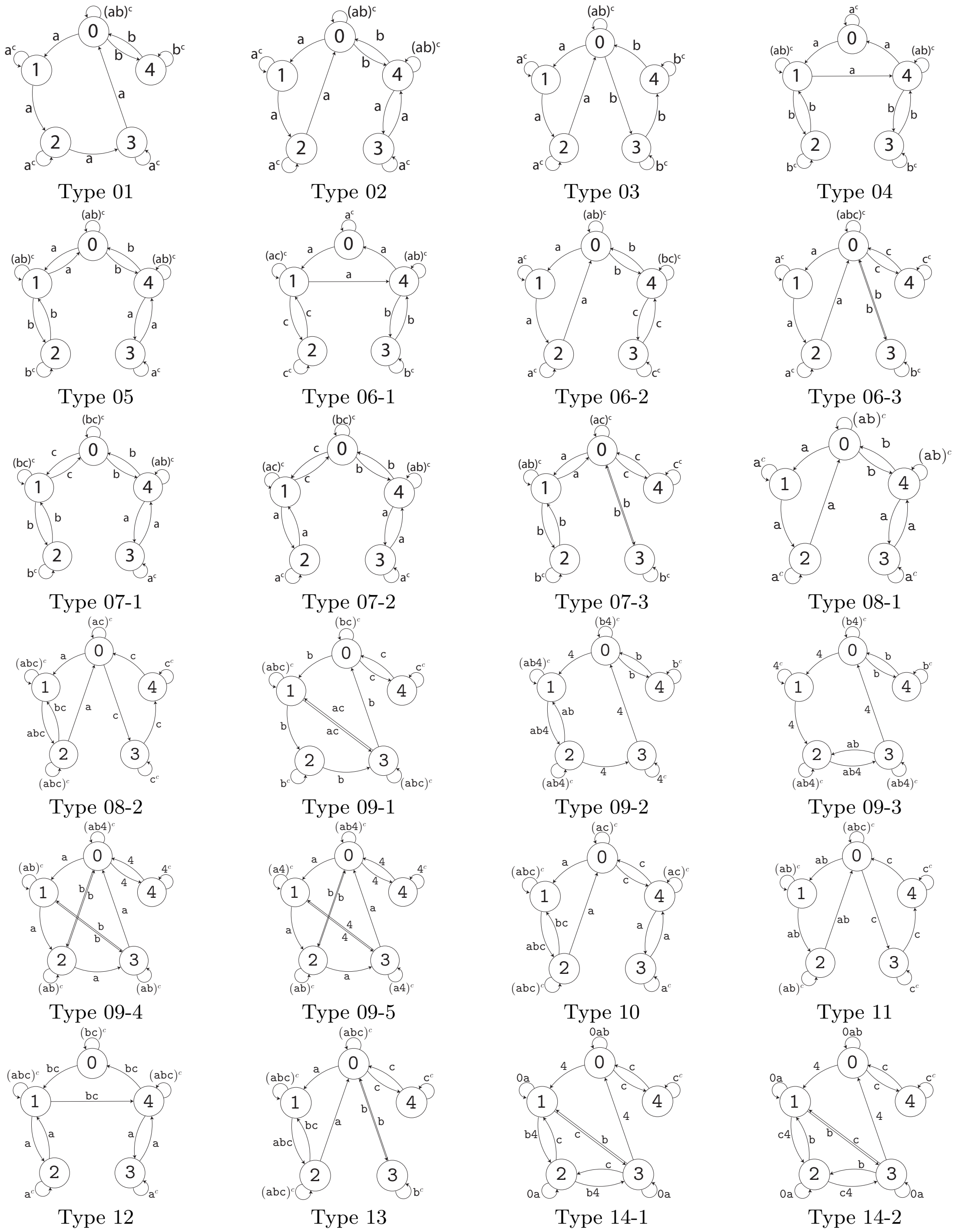

**Fig. 7** State transition diagrams of the ergodic rules for $k = 5$. The arrow's index shows the values of left nearest neighbor. An asterisk means *any* value. Superscript $c$ means complement. For example, $(12)^c$ means 034.

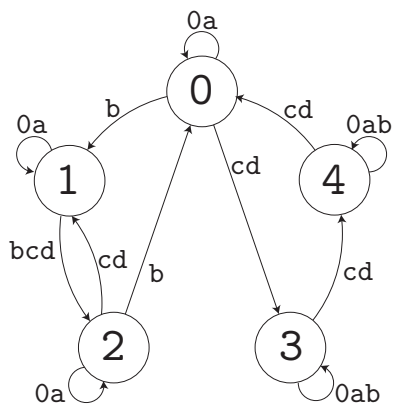

Type 15

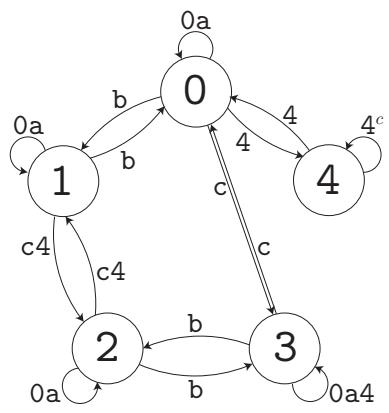

Type 16

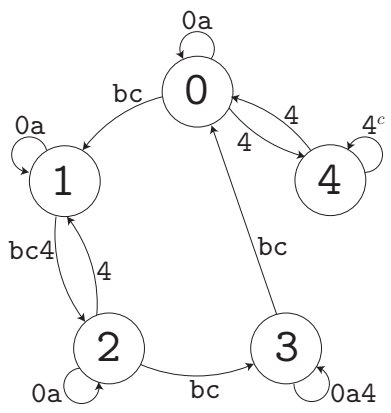

Type 17-1

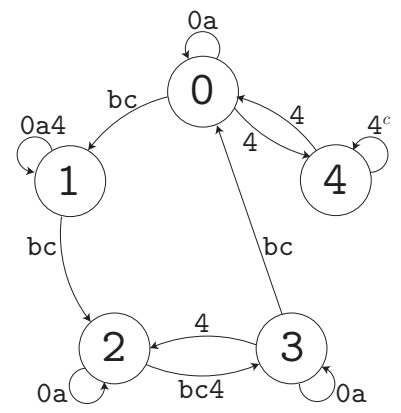

Type 17-2

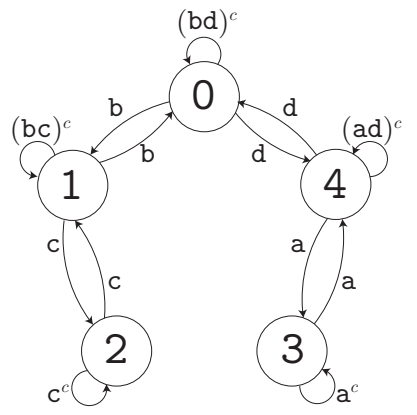

Type 18-1

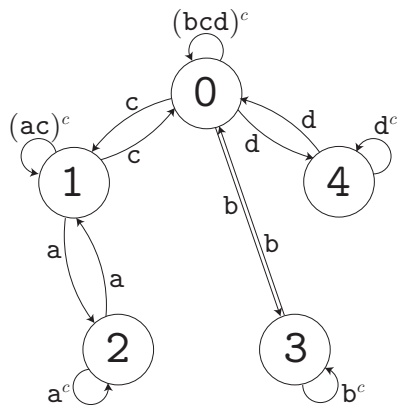

Type 18-2

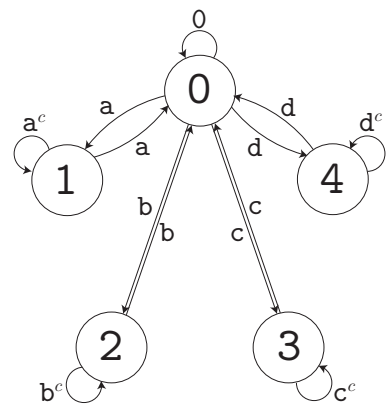

Type 18-3

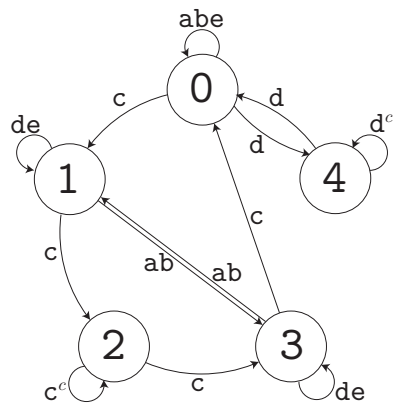

Type 19-1

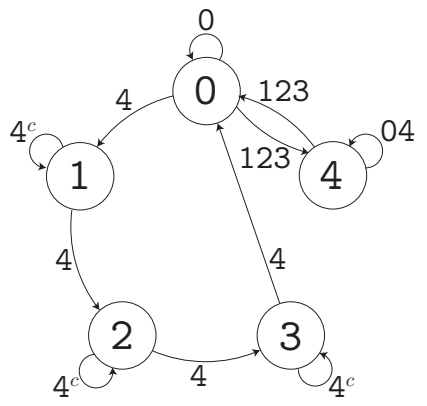

Type 19-2

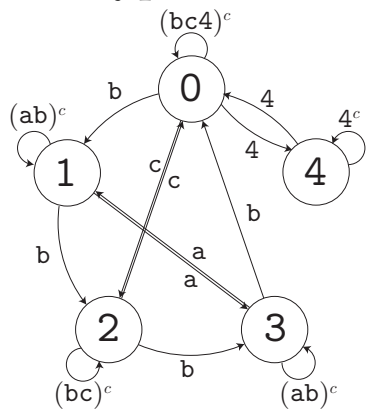

Type 19-3

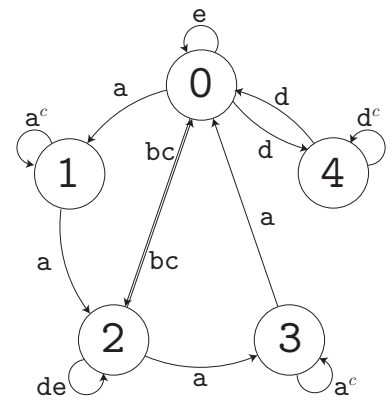

Type 19-4

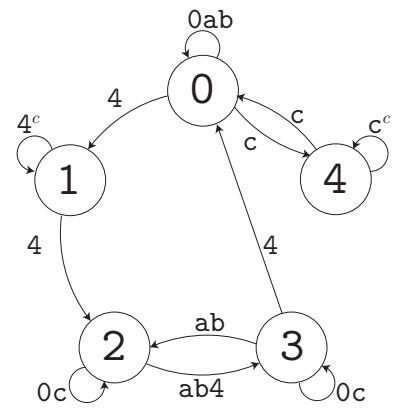

Type 19-5

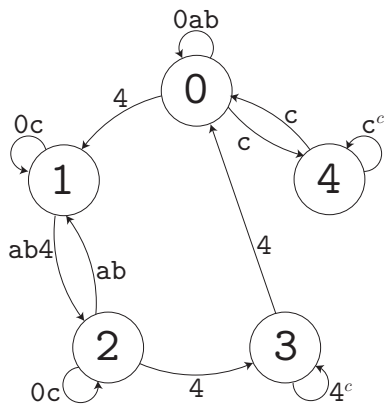

Type 19-6

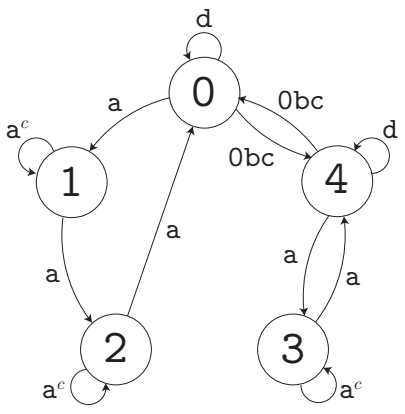

Type 20-1

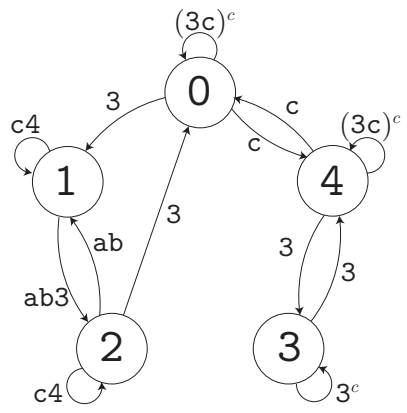

Type 20-2

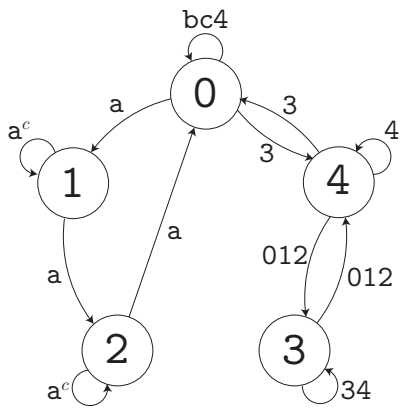

Type 20-3

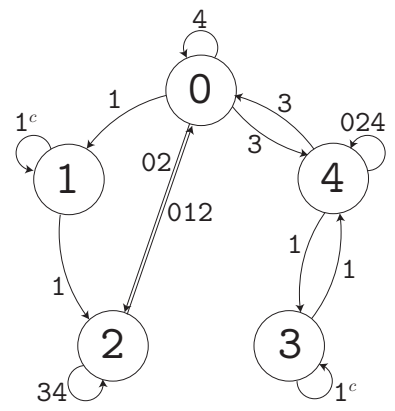

Type 20-4

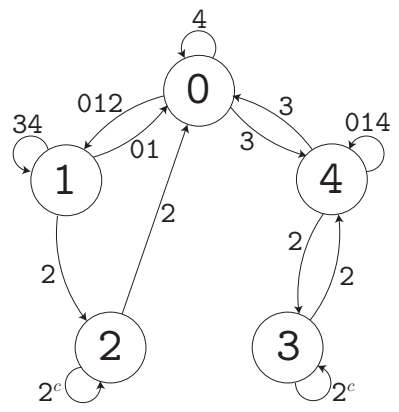

Type 20-5

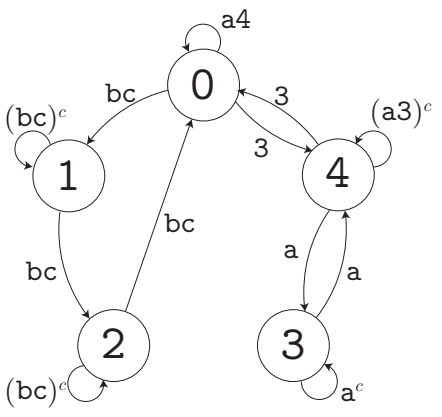

Type 21-1

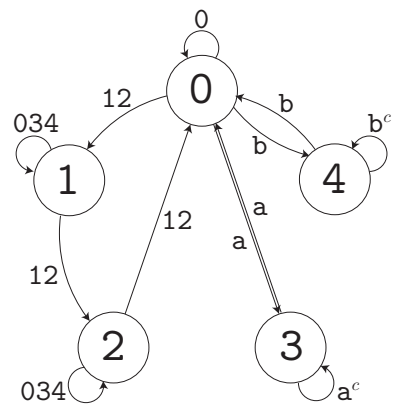

Type 21-2

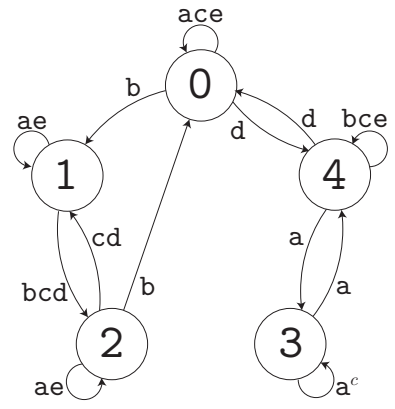

Type 22-1

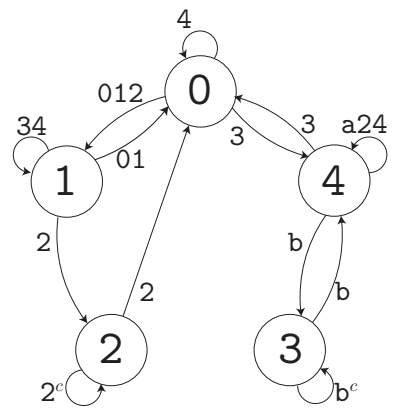

Type 22-2

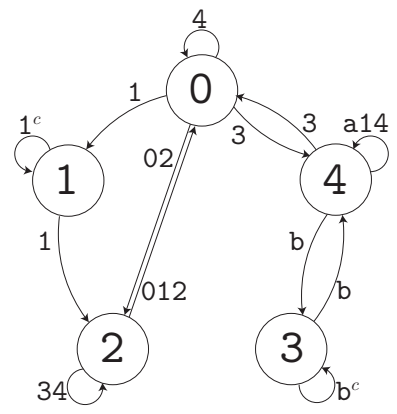

Type 22-3

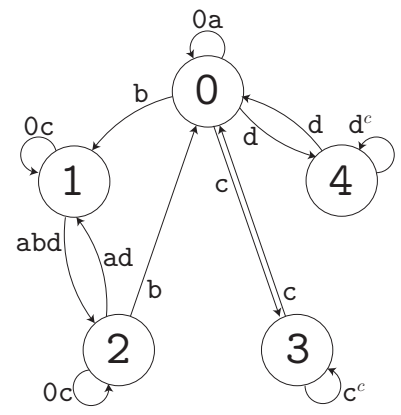

Type 22-4

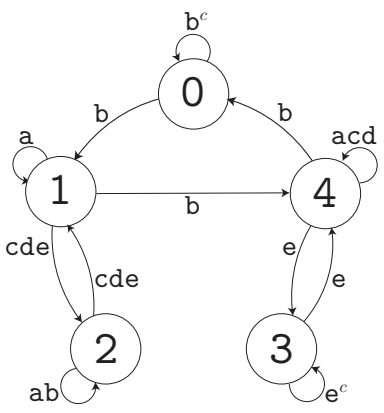

Type 22-5

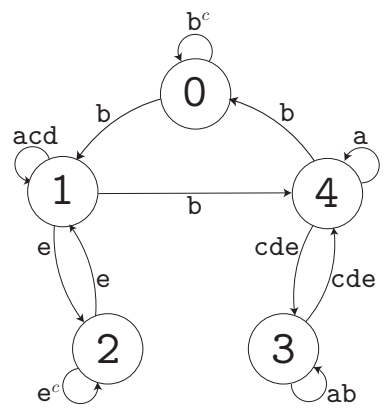

Type 22-6

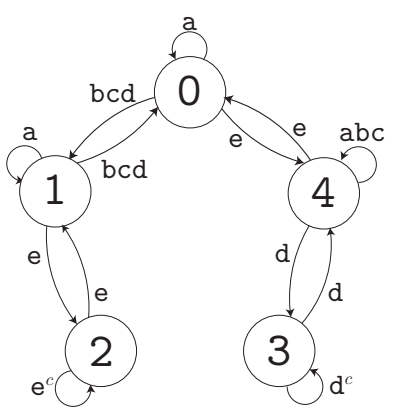

Type 23-1

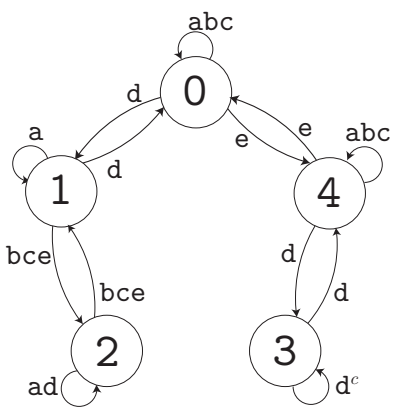

Type 23-2

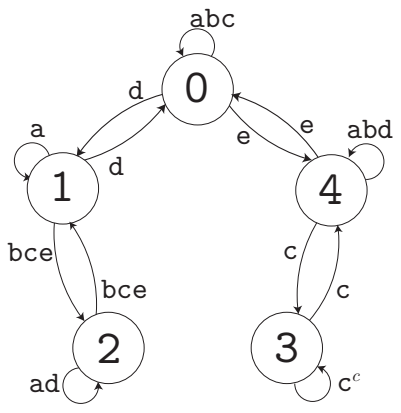

Type 23-3

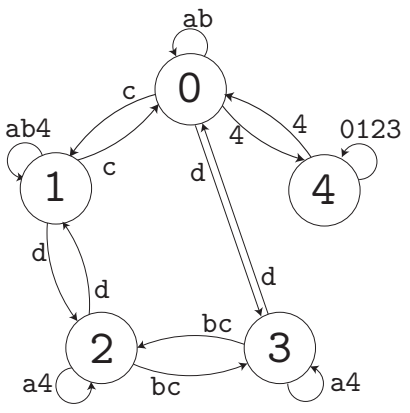

Type 23-4

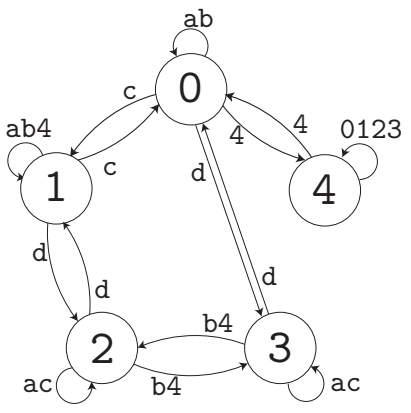

Type 23-5

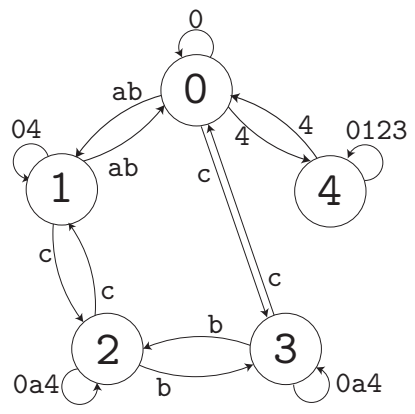

Type 23-6

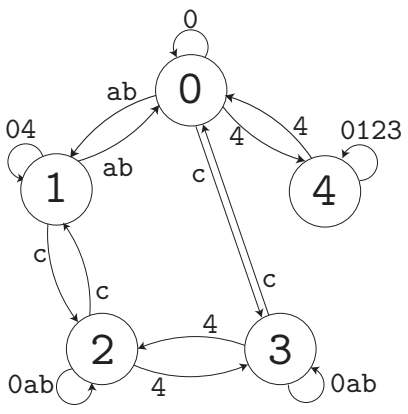

Type 23-7

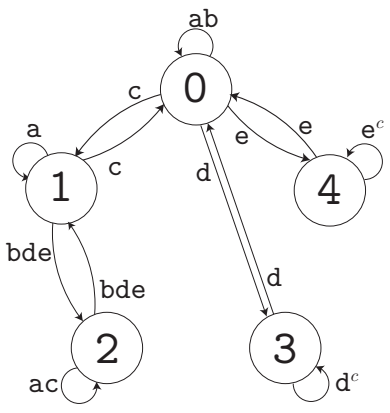

Type 23-8

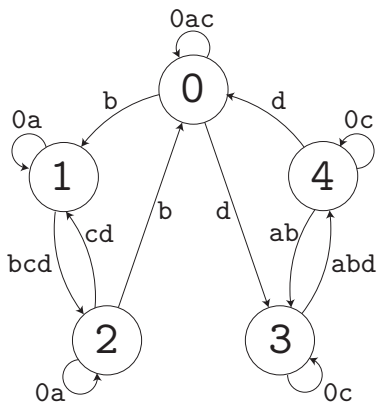

Type 24-1

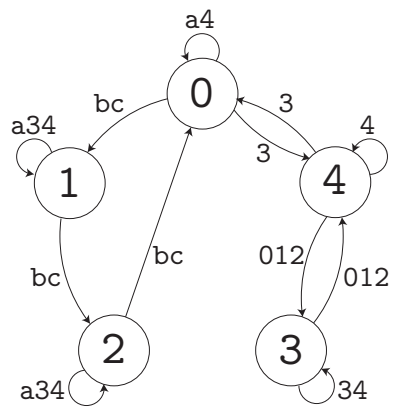

Type 24-2

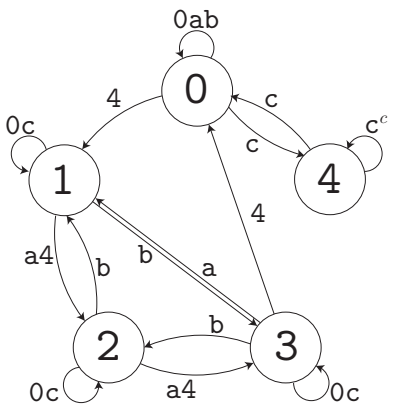

Type 25

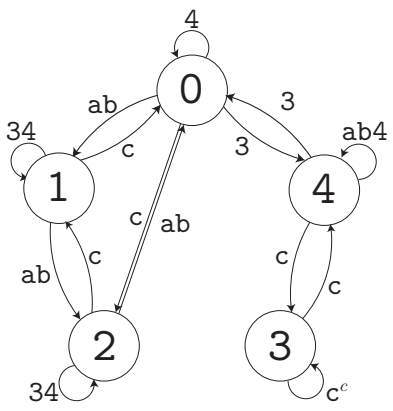

Type 26-1

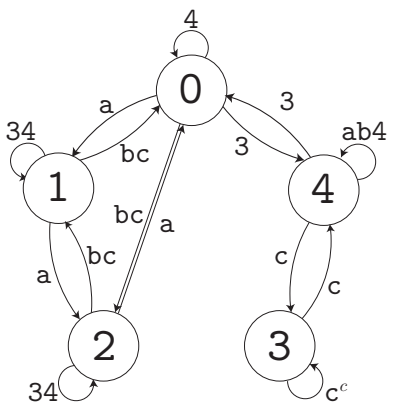

Type 26-2

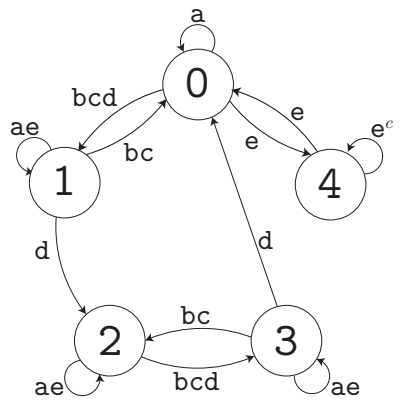

Type 27-1

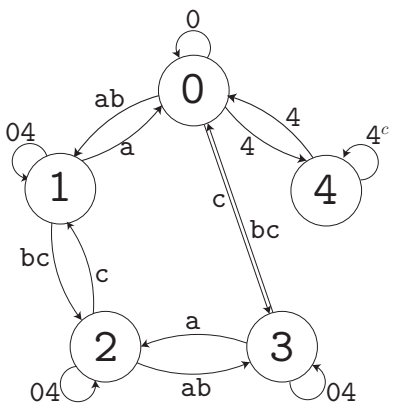

Type 27-2

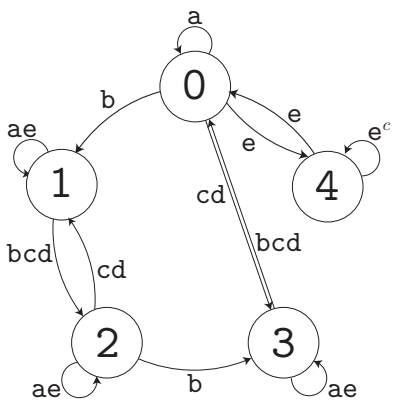

Type 27-3

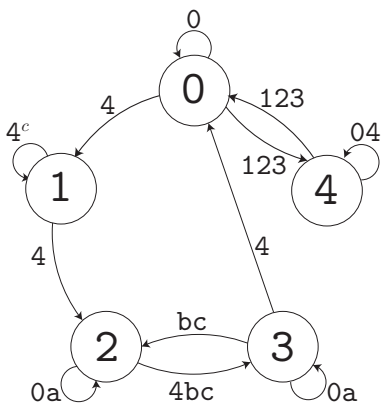

Type 27-4

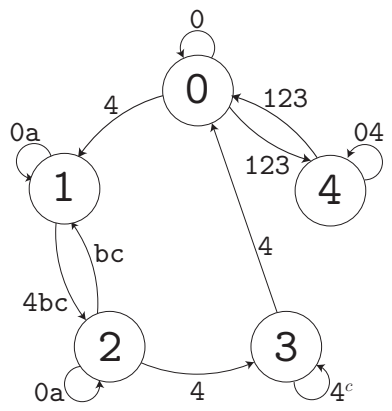

Type 27-5

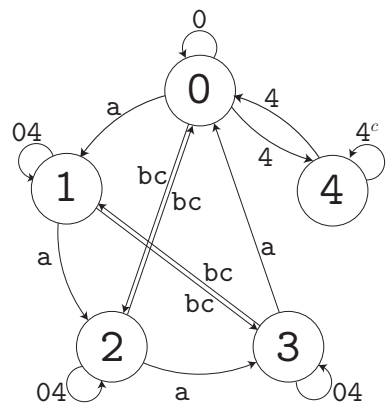

Type 27-6

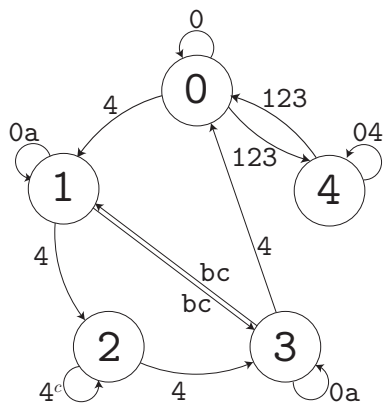

Type 27-7

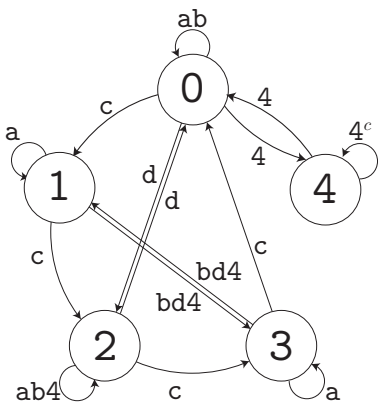

Type 27-8

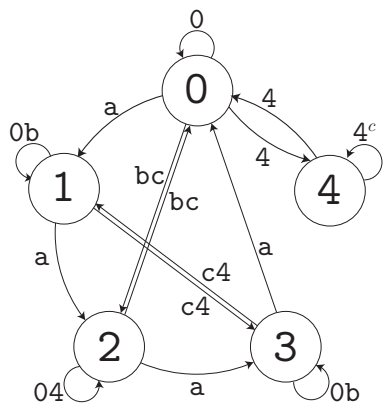

Type 27-9

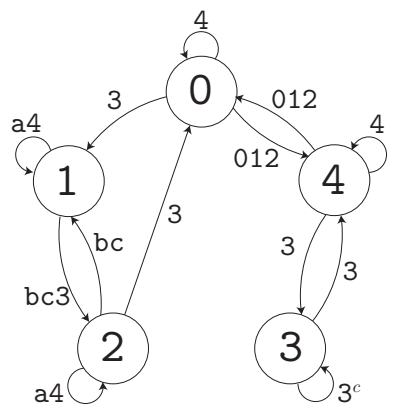
Type 28-1

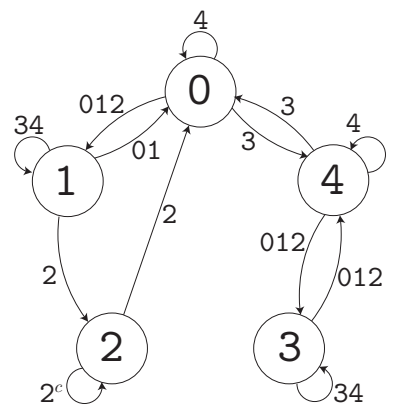
Type 28-2

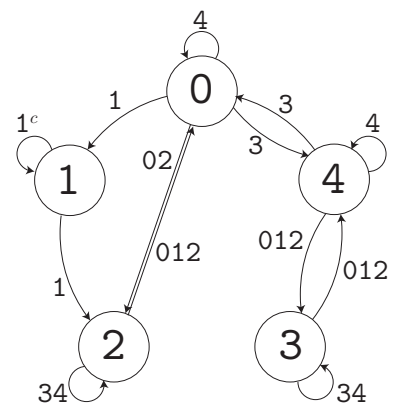
Type 28-3

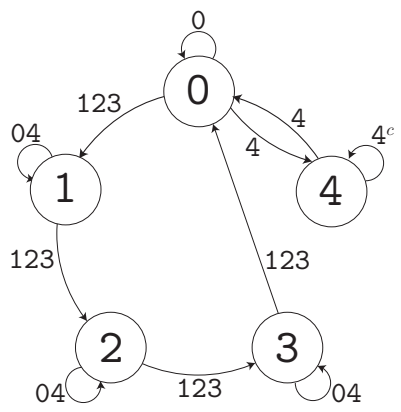
Type 29-1

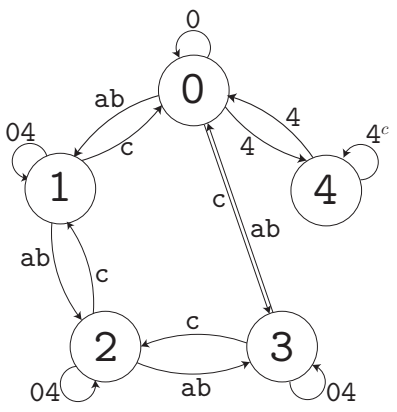
Type 29-2

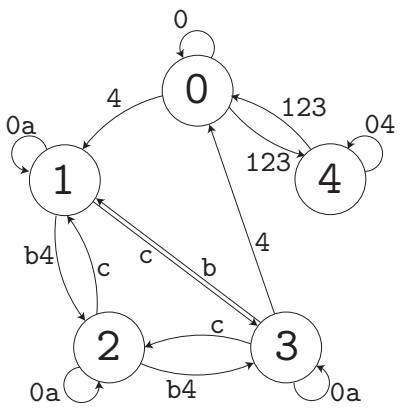
Type 30

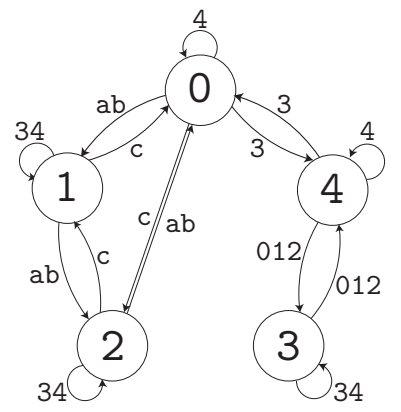
Type 31

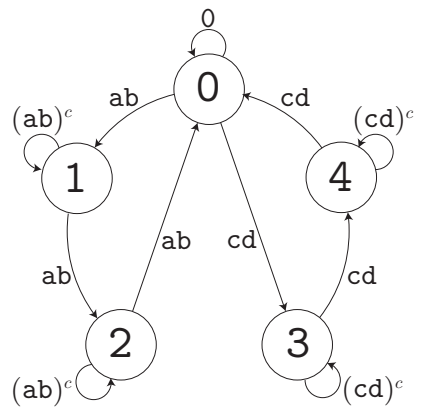
Type 32

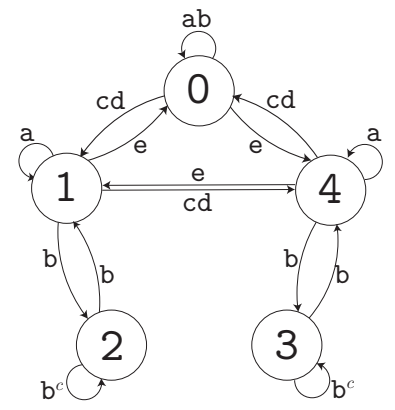
Type 33

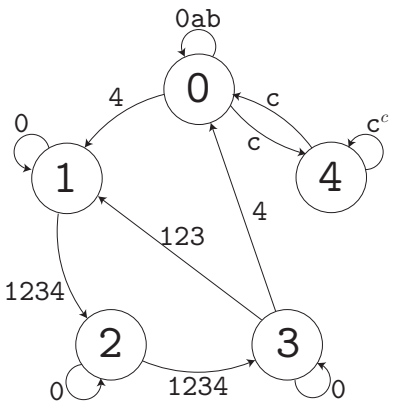
Type 34-1

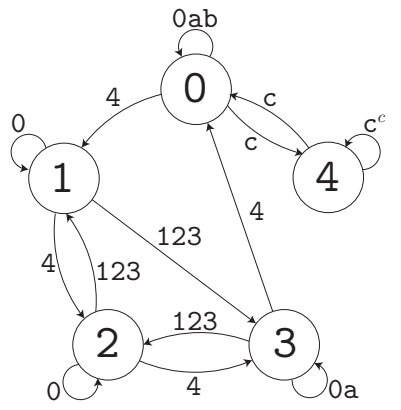
Type 34-2

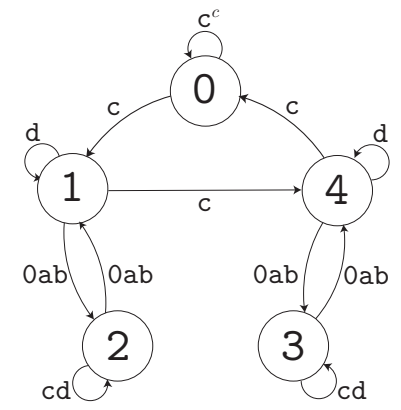
Type 35

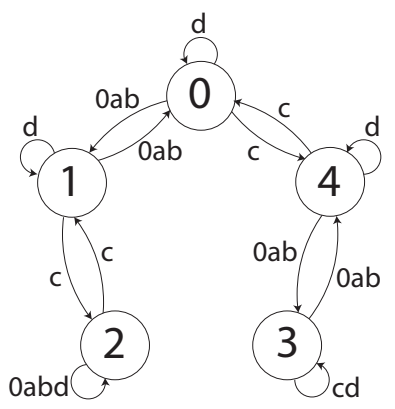
Type 36-1

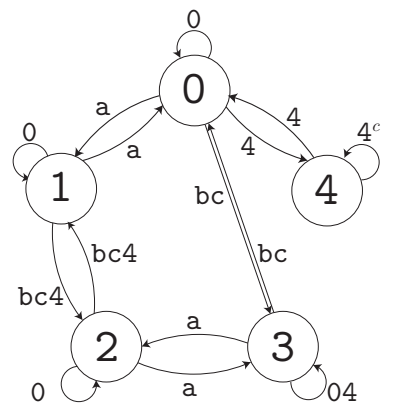
Type 36-2

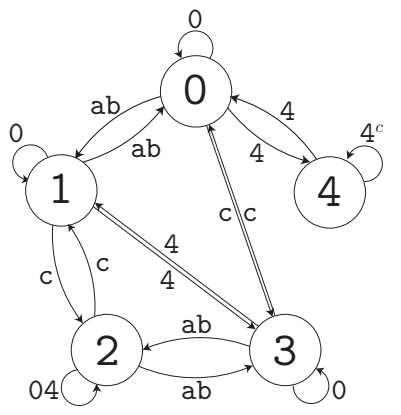
Type 36-3

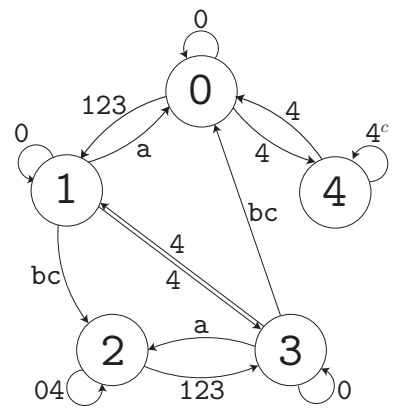
Type 37-1

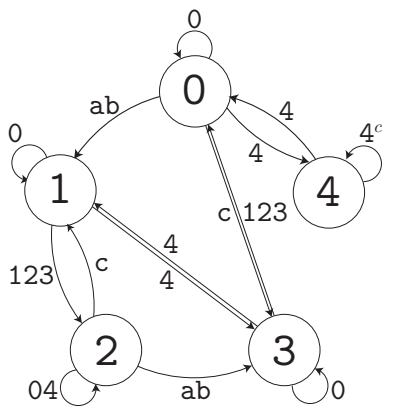
Type 37-2

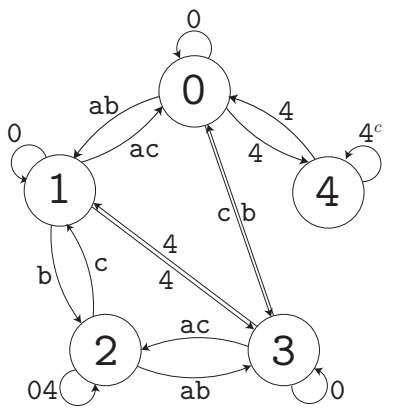
Type 37-3

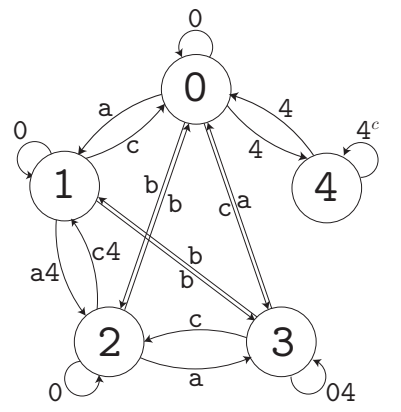
Type 37-4

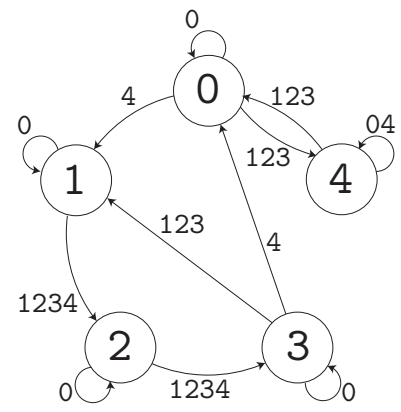
Type 38-1

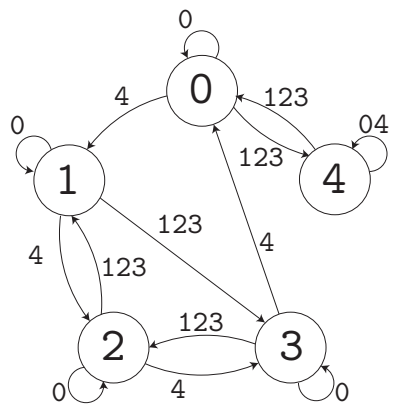
Type 38-2

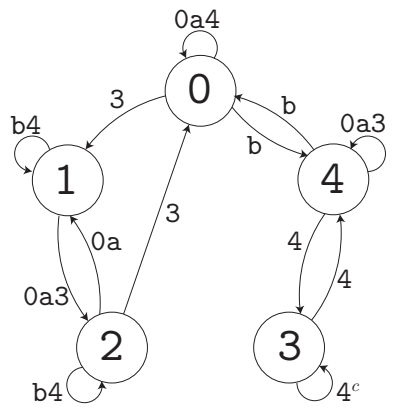
Type 39

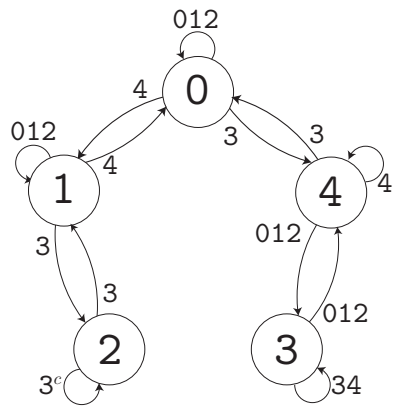
Type 40-1

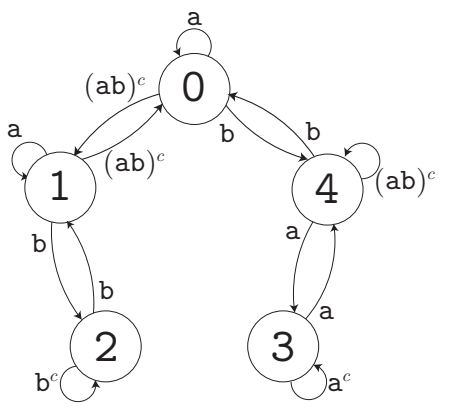
Type 40-2

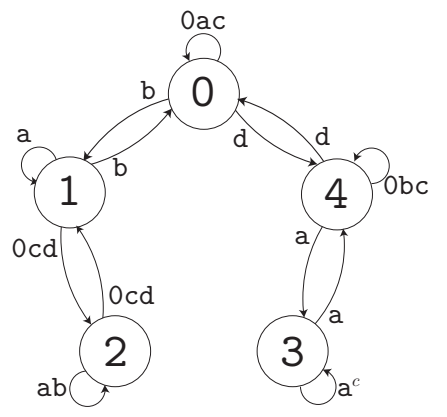

Type 40-3

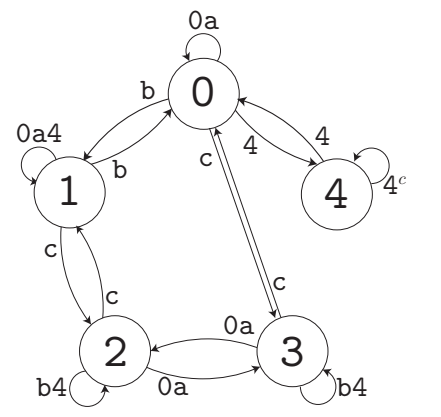

Type 40-4

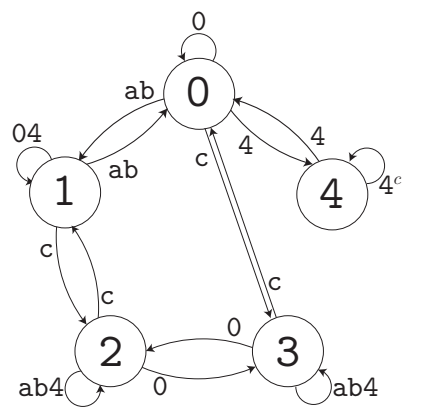

Type 40-5

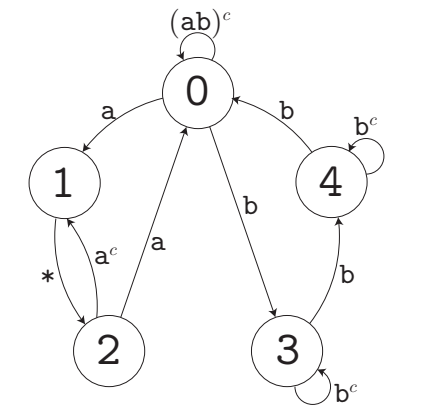

Type 41

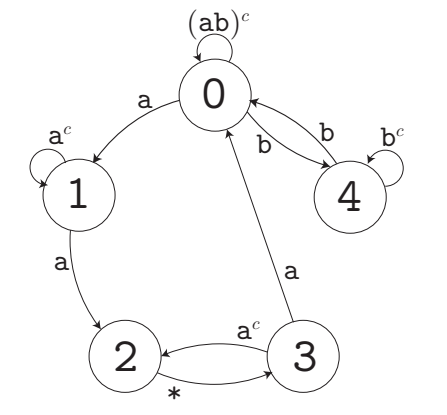

Type 42-1

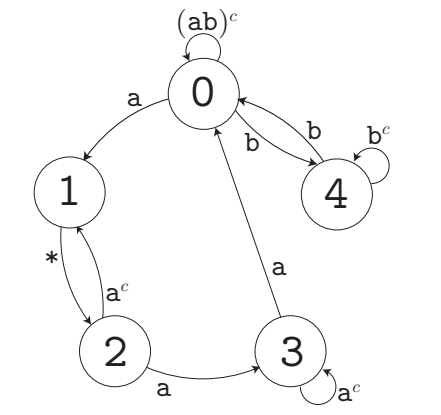

Type 42-2

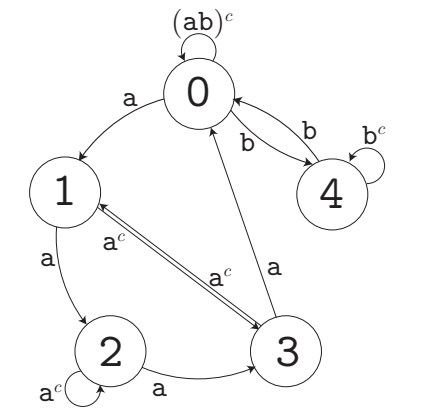

Type 42-3

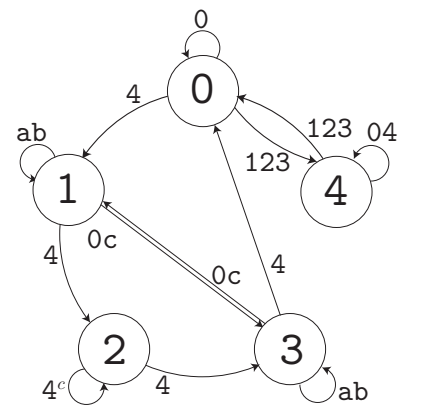

Type 42-4

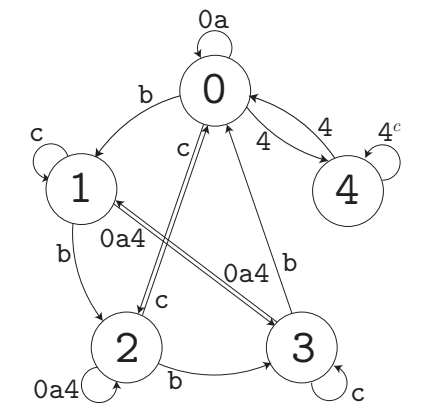

Type 42-5

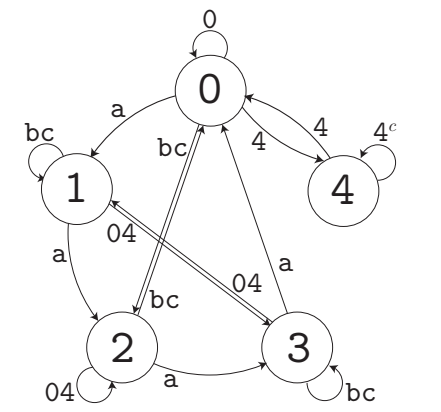

Type 42-6

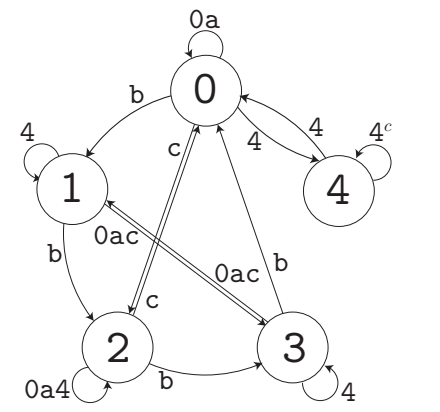

Type 42-7

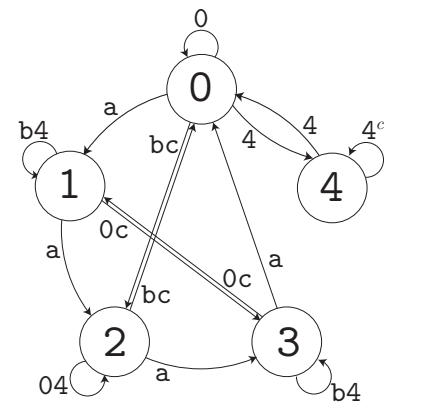

Type 42-8

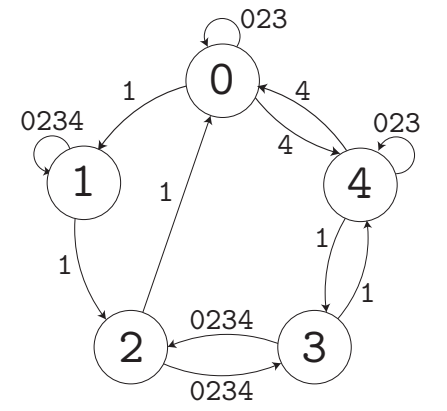

Type 43-1

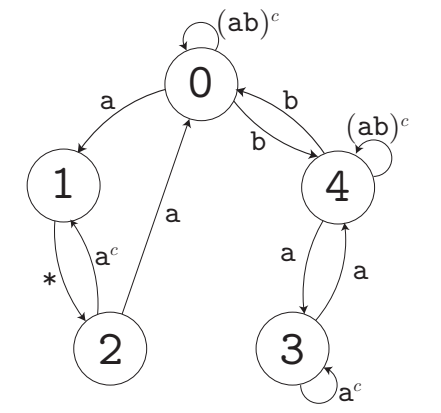

Type 43-2

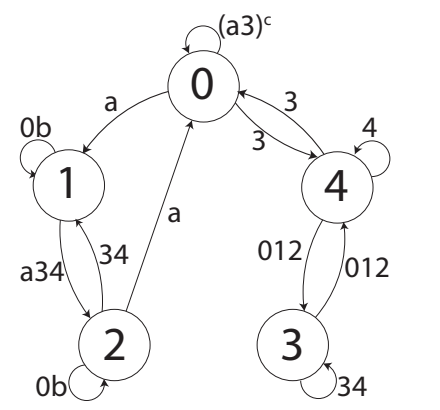

Type 43-3

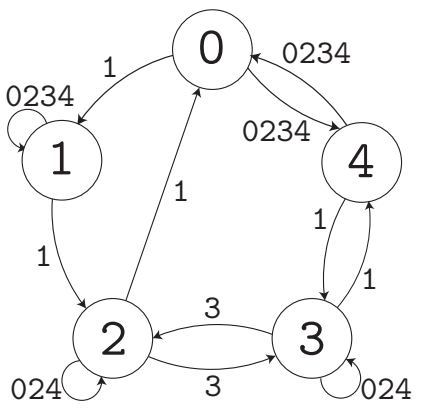

Type 43-4

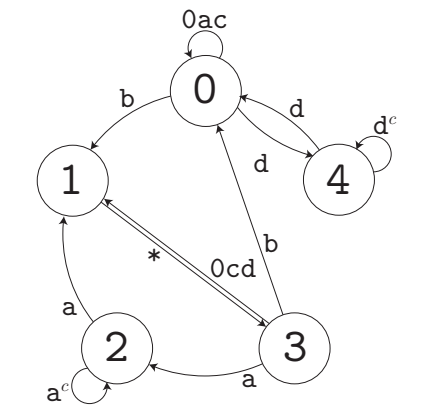

Type 44-1

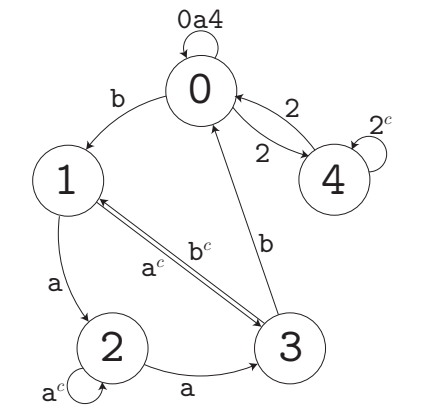

Type 44-2

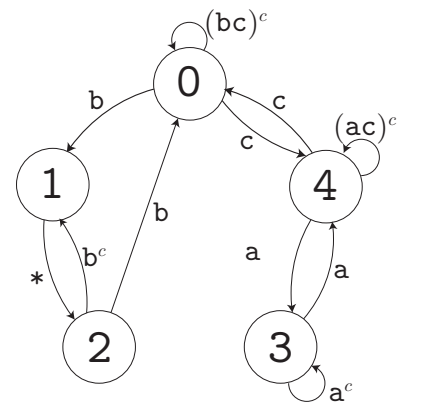

Type 45-1

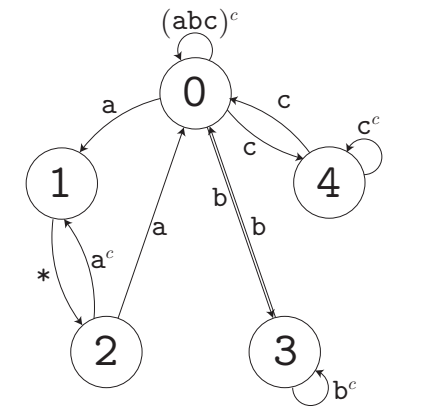

Type 45-2

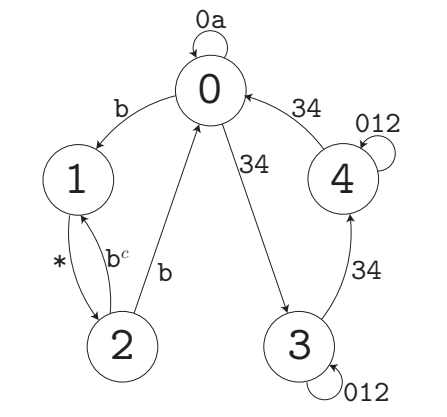

Type 46

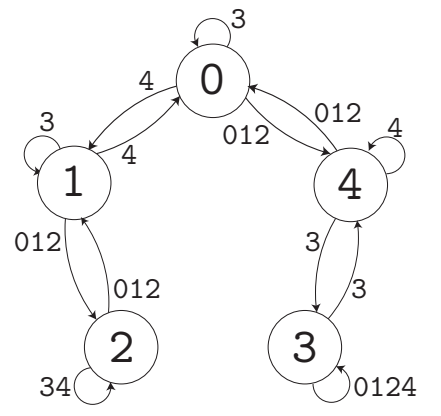

Type 47-1

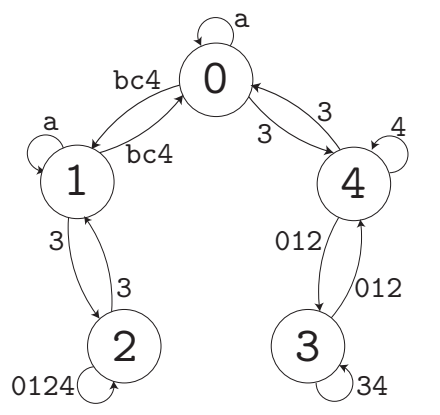

Type 47-2

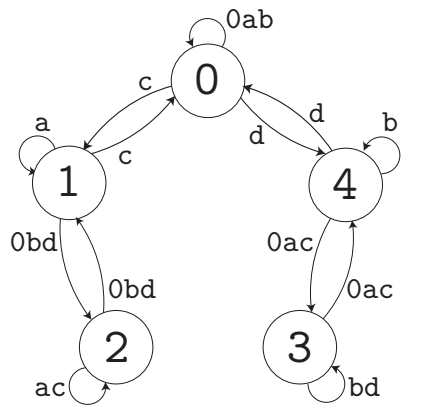

Type 47-3

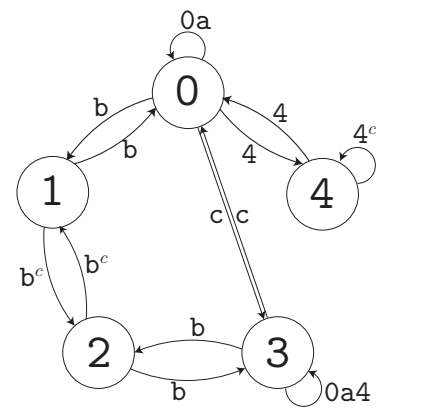
Type 47-4

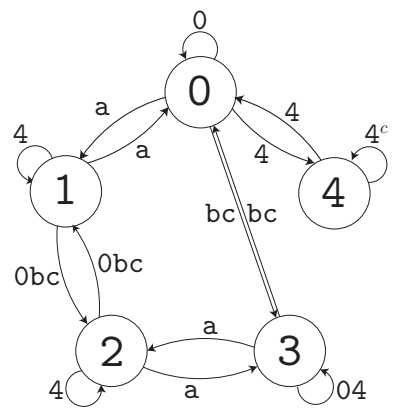
Type 47-5

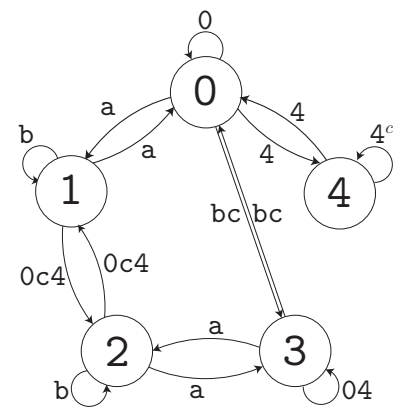
Type 47-6

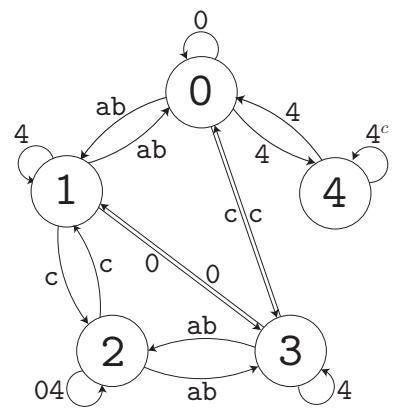
Type 47-7

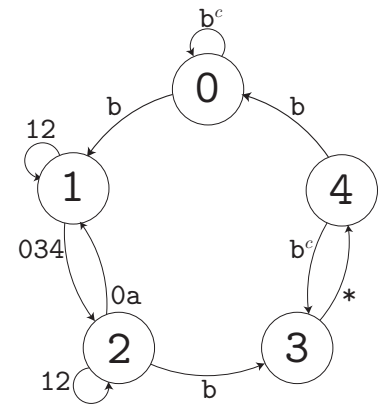
Type 48-1

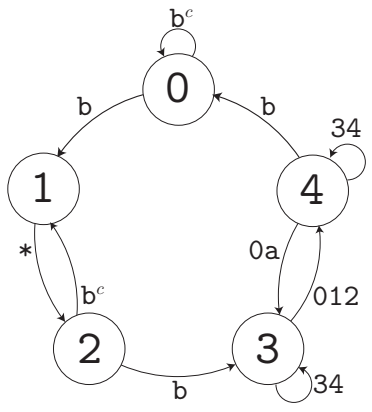
Type 48-2

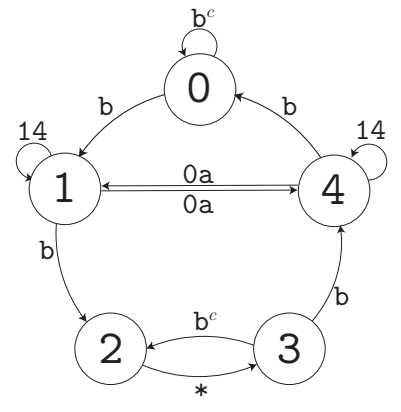
Type 48-3

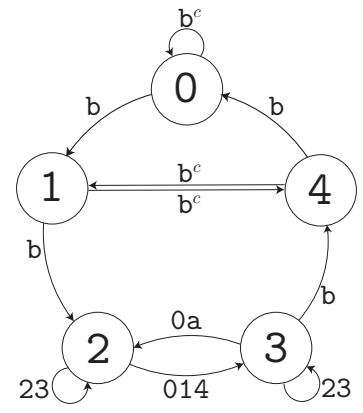
Type 48-4

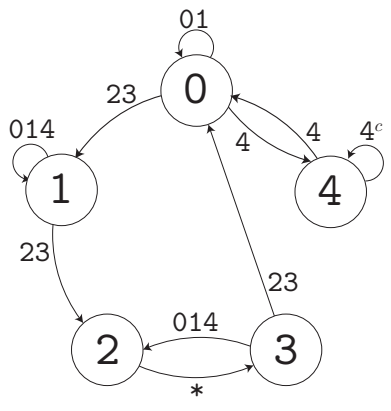
Type 49-1

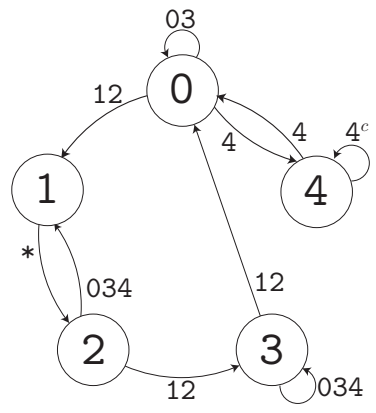
Type 49-2

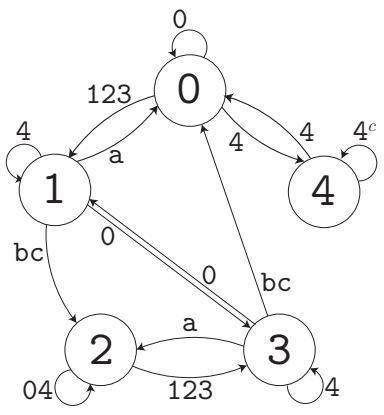
Type 49-3

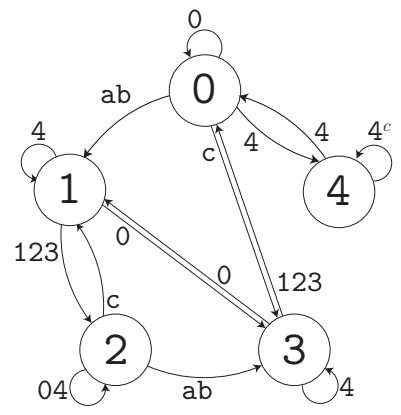
Type 49-4

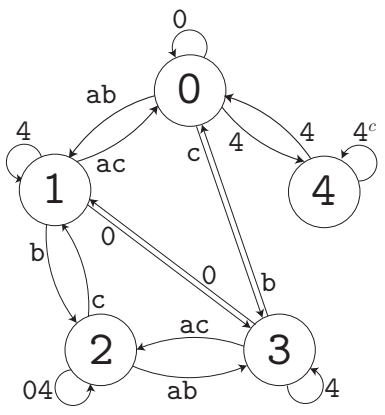
Type 49-5

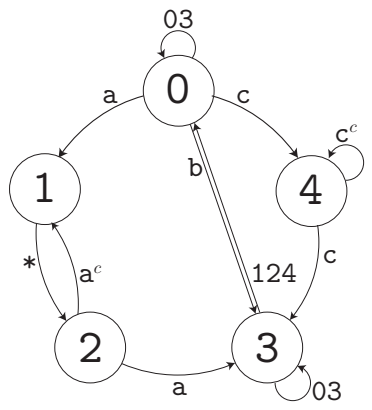
Type 50-1

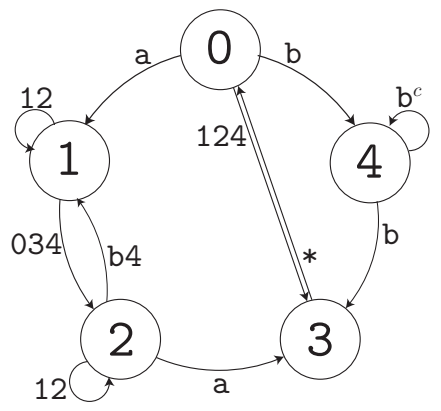
Type 50-2

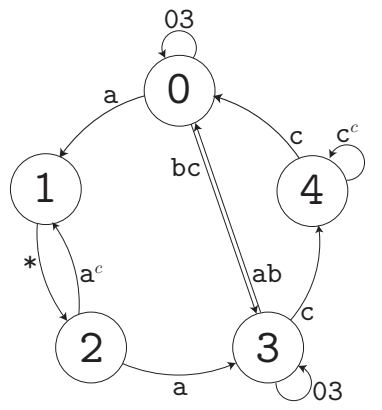
Type 50-3

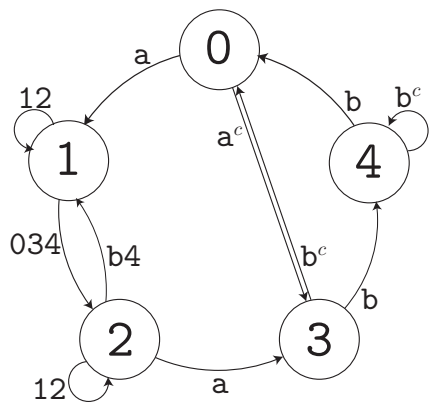
Type 50-4

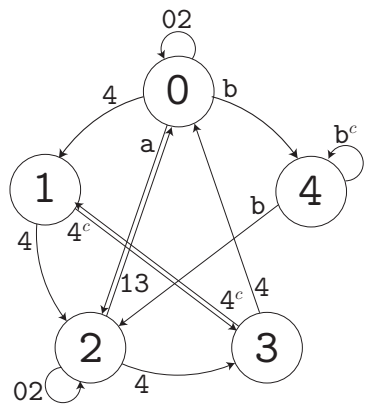
Type 50-5

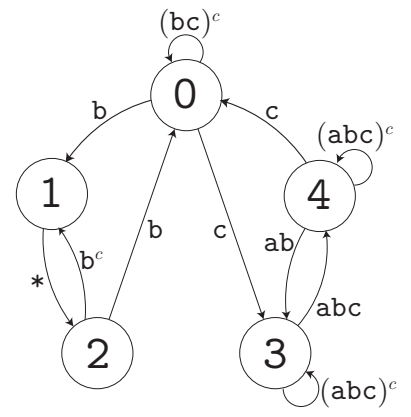
Type 51

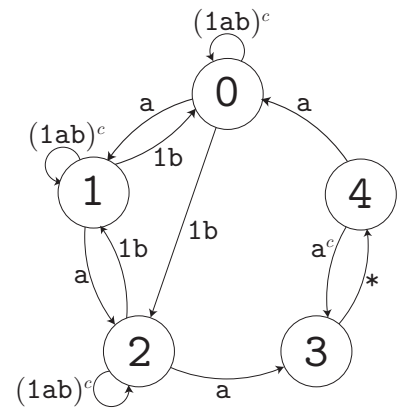
Type 52-1

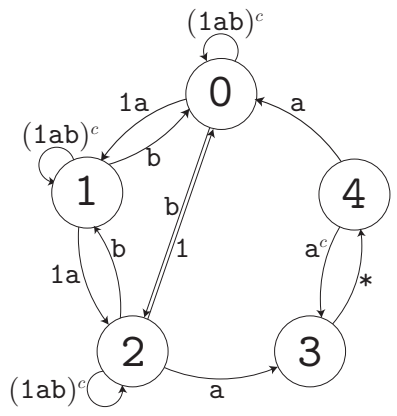
Type 52-2

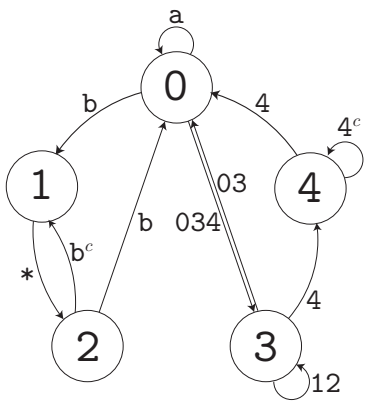
Type 53-1

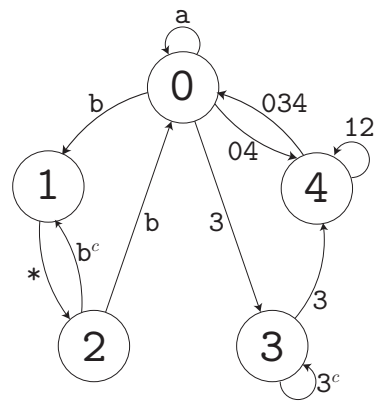
Type 53-2

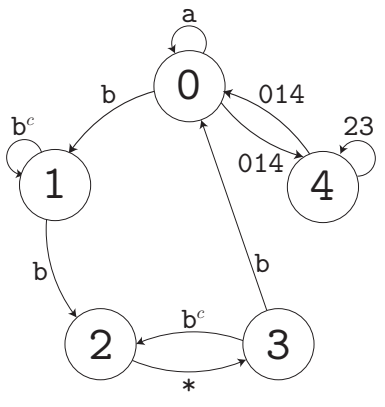
Type 54-1

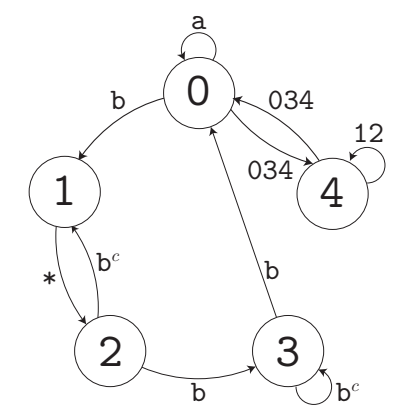

Type 54-2

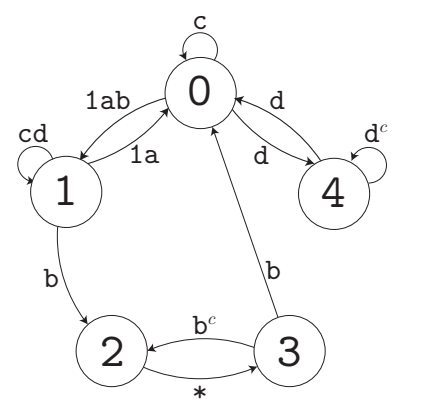

Type 54-3

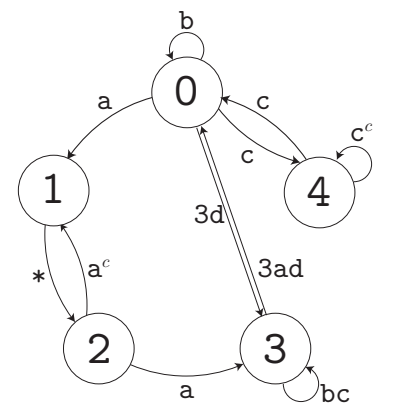

Type 54-4

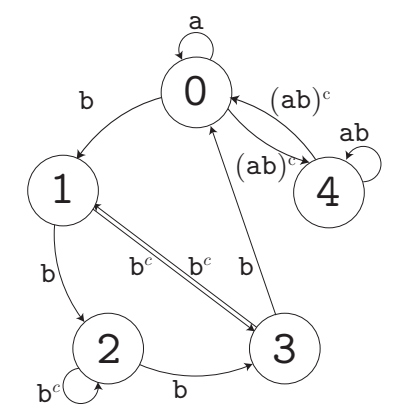

Type 54-5

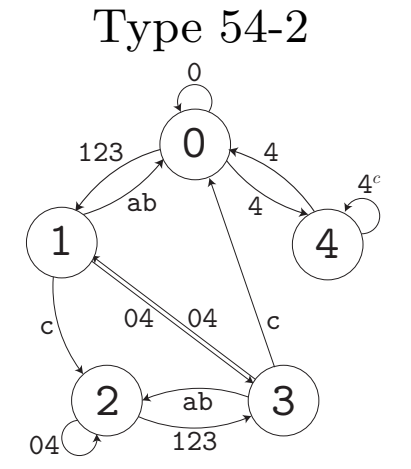

Type 54-6

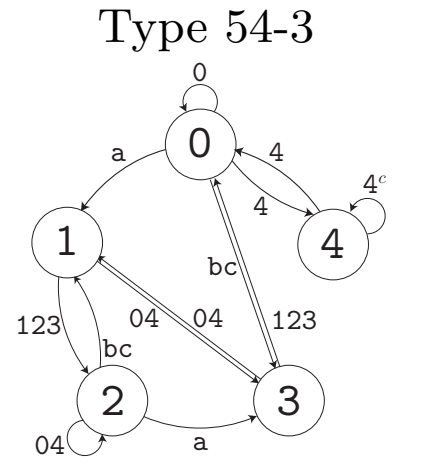

Type 54-7

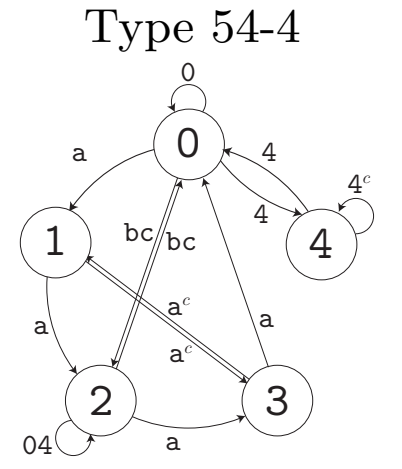

Type 54-8

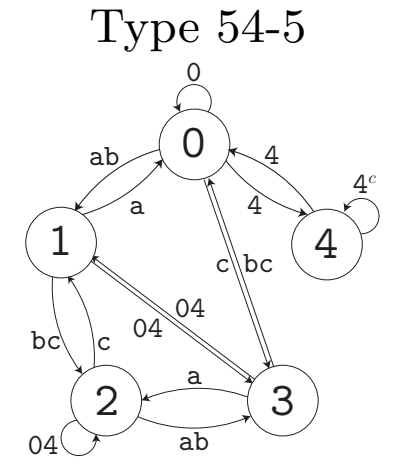

Type 54-9

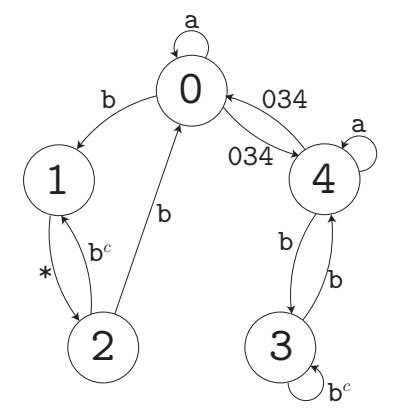

Type 55-1

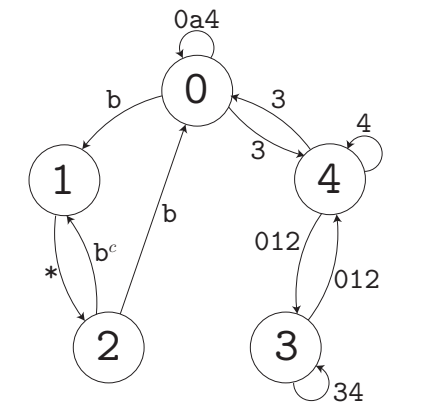

Type 55-2

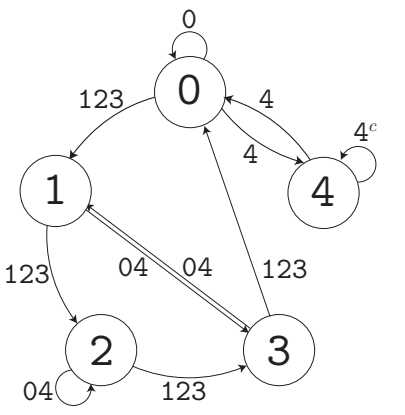

Type 56-1

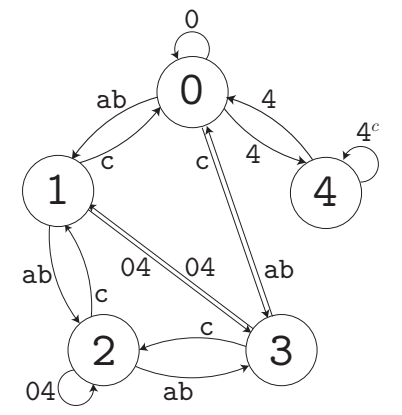

Type 56-2

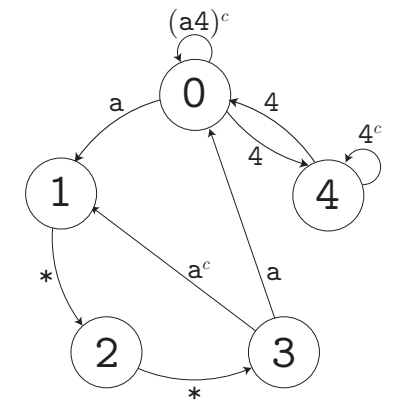

Type 57

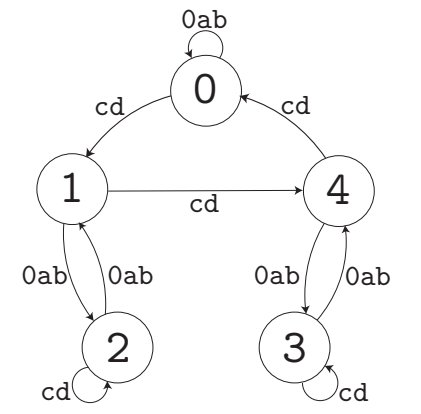

Type 58-1

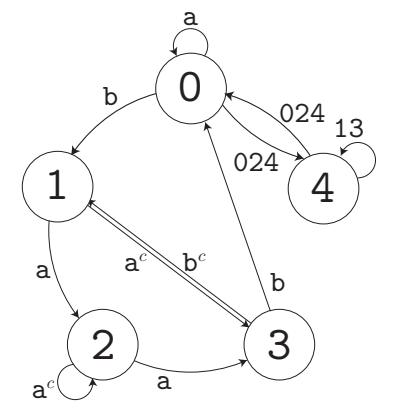

Type 58-2

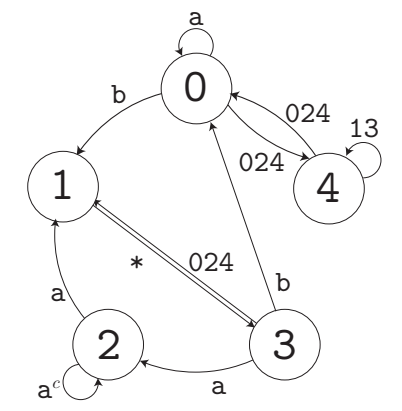

Type 58-3

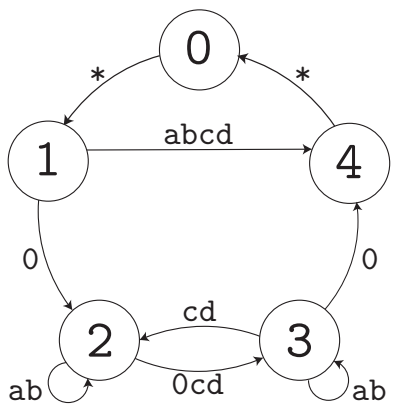

Type 59

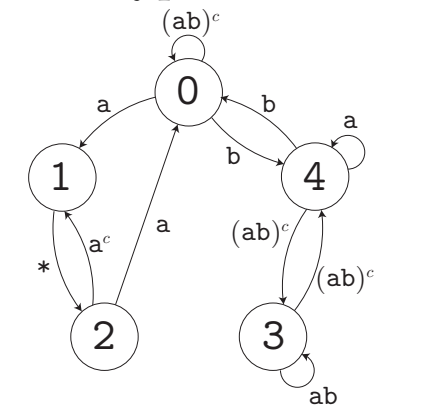

Type 60-1

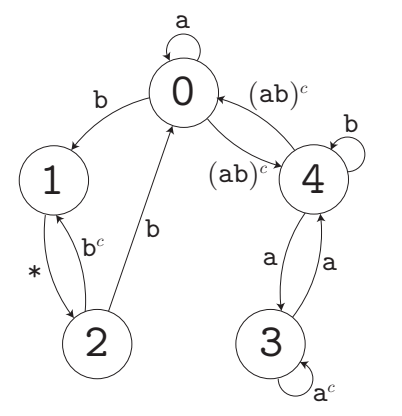

Type 60-2

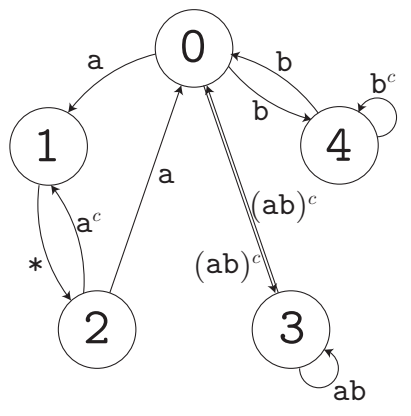

Type 60-3

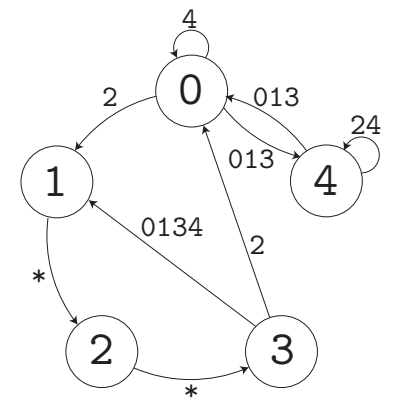

Type 61

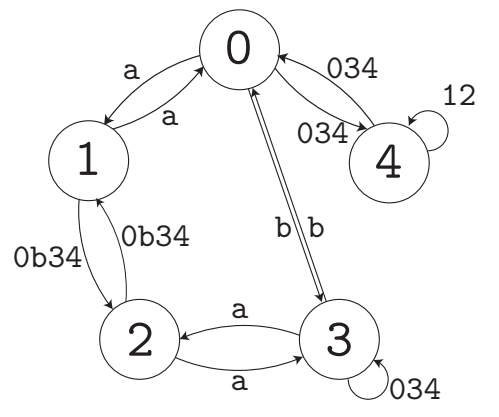

Type 62

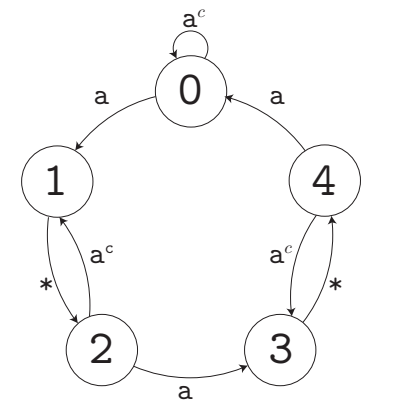

Type 63-1

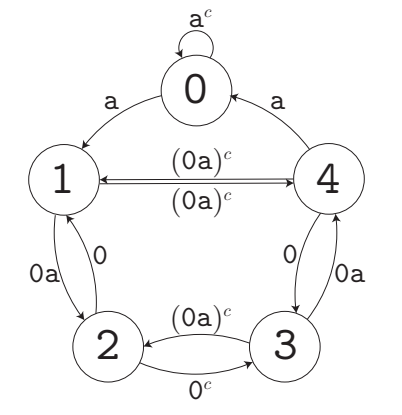

Type 63-2

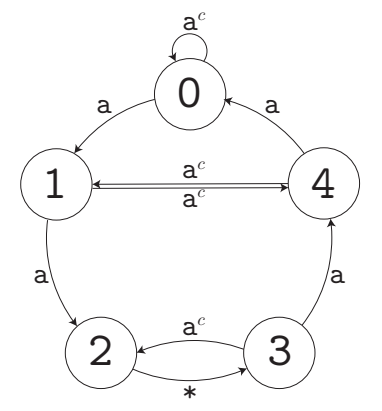

Type 63-4

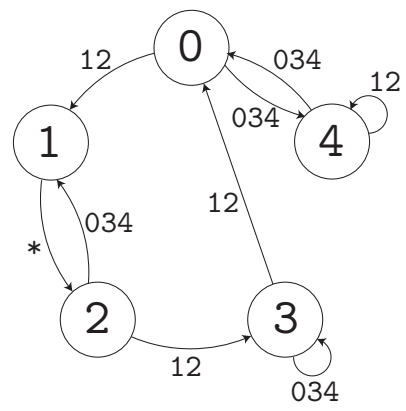

Type 64-1

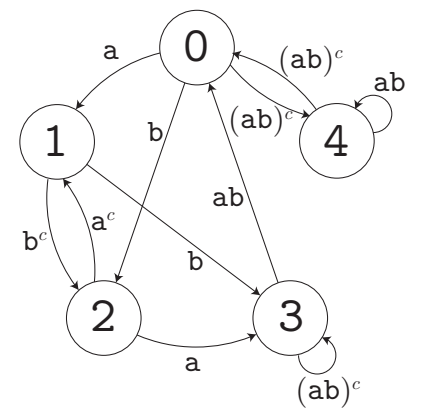

Type 64-2

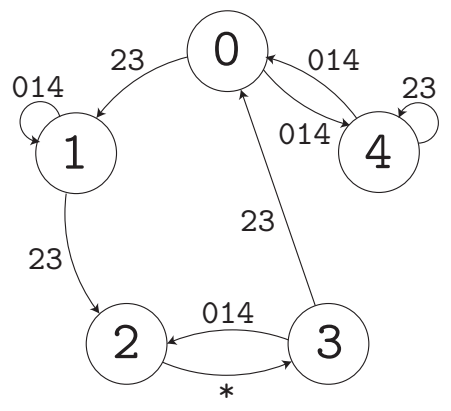

Type 64-3

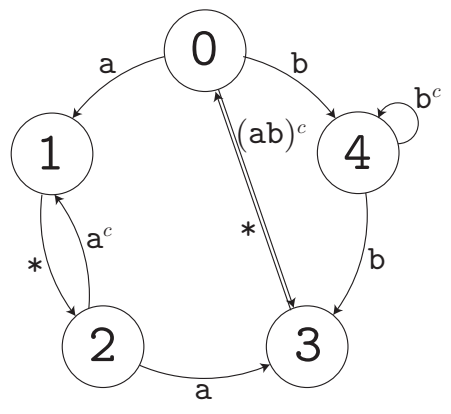

Type 65-1

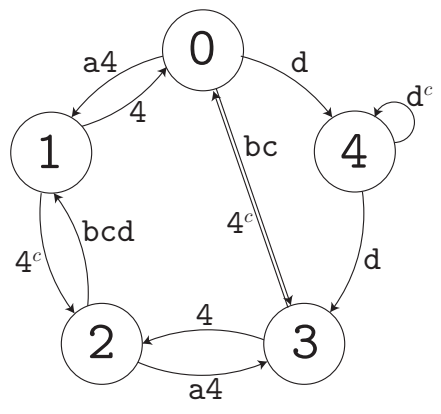

Type 65-2

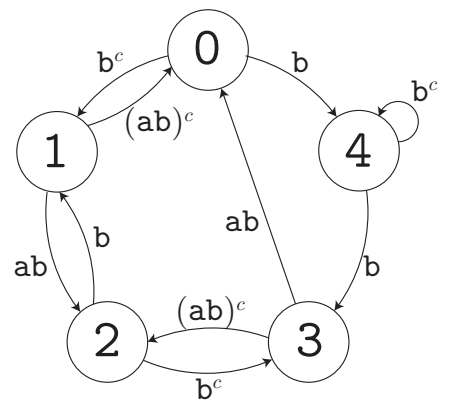

Type 65-3

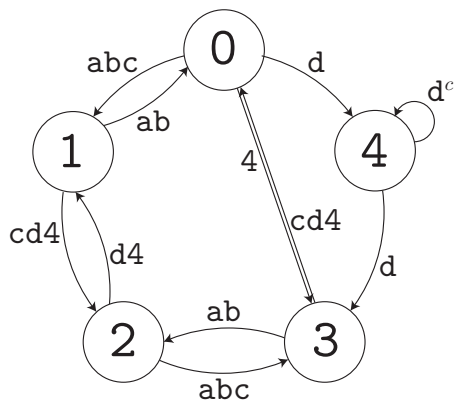

Type 65-4

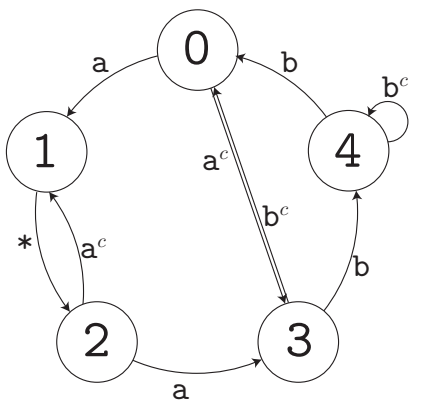

Type 65-5

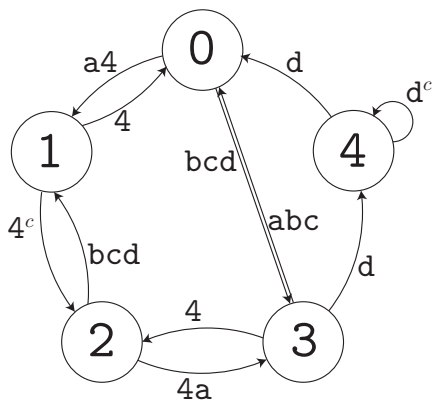

Type 65-6

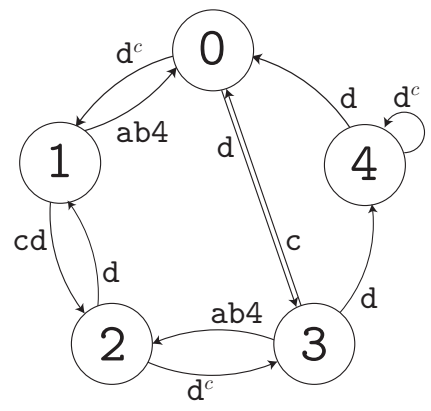

Type 65-7

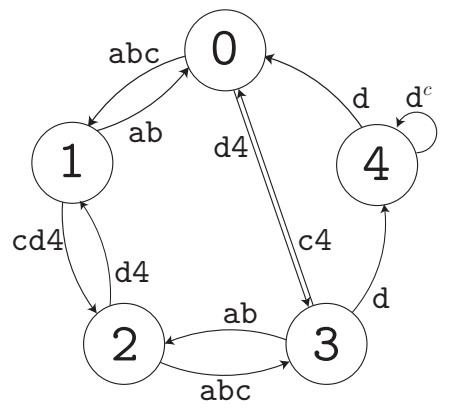

Type 65-8

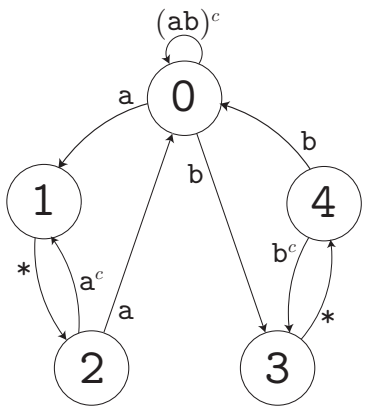

Type 66-1

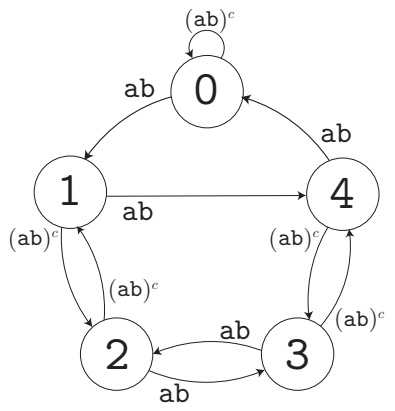

Type 66-2

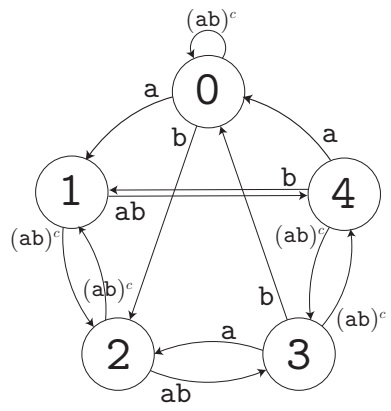

Type 66-3

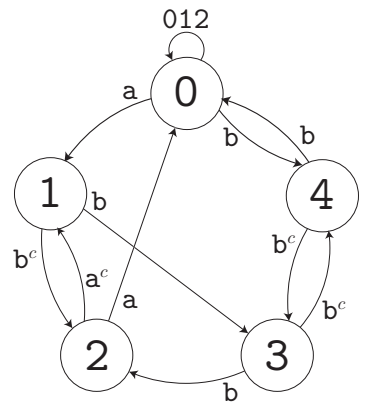

Type 66-4

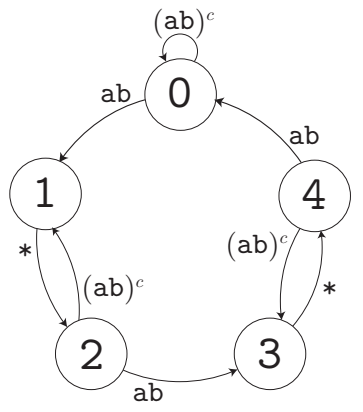

Type 67-1

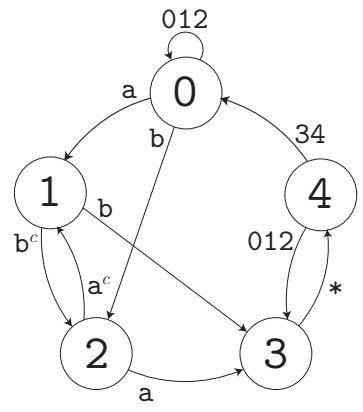

Type 67-2

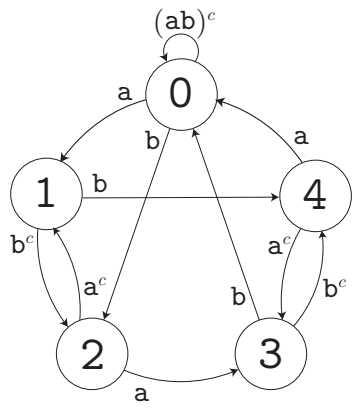

Type 67-3

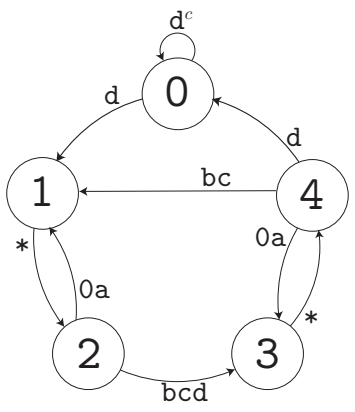

Type 68-1

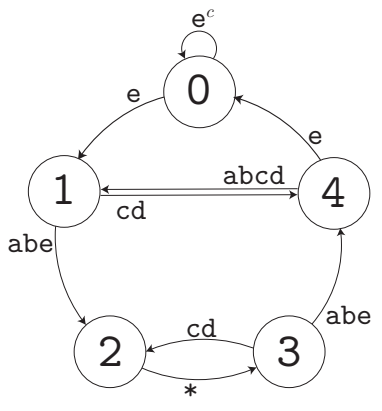

Type 68-2

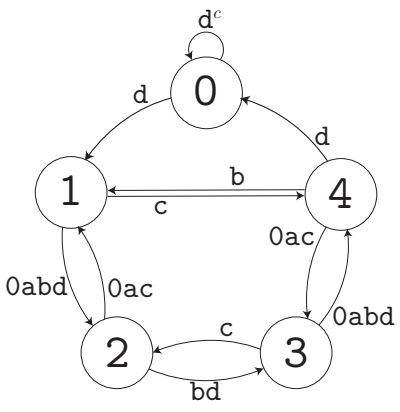

Type 68-3

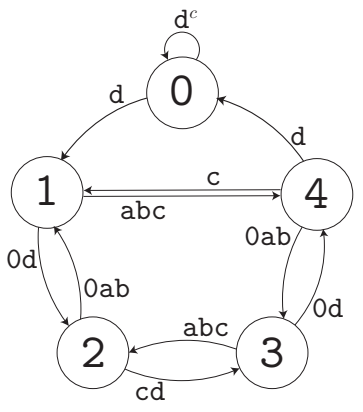

Type 68-4

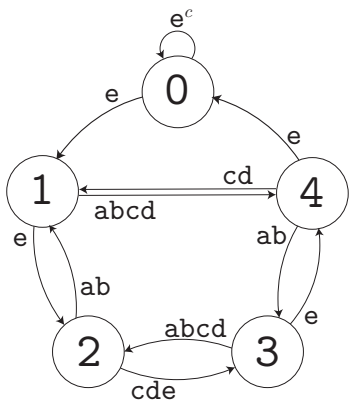

Type 68-5

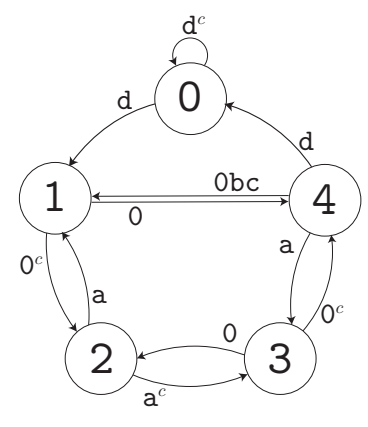

Type 68-6

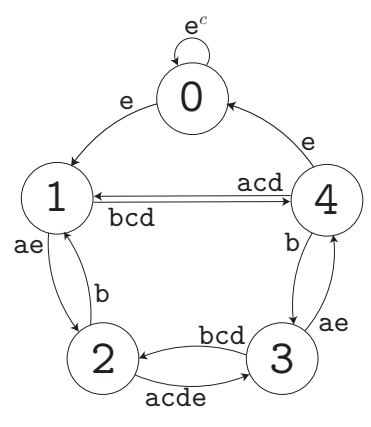

Type 68-7

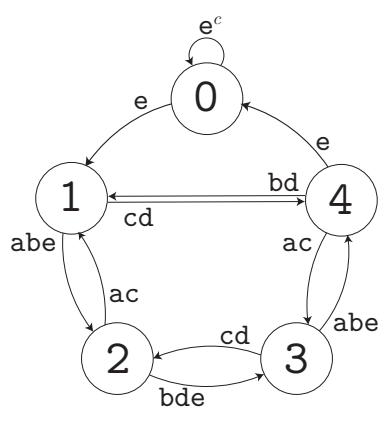

Type 68-8

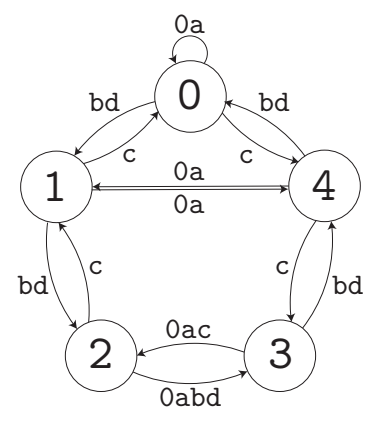

Type 69

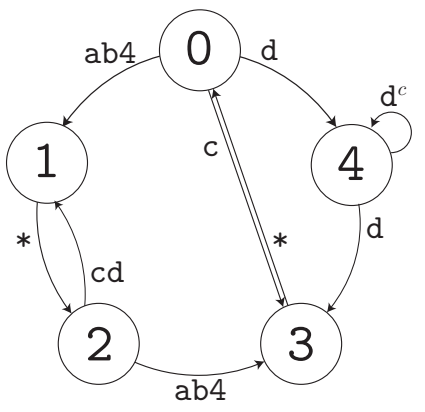

Type 70-1

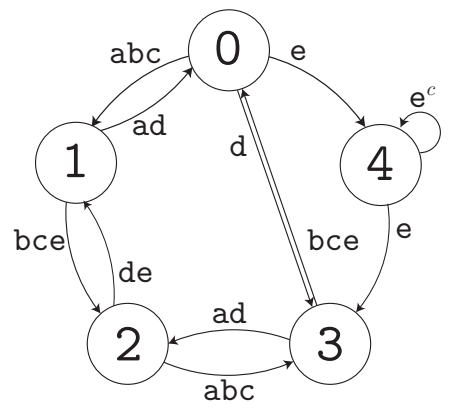

Type 70-2

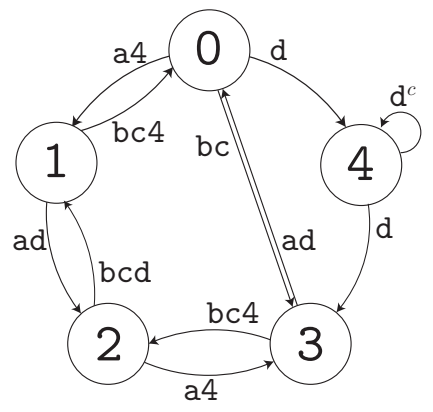

Type 70-3

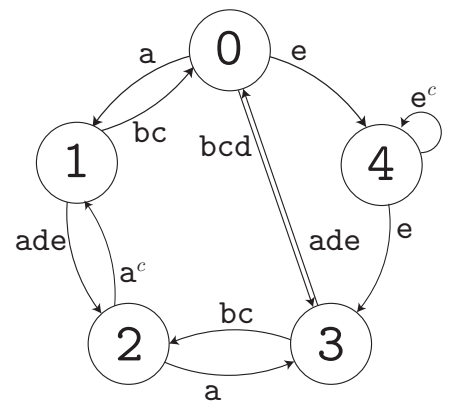

Type 70-4

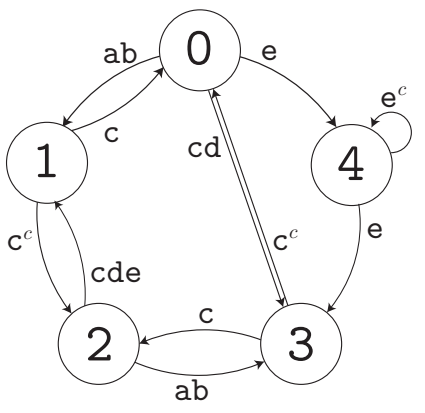

Type 70-5

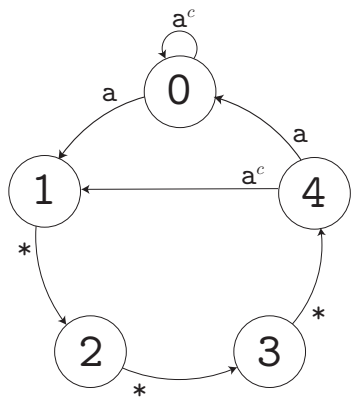

Type 71-1

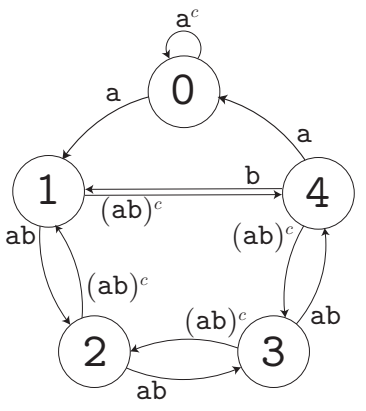

Type 71-2

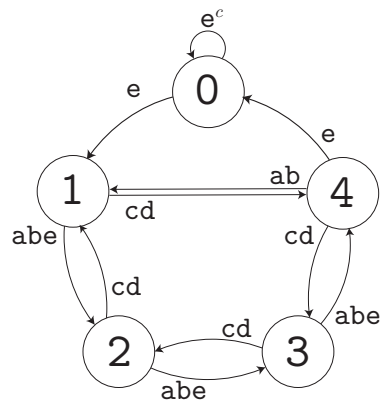

Type 71-3

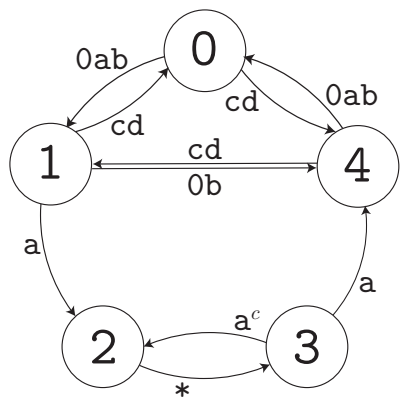

Type 72-1

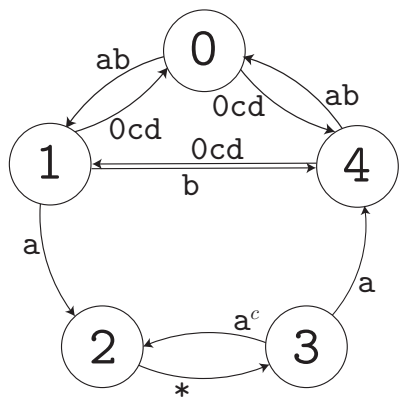

Type 72-2

**Table 4** Ergodic rules for $k = 5$. $\{i_1, i_2\}$ means that any order of the elements is accepted. The semicolon represents a separation of rules.

| Type | Number | Representative rules |
|---|---|---|
| 01 | 720 | $(\pi_a, \pi_b) = (\sigma_{32}, \sigma_{105})$, others are $\sigma_0$ |
| | | $(a, b) = \{1, 4\}, \{2, 4\}, \{3, 4\}$ |
| 02 | 1200 | $(\pi_a, \pi_b) = (\sigma_{31}, \sigma_{105})$, others are $\sigma_0$ |
| | | $(a, b) = \{0, 3\}, \{1, 3\}, \{1, 4\}, \{2, 3\}, \{2, 4\}$ |
| 03 | 480 | $(\pi_a, \pi_b) = (\sigma_{30}, \sigma_{81})$, others are $\sigma_0$ |
| | | $(a, b) = \{1, 3\}, \{1, 4\}, \{2, 3\}, \{2, 4\}$ |
| 04 | 1200 | $(\pi_a, \pi_b) = (\sigma_{45}, \sigma_7)$, others are $\sigma_0$ |
| | | $(a, b) = \{0, 2\}, \{0, 3\}\{1, 3\}, \{2, 3\}, \{2, 4\}$ |
| 05 | 720 | $(\pi_a, \pi_b) = (\sigma_{25}, \sigma_{111})$, others are $\sigma_0$ |
| | | $(a, b) = \{0, 2\}, \{0, 3\}\{1, 3\}, \{1, 4\}, \{2, 3\}, \{2, 4\}$ |
| 06-1 | 1680 | $(\pi_a, \pi_b, \pi_c) = (\sigma_{45}, \sigma_1, \sigma_6)$, others are $\sigma_0$ |
| | | $(a, b, c) = (0, \{2, 3\}), (2, \{0, 3\}), (2, \{0, 4\}), (2, \{3, 4\}), (3, \{0, 1\}), (3, \{0, 2\}), (3, \{1, 2\})$ |
| 06-2 | 1440 | $(\pi_a, \pi_b, \pi_c) = (\sigma_{30}, \sigma_{105}, \sigma_1)$, others are $\sigma_0$ |
| | | $b = 1$ or $2$, $(a, c) = \{3, 4\}$; $b = 3$, $(a, c) = \{0, 1\}, \{0, 2\}, \{1, 2\}$; $b = 4$, $(a, c) = \{1, 2\}$ |
| 06-3 | 720 | $(\pi_a, \pi_b, \pi_c) = (\sigma_{30}, \sigma_{80}, \sigma_{105})$, others are $\sigma_0$ |
| | | $(a, b, c) = \{1, 3, 4\}, \{2, 3, 4\}$ |
| 07-1 | 3120 | $(\pi_a, \pi_b, \pi_c) = (\sigma_1, \sigma_{111}, \sigma_{24})$, others are $\sigma_0$ |
| | | $(a, b, c) = \{0, 2, 3\}, (0, 2, 4), (0, 3, 1), (0, 3, 4), (1, 3, 0), (1, 3, 2), (1, 3, 4), (1, 4, 2), (2, 3, 1), (2, 3, 4), (2, 4, 1)$ |
| | | $(3, 0, 1), (3, 0, 4), (3, 1, 0), (3, 1, 2), (3, 1, 4), (3, 2, 1), (3, 2, 4), (4, 1, 3), (4, 2, 0), (4, 2, 3)$ |
| 07-2 | 360 | $\pi_1 = \pi_4 = \sigma_0$, $(\pi_a, \pi_b, \pi_c) = (\sigma_7, \sigma_{105}, \sigma_{24})$ |
| | | $(a, b, c) = \{0, 2, 3\}$ |
| 07-3 | 1680 | $\pi_1 = \pi_4 = \sigma_0$, $(\pi_a, \pi_b, \pi_c) = (\sigma_{24}, \sigma_{86}, \sigma_{105})$. |
| | | $(a, b, c) = \{1, 3, 4\}, \{2, 3, 4\}, (\{1, 2\}, 4)$ |
| 08-1 | 1440 | $(\pi_a, \pi_b, \pi_c) = (\sigma_{31}, \sigma_{30}, \sigma_{105})$, others are $\sigma_0$ |
| | | $(a, b, c) = (\{0, 1\}, 3), (\{0, 2\}, 3), (\{1, 2\}, 3), (\{1, 2\}, 4), (\{3, 4\}, 1), (\{3, 4\}, 2)$ |
| 08-2 | 960 | $(\pi_a, \pi_b, \pi_c) = (\sigma_{30}, \sigma_6, \sigma_{87})$ |
| | | $(a, b, c) = (1, \{3, 4\}), (2, \{3, 4\}), (3, \{1, 2\}), (4, \{1, 2\})$ |
| 09-1 | 2160 | $(\pi_a, \pi_b, \pi_c) = (\sigma_{14}, \sigma_{32}, \sigma_{119})$, others are $\sigma_0$ |
| | | $b$ or $c = 4$, and $b, c \neq 0$ |
| 09-2 | 720 | $(\pi_a, \pi_b, \pi_4) = (\sigma_6, \sigma_{111}, \sigma_{32})$, others are $\sigma_0$ |
| | | $(a, b) = \{1, 2\}, \{1, 3\}, \{2, 3\}$ |
| 09-3 | 720 | $(\pi_a, \pi_b, \pi_4) = (\sigma_2, \sigma_{107}, \sigma_{32})$, others are $\sigma_0$ |
| | | $(a, b) = \{1, 2\}, \{1, 3\}, \{2, 3\}$ |
| 09-4 | 720 | $(\pi_a, \pi_b, \pi_4) = (\sigma_{32}, \sigma_{60}, \sigma_{105})$, others are $\sigma_0$ |
| | | $(a, b) = \{1, 2\}, \{1, 3\}, \{2, 3\}$ |
| 09-5 | 720 | $(\pi_a, \pi_b, \pi_4) = (\sigma_{32}, \sigma_{54}, \sigma_{119})$, others are $\sigma_0$ |
| | | $(a, b) = \{1, 2\}, \{1, 3\}, \{2, 3\}$ |
| 10 | 1440 | $(\pi_a, \pi_b, \pi_c) = (\sigma_{31}, \sigma_6, \sigma_{111})$, others are $\sigma_0$ |
| | | $(a, b, c) = (2, \{3, 4\}), (3, \{0, 1\}), (3, \{0, 2\}), (3, \{1, 2\}), (4, \{1, 2\})$ |
| 11 | 480 | $(\pi_a, \pi_b, \pi_c) = (\sigma_{30}, \sigma_{30}, \sigma_{81})$, others are $\sigma_0$ |
| | | $(a, b, c) = (1, 2, 3), (1, 2, 4), (3, 4, 1), (3, 4, 2)$ |
| 12 | 720 | $(\pi_a, \pi_b, \pi_c) = (\sigma_7, \sigma_{45}, \sigma_{45})$, others are $\sigma_0$ |
| | | $(a, b, c) = (2, 0, 3), (2, 0, 4), (2, 3, 4), (3, 0, 1), (3, 0, 2), (3, 1, 2)$ |
| 13 | 720 | $(\pi_a, \pi_b, \pi_c) = (\sigma_{30}, \sigma_{86}, \sigma_{111})$, others are $\sigma_0$ |
| | | $(a, b, c) = \{1, 3, 4\}, \{2, 3, 4\}$ |
| 14-1 | 720 | $(\pi_0, \pi_a, \pi_b, \pi_c, \pi_4) = (\sigma_0, \sigma_0, \sigma_8, \sigma_{117}, \sigma_{32})$, $(a, b, c) = \{1, 2, 3\}$ |
| 14-2 | 720 | $(\pi_0, \pi_a, \pi_b, \pi_c, \pi_4) = (\sigma_0, \sigma_0, \sigma_{12}, \sigma_{113}, \sigma_{32})$, $(a, b, c) = \{1, 2, 3\}$ |
| 15 | 480 | $(\pi_0, \pi_a, \pi_b, \pi_c, \pi_d) = (\sigma_0, \sigma_0, \sigma_{30}, \sigma_{87}, \sigma_{87})$, $(a, b, c, d) = (\{1, 2\}, 3, 4), (\{3, 4\}, 1, 2)$ |
| 16 | 720 | $(\pi_0, \pi_a, \pi_b, \pi_c, \pi_4) = (\sigma_0, \sigma_0, \sigma_{26}, \sigma_{86}, \sigma_{111})$, $(a, b, c) = \{1, 2, 3\}$ |
| 17-1 | 360 | $(\pi_0, \pi_a, \pi_b, \pi_c, \pi_4) = (\sigma_0, \sigma_0, \sigma_{32}, \sigma_{32}, \sigma_{111})$, $(a, b, c) = (1, 2, 3), (2, 1, 3), (3, 1, 2)$ |
| 17-2 | 360 | $(\pi_0, \pi_a, \pi_b, \pi_c, \pi_4) = (\sigma_0, \sigma_0, \sigma_{32}, \sigma_{32}, \sigma_{107})$, $(a, b, c) = (1, 2, 3), (2, 1, 3), (3, 1, 2)$ |
| 18-1 | 1200 | $(\pi_a, \pi_b, \pi_c, \pi_d, \pi_e) = (\sigma_1, \sigma_{24}, \sigma_6, \sigma_{105}, \sigma_0)$, |
| | | $(a, b, c, d, e) = (\{\{1, 2\}, \{3, 4\}\}, 0), (0, 1, 4, 3, 2), (4, 3, 0, 1, 2), (0, 4, 1, 2, 3), (1, 2, 0, 4, 3)$ |
| | | $(a, d, b, c, e) = (\{0, 2\}, 3, 1, 4), (\{0, 3\}, 2, 4, 1), (1, 3, \{0, 2\}, 4), (4, 2, \{0, 3\}, 1)$ |
| 18-2 | 600 | $(\pi_a, \pi_b, \pi_c, \pi_d, \pi_e) = (\sigma_6, \sigma_{80}, \sigma_{24}, \sigma_{105}, \sigma_0)$ |
| | | $(a, b, c, d, e) = (\{1, 2\}, \{3, 4\}, 0)$, $(a, d, b, c, e) = (\{1, 2\}, \{3, 4\}, 0), (0, \{3, 4\}, 2, 1)$ |
| 18-3 | 120 | $(\pi_0, \pi_a, \pi_b, \pi_c, \pi_d) = (\sigma_0, \sigma_{24}, \sigma_{54}, \sigma_{80}, \sigma_{105})$, $(a, b, c, d) = \{1, 2, 3, 4\}$ |
| 19-1 | 2160 | $(\pi_a, \pi_b, \pi_c, \pi_d, \pi_e) = (\sigma_{14}, \sigma_{14}, \sigma_{32}, \sigma_{105}, \sigma_0)$ |
| | | $(e, a, b, c, d) = (\{0, 1, 2\}, \{3, 4\}), (\{0, 1, 3\}, \{2, 4\}), (\{0, 2, 3\}, \{1, 4\})$ |
| 19-2 | 120 | $(\pi_0, \pi_1, \pi_2, \pi_3, \pi_4) = (\sigma_0, \sigma_{105}, \sigma_{105}, \sigma_{105}, \sigma_{32})$ |
| 19-3 | 1440 | $(\pi_a, \pi_b, \pi_c, \pi_d, \pi_4) = (\sigma_{14}, \sigma_{32}, \sigma_{54}, \sigma_0, \sigma_{105})$ |
| | | $(a, b, c, d) = (\{1, 2, 3\}, 0), (0, \{1, 2\}, 3), (0, \{1, 3\}, 2), (0, \{2, 3\}, 1)$ |

| Type | Number | Representative rules |
|---|---|---|
| 19-4 | 600 | $(\pi_a,\pi_b,\pi_c,\pi_d,\pi_e)=(\sigma_{32},\sigma_{54},\sigma_{54},\sigma_{105},\sigma_0),\quad (a,b,c,d,e)=(\{1,2,3\},4,0),(4,0,2,\{1,3\})$ |
| 19-5 | 360 | $(\pi_0,\pi_a,\pi_b,\pi_c,\pi_4)=(\sigma_0,\sigma_2,\sigma_2,\sigma_{105},\sigma_{32}),\quad (a,b,c)=\{1,2,3\}$ |
| 19-6 | 360 | $(\pi_0,\pi_a,\pi_b,\pi_c,\pi_4)=(\sigma_0,\sigma_6,\sigma_6,\sigma_{105},\sigma_{32}),\quad (a,b,c)=\{1,2,3\}$ |
| 20-1 | 360 | $(\pi_0,\pi_a,\pi_b,\pi_c,\pi_d)=(\sigma_{105},\sigma_{31},\sigma_{105},\sigma_{105},\sigma_0),\quad (a,b,c,d)=(1,2,4,3),(2,1,4,3),(3,1,2,4)$ |
| 20-2 | 360 | $(\pi_a,\pi_b,\pi_c,\pi_3,\pi_4)=(\sigma_6,\sigma_6,\sigma_{105},\sigma_{31},\sigma_0),\quad (a,b,c)=\{0,1,2\}$ |
| 20-3 | 360 | $(\pi_a,\pi_b,\pi_c,\pi_3,\pi_4)=(\sigma_{31},\sigma_1,\sigma_1,\sigma_{105},\sigma_0),\quad (a,b,c)=\{0,1,2\}$ |
| 20-4 | 120 | $(\pi_0,\pi_1,\pi_2,\pi_3,\pi_4)=(\sigma_{54},\sigma_{31},\sigma_{54},\sigma_{105},\sigma_0)$ |
| 20-5 | 120 | $(\pi_0,\pi_1,\pi_2,\pi_3,\pi_4)=(\sigma_{24},\sigma_{24},\sigma_{31},\sigma_{105},\sigma_0)$ |
| 21-1 | 360 | $(\pi_a,\pi_b,\pi_c,\pi_3,\pi_4)=(\sigma_1,\sigma_{30},\sigma_{30},\sigma_{105},\sigma_0)\quad (a,b,c)=\{0,1,2\}$ |
| 21-2 | 120 | $(\pi_0,\pi_1,\pi_2,\pi_a,\pi_b)=(\sigma_0,\sigma_{30},\sigma_{30},\sigma_{80},\sigma_{105}),\quad (a,b)=\{3,4\}$ |
| 22-1 | 1440 | $(\pi_a,\pi_b,\pi_c,\pi_d,\pi_e)=(\sigma_1,\sigma_{30},\sigma_6,\sigma_{111},\sigma_0)$ |
| | | $(a,b,c,d,e)=(\{1,2\},\{3,4\},0),(0,1,4,3,2),(0,2,4,3,1),(4,3,0,\{1,2\})$ |
| 22-2 | 240 | $(\pi_a,\pi_b,\pi_2,\pi_3,\pi_4)=(\sigma_{24},\sigma_{25},\sigma_{30},\sigma_{105},\sigma_0),\quad (a,b)=\{0,1\}$ |
| 22-3 | 240 | $(\pi_a,\pi_1,\pi_b,\pi_3,\pi_4)=(\sigma_{54},\sigma_{30},\sigma_{55},\sigma_{105},\sigma_0),\quad (a,b)=\{0,2\}$ |
| 22-4 | 960 | $(\pi_0,\pi_a,\pi_b,\pi_c,\pi_d)=(\sigma_0,\sigma_6,\sigma_{30},\sigma_{80},\sigma_{111})$ |
| | | $(a,b,c,d)=(1,3,\{2,4\}),(1,4,\{2,3\}),(2,3,\{1,4\}),(2,4,\{1,3\})$ |
| 22-5 | 720 | $(\pi_a,\pi_b,\pi_c,\pi_d,\pi_e)=(\sigma_0,\sigma_{45},\sigma_6,\sigma_6,\sigma_7),\quad (a,b,c,d,e)=(1,2,\{0,3,4\}),(4,3,\{0,1,2\})$ |
| 22-6 | 720 | $(\pi_a,\pi_b,\pi_c,\pi_d,\pi_e)=(\sigma_0,\sigma_{45},\sigma_1,\sigma_1,\sigma_7),\quad (a,b,c,d,e)=(1,2,\{0,3,4\}),(4,3,\{0,1,2\})$ |
| 23-1 | 2400 | $(\pi_a,\pi_b,\pi_c,\pi_d,\pi_e)=(\sigma_0,\sigma_{24},\sigma_{24},\sigma_{25},\sigma_{111})$ |
| | | $(a,b,c,d,e)=(0,1,4,\{2,3\}),(4,0,1,\{2,3\}),(\{0,2,4\},\{1,3\}),(\{1,2,4\},\{0,3\}),(1,\{0,3,4\},2)$ |
| 23-2 | 720 | $(\pi_a,\pi_b,\pi_c,\pi_d,\pi_e)=(\sigma_0,\sigma_6,\sigma_6,\sigma_{25},\sigma_{111}),\quad (a,d,b,c,e)=(1,2,\{0,3,4\}),(4,3,\{0,1,2\})$ |
| 23-3 | 480 | $(\pi_a,\pi_b,\pi_c,\pi_d,\pi_e)=(\sigma_0,\sigma_6,\sigma_7,\sigma_{24},\sigma_{111}),\quad (a,b,d,c,e)=(1,4,2,\{0,3\}),(4,1,3,\{0,2\})$ |
| 23-4 | 1440 | $(\pi_a,\pi_b,\pi_c,\pi_d,\pi_4)=(\sigma_0,\sigma_2,\sigma_{26},\sigma_{86},\sigma_{105}),\quad (a,b,c,d)=(0,\{1,2,3\}),(b,a,c,d)=(0,\{1,2,3\})$ |
| 23-5 | 1440 | $(\pi_a,\pi_b,\pi_c,\pi_d,\pi_4)=(\sigma_0,\sigma_2,\sigma_{24},\sigma_{86},\sigma_{107}),\quad (a,b,c,d)=(0,\{1,2,3\}),(b,a,c,d)=(0,\{1,2,3\})$ |
| 23-6 | 720 | $(\pi_0,\pi_a,\pi_b,\pi_c,\pi_4)=(\sigma_0,\sigma_{24},\sigma_{26},\sigma_{86},\sigma_{105}),\quad (a,b,c)=\{1,2,3\}$ |
| 23-7 | 360 | $(\pi_0,\pi_a,\pi_b,\pi_c,\pi_4)=(\sigma_0,\sigma_{24},\sigma_{24},\sigma_{86},\sigma_{107}),\quad (a,b,c)=\{1,2,3\}$ |
| 23-8 | 600 | $(\pi_a,\pi_b,\pi_c,\pi_d,\pi_e)=(\sigma_0,\sigma_6,\sigma_{24},\sigma_{86},\sigma_{111})$ |
| | | $(a,b,c,d,e)=(1,0,2,\{3,4\}),(a,b,d,c,e)=(0,\{1,2\},\{3,4\}),(a,b,e,c,d)=(0,\{1,2\},\{3,4\})$ |
| 24-1 | 480 | $(\pi_0,\pi_a,\pi_b,\pi_c,\pi_d)=(\sigma_0,\sigma_1,\sigma_{31},\sigma_6,\sigma_{87}),\quad (a,b,c,d)=(\{1,2\},\{3,4\})$ |
| 24-2 | 360 | $(\pi_a,\pi_b,\pi_c,\pi_3,\pi_4)=(\sigma_1,\sigma_{31},\sigma_{31},\sigma_{105},\sigma_0),\quad (a,b,c)=\{0,1,2\}$ |
| 25 | 720 | $(\pi_0,\pi_a,\pi_b,\pi_c,\pi_4)=(\sigma_0,\sigma_8,\sigma_{12},\sigma_{105},\sigma_{32}),\quad (a,b,c)=\{1,2,3\}$ |
| 26-1 | 360 | $(\pi_a,\pi_b,\pi_c,\pi_3,\pi_4)=(\sigma_{30},\sigma_{30},\sigma_{49},\sigma_{105},\sigma_0)\quad (a,b,c)=\{0,1,2\}$ |
| 26-2 | 720 | $(\pi_a,\pi_b,\pi_c,\pi_3,\pi_4)=(\sigma_{30},\sigma_{48},\sigma_{49},\sigma_{105},\sigma_0)\quad (a,b,c)=\{0,1,2\}$ |
| 27-1 | 840 | $(\pi_a,\pi_b,\pi_c,\pi_d,\pi_e)=(\sigma_0,\sigma_{26},\sigma_{26},\sigma_{32},\sigma_{105})$ |
| | | $(a,b,c,d,e)=(0,1,2,3,4),(0,2,3,1,4),(0,3,1,2,4),(2,0,1,4,3),(2,1,4,0,3),(3,0,1,4,2),(3,1,4,0,2)$ |
| 27-2 | 720 | $(\pi_0,\pi_a,\pi_b,\pi_c,\pi_4)=(\sigma_0,\sigma_{26},\sigma_{32},\sigma_{86},\sigma_{105}),\quad (a,b,c)=\{1,2,3\}$ |
| 27-3 | 840 | $(\pi_a,\pi_b,\pi_c,\pi_d,\pi_e)=(\sigma_0,\sigma_{32},\sigma_{86},\sigma_{86},\sigma_{105})$ |
| | | $(a,b,c,d,e)=(0,1,2,3,4),(0,2,3,1,4),(0,3,1,2,4),(1,0,3,4,2),(1,4,0,3,2),(2,0,3,4,1),(2,4,0,3,1)$ |
| 27-4 | 360 | $(\pi_0,\pi_a,\pi_b,\pi_c,\pi_4)=(\sigma_0,\sigma_{105},\sigma_{107},\sigma_{107},\sigma_{32})$ |
| | | $(a,b,c)=\{1,2,3\}$ |
| 27-5 | 360 | $(\pi_0,\pi_a,\pi_b,\pi_c,\pi_4)=(\sigma_0,\sigma_{105},\sigma_{111},\sigma_{111},\sigma_{32})\quad (a,b,c)=\{1,2,3\}$ |
| 27-6 | 360 | $(\pi_0,\pi_a,\pi_b,\pi_c,\pi_4)=(\sigma_0,\sigma_{32},\sigma_{60},\sigma_{60},\sigma_{105})\quad (a,b,c)=\{1,2,3\}$ |
| 27-7 | 360 | $(\pi_0,\pi_a,\pi_b,\pi_c,\pi_4)=(\sigma_0,\sigma_{105},\sigma_{119},\sigma_{119},\sigma_{32})\quad (a,b,c)=\{1,2,3\}$ |
| 27-8 | 1440 | $(\pi_a,\pi_b,\pi_c,\pi_d,\pi_4)=(\sigma_0,\sigma_{14},\sigma_{32},\sigma_{60},\sigma_{119})$ |
| | | $(a,b,c,d)=(0,\{1,2,3\}),(b,a,c,d)=(0,\{1,2,3\})$ |
| 27-9 | 720 | $(\pi_0,\pi_a,\pi_b,\pi_c,\pi_4)=(\sigma_0,\sigma_{32},\sigma_{54},\sigma_{60},\sigma_{119})\quad (a,b,c)=\{1,2,3\}$ |
| 28-1 | 360 | $(\pi_a,\pi_b,\pi_c,\pi_3,\pi_4)=(\sigma_{105},\sigma_{111},\sigma_{111},\sigma_{31},\sigma_0)\quad (a,b,c)=\{0,1,2\}$ |
| 28-2 | 120 | $(\pi_0,\pi_1,\pi_2,\pi_3,\pi_4)=(\sigma_{25},\sigma_{25},\sigma_{31},\sigma_{105},\sigma_0)$ |
| 28-3 | 120 | $(\pi_0,\pi_1,\pi_2,\pi_3,\pi_4)=(\sigma_{55},\sigma_{31},\sigma_{55},\sigma_{105},\sigma_0)$ |
| 29-1 | 120 | $(\pi_0,\pi_1,\pi_2,\pi_3,\pi_4)=(\sigma_0,\sigma_{32},\sigma_{32},\sigma_{32},\sigma_{105})$ |
| 29-2 | 360 | $(\pi_0,\pi_a,\pi_b,\pi_c,\pi_4)=(\sigma_0,\sigma_{32},\sigma_{32},\sigma_{72},\sigma_{105})\quad (a,b,c)=\{1,2,3\}$ |
| 30 | 720 | $(\pi_0,\pi_a,\pi_b,\pi_c,\pi_4)=(\sigma_0,\sigma_{105},\sigma_{113},\sigma_{117},\sigma_{32})\quad (a,b,c)=\{1,2,3\}$ |
| 31 | 360 | $(\pi_a,\pi_b,\pi_c,\pi_3,\pi_4)=(\sigma_{31},\sigma_{31},\sigma_{49},\sigma_{105},\sigma_0)\quad (a,b,c)=\{0,1,2\}$ |
| 32 | 120 | $(\pi_0,\pi_a,\pi_b,\pi_c,\pi_d)=(\sigma_0,\sigma_{30},\sigma_{30},\sigma_{81},\sigma_{81})\quad (a,b,c,d)=(1,2,3,4),(3,4,1,2,)$ |
| 33 | 720 | $(\pi_a,\pi_b,\pi_c,\pi_d,\pi_e)=(\sigma_0,\sigma_7,\sigma_{45},\sigma_{45},\sigma_{99})\quad (a,b,c,d,e)=(1,2,\{0,3,4\}),(4,3,\{0,1,2\})$ |
| 34-1 | 360 | $(\pi_0,\pi_a,\pi_b,\pi_c,\pi_4)=(\sigma_0,\sigma_8,\sigma_8,\sigma_{113},\sigma_{32})\quad (a,b,c)=\{1,2,3\}$ |
| 34-2 | 360 | $(\pi_0,\pi_a,\pi_b,\pi_c,\pi_4)=(\sigma_0,\sigma_{12},\sigma_{12},\sigma_{117},\sigma_{32})\quad (a,b,c)=\{1,2,3\}$ |
| 35 | 240 | $(\pi_0,\pi_a,\pi_b,\pi_c,\pi_d)=(\sigma_7,\sigma_7,\sigma_7,\sigma_{45},\sigma_0)\quad (a,b,c,d)=(1,2,3,4),(3,4,2,1)$ |
| 36-1 | 240 | $(\pi_0,\pi_a,\pi_b,\pi_c,\pi_d)=(\sigma_{25},\sigma_{25},\sigma_{25},\sigma_{111},\sigma_0)\quad (a,b,c,d)=(1,2,3,4),(3,4,2,1)$ |
| 36-2 | 360 | $(\pi_0,\pi_a,\pi_b,\pi_c,\pi_4)=(\sigma_0,\sigma_{26},\sigma_{86},\sigma_{86},\sigma_{111})\quad (a,b,c)=\{1,2,3\}$ |
| 36-3 | 360 | $(\pi_0,\pi_a,\pi_b,\pi_c,\pi_4)=(\sigma_0,\sigma_{26},\sigma_{26},\sigma_{86},\sigma_{119})\quad (a,b,c)=\{1,2,3\}$ |
| 37-1 | 360 | $(\pi_0,\pi_a,\pi_b,\pi_c,\pi_4)=(\sigma_0,\sigma_{26},\sigma_{32},\sigma_{32},\sigma_{119})\quad (a,b,c)=\{1,2,3\}$ |
| 37-2 | 360 | $(\pi_0,\pi_a,\pi_b,\pi_c,\pi_4)=(\sigma_0,\sigma_{32},\sigma_{32},\sigma_{86},\sigma_{119})\quad (a,b,c)=\{1,2,3\}$ |
| 37-3 | 720 | $(\pi_0,\pi_a,\pi_b,\pi_c,\pi_4)=(\sigma_0,\sigma_{26},\sigma_{32},\sigma_{72},\sigma_{119})\quad (a,b,c)=\{1,2,3\}$ |
| 37-4 | 720 | $(\pi_0,\pi_a,\pi_b,\pi_c,\pi_4)=(\sigma_0,\sigma_{32},\sigma_{60},\sigma_{72},\sigma_{111})\quad (a,b,c)=\{1,2,3\}$ |

| Type | Number | Representative rules |
|---|---|---|
| 38-1 | 120 | $(\pi_0,\pi_1,\pi_2,\pi_3,\pi_4)=(\sigma_0,\sigma_{113},\sigma_{113},\sigma_{113},\sigma_{32})$ |
| 38-2 | 120 | $(\pi_0,\pi_1,\pi_2,\pi_3,\pi_4)=(\sigma_0,\sigma_{117},\sigma_{117},\sigma_{117},\sigma_{32})$ |
| 39 | 240 | $(\pi_0,\pi_a,\pi_b,\pi_3,\pi_4)=(\sigma_6,\sigma_6,\sigma_{105},\sigma_{30},\sigma_1)\quad (a,b)=\{1,2\}$ |
| 40-1 | 120 | $(\pi_0,\pi_1,\pi_2,\pi_3,\pi_4)=(\sigma_1,\sigma_1,\sigma_1,\sigma_{111},\sigma_{24})$ |
| 40-2 | 720 | $(\pi_a,\pi_b,\pi_c,\pi_d,\pi_4)=(\sigma_1,\sigma_{111},\sigma_{24},\sigma_{24},\sigma_{24})$ |
| | | $(a,b,c,d)=(\{0,3\},1,2),(\{1,3\},0,2),(\{2,3\},0,1)$ |
| 40-3 | 480 | $(\pi_0,\pi_a,\pi_b,\pi_c,\pi_d)=(\sigma_6,\sigma_1,\sigma_{24},\sigma_6,\sigma_{111})$ |
| | | $(a,b,c,d)=(1,2,\{3,4\}),(4,3,\{1,2\})$ |
| 40-4 | 720 | $(\pi_0,\pi_a,\pi_b,\pi_c,\pi_4)=(\sigma_2,\sigma_2,\sigma_{24},\sigma_{86},\sigma_{105})\quad (a,b,c)=\{1,2,3\}$ |
| 40-5 | 720 | $(\pi_0,\pi_a,\pi_b,\pi_c,\pi_4)=(\sigma_2,\sigma_{24},\sigma_{24},\sigma_{86},\sigma_{105})\quad (a,b,c)=\{1,2,3\}$ |
| 41 | 480 | $(\pi_0,\pi_a,\pi_b,\pi_c,\pi_d)=(\sigma_6,\sigma_{30},\sigma_{87},\sigma_6,\sigma_6)$ |
| | | $(a,b,c,d)=(1,3,2,4),(1,4,2,3),(2,3,1,4),(2,4,1,3)$ |
| 42-1 | 480 | $(\pi_0,\pi_a,\pi_b,\pi_c,\pi_d)=(\sigma_2,\sigma_{32},\sigma_{107},\sigma_2,\sigma_2)$ |
| | | $(a,b,c,d)=(2,1,3,4),(3,1,2,4),(4,2,1,3),(4,3,1,2)$ |
| 42-2 | 480 | $(\pi_0,\pi_a,\pi_b,\pi_c,\pi_d)=(\sigma_6,\sigma_{32},\sigma_{111},\sigma_6,\sigma_6)$ |
| | | $(a,b,c,d)=(1,3,2,4),(2,3,1,4),(4,1,2,3),(4,2,1,3)$ |
| 42-3 | 960 | $(\pi_0,\pi_a,\pi_b,\pi_c,\pi_d)=(\sigma_{14},\sigma_{32},\sigma_{119},\sigma_{14},\sigma_{14})$ |
| | | $(a,b,c,d)=(1,2,3,4),(3,2,1,4),(\{1,4\},2,3),(\{2,4\},1,3),(\{3,4\},1,2)$ |
| 42-4 | 360 | $(\pi_0,\pi_a,\pi_b,\pi_c,\pi_4)=(\sigma_{14},\sigma_{105},\sigma_{105},\sigma_{119},\sigma_{32})\quad (a,b,c)=\{1,2,3\}$ |
| 42-5 | 720 | $(\pi_0,\pi_a,\pi_b,\pi_c,\pi_4)=(\sigma_{14},\sigma_{14},\sigma_{32},\sigma_{54},\sigma_{119})\quad (a,b,c)=\{1,2,3\}$ |
| 42-6 | 360 | $(\pi_0,\pi_a,\pi_b,\pi_c,\pi_4)=(\sigma_{14},\sigma_{32},\sigma_{54},\sigma_{54},\sigma_{119})\quad (a,b,c)=\{1,2,3\}$ |
| 42-7 | 720 | $(\pi_0,\pi_a,\pi_b,\pi_c,\pi_4)=(\sigma_{14},\sigma_{14},\sigma_{32},\sigma_{60},\sigma_{105})\quad (a,b,c)=\{1,2,3\}$ |
| 42-8 | 720 | $(\pi_0,\pi_a,\pi_b,\pi_c,\pi_4)=(\sigma_{14},\sigma_{32},\sigma_{54},\sigma_{60},\sigma_{105})\quad (a,b,c)=\{1,2,3\}$ |
| 43-1 | 120 | $(\pi_0,\pi_1,\pi_2,\pi_3,\pi_4)=(\sigma_2,\sigma_{31},\sigma_2,\sigma_2,\sigma_{107})$ |
| 43-2 | 480 | $(\pi_0,\pi_a,\pi_b,\pi_c,\pi_d)=(\sigma_6,\sigma_{31},\sigma_{111},\sigma_6,\sigma_6)$ |
| | | $(a,b,c,d)=(1,3,2,4),(1,4,2,3),(2,3,1,4),(2,4,1,3)$ |
| 43-3 | 240 | $(\pi_0,\pi_a,\pi_b,\pi_3,\pi_4)=(\sigma_1,\sigma_{31},\sigma_1,\sigma_{111},\sigma_6),\quad (a,b)=\{1,2\}$ |
| 43-4 | 120 | $(\pi_0,\pi_1,\pi_2,\pi_3,\pi_4)=(\sigma_{105},\sigma_{31},\sigma_{105},\sigma_{107},\sigma_{105})$ |
| 44-1 | 480 | $(\pi_0,\pi_a,\pi_b,\pi_c,\pi_d)=(\sigma_{14},\sigma_{12},\sigma_{38},\sigma_{14},\sigma_{119})\quad (a,b,c,d)=(\{1,3\},\{2,4\})$ |
| 44-2 | 240 | $(\pi_0,\pi_a,\pi_2,\pi_b,\pi_4)=(\sigma_{14},\sigma_8,\sigma_{119},\sigma_{38},\sigma_{14})\quad (a,b)=\{1,3\}$ |
| 45-1 | 960 | $(\pi_a,\pi_b,\pi_c,\pi_d,\pi_e)=(\sigma_7,\sigma_{30},\sigma_{111},\sigma_6,\sigma_6)$ |
| | | $(a,b,c,d,e)=(\{1,2\},4,0,3),(\{1,2\},3,0,4),(0,\{1,2\},3,4),(0,1,3,2,4),(0,2,3,1,4)$ |
| 45-2 | 720 | $(\pi_0,\pi_a,\pi_b,\pi_c,\pi_d)=(\sigma_6,\sigma_{30},\sigma_{86},\sigma_{111},\sigma_6)$ |
| | | $(a,b,c,d)=(\{1,2\},3,4),(\{2,3\},4,1),(\{1,3\},4,2),(1,4,\{2,3\}),(2,4,\{1,3\}),(4,3,\{1,2\})$ |
| 46 | 240 | $(\pi_0,\pi_a,\pi_b,\pi_3,\pi_4)=(\sigma_6,\sigma_6,\sigma_{30},\sigma_{87},\sigma_{87})\quad (a,b)=\{1,2\}$ |
| 47-1 | 120 | $(\pi_0,\pi_1,\pi_2,\pi_3,\pi_4)=(\sigma_{111},\sigma_{111},\sigma_{111},\sigma_1,\sigma_{24})$ |
| 47-2 | 360 | $(\pi_a,\pi_b,\pi_c,\pi_3,\pi_4)=(\sigma_1,\sigma_{25},\sigma_{25},\sigma_{111},\sigma_{24}),\quad (a,b,c)=\{0,1,2\}$ |
| 47-3 | 120 | $(\pi_0,\pi_a,\pi_b,\pi_c,\pi_d)=(\sigma_7,\sigma_1,\sigma_6,\sigma_{25},\sigma_{111})$ |
| | | $(a,b,c,d)=(1,4,2,3),(4,1,3,2)$ |
| 47-4 | 720 | $(\pi_0,\pi_a,\pi_b,\pi_c,\pi_4)=(\sigma_6,\sigma_6,\sigma_{26},\sigma_{86},\sigma_{111}),\quad (a,b,c)=\{1,2,3\}$ |
| 47-5 | 360 | $(\pi_0,\pi_a,\pi_b,\pi_c,\pi_4)=(\sigma_6,\sigma_{26},\sigma_{86},\sigma_{86},\sigma_{105}),\quad (a,b,c)=\{1,2,3\}$ |
| 47-6 | 720 | $(\pi_0,\pi_a,\pi_b,\pi_c,\pi_4)=(\sigma_6,\sigma_{26},\sigma_{80},\sigma_{86},\sigma_{111}),\quad (a,b,c)=\{1,2,3\}$ |
| 47-7 | 360 | $(\pi_0,\pi_a,\pi_b,\pi_c,\pi_4)=(\sigma_{14},\sigma_{26},\sigma_{26},\sigma_{86},\sigma_{105}),\quad (a,b,c)=\{1,2,3\}$ |
| 48-1 | 240 | $(\pi_0,\pi_1,\pi_2,\pi_a,\pi_b)=(\sigma_7,\sigma_1,\sigma_1,\sigma_7,\sigma_{33}),\quad (a,b)=\{3,4\}$ |
| 48-2 | 240 | $(\pi_0,\pi_a,\pi_b,\pi_3,\pi_4)=(\sigma_7,\sigma_7,\sigma_{33},\sigma_6,\sigma_6),\quad (a,b)=\{1,2\}$ |
| 48-3 | 240 | $(\pi_0,\pi_1,\pi_a,\pi_b,\pi_4)=(\sigma_{23},\sigma_2,\sigma_{23},\sigma_{33},\sigma_2),\quad (a,b)=\{2,3\}$ |
| 48-4 | 240 | $(\pi_0,\pi_a,\pi_2,\pi_3,\pi_b)=(\sigma_{23},\sigma_{23},\sigma_{21},\sigma_{21},\sigma_{33}),\quad (a,b)=\{1,4\}$ |
| 49-1 | 120 | $(\pi_0,\pi_1,\pi_2,\pi_3,\pi_4)=(\sigma_2,\sigma_2,\sigma_{32},\sigma_{32},\sigma_{107})$ |
| 49-2 | 120 | $(\pi_0,\pi_1,\pi_2,\pi_3,\pi_4)=(\sigma_6,\sigma_{32},\sigma_{32},\sigma_6,\sigma_{111})$ |
| 49-3 | 360 | $(\pi_0,\pi_a,\pi_b,\pi_c,\pi_4)=(\sigma_{14},\sigma_{26},\sigma_{32},\sigma_{32},\sigma_{105}),\quad (a,b,c)=\{1,2,3\}$ |
| 49-4 | 360 | $(\pi_0,\pi_a,\pi_b,\pi_c,\pi_4)=(\sigma_{14},\sigma_{32},\sigma_{32},\sigma_{86},\sigma_{105}),\quad (a,b,c)=\{1,2,3\}$ |
| 49-5 | 720 | $(\pi_0,\pi_a,\pi_b,\pi_c,\pi_4)=(\sigma_{14},\sigma_{26},\sigma_{32},\sigma_{72},\sigma_{105}),\quad (a,b,c)=\{1,2,3\}$ |
| 50-1 | 720 | $(\pi_0,\pi_a,\pi_b,\pi_3,\pi_c)=(\sigma_6,\sigma_{32},\sigma_{86},\sigma_6,\sigma_{110}),\quad (a,b,c)=\{1,2,4\}$ |
| 50-2 | 720 | $(\pi_a,\pi_1,\pi_2,\pi_b,\pi_4)=(\sigma_{32},\sigma_{80},\sigma_{80},\sigma_{110},\sigma_{86}),\quad (a,b)=\{0,3\}$ |
| 50-3 | 720 | $(\pi_0,\pi_a,\pi_b,\pi_3,\pi_c)=(\sigma_6,\sigma_{32},\sigma_{86},\sigma_6,\sigma_{87}),\quad (a,b,c)=\{1,2,4\}$ |
| 50-4 | 720 | $(\pi_a,\pi_1,\pi_2,\pi_b,\pi_4)=(\sigma_{32},\sigma_{80},\sigma_{80},\sigma_{87},\sigma_{86}),\quad (a,b)=\{0,3\}$ |
| 50-5 | 240 | $(\pi_0,\pi_a,\pi_2,\pi_b,\pi_4)=(\sigma_{14},\sigma_{60},\sigma_{14},\sigma_{114},\sigma_{32}),\quad (a,b)=\{1,3\}$ |
| 51 | 720 | $(\pi_a,\pi_b,\pi_c,\pi_d,\pi_e)=(\sigma_7,\sigma_{31},\sigma_{87},\sigma_6,\sigma_6)$ |
| | | $(a,b,c,d,e)=(\{1,2\},\{3,4\},0),(0,\{1,2\},3,4)$ |
| 52-1 | 240 | $(\pi_0,\pi_1,\pi_2,\pi_a,\pi_b)=(\sigma_1,\sigma_{49},\sigma_1,\sigma_{33},\sigma_{49}),\quad (a,b)=\{3,4\}$ |
| 52-2 | 240 | $(\pi_0,\pi_1,\pi_2,\pi_a,\pi_b)=(\sigma_1,\sigma_{31},\sigma_1,\sigma_{33},\sigma_{49}),\quad (a,b)=\{3,4\}$ |
| 53-1 | 240 | $(\pi_0,\pi_a,\pi_b,\pi_3,\pi_4)=(\sigma_{86},\sigma_6,\sigma_{30},\sigma_{86},\sigma_{87}),\quad (a,b)=\{1,2\}$ |
| 53-2 | 240 | $(\pi_0,\pi_a,\pi_b,\pi_3,\pi_4)=(\sigma_{111},\sigma_6,\sigma_{30},\sigma_{87},\sigma_{111}),\quad (a,b)=\{1,2\}$ |

| Type | Number | Representative rules |
|---|---|---|
| 54-1 | 240 | $(\pi_0,\pi_1,\pi_a,\pi_b,\pi_4)=(\sigma_{107},\sigma_{107},\sigma_2,\sigma_{32},\sigma_{107}),\quad (a,b)=\{2,3\}$ |
| 54-2 | 240 | $(\pi_0,\pi_a,\pi_b,\pi_3,\pi_4)=(\sigma_{111},\sigma_6,\sigma_{32},\sigma_{111},\sigma_{111}),\quad (a,b)=\{1,2\}$ |
| 54-3 | 480 | $(\pi_a,\pi_1,\pi_b,\pi_c,\pi_d)=(\sigma_{26},\sigma_{26},\sigma_{32},\sigma_2,\sigma_{107})\quad (a,b,c,d)=(\{0,4\},\{2,3\})$ |
| 54-4 | 480 | $(\pi_a,\pi_b,\pi_c,\pi_3,\pi_d)=(\sigma_{32},\sigma_6,\sigma_{111},\sigma_{86},\sigma_{86})\quad (a,d,b,c)=(\{0,4\},\{1,2\})$ |
| 54-5 | 360 | $(\pi_a,\pi_b,\pi_2,\pi_c,\pi_d)=(\sigma_{14},\sigma_{32},\sigma_{119},\sigma_{119},\sigma_{119})$<br>$(a,b,c,d)=(0,4,1,2),(\{1,3\},0,4)$ |
| 54-6 | 360 | $(\pi_0,\pi_a,\pi_b,\pi_c,\pi_4)=(\sigma_{14},\sigma_{26},\sigma_{26},\sigma_{32},\sigma_{119}),\quad (a,b,c)=\{1,2,3\}$ |
| 54-7 | 360 | $(\pi_0,\pi_a,\pi_b,\pi_c,\pi_4)=(\sigma_{14},\sigma_{32},\sigma_{86},\sigma_{86},\sigma_{119}),\quad (a,b,c)=\{1,2,3\}$ |
| 54-8 | 360 | $(\pi_0,\pi_a,\pi_b,\pi_c,\pi_4)=(\sigma_{14},\sigma_{32},\sigma_{60},\sigma_{60},\sigma_{119}),\quad (a,b,c)=\{1,2,3\}$ |
| 54-9 | 720 | $(\pi_0,\pi_a,\pi_b,\pi_c,\pi_4)=(\sigma_{14},\sigma_{26},\sigma_{32},\sigma_{86},\sigma_{119}),\quad (a,b,c)=\{1,2,3\}$ |
| 55-1 | 240 | $(\pi_0,\pi_a,\pi_b,\pi_3,\pi_4)=(\sigma_{111},\sigma_6,\sigma_{31},\sigma_{111},\sigma_{111}),\quad (a,b)=\{1,2\}$ |
| 55-2 | 240 | $(\pi_0,\pi_a,\pi_b,\pi_3,\pi_4)=(\sigma_7,\sigma_7,\sigma_{31},\sigma_{111},\sigma_6),\quad (a,b)=\{1,2\}$ |
| 56-1 | 120 | $(\pi_0,\pi_1,\pi_2,\pi_3,\pi_4)=(\sigma_{14},\sigma_{32},\sigma_{32},\sigma_{32},\sigma_{119})$ |
| 56-2 | 360 | $(\pi_0,\pi_a,\pi_b,\pi_c,\pi_4)=(\sigma_{14},\sigma_{32},\sigma_{32},\sigma_{72},\sigma_{119}),\quad (a,b,c)=\{1,2,3\}$ |
| 57 | 360 | $(\pi_0,\pi_a,\pi_b,\pi_c,\pi_4)=(\sigma_8,\sigma_{32},\sigma_8,\sigma_8,\sigma_{113}),\quad (a,b,c)=\{1,2,3\}$ |
| 58-1 | 240 | $(\pi_0,\pi_a,\pi_b,\pi_c,\pi_d)=(\sigma_7,\sigma_7,\sigma_7,\sigma_{45},\sigma_{45})$<br>$(a,b,c,d)=(1,2,3,4),(3,4,1,2)$ |
| 58-2 | 240 | $(\pi_0,\pi_a,\pi_2,\pi_b,\pi_4)=(\sigma_{119},\sigma_8,\sigma_{119},\sigma_{38},\sigma_{119}),\quad (a,b)=\{1,3\}$ |
| 58-3 | 240 | $(\pi_0,\pi_a,\pi_2,\pi_b,\pi_4)=(\sigma_{119},\sigma_{12},\sigma_{119},\sigma_{38},\sigma_{119}),\quad (a,b)=\{1,3\}$ |
| 59 | 240 | $(\pi_0,\pi_a,\pi_b,\pi_c,\pi_d)=(\sigma_{33},\sigma_{45},\sigma_{45},\sigma_{47},\sigma_{47})$<br>$(a,b,c,d)=(1,3,2,4),(2,4,1,3)$ |
| 60-1 | 240 | $(\pi_0,\pi_a,\pi_b,\pi_3,\pi_4)=(\sigma_7,\sigma_{30},\sigma_{111},\sigma_7,\sigma_7),\quad (a,b)=\{1,2\}$ |
| 60-2 | 240 | $(\pi_0,\pi_a,\pi_b,\pi_3,\pi_4)=(\sigma_{111},\sigma_7,\sigma_{30},\sigma_{111},\sigma_{111}),\quad (a,b)=\{1,2\}$ |
| 60-3 | 240 | $(\pi_0,\pi_a,\pi_b,\pi_3,\pi_4)=(\sigma_{86},\sigma_{30},\sigma_{111},\sigma_{86},\sigma_{86}),\quad (a,b)=\{1,2\}$ |
| 61 | 120 | $(\pi_0,\pi_1,\pi_2,\pi_3,\pi_4)=(\sigma_{113},\sigma_{113},\sigma_{32},\sigma_{113},\sigma_8)$ |
| 62 | 240 | $(\pi_0,\pi_a,\pi_b,\pi_3,\pi_4)=(\sigma_{111},\sigma_{26},\sigma_{86},\sigma_{111},\sigma_{111}),\quad (a,b)=\{1,2\}$ |
| 63-1 | 480 | $\pi_a=\sigma_{33}$, others are $\sigma_7,\quad a=1,2,3,4$ |
| 63-2 | 480 | $(\pi_0,\pi_a)=(\sigma_7,\sigma_{33})$, others are $\sigma_{23},\quad a=1,2,3,4$ |
| 63-3 | 480 | $(\pi_0,\pi_a)=(\sigma_{23},\sigma_{33})$, others are $\sigma_7,\quad a=1,2,3,4$ |
| 63-4 | 480 | $\pi_a=\sigma_{33}$, others are $\sigma_{23},\quad a=1,2,3,4$ |
| 64-1 | 120 | $(\pi_0,\pi_1,\pi_2,\pi_3,\pi_4)=(\sigma_{111},\sigma_{32},\sigma_{32},\sigma_{111},\sigma_{111})$ |
| 64-2 | 240 | $(\pi_0,\pi_a,\pi_b,\pi_3,\pi_4)=(\sigma_{111},\sigma_{32},\sigma_{62},\sigma_{111},\sigma_{111}),\quad (a,b)=\{1,2\}$ |
| 64-3 | 120 | $(\pi_0,\pi_1,\pi_2,\pi_3,\pi_4)=(\sigma_{107},\sigma_{107},\sigma_{32},\sigma_{32},\sigma_{107})$ |
| 65-1 | 1440 | $(\pi_a,\pi_b)=(\sigma_{32},\sigma_{110})$, others are $\sigma_{86}$<br>$(a,b)=\{0,1\},\{0,3\},\{1,2\},\{1,4\},\{2,3\},\{2,4\}$ |
| 65-2 | 960 | $(\pi_a,\pi_b,\pi_c,\pi_d,\pi_4)=(\sigma_{32},\sigma_{86},\sigma_{86},\sigma_{110},\sigma_{26})$<br>$(a,d,b,c)=(\{0,1\},2,3),(\{0,3\},1,2),(\{1,2\},0,3),(\{2,3\},0,1)$ |
| 65-3 | 480 | $(\pi_a,\pi_b)=(\sigma_{32},\sigma_{110})$, others are $\sigma_{26}$<br>$(a,b)=\{0,2\},\{1,3\}$ |
| 65-4 | 480 | $(\pi_a,\pi_b,\pi_c,\pi_d,\pi_4)=(\sigma_{26},\sigma_{26},\sigma_{32},\sigma_{110},\sigma_{86})$<br>$(a,b,c,d)=(0,2,\{1,3\}),(1,3,\{0,2\})$ |
| 65-5 | 1440 | $(\pi_a,\pi_b)=(\sigma_{32},\sigma_{87})$, others are $\sigma_{86}$<br>$(a,b)=\{0,2\},\{0,3\},\{1,2\},\{1,3\},\{1,4\},\{2,4\}$ |
| 65-6 | 480 | $(\pi_a,\pi_b,\pi_c,\pi_d,\pi_4)=(\sigma_{32},\sigma_{86},\sigma_{86},\sigma_{87},\sigma_{26})$<br>$(a,d,b,c)=(\{0,2\},1,3),(\{1,3\},0,2)$ |
| 65-7 | 960 | $(\pi_a,\pi_b,\pi_c,\pi_d,\pi_4)=(\sigma_{26},\sigma_{26},\sigma_{32},\sigma_{87},\sigma_{26})$<br>$(a,b,c,d)=(0,1,\{2,3\}),(0,3,\{1,2\}),(1,2,\{0,3\}),(2,3,\{0,1\})$ |
| 65-8 | 960 | $(\pi_a,\pi_b,\pi_c,\pi_d,\pi_4)=(\sigma_{26},\sigma_{26},\sigma_{32},\sigma_{87},\sigma_{86})$<br>$(a,b,c,d)=(0,1,\{2,3\}),(0,3,\{1,2\}),(1,2,\{0,3\}),(2,3,\{0,1\})$ |
| 66-1 | 720 | $(\pi_0,\pi_a,\pi_b,\pi_c,\pi_d)=(\sigma_7,\sigma_{31},\sigma_{87},\sigma_7,\sigma_7)$<br>$(a,b,c,d)=\{1,2,3,4\}$ |
| 66-2 | 240 | $(\pi_0,\pi_a,\pi_b,\pi_c,\pi_d)=(\sigma_7,\sigma_{47},\sigma_{47},\sigma_7,\sigma_7)$<br>$(a,b,c,d)=(1,3,2,4),(2,4,1,3)$ |
| 66-3 | 480 | $(\pi_a,\pi_b)=(\sigma_{47},\sigma_{70})$, others are $\sigma_7$<br>$(a,b)=\{1,2\},\{1,4\},\{2,3\},\{3,4\}$ |
| 66-4 | 240 | $(\pi_a,\pi_b)=(\sigma_{31},\sigma_{117})$, others are $\sigma_7,\quad (a,b)=\{3,4\}$ |
| 67-1 | 240 | $(\pi_a,\pi_b)=(\sigma_{33},\sigma_{33})$, others are $\sigma_7,\quad (a,b)=(1,3),(2,4)$ |
| 67-2 | 240 | $(\pi_a,\pi_b)=(\sigma_{33},\sigma_{63})$, others are $\sigma_7,\quad (a,b)=\{3,4\}$ |
| 67-3 | 480 | $(\pi_a,\pi_b)=(\sigma_{33},\sigma_{68})$, others are $\sigma_7$<br>$(a,b)=(1,4),(2,1),(2,3),(3,4)$ |
| 68-1 | 240 | $(\pi_0,\pi_a,\pi_b,\pi_c,\pi_d)=(\sigma_7,\sigma_7,\sigma_9,\sigma_9,\sigma_{33})$<br>$(a,b,c,d)=(\{1,2\},\{3,4\})$ |
| 68-2 | 720 | $(\pi_a,\pi_b,\pi_c,\pi_d,\pi_e)=(\sigma_9,\sigma_9,\sigma_{23},\sigma_{23},\sigma_{33})$<br>$(a,b,c,d,e)=(0,1,2,4,3),(0,4,1,3,2),(1,2,0,\{3,4\}),(3,4,0,\{1,2\})$ |
| 68-3 | 960 | $(\pi_0,\pi_a,\pi_b,\pi_c,\pi_d)=(\sigma_7,\sigma_7,\sigma_9,\sigma_{18},\sigma_{33})$<br>$(a,b,c,d)=(\{1,3\},\{2,4\}),(\{2,4\},\{1,3\})$ |
| 68-4 | 480 | $(\pi_0,\pi_a,\pi_b,\pi_c,\pi_d)=(\sigma_7,\sigma_{18},\sigma_{18},\sigma_{23},\sigma_{33})$<br>$(a,b,c,d)=(1,3,\{2,4\}),(2,4,\{1,3\})$ |
| 68-5 | 960 | $(\pi_a,\pi_b,\pi_c,\pi_d,\pi_e)=(\sigma_{18},\sigma_{18},\sigma_{23},\sigma_{23},\sigma_{33})$<br>$a=0$ and "$e=2$ and $4\in\{c,d\}$" or "$e=3$ and $1\in\{c,d\}$";<br>$(a,b,c,d,e)=(1,3,0,\{2,4\}),(2,4,0,\{1,3\})$ |
| 68-6 | 480 | $(\pi_0,\pi_a,\pi_b,\pi_c,\pi_d)=(\sigma_{23},\sigma_7,\sigma_9,\sigma_9,\sigma_{33})$<br>$(a,d,b,c)=(\{1,3\},2,4),(\{2,4\},1,3)$ |
| 68-7 | 1680 | $(\pi_a,\pi_b,\pi_c,\pi_d,\pi_e)=(\sigma_9,\sigma_{18},\sigma_{23},\sigma_{23},\sigma_{33})$<br>$(a,b,e,c,d)=(0,1,2,3,4),(0,\{2,3\},1,4),(0,4,3,1,2)$;<br>$(c,a,e,b,d)=(0,\{1,3\},\{2,4\}),(0,\{2,4\},\{1,3\})$ |
| 68-8 | 1920 | $(\pi_a,\pi_b,\pi_c,\pi_d,\pi_e)=(\sigma_7,\sigma_9,\sigma_{18},\sigma_{23},\sigma_{33})$<br>$(a,b,e,c,d)=(0,\{1,3\},\{2,4\}),(0,\{2,4\},\{1,3\})$;<br>$(d,a,b,c,e)=(0,\{1,3\},\{2,4\}),(0,\{2,4\},\{1,3\})$ |
| 69 | 480 | $(\pi_0,\pi_a,\pi_b,\pi_c,\pi_d)=(\sigma_{23},\sigma_{23},\sigma_{33},\sigma_{33},\sigma_{96})$<br>$\{a,c\}=\{1,3\}$ or $\{a,c\}=\{2,4\}$ |
| 70-1 | 240 | $(\pi_a,\pi_b,\pi_c,\pi_d,\pi_4)=(\sigma_{32},\sigma_{32},\sigma_{86},\sigma_{110},\sigma_{32})$<br>$(a,b,c,d)=(0,2,3,1),(1,3,0,2)$ |
| 70-2 | 720 | $(\pi_a,\pi_b,\pi_c,\pi_d,\pi_e)=(\sigma_{26},\sigma_{32},\sigma_{32},\sigma_{72},\sigma_{110})$<br>$(b,c)=(0,2)$ and $(a,d,e)=(1,3,4),(4,1,3),(4,3,1)$<br>$(b,c)=(1,3)$ and $(a,d,e)=(2,0,4),(4,0,2),(4,2,0)$ |
| 70-3 | 480 | $(\pi_a,\pi_b,\pi_c,\pi_d,\pi_4)=(\sigma_{32},\sigma_{72},\sigma_{72},\sigma_{110},\sigma_{26})$<br>$(a,d,b,c)=(\{1,3\},0,2),(\{0,2\},1,3)$ |
| 70-4 | 960 | $(\pi_a,\pi_b,\pi_c,\pi_d,\pi_e)=(\sigma_{32},\sigma_{72},\sigma_{72},\sigma_{86},\sigma_{110})$<br>$(c,a,d,b,e)=(4,\{0,2\},3,1),(4,\{1,3\},0,2)$;<br>$(d,a,e,b,c)=(4,\{0,2\},3,1),(4,\{1,3\},0,2)$; |
| 70-5 | 1440 | $(\pi_a,\pi_b,\pi_c,\pi_d,\pi_e)=(\sigma_{32},\sigma_{32},\sigma_{72},\sigma_{86},\sigma_{110})$<br>$(d,a,b,c,e)=(4,0,2,\{1,3\}),(4,1,3,\{0,2\})$;<br>$(a,b,d,c,e)=(4,\{0,2\},3,1),(4,\{1,3\},0,2)$;<br>$(c,a,b,d,e)=(4,0,2,3,1),(4,1,3,0,2)$ ;<br>$(e,a,b,c,d)=(4,0,2,3,1),(4,1,3,0,2)$. |
| 71-1 | 480 | $\pi_a=\sigma_{33}$ and others are $\sigma_9$, $a=1,2,3,4$ |
| 71-2 | 240 | $(\pi_a,\pi_b)=(\sigma_{33},\sigma_9)$ and others are $\sigma_{18}$, $(a,b)=(2,4),(3,1)$ |
| 71-3 | 240 | $(\pi_a,\pi_b,\pi_c,\pi_d,\pi_e)=(\sigma_9,\sigma_9,\sigma_{18},\sigma_{18},\sigma_{33})$<br>$(a,b,c,d,e)=(0,1,2,4,3),(0,4,1,3,2)$ |
| 72-1 | 240 | $(\pi_0,\pi_a,\pi_b,\pi_c,\pi_d)=(\sigma_{47},\sigma_{33},\sigma_{47},\sigma_{101},\sigma_{101})$<br>$(a,b,c,d)=(2,4,1,3),(3,1,2,4)$ |
| 72-2 | 480 | $(\pi_0,\pi_a,\pi_b,\pi_c,\pi_d)=(\sigma_{101},\sigma_{33},\sigma_{47},\sigma_{101},\sigma_{101})$<br>$(a,b,c,d)=(2,1,3,4),(3,1,2,4),(2,4,1,3),(3,4,1,2)$ |

## 5.2 Proof strategy

Our proof of ergodicity takes the form of mathematical induction. Given the sequence in site $n$ with period $5^n$. Let $\Sigma = T_{n+1}^{0\to 5^n}$ be the permutation in site $n+1$ through $5^n$ steps (one period in site $n$), which is induced by the sequence in site $n$. In our mathematical induction, we show that

(a) Site $n$ has period $5^n$ (i.e., $\Sigma$ is a 5-cycle).
(b) The sequence of site $n$ satisfies some *structure*.

The former fact implies that site $n$ is ergodic. The form of the structure is specified on each rule of the CA, which will be explained later.

The above two properties are shown to be satisfied in site $n = 2$ by a direct computation, and therefore the remainder of this paper is devoted to proving the aforementioned two properties in site $n+1$ under the assumption that they are satisfied in site $n$. Since it requires extremely many pages to fill the proofs for all possible rules, and most of the proofs employ similar techniques, we classify the structures of all rules into several *patterns* and provide proofs only for some representative rules. We list structures of all rules explicitly in Appendix. A , with which a reader can prove the ergodicity of each rule without much effort if one needs it.

## 5.3 Symbols

To present the structures of CA, We use the following terminology:

- period sequence: a sequence of $5^n$ symbols (i.e., one period) in site $n$.
- island: We decompose a period sequence into several large regions. A single long region is called *island*. See the explanation on Pattern A (Sec. 6.2) for details and examples.
- unit : A sequence of a few symbols (usually two or three symbols) is called *unit*. See the explanation on Pattern B (Sec. 6.3) for details and examples.
- block unit: A finite length of symbol sequence is called *block unit*, if what symbols or units can appear in this sequence is determined while its order is not determined (see the explanation on $\lfloor\cdot\rfloor$ below). In several cases, we need not to distinguish units and block units. In these cases, we call block units also as simply *unit*.
- queue: A finite length of symbol sequence is called *queue*, if the order of symbols or units in this sequence is determined (see the explanation on $\langle\cdot\rangle$ below).

In addition, we employ the following symbols to describe structures in shorthand:

- $[\cdot]$ : representing a unit. The symbols in the bracket appear in its order.
- $(\cdot)$ : representing an island. The symbols in the bracket can appear in the island. For example, $(0,1,2)$ means that only 0, 1, and 2 may appear in this island (i.e., 3 and 4 never appear). The island $(0,[12])$ can be 001200 and 1212120, but neither 001020 nor 0112200.
- $\{\cdot\}$: representing a symbol which is one of the symbols in the bracket. For example, $\{0,2,3\}$ is a single symbol of 0, 2, or 3. $\{0,[12]\}$ is one of symbol 0 or unit 12.
- $\lfloor\cdot\rfloor_{\text{even}}$ : representing a block with even length. For example, $\lfloor 0,3\rfloor_{\text{even}}$ might be 00, 0303, 333033.
- $\lfloor\cdot\rfloor_{\text{odd}}$ : representing a block with odd length. For example, $\lfloor 0,3\rfloor_{\text{odd}}$ might be 033, 0, 303, 00300.
- $\lfloor\cdot\rfloor^1_{\text{odd}}$ : meaning that this block appears only once in one period sequence with $5^n$ symbols. The superscript 1 in $\lfloor\cdot\rfloor^1_{\text{even}}$, $\langle\cdot\rangle^1_{\text{even}}$, and $\langle\cdot\rangle^1_{\text{odd}}$ represents the same property.
- $\langle\cdot\rangle_{\text{even}}$ : representing a queue with even number of symbols. The "+" sign concatenates symbols. For example, $\langle 2+\{3,4\}\rangle_{\text{even}}$ might be 2324242423, 4232323232, where 2 and "3 or 4" appears alternately. $\langle\cdot\rangle_{\text{odd}}$ represents a queue with an odd number of symbols.
- $*$: The superscript $*$ (e.g., $\lfloor 0,1\rfloor^*_{\text{even}}$), which is used only in Pattern D-2-2, means that the rule (even or odd) might be violated at the edge of islands. See the section for Pattern D-2-2 for details.

Furthermore, to explain the dynamics, we distinguish *transition* and *permutation* itself. Let $\pi_1 = (12)(34)$ and $\pi_4 = (23)$. Then, the permutation $\pi_{4141} = \pi_4\pi_1\pi_4\pi_1$ simply means $(14)(23)$. If the initial state is 1, then the permutation $\pi_{4141}$ conveys state 1 to state 4. On the other hand, if the initial state is 1, the transition induced by $\pi_{4141}$ is $1 \to 2 \to 3 \to 4$, where we take the path into account. Thus, the permutation $\pi_{4141}$ is inside $\{1,4\}$ and $\{2,3\}$, while the transition induced by $\pi_{4141}$ is not inside $\{1,4\}$ and $\{2,3\}$ (because a state inside $\{1,4\}$ goes out from this set in the middle of the transition).

## 6 CA with 5 states: Patterns and proofs

### 6.1 Classification

We here briefly summarize the classification of the rules of ergodic CA with 5 states. Details are described in the corresponding subsections.

We classify the rules of ergodic CA into five patterns from Pattern A to Pattern E. Pattern A employs several islands in its structure, which can be understood as a generalization of type-2 in 3-state CA. Pattern B employs several units in its structure, which can be understood as a generalization of type-1 in 3-state CA. Pattern C newly appears in 5-state CA, where states are decomposed into two sets, and states in these two sets appear alternatingly. Pattern D is a hybrid of Pattern A and Pattern B. Pattern E is highly exceptional, and few rules belong to it.

Patterns A, B, and D have several subclasses. Even in Patterns A and B, some subclasses have complicated structures.

### 6.2 Pattern A: divided by several islands

#### *6.2.1 A-1: The simplest case*

The structure in pattern A-1 consists of several islands $I_1, I_2, \dots$ on 5 states. The rule in pattern A-1 satisfies the following properties:

(a). States $a, b$ belonging to the same island (i.e., $a, b \in I_i$ for some $i$) provide commutative permutations; $\pi_a \pi_b = \pi_b \pi_a$.

(b). State $x$ can be mapped onto state $y$ by some permutation $\pi_a$ only if $x$ and $y$ belong to the same island.

Thanks to the commuting condition (a), the computation of the permutation with one period $\Sigma$ becomes simple.

We shall demonstrate how to prove ergodicity and the structure by taking several representatives. The structures of the remaining rules are presented in Appendix. A .

Type03; $(a, b) = (2, 4)$: ($\pi_0 = \pi_1 = \pi_3 = \mathrm{id}$, $\pi_2 = (012)$, $\pi_4 = (034)$)

The structure of this rule is $(0,1,2)(0,3,4)$, which consists of two islands, $I_1 = (0,1,2)$ and $I_2 = (0,3,4)$. We can confirm condition (b) by just drawing its transition map (see Figure. 7 in Sec. 5.1). Since $\pi_0, \pi_1, \pi_3$ are identity maps, we see that the commuting condition (a) is satisfied.

By the assumption from induction, the period sequence with length $5^n$ in site $n$ consists of $I_1^{[n]}$ and $I_2^{[n]}$, which are sequences of symbols only in $I_1$ and $I_2$, respectively. We denote by $\pi_{I_1^{[n]}}$ and $\pi_{I_2^{[n]}}$ the transitions on site $n+1$ induced by islands $I_1$ and $I_2$ in site $n$. The total transition in site $n+1$ in a single period $5^n$, $\Sigma = T_{n+1}^{0 \to 5^n}$, can be written as $(\pi_4)^{5^{n-1}} (\pi_2)^{5^{n-1}} = (034)^{5^{n-1}} (012)^{5^{n-1}}$, where $(034)$ and $(012)$ represent 3-cycles $\begin{pmatrix} 0 & 3 & 4 \\ 3 & 4 & 0 \end{pmatrix}$ and $\begin{pmatrix} 0 & 1 & 2 \\ 1 & 2 & 0 \end{pmatrix}$, respectively.

Since we have

$$\pi_{I_2^{[n]}} = (\pi_4)^{5^{n-1}} = (034)^{5^{n-1}} = \begin{cases} (043) & n \text{ is even,} \\ (034) & n \text{ is odd,} \end{cases} \tag{13}$$

$$\pi_{I_1^{[n]}} = (\pi_2)^{5^{n-1}} = (012)^{5^{n-1}} = \begin{cases} (021) & n \text{ is even,} \\ (012) & n \text{ is odd,} \end{cases} \tag{14}$$

the permutation on site $n+1$ for one period of site $n$ is a 5-cycle:

$$\Sigma = \pi_{I_2^{[n]}} \pi_{I_1^{[n]}} = \begin{cases} (043)(021) = (02143) & n \text{ is even,} \\ (034)(012) = (01234) & n \text{ is odd,} \end{cases} \tag{15}$$

which guarantees the ergodicity in site $n+1$. Furthermore, the structure $(0,1,2)(0,3,4)$ is inherited from site $n$ to site $n+1$, because the state in island $I_1$ is transferred to island $I_2$ only when the state after $\pi_{I_1^{[n]}}$ is 0 (see Fig. 8).

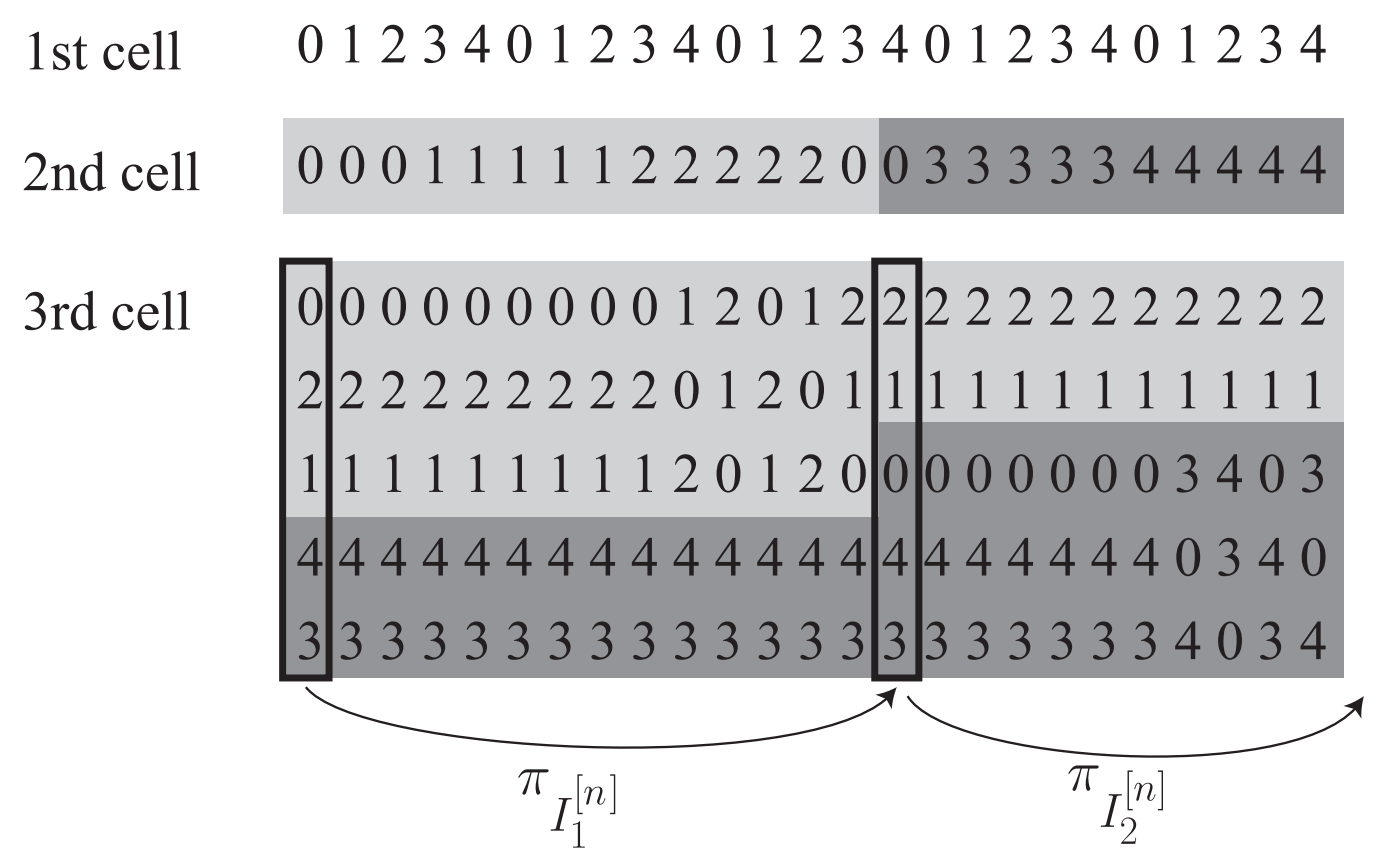

**Fig. 8** The sequence of first, second, and third cells in Type03 with $(a,b)=(2,4)$. The sequence in the third cell is divided into 5 lines that go from the top to the bottom. The light gray and dark gray regions represent islands $I_1$ and $I_2$, respectively. Two solid rectangles represent the initial (resp. final) state and the final (resp. initial) state of permutation $\pi_{I_1^{[n]}}$ (resp. $\pi_{I_2^{[n]}}$) for $n=2$.

Since this is our first example, we shall examine the last point (inheritance of structure) in detail. Owing to ergodicity, $\pi_{I_1^{[n]}}$ restricted to $I_1$ serves as a cyclic permutation in $I_1$, and $\pi_{I_2^{[n]}}$ restricted to $I_2$ serves as a cyclic permutation in $I_2$. In addition, $\pi_{I_1^{[n]}}$ acts trivially on states 3 and 4, and $\pi_{I_2^{[n]}}$ acts trivially on states 1 and 2. Now we can demonstrate the inheritance of structure as follows: Suppose that the initial state is 0. First, $\pi_{I_1^{[n]}}$ applies, which induces a transition inside island $I_1$, and the resulting state is not 0 (i.e., 1 or 2) due to ergodicity. Next, $\pi_{I_2^{[n]}}$ applies, while it keeps state 1 as it is, and thus the state remains inside island $I_1$. Then, $\pi_{I_1^{[n]}}$ applies, which induces a transition inside island $I_1$, and the resulting state is still 1 or 2 due to ergodicity. Further, $\pi_{I_2^{[n]}}$ applies again, while it keeps state 1 as it is, and thus the state remains inside island $I_1$. Finally, $\pi_{I_1^{[n]}}$ applies, which induces a transition inside island $I_1$, and the resulting state is now 0. Thus, the next application of $\pi_{I_2^{[n]}}$ induces a nontrivial transition inside $I_2$, and the resulting state is not 0 (i.e., 3 or 4) due to ergodicity. The remaining argument is similar to the above, with exchanging the role of $I_1$ and $I_2$. In summary, the state remains inside $I_1$ for two and a half periods, and remains inside $I_2$ for another two and half periods, which implies the inheritance of the structure.

In most of type-A rules, the inheritance of the structure can be shown in the aforementioned manner with the help of ergodicity. Therefore, in the following, we mainly investigate the ergodicity, and briefly comment on the inheritance of the structure.

Type09-2; $(a,b)=(2,3)$: ($\pi_0=\pi_1=\mathrm{id}$, $\pi_2=(12)$, $\pi_3=(04)(12)$, $\pi_4=(0123)$)

The structure of this rule is $(0,1,2,3)(0,4)$, which consists of two islands, $I_1=(0,1,2,3)$ and $I_2=(0,4)$. We can confirm condition (b) by just drawing its transition map (see Figure. 7 in Sec. 5.1). Observe that $\pi_0,\pi_1$ are identity maps and $\pi_2=\pi_3$ behave as (12) in $I_1$, and that $\pi_0,\pi_1,\pi_2,\pi_4$ are identity maps in $I_2$, which ensure the commuting condition (a).

The maps $\pi_{I_1^{[n]}}$ and $\pi_{I_2^{[n]}}$ are calculated as

$$\pi_{I_1^{[n]}}=(\pi_2)^{5^{n-1}}(\pi_3)^{5^{n-1}}=(12)^{2\cdot 5^{n-1}}(04)^{5^{n-1}}=(04), \tag{16}$$

$$\pi_{I_2^{[n]}}=(\pi_4)^{5^{n-1}}=(0123)^{5^{n-1}}=(0123). \tag{17}$$

Hence, the permutation with one period is a 5-cycle:

$$\pi_{I_2^{[n]}}\pi_{I_1^{[n]}} = (0123)(04) = (04123), \tag{18}$$

which guarantees the ergodicity in site $n+1$. Furthermore, the structure $(0,1,2,3)(0,4)$ is kept, because the state in island $I_1$ is transferred to island $I_2$ only when the state after $\pi_{I_1^{[n]}}$ is 0 (see Fig. 9).

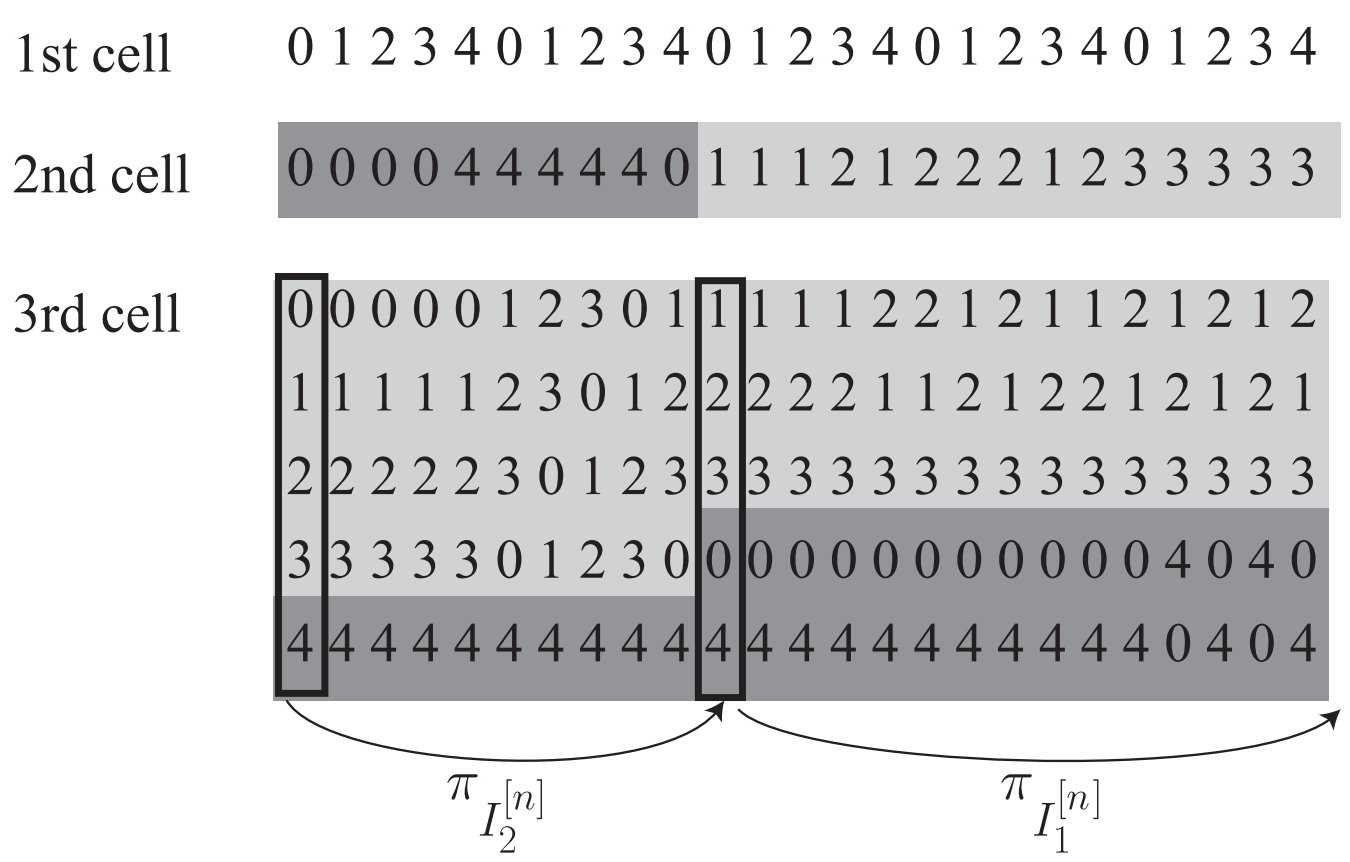

**Fig. 9** The sequence of first, second, and third cells in Type09-2 with $(a,b) = (2,3)$. The sequence in the third cell is divided into 5 lines that go from the top to the bottom. The light gray and dark gray regions represent islands $I_1$ and $I_2$, respectively. Two solid rectangles represent the initial (resp. final) state and the final (resp. initial) state of permutation $\pi_{I_2^{[n]}}$ (resp. $\pi_{I_1^{[n]}}$) for $n = 2$.

*6.2.2 A-1-2: The case where some islands appear twice*

The structure and the proof technique for pattern A-1-2 are very similar to those for pattern A-1. The only difference lies in the fact that the same island appears twice in the structure.

Let us take the case with three islands, $I_1$, $I_2$, and $I_3$, as an example. The case with four or more islands can be treated similarly. Suppose that these three islands sit in the order $I_1 - I_2 - I_3$ in the transition map (see Fig. **??**). Then, to cover all possible states, the state should pass $I_2$ at least twice, e.g., as $I_1 \to I_2 \to I_3 \to I_2 (\to I_1)$. In this case, the rule in pattern A-1-2 should satisfy the following property in addition to properties (a) and (b) for pattern A-1:

(c). For any state $a \in I_2$, $\pi_a$ commutes with all permutations induced by states in island $I_1$ or those induced by states in island $I_3$. Here, which island is employed depends on the state $a$.

Using this commutative condition, we can sort the permutation $\pi_0, \ldots, \pi_4$ in a tractable form.

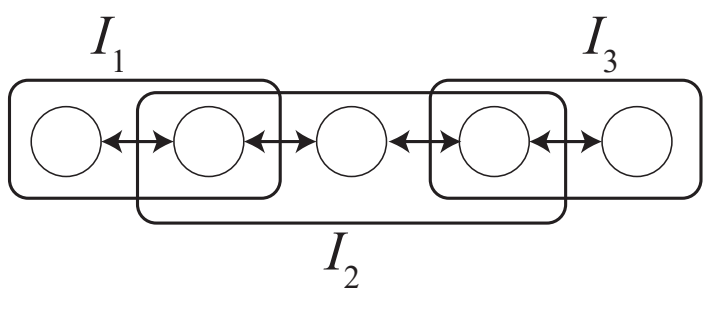

**Fig. 10** An example of the state space of pattern A-1-2.

<u>Type40-3; $(a,b,c,d) = (1,2,3,4)$: $(\pi_0 = \pi_3 = (12)$, $\pi_1 = (34)$, $\pi_2 = (01)$, $\pi_4 = (04)(12))$</u>

The structure of this rule is $(1,2)(0,1)(0,3,4)(0,1)$, which consists of three islands, $I_1 = (1,2)$, $I_2 = (0,1)$, and $I_3 = (0,3,4)$. We can confirm condition (b) by just drawing its transition map (see

Figure. 7 in Sec. 5.1). Five permutations are written as $\pi_0 = \pi_3 = (12)$, $\pi_1 = (34)$, $\pi_2 = (01)$, and $\pi_4 = (12)(04)$, which directly suggests condition (a). In addition, $\pi_1$ commutes with all permutations $\pi_a$ with $a \in I_1$, and $\pi_0$ commutes with all permutations $\pi_b$ with $b \in I_3$, which confirm condition (c).

Let us denote by $\pi_{I_{2,1}^{[n]}}$ and $\pi_{I_{2,2}^{[n]}}$ the permutations induced by the first $I_2$ and the second $I_2$ in site $n$. Then, we can see that the structure $(1,2)(0,1)(0,3,4)(0,1)$ is inherited from site $n$ to site $n+1$ as follows. First, $\pi_{I_{2,2}^{[n]}}\pi_{I_3^{[n]}}\pi_{I_{2,1}^{[n]}}$ induces transitions inside $(1,2)(= I_1)$ and $(0,3,4)(= I_3)$, and $\pi_{I_1^{[n]}}$ induces transitions inside $(0,1)(= I_2)$. Here, a transition is said to be inside an island when all the $\left|I_{2,1}^{[n]}\right| + \left|I_3^{[n]}\right| + \left|I_{2,2}^{[n]}\right|$ states in the middle of the path are kept in this island. Combining this observation with the ergodicity shown below, we conclude the inheritance of structure (A similar argument to Type03 in Pattern A-1 holds).

We remark that the time to start the structure varies depending on the sites. For example, as seen in Fig. 11, the structure in the second cell starts at 5 steps (the left end of white cells), while the structure in the third cell starts at 13 steps, not 5 steps.

Now we shall demonstrate the ergodicity of this CA. The total permutation in a single period is computed as

$$\pi_{I_{2,2}^{[n]}}\pi_{I_3^{[n]}}\pi_{I_{2,1}^{[n]}}\pi_{I_1^{[n]}} = (\pi_1)^p(\pi_0)^{5^{n-1}}(\pi_3)^{5^{n-1}}(\pi_4)^{5^{n-1}}(\pi_2)^{5^{n-1}}(\pi_1)^q = (\pi_1)^p\pi_4\pi_2(\pi_1)^q \tag{19}$$

with $p+q = 5^{n-1}$. Noticing that one of $p$ or $q$ is even and the other is odd, we compute this permutation explicitly. In the case with even $p$ and odd $q$, we have

$$(\pi_1)^p\pi_4\pi_2(\pi_1)^q = \pi_4\pi_2\pi_1 = (02143), \tag{20}$$

and in the case with odd $p$ and even $q$, we have

$$(\pi_1)^p\pi_4\pi_2(\pi_1)^q = \pi_1\pi_4\pi_2 = (02134), \tag{21}$$

both of which are 5-cycles.

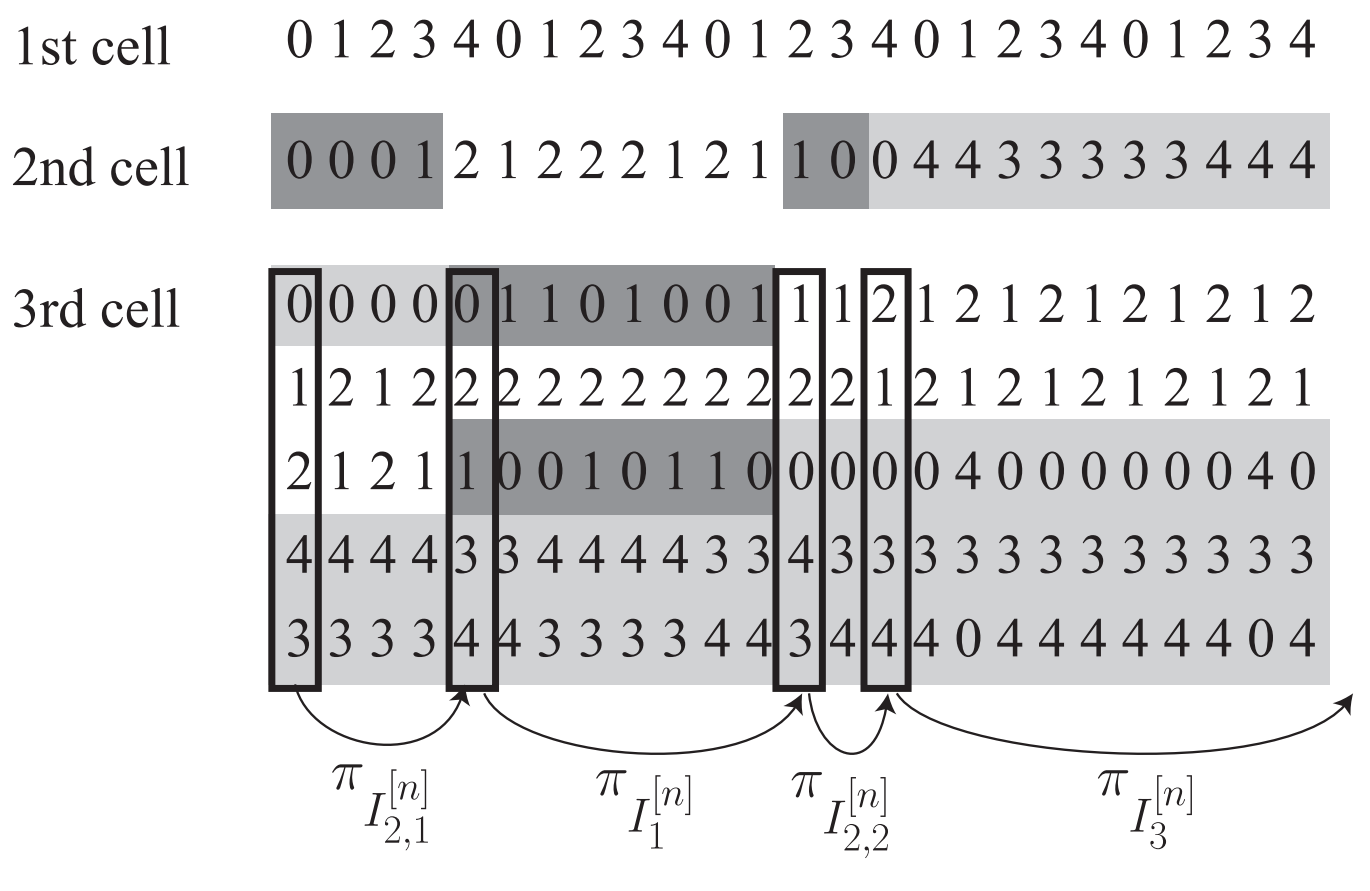

**Fig. 11** The sequence of first, second, and third cells in Type40-3 with $(a,b,c,d) = (1,2,3,4)$. The sequence in the third cell is divided into 5 lines that go from the top to the bottom. The white, dark gray, and light gray regions represent islands $I_1$, $I_2$, and $I_3$, respectively. We draw permutations $\pi_{I_{2,1}^{[n]}}$, $\pi_{I_1^{[n]}}$, $\pi_{I_{2,2}^{[n]}}$, and $\pi_{I_3^{[n]}}$ for $n = 2$.

### *6.2.3 A-2: non-commutative permutations exist in some island*

In pattern A-2, some permutations in one island are non-commutative. On the other hand, we have a coarse-grained description under which permutations behave as commutative, and this hidden commutative property confirms ergodicity.

The structure of pattern A-2 takes the form of (0,1,2,3)(0,4) (or its permutation). One of $\{0,1,2,3\}$ and $\{0,4\}$ accompanies commutative permutations, and the other (the noncommutative one) acts nontrivially on $\{0,1,2,3\}$ but trivially on $\{0,4\}$. For explanation, we take the case that $\{0,1,2,3\}$ accompanies noncommutative permutations (i.e., some of $\pi_0$, $\pi_1$, $\pi_2$, and $\pi_3$ are not commutative with each other), and only $\pi_4$ acts nontrivially on state 4. In spite of non-commutativity, we suppose that by decomposing the set (0,1,2,3) into two subsets $X$ and $Y$ each of which consists of two states, then with this coarse-grained view permutations $\pi_0$, $\pi_1$, $\pi_2$, $\pi_3$, (and $\pi_4$) are identity or transposition of $X$ and $Y$ at any microscopic states. Thanks to this coarse-grained description, we can prove the ergodicity.

The inheritance of structure from site $n$ to site $n+1$ is directly confirmed by the fact that only one of $\{0,1,2,3\}$ and $\{0,4\}$ accompanies permutations nontrivially acts on $\{0,4\}$, with combining the ergodicity. In fact, except for the hardness to show the ergodicity Pattern A-1 and Pattern A-2 have the same structure (decomposition into islands). Thus, the inheritance of the structure can be shown in a similar manner to Type03 in Pattern A-1, and therefore in the following we only discuss the ergodicity.

Type 42-5; $(a,b,c)=(1,2,3)$: $(\pi_0=\pi_1=(13)$, $\pi_2=(0123)$, $\pi_3=(02)$, $\pi_4=(04)(13))$

The structure is $(0,1,2,3)(0,4)$. We denote by $I_1=(0,1,2,3)$ and $I_2=(0,4)$. Noticing $\pi_{I_2^{[n]}}=(04)^{5^{n-1}}(13)^N$ with $N=\left|I_2^{[n]}\right|$ as the length of $I_2^{[n]}$, we find that the dynamics in island $(0,4)$ can be treated separately and thus it is convenient to treat only $(0,1,2,3)$. By restricting permutations to space (0,1,2,3), permutations act as $\pi_0=\pi_1=\pi_4=(13)$, $\pi_2=(0123)$, and $\pi_3=(02)$, and hence $\pi_2$ and others are not commutative. On the other hand, by setting $X=\{0,2\}$ and $Y=\{1,3\}$ and viewing these permutations in a coarse-grained state space $\{X,Y\}$, $\pi_0$, $\pi_1$, $\pi_3$, and $\pi_4$ are identity in this state space, and $\pi_2$ is a transposition $(XY)$. Hence, in the coarse-grained description, $\pi_{I_1^{[n]}}=(XY)^{5^n}=(XY)$ is obtained.

We introduce modified permutation $\tilde{\pi}_i$ with $\tilde{\pi}_j=\pi_j$ for $j=0,1,2,3$, and $\tilde{\pi}_4=(13)$, which is obtained from $\pi_i$ by removing all transitions between 0 and 4. $\tilde{\pi}_0$, $\tilde{\pi}_1$, and $\tilde{\pi}_4$ have nontrivial actions only when the present state is in $Y$, and $\tilde{\pi}_3$ has nontrivial action only when the present state is in $X$. Denoting $\tilde{\Sigma}:=\tilde{\pi}_{I_2^{[n]}}\tilde{\pi}_{I_1^{[n]}}$, and consider transitions in two periods $\tilde{\Sigma}^2=\tilde{\pi}_{I_2^{[n]}}\tilde{\pi}_{I_1^{[n]}}\tilde{\pi}_{I_2^{[n]}}\tilde{\pi}_{I_1^{[n]}}$ with length $2\cdot 5^n$. A key observation is that if and only if $p$-th $\tilde{\pi}_1$ in the first $5^n$ sequel permutations act nontrivially (i.e., the state is in $X$), then the same $p$-th $\tilde{\pi}_1$ in the second period act trivially (i.e., the state is in $Y$). Similar observations are valid for $\tilde{\pi}_0$ and $\tilde{\pi}_3$. These observations lead to an important fact that through two periods $\tilde{\pi}_1$, $\tilde{\pi}_0$ and $\tilde{\pi}_4$ act on $X$ by $5^n$ times, and $\tilde{\pi}_3$ acts on $Y$ by $5^n$ times.

We now argue the ergodicity. In the coarse-grained description, $\tilde{\pi}_{I_2^{[n]}}\tilde{\pi}_{I_1^{[n]}}\tilde{\pi}_{I_2^{[n]}}\tilde{\pi}_{I_1^{[n]}}$ is identity (i.e., $X\to X$ and $Y\to Y$). Hence, $\tilde{\pi}_{I_2^{[n]}}\tilde{\pi}_{I_1^{[n]}}\tilde{\pi}_{I_2^{[n]}}\tilde{\pi}_{I_1^{[n]}}$ belongs to the set $\{(02),\mathrm{id}\}\otimes\{(13),\mathrm{id}\}$. Using the fact that $\tilde{\pi}_1$, $\tilde{\pi}_0$ and $\tilde{\pi}_4$ act on $X$ by $5^n$ times, and $\tilde{\pi}_3$ acts on $Y$ by $5^n$ times throughout these two periods, we find that $\tilde{\pi}_{I_2^{[n]}}\tilde{\pi}_{I_1^{[n]}}\tilde{\pi}_{I_2^{[n]}}\tilde{\pi}_{I_1^{[n]}}$ is the product of two transpositions (02)(13). Since $\tilde{\pi}_{I_2^{[n]}}\tilde{\pi}_{I_1^{[n]}}=(XY)$ in the coarse-grained description, we see that $\tilde{\pi}_{I_2^{[n]}}\tilde{\pi}_{I_1^{[n]}}$ behaves as (0123) or (0321).

Finally, taking the effect of transposition (04) in $\pi_4$ into account, we conclude that

$$\pi_{I_2^{[n]}}\pi_{I_1^{[n]}}=\begin{cases}(04)^{5^{n-1}}(0123)=(04)(0123)=(01234) & \text{or}\\ (04)^{5^{n-1}}(0321)=(04)(0321)=(03214), & \end{cases} \tag{22}$$

both of which are 5-cycles.

Type 40-1: $(\pi_0=\pi_1=\pi_2=(34)$, $\pi_3=(04)(12)$, $\pi_4=(01))$

The structure is $(0,1,2,4)(3,4)$. We denote by $I_1=(0,1,2,4)$ and $I_2=(3,4)$. Two permutations yielded by $I_2$, $\pi_3=(04)(12)$ and $\pi_4=(01)$, are not commutative. On the other hand, by setting $X=\{0,1\}$ and $Y=\{2,4\}$ and viewing these permutations in this coarse-grained state space $\{X,Y\}$, $\pi_4$ is identity in this state space, and $\pi_3$ is a transposition $(XY)$. In this coarse-grained description, we have $\pi_{I_2^{[n]}}=(XY)^{5^{n-1}}=(XY)$.

To analyze the dynamics in $(0,1,2,4)$ in detail, we introduce modified permutation $\tilde{\pi}_i$ with $\tilde{\pi}_4=\pi_4$, $\tilde{\pi}_3=\pi_3$, while $\tilde{\pi}_0=\tilde{\pi}_1=\tilde{\pi}_2=\mathrm{id}$, which is obtained from $\pi_i$ by removing all transitions between 3 and 4. We notice that $\pi_4=\tilde{\pi}_4$ has nontrivial action only when the present state is in $X$. Recalling $\pi_0 = \pi_1 = \pi_2 = (34)$, we find that $\tilde{\Sigma} = \tilde{\pi}_{I_2^{[n]}} \tilde{\pi}_{I_1^{[n]}}$ and $\Sigma = \pi_{I_2^{[n]}} \pi_{I_1^{[n]}}$ induce the same transition if the initial state is 0, 1, and 2.

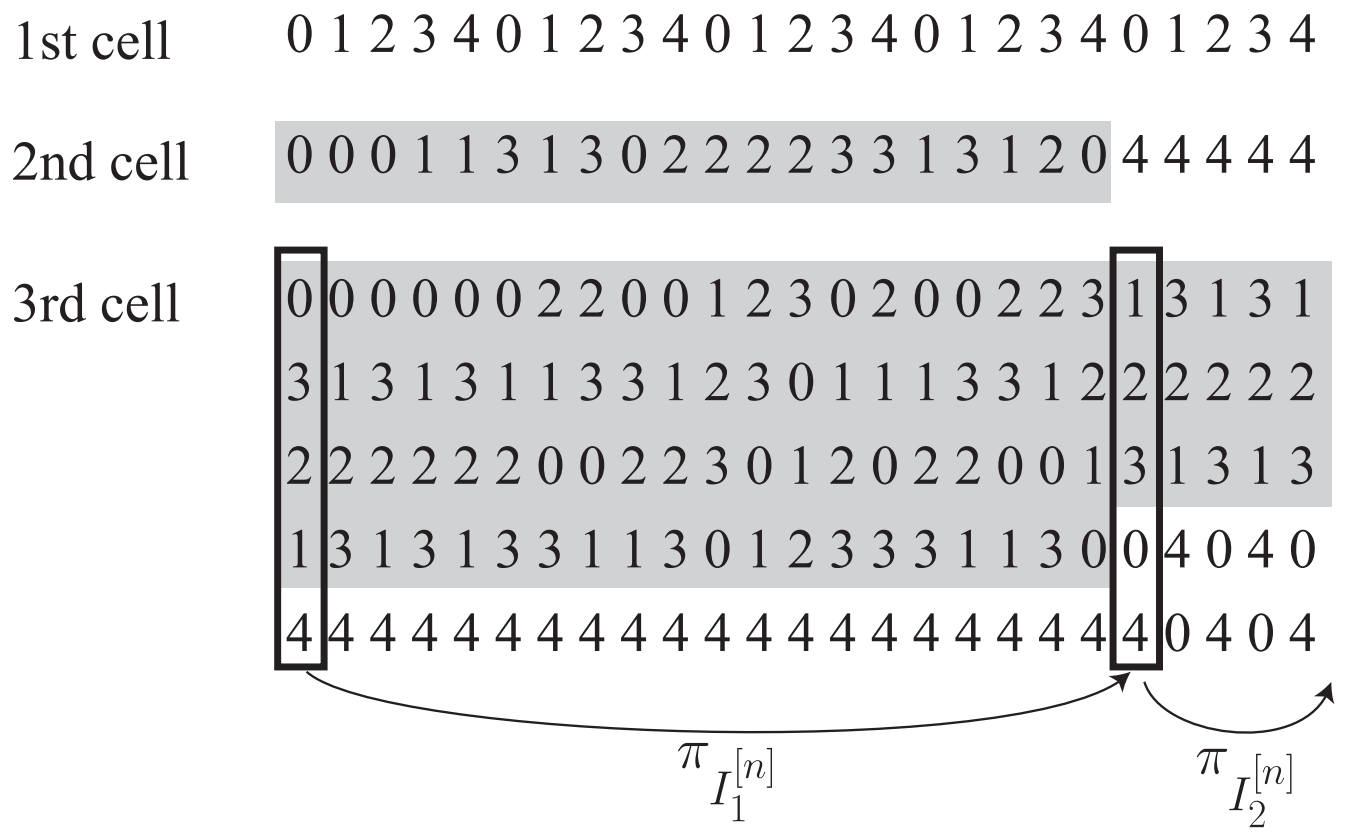

**Fig. 12** The sequence of first, second, and third cells in Type42-5 with $(a, b, c) = (1, 2, 3)$. The sequence in the third cell is divided into 5 lines that go from the top to the bottom. We draw island $I_1$ in light gray. We draw permutations for $n = 2$.

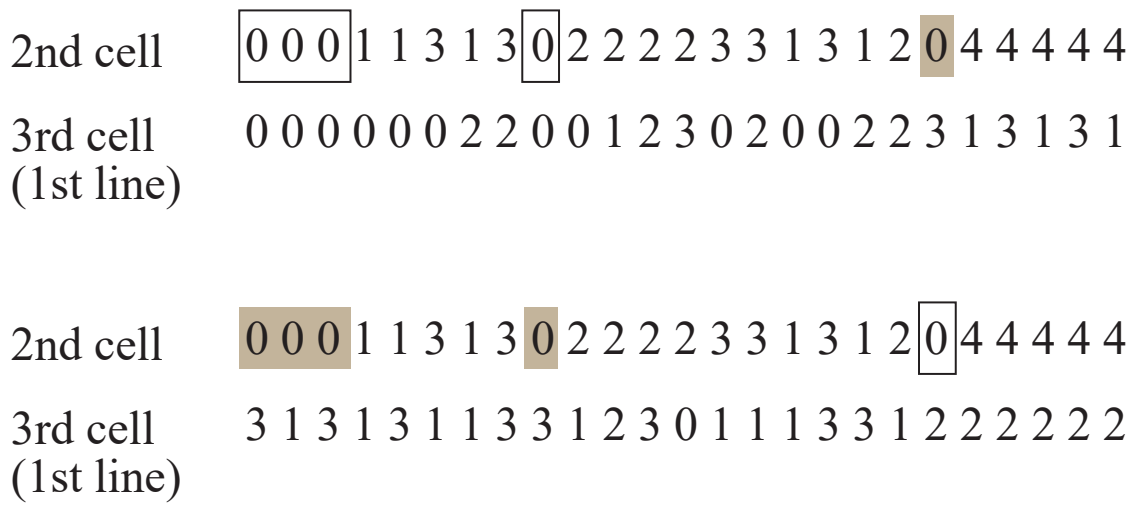

**Fig. 13** We pick up the first and the second line of the third cell. We draw $\pi_0$ in the second cell acting nontrivially in light brown and that acting trivially white with a bold rectangle. As clearly seen, if $\pi_0$ acts nontrivially in the first line, it acts trivially in the second line, and vice versa.

Consider transitions under the modified permutation in two periods $\tilde{\Sigma}^2 = \tilde{\pi}_{I_2^{[n]}} \tilde{\pi}_{I_1^{[n]}} \tilde{\pi}_{I_2^{[n]}} \tilde{\pi}_{I_1^{[n]}}$ with length $2 \cdot 5^n$. A key observation is that if and only if $p$-th $\tilde{\pi}_4$ in the first $5^n$ sequel permutations acts nontrivially (i.e., the state is in $X$), in the second period (i.e., the second $5^n$ sequel permutations) the same $p$-th $\tilde{\pi}_4$ acts trivially (i.e., the state is in $Y$) (see Fig. 15). This is confirmed by the fact that between two $p$-th $\tilde{\pi}_4$'s in the first and the second periods, $\tilde{\pi}_3 = (XY)$ acts $5^{n-1}$ times, leading to permutation $(XY)^{5^{n-1}} = (XY)$. Owing to this observation, exactly half of $\tilde{\pi}_4$'s in these two periods act on $X$, and hence the total number of actions of $\tilde{\pi}_4$ on $X$ through these two periods is $5^n$ times.

We now argue the ergodicity. In the coarse-grained description, $\tilde{\pi}_{I_2^{[n]}} \tilde{\pi}_{I_1^{[n]}} \tilde{\pi}_{I_2^{[n]}} \tilde{\pi}_{I_1^{[n]}}$ is identity (i.e., $X \to X$ and $Y \to Y$). Hence, $\tilde{\pi}_{I_2^{[n]}} \tilde{\pi}_{I_1^{[n]}} \tilde{\pi}_{I_2^{[n]}} \tilde{\pi}_{I_1^{[n]}}$ belongs to the set $\{(01), \mathrm{id}\} \otimes \{(24), \mathrm{id}\}$ (i.e., a product of the transposition in $X$ and/or that in $Y$ and/or identity). In addition, observe that $\tilde{\pi}_4$ transpose states in $Z := \{0, 4\}$ and states in $U := \{1, 2\}$ if and only if the state is in $X$, and $\pi_3$ keeps $Z$ and $U$ unchanged. Since $\tilde{\pi}_4$ acts on $X$ by $5^n$ times in these two periods, we find that $\tilde{\pi}_{I_2^{[n]}} \tilde{\pi}_{I_1^{[n]}} \tilde{\pi}_{I_2^{[n]}} \tilde{\pi}_{I_1^{[n]}}$ swaps states in $Z$ and states in $U$, which implies $\tilde{\pi}_{I_2^{[n]}} \tilde{\pi}_{I_1^{[n]}} \tilde{\pi}_{I_2^{[n]}} \tilde{\pi}_{I_1^{[n]}} = (01)(24)$. Since $\tilde{\pi}_{I_2^{[n]}} \tilde{\pi}_{I_1^{[n]}} = (XY)$ in the coarse-grained description, we find that $\tilde{\pi}_{I_2^{[n]}} \tilde{\pi}_{I_1^{[n]}}$ behaves as (0214) or (0412). Finally, going back to the original permutation $\pi$ and taking back the transposition (34) in $\pi_{I_1^{[n]}}$ into account, we conclude that

$$\pi_{I_2^{[n]}} \pi_{I_1^{[n]}} = \begin{cases} (0214)(34)^{5^{n-1}} = (0214)(34) = (02143), & \text{or} \\ (0412)(34)^{5^{n-1}} = (0412)(34) = (04312), & \end{cases} \tag{23}$$

both of which are 5-cycles. The obtained permutation can be seen in the right rectangle in Fig. 14. The right rectangle visualize the move of the permutation $\pi_{I_2^{[n]}}\pi_{I_1^{[n]}}$, which reads $2 \to 1 \to 4 \to 3 \to 0 \to 2$, corresponding to (02143).

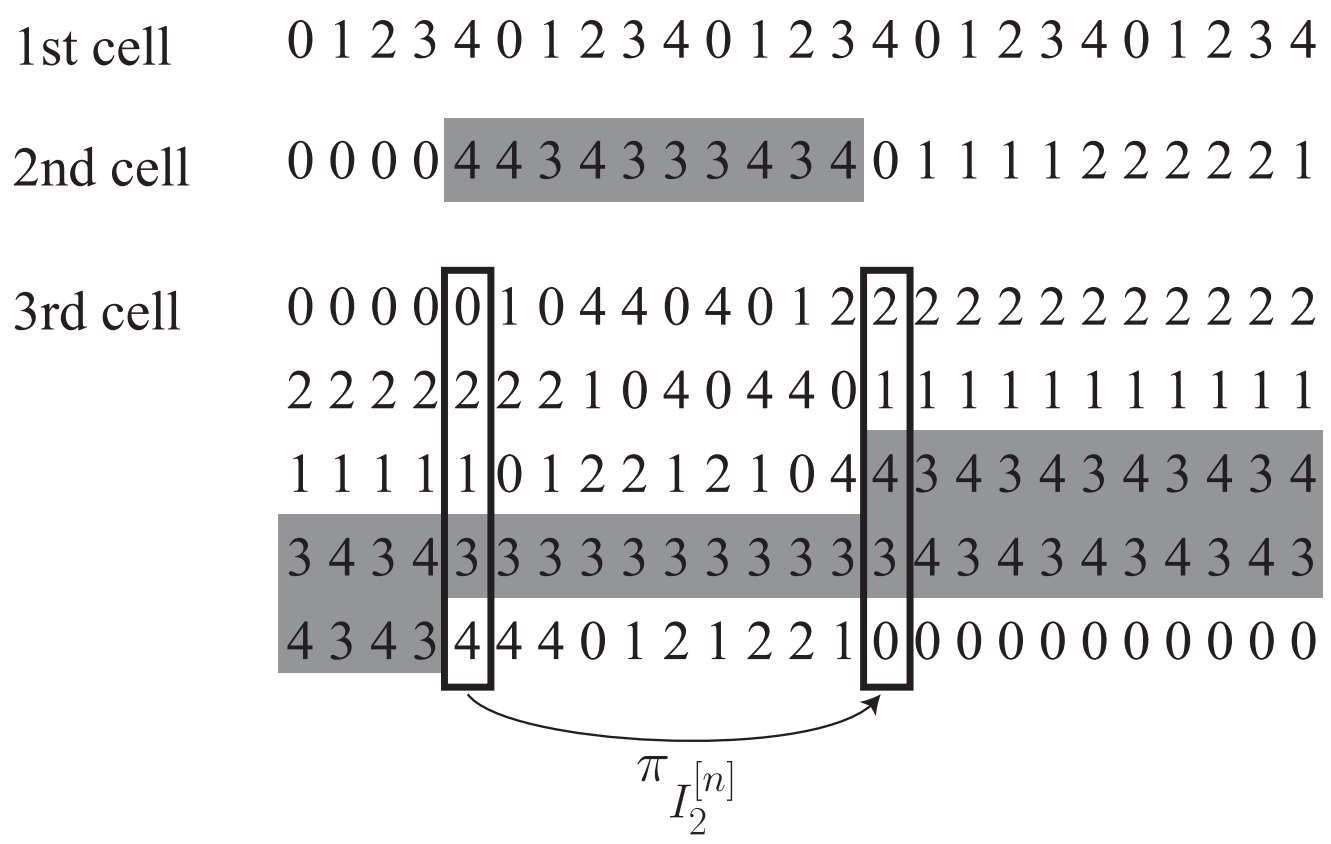

**Fig. 14** The sequence of first, second, and third cells in Type40-1. The sequence in the third cell is divided into 5 lines that go from the top to the bottom. We draw island $I_2$ in dark gray. We draw permutations for $n = 2$.

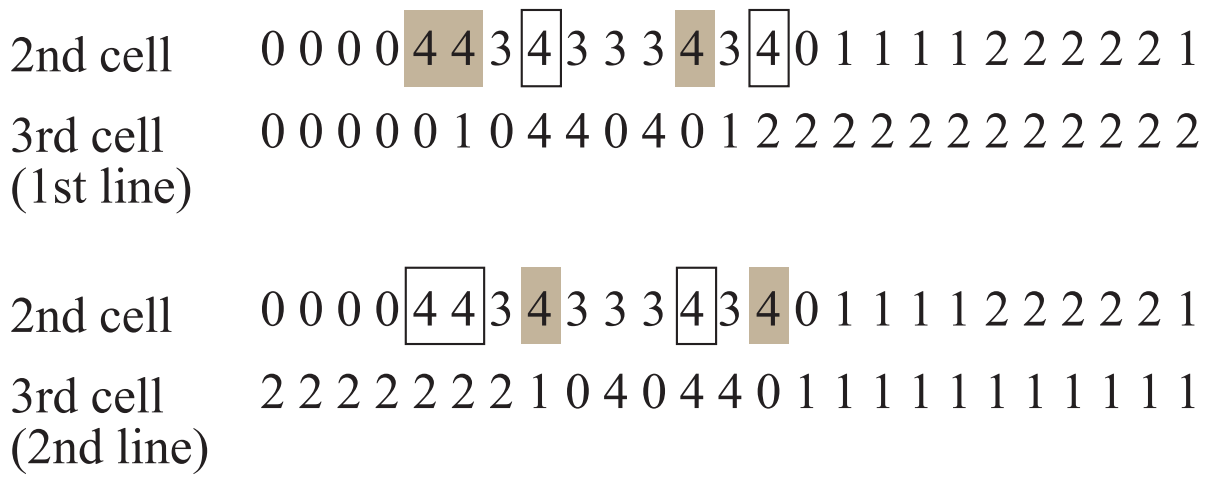

**Fig. 15** We pick up the first and the second line of the third cell. We draw $\pi_4$ in the second cell acting nontrivially in light brown and that acting trivially white with a bold rectangle. As clearly seen, if $\pi_4$ acts nontrivially in the first line, it acts trivially in the second line, and vice versa.

Type27-1; $(a, b, c, d, e) = (0, 1, 2, 3, 4)$: ($\pi_0 = \text{id}$, $\pi_1 = \pi_2 = (01)(23)$, $\pi_3 = (0123)$, $\pi_4 = (04)$)

The structure is $(0, 1, 2, 3)(0, 4)$. We denote by $I_1 = (0, 1, 2, 3)$ and $I_2 = (0, 4)$. In (0,1,2,3), permutations act as $\pi_1 = \pi_2 = (01)(23)$ and $\pi_3 = (0123)$, and hence $\pi_3$ and others are not commutative. On the other hand, by setting $X = \{0, 2\}$ and $Y = \{1, 3\}$ and viewing these permutations in a coarse-grained state space $\{X, Y\}$, all of $\pi_1$, $\pi_2$ and $\pi_3$ are transposition $(XY)$. Hence, in the coarse-grained description, $\pi_{I_1^{[n]}} = (XY)^{3\cdot 5^n} = (XY)$ is obtained. In addition, since $\pi_0$ and $\pi_4$ commute and $I_2$ has odd number of 4, we have $\pi_{I_2^{[n]}} = (04)$.

Now we regard states 0, 1, 2, and 3 as elements of $\mathbb{Z}_4$. Then, $\pi_3$ serves as an addition of $+1$. In contrast, $\pi_1$ and $\pi_2$ play as addition of $+1$ if the state is in $X = \{0, 2\}$ and as addition of $-1$ if the state is in $Y = \{1, 3\}$. As following a similar argument to previous ones, we introduce modified transition $\tilde{\pi}_i$ with $\tilde{\pi}_j = \pi_j$ for $j = 0, 1, 2, 3$ and $\tilde{\pi}_4 = \text{id}$, and consider transitions in two periods $\tilde{\Sigma}^2 = (\tilde{\pi}_{I_1^{[n]}})^2$. We observe that if $p$-th $\tilde{\pi}_1$ in the $5^n$ sequel permutations act as $+1$ (i.e., the state is in $X$) in the first period, then the same $p$-th $\tilde{\pi}_1$ in the second period act as $-1$ (i.e., the state is in $Y$), and vice versa. Similar observations hold for $\tilde{\pi}_2$. Hence, through two periods, $\tilde{\pi}_1$ and $\tilde{\pi}_2$ act on $X$ by $5^n$ times and on $Y$ by $5^n$ times. This means that through these two periods $\tilde{\pi}_1$ and $\tilde{\pi}_2$ have no total contribution, since $\tilde{\pi}_1$ acts as $+1$ by $5^n$ times and acts as $-1$ by $5^n$ times, whose sum is zero. In summary, $(\tilde{\pi}_{I_1^{[n]}})^2$

is computed as

$$(\tilde{\pi}_{I_1^{[n]}})^2 = (\tilde{\pi}_3)^{2\cdot 5^n} = (02)(13), \tag{24}$$

and hence $\tilde{\pi}_{I_1^{[n]}}$ is (0123) or (0321).

Using these facts, we can easily show the ergodicity. The total permutation in a single period with length $5^n$ denoted by $\Sigma$ is computed as

$$\Sigma = \pi_{I_2^{[n]}}\pi_{I_1^{[n]}} = \begin{cases} (04)(0123) = & (01234) \text{ or} \\ (04)(0321) = & (03214), \end{cases} \tag{25}$$

both of which are 5-cycles.

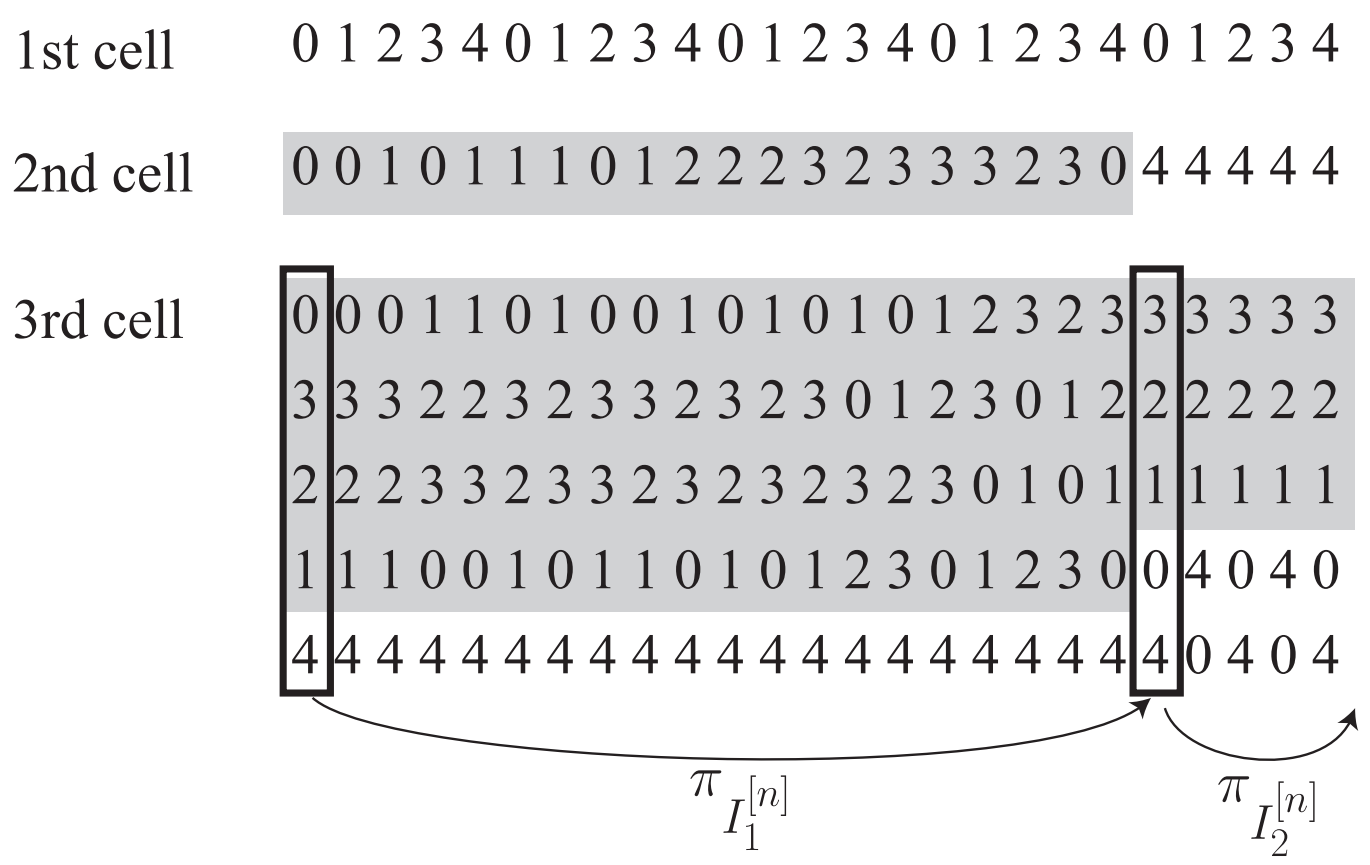

**Fig. 16** The sequence of first, second, and third cells in Type27-1 with $(a, b, c, d, e) = (0, 1, 2, 3, 4)$. The sequence in the third cell is divided into 5 lines that go from the top to the bottom. We draw island $I_1$ in light gray. We draw permutations for $n = 2$.

## 6.3 Pattern B: Constructed by some units

### *6.3.1 B-1: The simplest case*

The structure in pattern B-1 consists of several units and block units, e.g., [01] and $\lfloor 2, 3, 4 \rfloor_{\text{even}}$. In case of no confusion, we call collectively them simply "units". In most rules in pattern B-1, the structure has the following components:

– Several *driving unit* [*ab*] (or a longer one) which induces a nontrivial permutation $\pi_b\pi_a$.
– Several *non-driving unit* with even length $\lfloor c, d, e \rfloor_{\text{even}}$ such that they induce trivial permutations (e.g., $\pi_d\pi_e$ and $\pi_c\pi_c\pi_d\pi_c$ are identity permutations).
– A single *exceptional unit* with odd length $\lfloor c, d, e \rfloor_{\text{odd}}$.

In addition, transitions induced by driving units [ab] and non-driving units with even length $\lfloor c, d, e \rfloor_{\text{even}}$ reproduce [ab] and $\lfloor c, d, e \rfloor_{\text{even}}$, which guarantees the structure.

Type54-1; $(a, b) = (3, 2)$: $(\pi_0 = \pi_1 = \pi_4 = (04)(23),\ \pi_2 = (0123),\ \pi_3 = (23))$

The structure is [23], $\lfloor 0, 1, 4 \rfloor_{\text{even}}$, $\lfloor 0, 1, 4 \rfloor^1_{\text{odd}}$. The driving unit is [23], and the non-driving unit is $\lfloor 0, 1, 4 \rfloor_{\text{even}}$.

We first observe the following fact: The permutation $\pi_{[23]} = \pi_3\pi_2$ yields transitions $(0 \to)1 \to 1$, $(3 \to)0 \to 0$, and $(4 \to)4 \to 4$, which reproduce the structure $\lfloor 0, 1, 4 \rfloor_{\text{even}}$, and $(1 \to)2 \to 3$ and $(2 \to)3 \to 2$, which reproduce the structure [23]. In addition, the permutation $\pi_b\pi_a$ with $a, b \in \{0, 1, 4\}$ yields transition $(0 \to)4 \to 0$, $(4 \to)0 \to 4$, and $(1 \to)1 \to 1$ , which reproduce the structure $\lfloor 0, 1, 4 \rfloor_{\text{even}}$, and $(3 \to)2 \to 3$ and $(2 \to)3 \to 2$, which reproduce the structure [23]. We also observe that the permutation $\pi_b \pi_a$ with $a, b \in \{0, 1, 4\}$ is an identity permutation. Hence, permutations $\pi_3\pi_2$ and $\pi_b\pi_a$ with $a, b \in \{0, 1, 4\}$ exchange states in $\{0, 1, 3\}$ (which corresponds to the second number in the rectangles filled with two numbers in the main transition in Fig. 17), and keep states 2 and 4 unchanged (which corresponds to that in the sub-transition in Fig. 17). In addition, $\pi_{\lfloor 0,1,4 \rfloor^1_{\mathrm{odd}}}$ is equal to (23)(04), which plays the role of switching the main transition and sub-transition (see Fig. 17) unless the state is 1. If the state is 1, $\pi_{\lfloor 0,1,4 \rfloor^1_{\mathrm{odd}}}$ plays the role of addition of a single 1, which results in the odd length sequence $\lfloor 0, 1, 4 \rfloor^1_{\mathrm{odd}}$ in site $n+1$. This mechanism ensures the inheritance of the structure from site $n$ to site $n+1$.

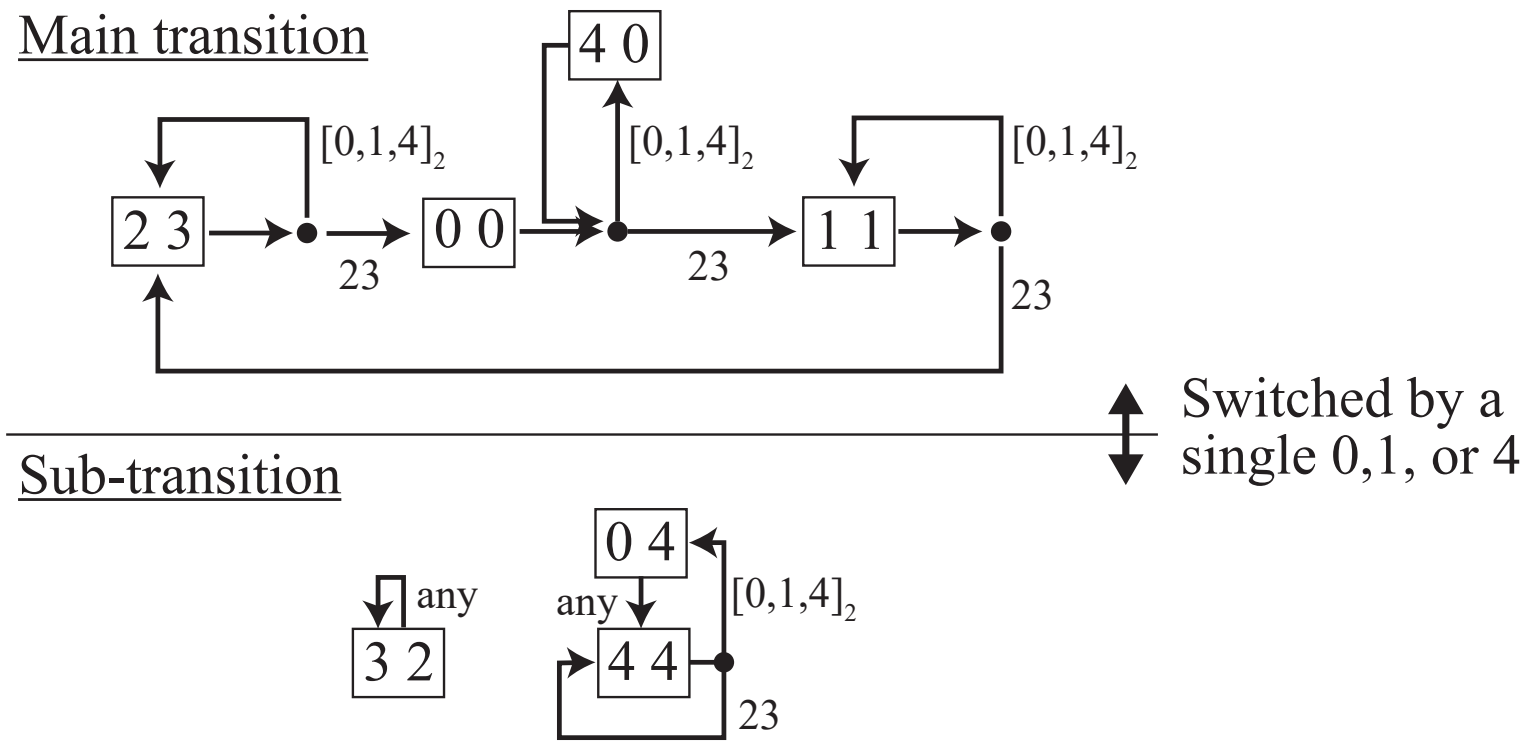

**Fig. 17** A flow chart of the transition of Type54-1 with $(a, b) = (3, 2)$. When permutations induced by [23] or $\lfloor 0, 1, 4 \rfloor_{\mathrm{even}}$ apply, the time evolution of the cell follows the flow chart of the main transition (above) or the sub transition (below). If a permutation induced by $\lfloor 0, 1, 4 \rfloor_{\mathrm{odd}}$ (more precisely, a single 0, 1, or 4) applies, the main transition and the sub transition are switched unless the state is 1. When the state is 1, the application of $\pi_1$ plays the role of just adding a single 1 to the sequence in site $n+1$.

We now confirm the ergodicity. The total permutation in a single period with length $5^n$ denoted by $\Sigma$ is computed as

$$\Sigma = (\pi_3\pi_2)^p (23)(04)(\pi_3\pi_2)^q \tag{26}$$

with $p + q = 5^{n-1}$. By setting the initial state of the period at just after $\lfloor 0, 1, 4 \rfloor^1_{\mathrm{odd}}$, this permutation denoted by $\Sigma'$ reads

$$\Sigma' = (23)(04)(\pi_3\pi_2)^{5^{n-1}} = (23)(04)(013)^{5^{n-1}} = \begin{cases} (23)(04)(013) & = (01234), \quad n \text{ is odd} \\ (23)(04)(031) & = (02314), \quad n \text{ is even}, \end{cases} \tag{27}$$

both of which are 5-cycles.

| | |
|---|---|
| 1st cell | 0 1 2 3 4 0 1 2 3 4 0 1 2 3 4 0 1 2 3 4 0 1 2 3 4 |
| 2nd cell | 0 4 0 1 1 1 1 1 2 3 2 3 2 3 2 3 2 3 0 0 4 0 4 4 4 |
| 3rd cell | 0 4 0 4 0 4 0 4 0 1 1 2 3 0 0 1 1 2 3 2 3 2 3 2 3 |
| | 2 3 2 3 2 3 2 3 2 3 2 3 2 3 2 3 2 3 2 3 2 3 2 3 2 |
| | 3 2 3 2 3 2 3 2 3 0 0 1 1 2 3 0 0 1 1 1 1 1 1 1 1 |
| | 1 1 1 1 1 1 1 1 1 2 3 0 0 1 1 2 3 0 0 4 0 4 0 4 0 |
| | 4 0 4 0 4 0 4 0 4 4 4 4 4 4 4 4 4 4 4 0 4 0 4 0 4 |

**Fig. 18** The sequence of first, second, and third cells in Type54-1 with $(a, b) = (3, 2)$. The sequence in the third cell is divided into 5 lines that go from the top to the bottom. We draw [23] in light yellow and $\lfloor 0, 1, 4 \rfloor^1_{\mathrm{odd}}$ in brown (with length 15 in the 2nd cell and length 19 in the 3rd cell).

Type48-2; $(a,b)=(2,1)$: $(\pi_0=\pi_2=(12)(34),\ \pi_1=(01234),\ \pi_3=\pi_4=(12))$

The structure is $[12],[00],\lfloor 3,4\rfloor_{\text{even}},\lfloor 0\rfloor^1_{\text{odd}}$. The driving unit is $[12]$, and the non-driving units are $[00]$ and $\lfloor 3,4\rfloor_{\text{even}}$.

We first observe the following facts: The permutation $\pi_2\pi_1$ yields transition $(0\to)1\to 2$ and $(1\to)2\to 1$, which belong to $[12]$, and $(2\to)3\to 4$ and $(3\to)4\to 3$, which reproduce $\lfloor 3,4\rfloor_{\text{even}}$, and $(4\to)0\to 0$ , which reproduces $[00]$. In addition, the permutations $\pi_b\pi_a$ with $a,b\in\{3,4\}$ and $\pi_0\pi_0$ yield transitions $(1\to)2\to 1$ and $(2\to)1\to 2$, which reproduce $[12]$, and act as identities on 3, 4, and 0, which belong to $\lfloor 3,4\rfloor_{\text{even}}$ and $[00]$. We also observe that the permutations $\pi_b\pi_a$ with $a,b\in\{3,4\}$ and $\pi_0\pi_0$ are an identity permutation. In other words, all of permutations $\pi_2\pi_1$, $\pi_0\pi_0$, and $\pi_b\pi_a$ with $a,b\in\{3,4\}$ act as permutations only on $\{0,2,4\}$, and keep states 1 and 3 unchanged. In addition, $\pi_0$ is equal to $(12)(34)$. Following a similar argument to the case of Type 54-1, we conclude that if the structure holds in site $n$, the structure also holds in site $n+1$.

Furthermore, the total permutation in a single period with length $5^n$ denoted by $\Sigma$ is computed as

$$\Sigma=(\pi_2\pi_1)^p(12)(34)(\pi_2\pi_1)^q \tag{28}$$

with $p+q=5^{n-1}$. By setting the initial state of the period at just after the single $\pi_0$, the total permutation denoted by $\Sigma'$ reads

$$\Sigma'=(12)(34)(\pi_2\pi_1)^{5^{n-1}}=(12)(34)(024)^{5^{n-1}}=\begin{cases}(12)(34)(024)= & (01234),\quad n\text{ is odd}\\ (12)(34)(042)= & (03412),\quad n\text{ is even}\end{cases} \tag{29}$$

both of which are 5-cycles.

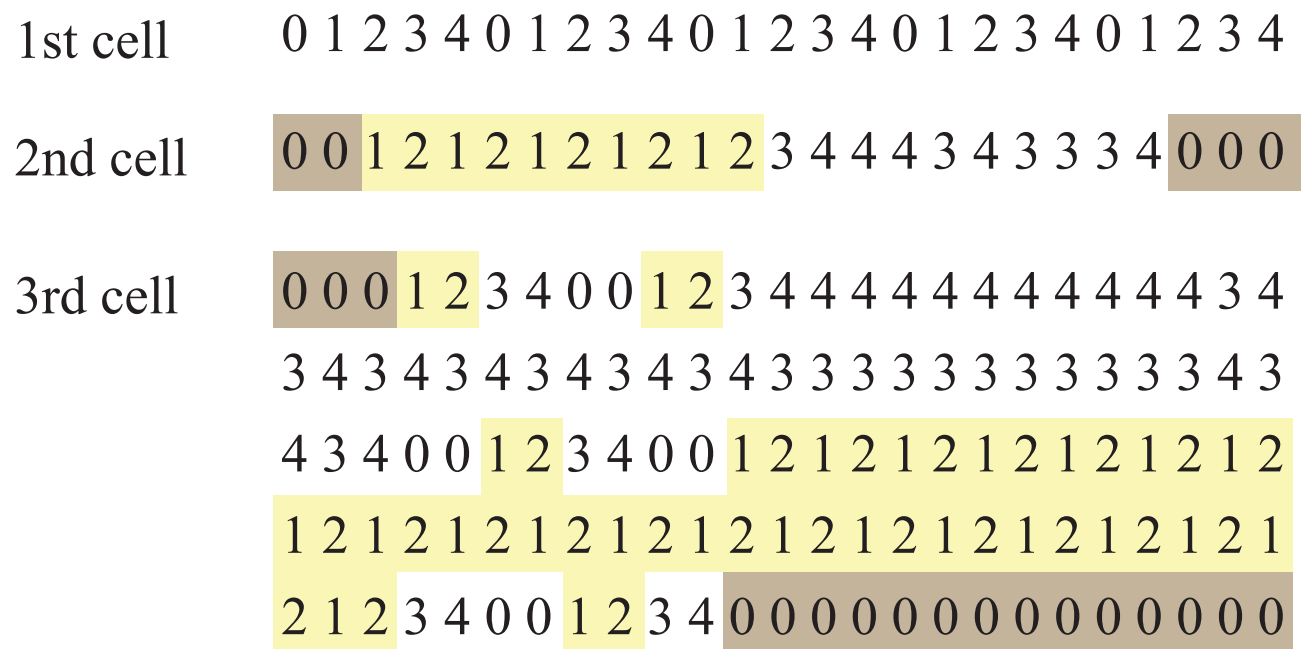

**Fig. 19** The sequence of first, second, and third cells in Type48-2 with $(a,b)=(2,1)$. The sequence in the third cell is divided into 5 lines that go from the top to the bottom. We draw $[12]$ in light yellow and $\lfloor 0\rfloor^1_{\text{odd}}$ in brown.

Type59; $(a,b,c,d)=(1,3,2,4)$: $(\pi_0=(01234),\ \pi_1=\pi_3=(014),\ \pi_2=\pi_4=(014)(23))$

The structure is $[401],[223],\lfloor 2,3\rfloor^1_{l_n}$ with $l_n\equiv 2\cdot 5^{n-1}\mod 3$, where $\lfloor 2,3\rfloor^1_{l_n}$ is a block unit of 2 and 3 with length $l_n\equiv 2\cdot 5^{n-1}\mod 3$ appearing just once in a single period. The driving unit is $[401]$, and the non-driving unit is $[223]$.

We first notice that state change from set $\{0,1,4\}$ to $\{2,3\}$ occurs only when $\pi_0$ acts on state 0, and the opposite occurs only when $\pi_0$ acts on 3. Under the structure, this means a sequence composed of 2 and 3 only begins when $\pi_1\pi_0\pi_4$ acts on state 0, and is terminated when $\pi_1\pi_0\pi_4$ acts on state 2. All the other transitions induced by $\pi_1\pi_0\pi_4$, $\pi_3\pi_2\pi_2$ or $\pi_{\lfloor 2,3\rfloor^1_{l_n}}$ occur in set $\{0,1,4\}$ or $\{2,3\}$. Thus, the transition from initial value 0 induced by $401(223)^{\otimes m}401$ generates $01(223)^{\otimes m+1}40$ and that from initial value 0 induced by $401\lfloor 2,3\rfloor^1_{l_n}401$ results in 01 appended by a sequence composed of 2 and 3 only. The sequence of 2 and 3 is not terminated in that transition, because $\lfloor 2,3\rfloor^1_{l_n}$ contains an odd number of 2, and so the last $\pi_1\pi_0\pi_4$ acts on state 3, which causes a transition in $\{2,3\}$. The sequence is terminated after 5n further steps because $\pi_1\pi_0\pi_4$ operates on state 2 after the second $\lfloor 2,3\rfloor^1_{l_n}$. As the result, sequence $\lfloor 2,3 \rfloor^1_{l_{n+1}}$ is generated, where $l_{n+1} = l_n + 5^n + 3$. These observations indicate the inheritance of the structure.

We can see that $\pi_1\pi_0\pi_4 = (02)$ and $\pi_3\pi_2\pi_2 = \mathrm{id}$. We further notice that if site $n$ is ergodic, $\lfloor 2,3 \rfloor^1_{l_n}$ in site $n$ contains an odd number of 2. Thus, recalling $l_n \equiv 2 \cdot 5^{n-1} \mod 3$, we have $\pi_{\lfloor 2,3 \rfloor^1_{l_n}} = (23)(014)^{l_n} = (23)(041)$ for odd $n$ and $(23)(014)$ for even $n$. With noting $\pi_{[223]} = \mathrm{id}$, the total permutation in a single period with length $5^n$ denoted by $\Sigma$ is computed as

$$\Sigma = \begin{cases} (\pi_1\pi_0\pi_4)^p(23)(041)(\pi_1\pi_0\pi_4)^q, & n \text{ is odd}, \\ (\pi_1\pi_0\pi_4)^p(23)(014)(\pi_1\pi_0\pi_4)^q, & n \text{ is even}, \end{cases} \tag{30}$$

with $p+q = 5^{n-1}$. By setting the initial state of the period at just after $\lfloor 2,3 \rfloor^1_{l_n}$, the total permutation denoted by $\Sigma'$ reads

$$\Sigma' = \begin{cases} (23)(041)(\pi_1\pi_0\pi_4)^{5^{n-1}} & = (23)(041)(02)^{5^{n-1}} = (23)(041)(02) = (03241), \quad n \text{ is odd}, \\ (23)(014)(\pi_1\pi_0\pi_4)^{5^{n-1}} & = (23)(014)(02)^{5^{n-1}} = (23)(014)(02) = (03214). \quad n \text{ is even}, \end{cases} \tag{31}$$

both of which are 5-cycles.

| | |
|---|---|
| 1st cell | 0 1 2 3 4 0 1 2 3 4 0 1 2 3 4 0 1 2 3 4 0 1 2 3 4 |
| 2nd cell | 0 1 4 0 1 4 0 1 4 0 1 2 2 3 3 2 3 3 2 2 3 4 0 1 4 |
| 3rd cell | 0 1 4 0 1 4 0 1 4 0 1 4 0 1 4 0 1 4 0 1 4 0 1 2 2 |
| | 3 4 0 1 2 2 3 4 0 1 2 2 3 2 2 2 3 3 3 2 2 3 2 3 3 |
| | 2 3 3 2 3 3 2 3 3 2 3 3 2 3 3 3 2 2 2 3 3 2 3 4 0 |
| | 1 4 0 1 4 0 1 4 0 1 4 0 1 4 0 1 4 0 1 4 0 1 4 0 1 |
| | 4 0 1 4 0 1 4 0 1 4 0 1 4 0 1 4 0 1 4 0 1 4 0 1 4 |

**Fig. 20** The sequence of first, second, and third cells in Type59 with $(a,b,c,d) = (1,3,2,4)$. The sequence in the third cell is divided into 5 lines that go from the top to the bottom. We draw [401] in light yellow and $\lfloor 2,3 \rfloor^1_{l_n}$ with $l_n \equiv 2 \cdot 5^{n-1} \mod 3$, in brown. Although most of 2 and 3 belong to $\lfloor 2,3 \rfloor^1_{l_n}$ in this figure, in site $n$ with large $n$, most of 2 and 3 belong to [223].

Type58-1; $(a,b,c,d) = (1,2,3,4)$: $(\pi_0 = \pi_1 = \pi_2 = (12)(34),\ \pi_3 = \pi_4 = (014))$

The structure of this rule is $3, 4, \lfloor 0,1,2 \rfloor_{\mathrm{even}}, \lfloor 0,1,2 \rfloor^1_{\mathrm{odd}}$, where we put no restriction on the positions of states 3 and 4. These two states 3 and 4 serve as driving units, and $\lfloor 0,1,2 \rfloor_{\mathrm{even}}$ is a non-driving unit. Since $\lfloor 0,1,2 \rfloor_{\mathrm{even}}$ induces transitions $4 \to 3 \to 4$, $1 \to 2 \to 1$, and $0 \to 0 \to 0$, if the present state is 0, 1, or 4, a possible transition takes the form of repeats of

$$4(\to 3 \to 4)^{\otimes k} \to 0(\to 0 \to 0)^{\otimes l} \to 1(\to 2 \to 1)^{\otimes m} \to 4 \tag{32}$$

with some $k$, $l$, and $m$, which satisfies the structure.

In addition, since $\pi_3 = \pi_4 = (014)$ and $\pi_{\lfloor 0,1,2 \rfloor_{\mathrm{even}}} = \mathrm{id}$, the total permutation in a single period with length $5^n$ denoted by $\Sigma$ is computed as

$$\Sigma = (014)^p \pi_{\lfloor 0,1,2 \rfloor_{\mathrm{odd}}} (014)^q = (014)^p(12)(34)(014)^q \tag{33}$$

with $p+q = 2 \cdot 5^{n-1}$. Setting the initial state of the period at just after $(12)(34)$, the total permutation denoted by $\Sigma'$ reads

$$\Sigma' = (12)(34)(014)^{2 \cdot 5^{n-1}} = \begin{cases} (12)(34)(041) = (03421) & n \text{ is odd}, \\ (12)(34)(014) = (02134) & n \text{ is even}, \end{cases} \tag{34}$$

both of which are 5-cycles.

| | |
|---|---|
| 1st cell | 0 1 2 3 4 0 1 2 3 4 0 1 2 3 4 0 1 2 3 4 0 1 2 3 4 |
| 2nd cell | 0 0 0 0 1 4 3 4 3 3 3 4 3 4 0 1 2 1 2 2 2 1 2 1 4 |
| 3rd cell | 0 0 0 0 0 0 1 4 0 1 4 0 1 4 0 0 0 0 0 0 0 0 0 0 0 |
| | 1 2 1 2 1 2 2 2 2 2 2 2 2 2 2 1 2 1 2 1 2 1 2 1 2 |
| | 2 1 2 1 2 1 4 0 1 4 0 1 4 0 1 2 1 2 1 2 1 2 1 2 1 |
| | 4 3 4 3 4 3 3 3 3 3 3 3 3 3 3 4 3 4 3 4 3 4 3 4 3 |
| | 3 4 3 4 3 4 0 1 4 0 1 4 0 1 4 3 4 3 4 3 4 3 4 3 4 |

**Fig. 21** The sequence of first, second, and third cells in Type58-1 with $(a, b, c, d) = (1, 2, 3, 4)$. The sequence in the third cell is divided into 5 lines that go from the top to the bottom. We draw $\lfloor 0, 1, 2 \rfloor^1_{\text{odd}}$ in brown.

### *6.3.2 B-2: Alternating sequence appears with removing even length units*

The structure in pattern B-2 takes the form of $a + * + b + *$, where * is some units, usually with even length. The unit in * yields an identity permutation $\pi_* = \text{id}$ (except for the unique unit with odd length). The permutation induced by the sequence with this structure is thus equivalent to $ababab\cdots$, with which we prove the ergodicity.

Type58-2; $(a, b) = (1, 3)$: $(\pi_0 = \pi_2 = \pi_4 = (04)(13),\ \pi_1 = (123),\ \pi_3 = (013))$

The structure of this rule is $1 + \{\lfloor 2 \rfloor_{\text{even}}, \lfloor 2 \rfloor^1_{\text{odd}}\} + 3 + \lfloor 0, 4 \rfloor_{\text{even}}$, where $\{\lfloor 2 \rfloor_{\text{even}}, \lfloor 2 \rfloor^1_{\text{odd}}\}$ means one of $\lfloor 2 \rfloor_{\text{even}}$ or $\lfloor 2 \rfloor^1_{\text{odd}}$.

We first observe that $\lfloor 0, 4 \rfloor_{\text{even}}$ and $\lfloor 2 \rfloor_{\text{even}}$ yield transitions $1 \to 3 \to 1 \to 3 \cdots$, $2 \to 2 \to 2$, and $0 \to 4 \to 0 \to 4 \cdots$, all of which keep the structure $1 + \{\lfloor 2 \rfloor_{\text{even}}, \lfloor 2 \rfloor^1_{\text{odd}}\} + 3 + \lfloor 0, 4 \rfloor_{\text{even}}$. In addition, we have $\pi_{\lfloor 0,4 \rfloor_{\text{even}}} = \pi_{\lfloor 2 \rfloor_{\text{even}}} = \text{id}$. Hence, the permutation induced by $1 + \lfloor 2 \rfloor_{\text{even}} + 3 + \lfloor 0, 4 \rfloor_{\text{even}}$ is equivalent to that induced by $1313\cdots$, which yields a transition $(0 \to)0 \to 1 \to 2 \to 2 \to 3 \to 0$ if the initial state is 0, 1, or 2 (see the main transition in Fig. 22) and keeps the state at the same one if the initial state is 3 or 4 (see the sub-transition in Fig. 22). Furthermore, a single $\pi_2$ is permutation (13)(04), which switches the main transition and sub-transition (see Fig. 22) unless the state is 2. In summary, the alternating emergence of 1 and 3 is inherited from site $n$ to site $n + 1$.

Note that in the main transition (i.e, the structure in site $n$ is $1 + \lfloor 2 \rfloor_{\text{even}} + 3 + \lfloor 0, 4 \rfloor_{\text{even}}$), both the length of the sequence of 2 and that of 0 and 4 are even for the following reason: The sequence of 2 starts when $\pi_1$ applies, and it ends also when $\pi_1$ applies. The number of steps between state 1 and another state 1 in site $n$ is even, which guarantees that the length of the sequence of 2 is even. A similar argument holds for the sequence of 0 and 4. We then consider the role of single $\pi_2$. If the present state is 1 or 3, single $\pi_2$ switches to the sub-transition where the state remains the sequence $\lfloor 13 \rfloor$, and single $\pi_2$ appearing after a single period ($5^n$ steps) moves back to the main transition. If the present state is 2, then single $\pi_2$ simply adds 2 with keeping the transition in the main transition, which results in the unique odd length sequence of 2 in one period. If the present state is 0 or 4, single $\pi_2$ switches to the sub-transition where the state remains in the sequence of 0 and 4, and single $\pi_2$ appearing after a single period ($5^n$ steps) moves back to the main transition, which results in a sequence of 0 and 4 with even length. In summary, the structure is inherited from site $n$ to site $n + 1$ including the length of the sequence of 2 and that of 0 and 4.

Then, the total permutation in a single period with length $5^n$ denoted by $\Sigma$ is computed as

$$\Sigma = (\pi_1 \pi_3)^p (13)(04) (\pi_1 \pi_3)^q \tag{35}$$

with $p + q = 5^{n-1}$. Setting the initial state of the period at just after (13)(04), the total permutation denoted by $\Sigma'$ reads

$$\Sigma' = (13)(04)(\pi_1 \pi_3)^{5^{n-1}} = (13)(04)(023) = (02134) \tag{36}$$

which is a 5-cycle.

Type62; $(a, b) = (1, 2)$: $(\pi_0 = \pi_3 = \pi_4 = (04)(12),\ \pi_1 = (01)(23),\ \pi_2 = (03)(12))$

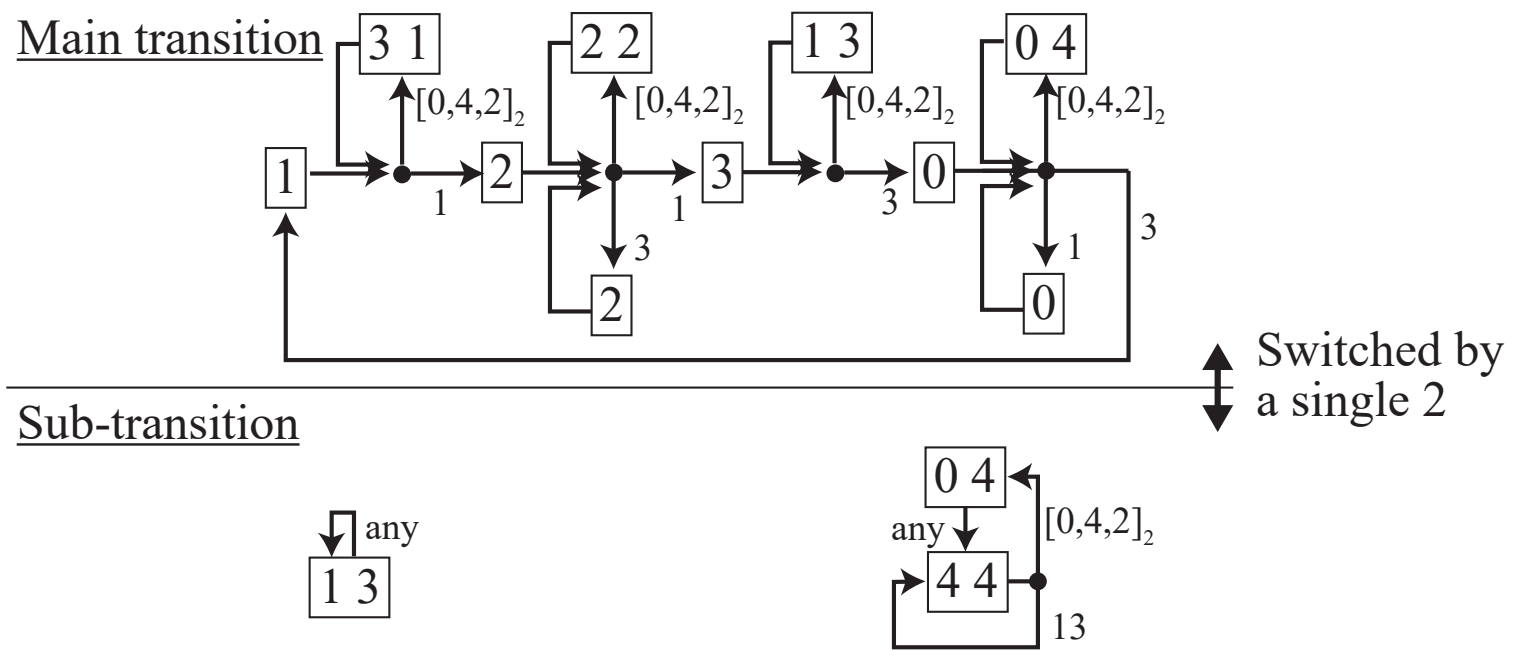

**Fig. 22** A flow chart of the transition of Type58-2 with $(a, b) = (1, 3)$. When permutations induced by $1 + \lfloor 2 \rfloor_{\text{even}} + 3 + \lfloor 0, 4 \rfloor_{\text{even}}$ apply, the time evolution of the cell follows the flow chart of the main transition (above) or the sub transition (below). Here, we treat $\lfloor 2 \rfloor_{\text{even}}$ and $\lfloor 0, 4 \rfloor_{\text{even}}$ together. If a permutation $\pi_2$ applies, the main transition and the sub transition are switched unless the state is 2. When the state is 2, the application of $\pi_2$ plays the role of just adding a single 2 to the sequence in site $n + 1$.

1st cell 0 1 2 3 4 0 1 2 3 4 0 1 2 3 4 0 1 2 3 4 0 1 2 3 4

2nd cell 0 4 4 0 1 3 1 2 2 2 2 2 3 1 3 1 3 1 3 0 4 0 0 4 4

3rd cell 0 4 0 4 0 0 1 2 2 2 2 2 2 2 3 0 0 1 2 2 2 2 2 2 2
2 2 2 2 2 3 0 0 4 0 4 0 4 4 4 4 4 4 4 4 0 4 0 4 0
4 0 4 0 4 4 4 4 0 4 0 4 0 1 2 2 3 0 0 1 3 1 3 1 3
1 3 1 3 1 2 2 3 1 3 1 3 1 3 1 3 1 3 1 3 1 3 1 3 1
3 1 3 1 3 1 3 1 3 1 3 1 3 0 0 1 2 2 3 0 4 0 4 0 4

**Fig. 23** The sequence of first, second, and third cells in Type58-2 with $(a, b) = (1, 3)$. The sequence in the third cell is divided into 5 lines that go from the top to the bottom. We draw 1 and 3 in light yellow and $\lfloor 2 \rfloor^1_{\text{odd}}$ in brown. As clearly seen, the alternation of 1 and 3 is kept anywhere.

The structure is $1 + * + 2 + *$ ($* = \lfloor 0, 3, 4 \rfloor_{\text{even}}, \lfloor 0, 3, 4 \rfloor^1_{\text{odd}}$), i.e., the sequence is basically $1 + \lfloor 0, 3, 4 \rfloor_{\text{even}} + 2 + \lfloor 0, 3, 4 \rfloor_{\text{even}}$, while one of $\lfloor 0, 3, 4 \rfloor_{\text{even}}$ in a single period is replaced by $\lfloor 0, 3, 4 \rfloor_{\text{odd}}$.

We first observe that $\lfloor 0, 3, 4 \rfloor_{\text{even}}$ yields transitions $(1 \to)2 \to 1$ and $(2 \to)1 \to 2$, and $(3 \to)3 \to 3$, $(0 \to)4 \to 0$, and $(4 \to)0 \to 4$, which keep the structure $1 + \lfloor 0, 3, 4 \rfloor_{\text{even}} + 2 + \lfloor 0, 3, 4 \rfloor_{\text{even}}$. With noting $\pi_{\lfloor 0,3,4 \rfloor_{\text{even}}} = \text{id}$, the permutation induced by $1 + \lfloor 0, 3, 4 \rfloor_{\text{even}} + 2 + \lfloor 0, 3, 4 \rfloor_{\text{even}}$ is equivalent to that induced by $1212\cdots$, which yields a transition $1 \to 0 \to 3 \to 2 \to 1$ or $2 \to 3 \to 0 \to 1 \to 2$. Which transition realizes is determined by the initial state. The former realizes when the initial state is 1 or 3, and the latter realizes when the initial state is 2 or 0. In summary, the alternating emergence of 1 and 2 is inherited from site $n + 1$ to site $n$.

A single $\lfloor 0, 3, 4 \rfloor_{\text{odd}}$ induces permutation (12)(04), which switches the direction of the aforementioned global rotations, $1 \to 0 \to 3 \to 2 \to 1$ and $2 \to 3 \to 0 \to 1 \to 2$. Note that this switch still keeps the structure.

Then, the total permutation in a single period with length $5^n$ denoted by $\Sigma$ is computed as

$$\Sigma = (\pi_2 \pi_1)^p (12)(04)(\pi_2 \pi_1)^q \tag{37}$$

with $p + q = 5^{n-1}$. By setting the initial state of the period at just after (12)(04), the total permutation denoted by $\Sigma'$ reads

$$\Sigma' = (12)(04)(\pi_2 \pi_1)^{5^{n-1}} = (12)(04)(13)(02) = (01324) \tag{38}$$

which is a 5-cycle.

Type68-2; $(a, b, c, d, e) = (3, 4, 0, 1, 2)$: $(\pi_0 = \pi_1 = (14)(23),\ \pi_2 = (01234),\ \pi_3 = \pi_4 = (1234))$

The structure of this rule is $4 + \{\lfloor 0 \rfloor_{\text{even}}, \lfloor 0 \rfloor^1_{\text{odd}}\} + 1 + [23]^{\otimes k}$, where $[23]^{\otimes k}$ represents a repeat of 23 by $k$ times (i.e., 232323...).

| | |
|---|---|
| 1st cell | 0 1 2 3 4 0 1 2 3 4 0 1 2 3 4 0 1 2 3 4 0 1 2 3 4 |
| 2nd cell | 0 4 4 4 0 4 0 1 2 1 2 1 0 3 3 3 3 2 1 2 1 2 3 0 4 |
| 3rd cell | 0 4 0 4 0 4 0 4 4 4 4 4 4 0 4 0 4 0 3 2 1 0 3 3 3 |
| | 3 3 3 3 3 3 3 3 2 1 0 3 2 1 2 1 2 1 2 3 0 1 2 1 2 |
| | 1 2 1 2 1 2 1 2 3 0 1 2 3 3 3 3 3 3 0 1 2 3 0 4 0 |
| | 4 0 4 0 4 0 4 0 1 2 3 0 1 2 1 2 1 2 1 0 3 2 1 2 1 |
| | 2 1 2 1 2 1 2 1 0 3 2 1 0 4 0 4 0 4 4 4 4 4 4 0 4 |

**Fig. 24** The sequence of first, second, and third cells in Type62 with $(a,b)=(1,2)$. The sequence in the third cell is divided into 5 lines that go from the top to the bottom. We draw 1 and 2 in light yellow and $\lfloor 0,3,4 \rfloor^1_{\mathrm{odd}}$ in brown. As clearly seen, the alternation of 1 and 2 is kept anywhere. In addition, we can see that the direction of the global rotation, $1 \to 0 \to 3 \to 2 \to 1$ and $2 \to 3 \to 0 \to 1 \to 2$, switches when $\pi_{\lfloor 0,3,4 \rfloor^1_{\mathrm{odd}}}$ (i.e., brown region in site $n$) applies.

We first observe that $\lfloor 0 \rfloor_{\mathrm{even}}$ yields transitions $4 \to 1 \to 4$, $2 \to 3 \to 2$, and $0 \to 0 \to 0$, which satisfy the structure. In addition, $[23]^{\otimes k}$ yields transitions $4 \to 0 \to 0 \to 1 \to 2 \to 3 \to 4$ and $3 \to 4 \to 1 \to 2 \to 3$, which also satisfy the structure. Furthermore, $\pi_1\pi_4$ (which is obtained from the structure by removing $\lfloor 0 \rfloor_{\mathrm{even}}$) yields transitions $(0 \to)0 \to 0$, $(1 \to)2 \to 3$, $(2 \to)3 \to 2$, $(3 \to)4 \to 1$, and $(4 \to)1 \to 4$, all of which satisfy the structure. The above observation, together with the ergodicity shown below, confirms the inheritance of the structure from site $n$ to site $n+1$.

Note that $\pi_{\lfloor 0 \rfloor_{\mathrm{even}}} = \mathrm{id}$, $\pi_1\pi_4 = (13)$, and $\pi_{[23]} = (024)(13)$. By setting the initial state of a single period with length $5^n$ at just after $\lfloor 0 \rfloor^1_{\mathrm{odd}}$, the total permutation in a single period $\Sigma'$ is computed as

$$\begin{aligned}\Sigma' =& \pi_0\pi_4(\pi_1\pi_4)^{5^{n-1}-1}\pi_{[23]}^{5^{n-1}}\pi_1 = (13)(13)^{5^{n-1}-1}((024)(13))^{5^{n-1}}(14)(23) \\ =& \begin{cases} (024)(14)(23) = (02341) & n \text{ is odd,} \\ (042)(14)(23) = (04123) & n \text{ is even,} \end{cases}\end{aligned} \tag{39}$$

both of which are 5-cycles.

| | |
|---|---|
| 1st cell | 0 1 2 3 4 0 1 2 3 4 0 1 2 3 4 0 1 2 3 4 0 1 2 3 4 |
| 2nd cell | 0 0 0 1 2 3 2 3 4 1 2 3 2 3 4 1 4 1 2 3 4 1 4 0 0 |
| 3rd cell | 0 0 0 0 0 1 2 3 4 1 4 0 0 1 2 3 2 3 2 3 4 1 4 1 4 |
| | 1 4 1 4 1 2 3 4 1 2 3 4 1 2 3 4 1 2 3 4 1 2 3 4 1 |
| | 4 1 4 1 4 0 0 1 2 3 2 3 4 0 0 0 0 0 0 1 2 3 2 3 2 |
| | 3 2 3 2 3 4 1 2 3 4 1 2 3 4 1 2 3 4 1 2 3 4 1 2 3 |
| | 2 3 2 3 2 3 4 0 0 0 0 1 2 3 4 1 4 1 4 0 0 0 0 0 0 |

**Fig. 25** The sequence of first, second, and third cells in Type68-2 with $(a,b,c,d,e) = (3,4,0,1,2)$. The sequence in the third cell is divided into 5 lines that go from the top to the bottom. We draw 1 and 4 in yellow and $\lfloor 0 \rfloor^1_{\mathrm{odd}}$ in brown.

Type 67-3; $(a,b)=(3,4)$: ($\pi_0 = \pi_1 = \pi_2 = (12)(34)$, $\pi_3 = (01234)$, $\pi_4 = (02143)$)

The structure of this rule is $3 + * + 4 + *$ with $* = \{\lfloor 0,1,2 \rfloor_{\mathrm{even}}, \lfloor 0,1,2 \rfloor^1_{\mathrm{odd}}\}$.

We first observe that $\lfloor 0,1,2\rfloor_{\text{even}}$ induces transitions $1 \to 2 \to 1$, $3 \to 4 \to 3$, and $0 \to 0 \to 0$, all of which satisfy the structure. As its corollary, we find $\pi_{\lfloor 0,1,2\rfloor_{\text{even}}} = \text{id}$, and hence $3 + * + 4 + *$ with $* = \{\lfloor 0,1,2\rfloor_{\text{even}}, \lfloor 0,1,2\rfloor^1_{\text{odd}}\}$ induces the same permutation as $3434\cdots$, which yields a transition $0 \to 1 \to 4 \to 0 \to 2 \to 3 \to 0$, satisfying the structure ($\lfloor 0,1,2\rfloor_{\text{even}}$ can be inserted between two numbers, though it does not disturb the structure). In summary, the structure is inherited from site $n$ to site $n+1$.

By setting the initial state of the period at just after applying $\lfloor 0,1,2\rfloor^1_{\text{odd}}$, the total permutation $\Sigma'$ reads

$$\Sigma' = (12)(34)(042)^{5^{n-1}} = \begin{cases} (12)(34)(042) = (03412) & n \text{ is odd,} \\ (12)(34)(024) = (01234) & n \text{ is even,} \end{cases} \tag{40}$$

if the state after $\lfloor 0,1,2\rfloor^1_{\text{odd}}$ is 3, and

$$\Sigma' = (12)(34)(031)^{5^{n-1}} = \begin{cases} (12)(34)(031) = (04321) & n \text{ is odd,} \\ (12)(34)(013) = (02143) & n \text{ is even,} \end{cases} \tag{41}$$

if the state after $\lfloor 0,1,2\rfloor^1_{\text{odd}}$ is 4. All of these four permutations are 5-cycles.

| | |
|---|---|
| 1st cell | 0 1 2 3 4 0 1 2 3 4 0 1 2 3 4 0 1 2 3 4 0 1 2 3 4 |
| 2nd cell | 0 0 0 0 1 4 3 4 3 4 3 4 3 4 0 2 1 2 1 2 1 2 1 2 3 |
| 3rd cell | 0 0 0 0 0 0 2 3 0 1 4 0 2 3 0 0 0 0 0 0 0 0 0 0 0 |
| | 1 2 1 2 1 2 1 2 1 2 1 2 1 2 1 2 1 2 1 2 1 2 1 2 1 |
| | 2 1 2 1 2 1 4 0 2 3 0 1 4 0 2 1 2 1 2 1 2 1 2 1 2 |
| | 3 4 3 4 3 4 3 4 3 4 3 4 3 4 3 4 3 4 3 4 3 4 3 4 3 |
| | 4 3 4 3 4 3 0 1 4 0 2 3 0 1 4 3 4 3 4 3 4 3 4 3 4 |

**Fig. 26** The sequence of first, second, and third cells in Type67-3 with $(a,b) = (3,4)$. The sequence in the third cell is divided into 5 lines that go from the top to the bottom. We draw 3 and 4 in yellow and $\lfloor 0,1,2\rfloor^1_{\text{odd}}$ in brown.

### *6.3.3 B-3: Complex sequence appears with removing even length units*

Type45-1; $(a,b,c) = (0,1,2)$: ($\pi_0 = (12)(34)$, $\pi_1 = (012)$, $\pi_2 = (04)(12)$, $\pi_3 = \pi_4 = (12)$)

The structure of this rule is $[12]^{\otimes k} + 0 + \lfloor 4,3\rfloor_{\text{even}} + \{\lfloor 0\rfloor_{\text{odd}}, \lfloor 0\rfloor^1_{\text{even}}\}$.

We first observe that $[12]^{\otimes k}$ yields transition $(1 \to)2 \to 0 \to 4 \to 4 \to 0 \to 1$, which satisfies the above structure. In addition, $0 + \lfloor 4,3\rfloor_{\text{even}} + \lfloor 0\rfloor_{\text{odd}}$ yields transitions inside $\{1,2\}$ or inside $\{3,4\}$ or staying at 0, which also keeps the structure. A key fact is that $0 + \lfloor 4,3\rfloor_{\text{even}} + \lfloor 0\rfloor_{\text{odd}}$ contains an even number of 0's. Noting $\pi_{\lfloor 4,3\rfloor_{\text{even}}} = \text{id}$ and $(\pi_0)^2 = \text{id}$, we easily have $\pi_{[12]^{\otimes k}+0+\lfloor 4,3\rfloor_{\text{even}}+\lfloor 0\rfloor_{\text{odd}}} = \pi_{[12]^{\otimes k}+0\lfloor 0\rfloor_{\text{odd}}} = (\pi_2\pi_1)^k$.

by setting the initial state of the period at just after applying $\lfloor 0\rfloor^1_{\text{even}}$, the total permutation $\Sigma'$ for a single period with length $5^n$ reads

$$\Sigma' = \pi_0(\pi_2\pi_1)^{5^{n-1}} = (12)(34)(024)^{5^{n-1}} = \begin{cases} (01234) & n \text{ is odd,} \\ (03412) & n \text{ is even,} \end{cases} \tag{42}$$

both of which are 5-cycles.

Type69; $(a,b,c,d) = (1,2,3,4)$: ($\pi_0 = \pi_1 = (12)(34)$, $\pi_2 = \pi_4 = (01234)$, $\pi_3 = (04321)$)

The structure of this rule is $\{2,4\} + \{3, \lfloor 0,1\rfloor_{\text{odd}}, \lfloor 0,1\rfloor^1_{\text{even}}\}$, which means that (i) 2 or 4, and (ii) 3 or "odd length sequence with 0 and 1" (this length is even once in a single period) appear alternately.

1st cell 0 1 2 3 4 0 1 2 3 4 0 1 2 3 4 0 1 2 3 4 0 1 2 3 4

2nd cell 0 0 1 2 1 2 1 2 1 2 1 2 0 4 4 4 3 3 3 3 3 4 4 0 0

3rd cell 0 0 0 1 2 0 4 4 0 1 2 0 4 3 3 3 3 3 3 3 3 3 3 3 4
3 4 3 3 3 3 3 3 3 3 3 3 3 4 4 4 4 4 4 4 4 4 4 4 4
4 4 4 4 0 1 2 0 4 4 0 1 2 1 2 1 2 1 2 1 2 1 2 1 2
1 2 1 2 1 2 1 2 1 2 1 2 1 2 1 2 1 2 1 2 1 2 1 2 1
2 1 2 0 4 4 0 1 2 0 4 4 0 0 0 0 0 0 0 0 0 0 0 0 0

**Fig. 27** The sequence of first, second, and third cells in Type45-1 with $(a, b, c) = (0, 1, 2)$. The sequence in the third cell is divided into 5 lines that go from the top to the bottom. We draw units [12] in yellow, and $\lfloor 0 \rfloor^1_{\mathrm{even}}$ in brown.

Examples are 234341010141 and 21412320, while 21433320 is not an example, since the symbol after 3 must be 2 or 4.

We first observe that $\{2, 4\} + 3$ induces transitions $(3 \to)4 \to 3$, $(4 \to)0 \to 4$, $(0 \to)1 \to 0$, $(1 \to)2 \to 1$, and $(2 \to)3 \to 2$, which keep the structure. In addition, $\{2, 4\} + \lfloor 0, 1 \rfloor_{\mathrm{odd}}$ induces transitions $(3 \to)4 \to 1 \to 4 \to 1 \to \cdots$, $(1 \to)2 \to 3 \to 2 \to 3 \to \cdots$, and $(4 \to)0 \to 0 \to \cdots$, $(0 \to)1 \to 4 \to 1 \to \cdots$, $(2 \to)3 \to 2 \to 3 \to \cdots$, which also keep the structure. Thus, the structure is inherited from site $n$ to site $n + 1$.

In site $n$, the alternation of $\{2, 4\}$ and $\{3, \lfloor 0, 1 \rfloor_{\mathrm{odd}}, \lfloor 0, 1 \rfloor^1_{\mathrm{even}}\}$ appears $2 \cdot 5^{n-1}$ times in a single period with length $5^n$. Hence, in a single period, $\pi_3 \pi_{\{2,4\}}$ appears $5^{n-1}$ times (since 3 appears $5^{n-1}$ times in a single period), $\pi_{\lfloor 0,1 \rfloor_{\mathrm{odd}}} \pi_{\{2,4\}}$ appears $5^{n-1} - 1$ times, and $\pi_{\lfloor 0,1 \rfloor_{\mathrm{even}}} \pi_{\{2,4\}}$ appears once. With keeping in mind $\pi_{\{2,4\}+3} = \mathrm{id}$ and $\pi_{\{2,4\}+\lfloor 0,1 \rfloor_{\mathrm{odd}}} = (04)(13)$, the total permutation $\Sigma'$ in a single period with setting the initial state at just after $\lfloor 0, 1 \rfloor^1_{\mathrm{even}}$ is calculated as

$$\Sigma' = (\pi_{\lfloor 0,1 \rfloor_{\mathrm{even}}} \pi_{\{2,4\}})(\pi_{\lfloor 0,1 \rfloor_{\mathrm{odd}}} \pi_{\{2,4\}})^{5^{n-1}-1} = (01234)((04)(13))^{5^{n-1}-1} = (01234), \tag{43}$$

which is a 5-cycle.

1st cell 0 1 2 3 4 0 1 2 3 4 0 1 2 3 4 0 1 2 3 4 0 1 2 3 4

2nd cell 0 0 0 1 0 1 4 1 2 1 2 3 2 3 2 3 2 3 4 3 4 1 4 0 4

3rd cell 0 0 0 0 0 0 0 1 4 0 0 1 0 1 0 1 0 1 0 1 0 1 4 0 0
1 4 1 4 1 4 1 2 3 4 1 2 1 2 1 2 1 2 1 2 1 2 3 4 1
2 3 2 3 2 3 2 3 2 3 2 3 2 3 2 3 2 3 2 3 2 3 2 3 2
3 2 3 2 3 2 3 4 1 2 3 4 3 4 3 4 3 4 3 4 1 4 1 2 3
4 1 4 1 4 1 4 0 0 1 4 0 4 0 4 0 4 0 4 0 4 0 0 1 4

**Fig. 28** The sequence of first, second, and third cells in Type69 with $(a, b, c, d) = (1, 2, 3, 4)$. The sequence in the third cell is divided into 5 lines that go from the top to the bottom. We draw 2 and 4 in light gray and $\lfloor 0, 1 \rfloor_{\mathrm{even}}$ in dark gray.

## 6.4 Pattern C: Even cells and odd cells take specific states

The structure in pattern C takes the form of $[\{1, 3\} + \{2, 4\}]$, $[00]$, $\lfloor 0 \rfloor^1_{\mathrm{odd}}$ (and its permutation), i.e., the sequence is constructed by [12], [14], [32], [34], [00], and a single [0]. This happens when the transition map is drawn as a composition of triangle 014 and square 1234 sharing edge 14 (e.g., the transition map of Type63-2 in Fig. **??** in Section. 5.1). In addition, for $x \neq 0$, $\pi_b(x) \neq x$ is satisfied for any $b$.

In pattern C, the following conditions are satisfied in order to keep the structure and to satisfy ergodicity:

- $\pi_0(0) = 0$, $\pi_0(a) \neq a$ for $a \neq 0$, and $\pi_0\pi_0$ is the identity permutation.
- $\pi_2\pi_1$, $\pi_4\pi_1$, $\pi_2\pi_3$, and $\pi_4\pi_3$ commute with each other.
- The following permutation

$$\pi_0(\pi_2\pi_1)^p(\pi_4\pi_1)^q(\pi_2\pi_3)^q(\pi_4\pi_3)^p \tag{44}$$

  with $p + q = 5^{n-1}$ is a 5-cycle for any $p$.

We shall briefly explain why these conditions confirm ergodicity. Thanks to the structure, the total permutation $\Sigma$ is a product of $\pi_2\pi_1$, $\pi_4\pi_1$, $\pi_2\pi_3$, $\pi_4\pi_3$, and $\pi_0\pi_0 = \text{id}$ (except for a single $\pi_0$). Since these four permutations commute with each other, we have $\Sigma = \pi_0(\pi_2\pi_1)^p(\pi_4\pi_1)^q(\pi_2\pi_3)^q(\pi_4\pi_3)^p$ with $p + q = 5^{n-1}$, which is ergodic by assumption. The last constraint $p + q = 5^{n-1}$ comes from the fact that $\pi_1$, $\pi_2$, $\pi_3$, and $\pi_4$ respectively appear $5^{n-1}$ times in one period.

Type65-2; $(a, b, c, d) = (2, 0, 1, 3)$: ($\pi_0 = \pi_1 = (03)(12)$, $\pi_2 = (0123)$, $\pi_3 = (043)(12)$, $\pi_4 = (01)(23)$)

The structure is $[\{1, 3\} + \{0, 2\}]$, $[44]$, $\lfloor 4 \rfloor^1_{\text{odd}}$. Here, state 4 plays the role of state 0 in the above explanation. It is easy to confirm that the transition map takes the aforementioned form (a composition of a triangle and a square: see Fig. **??** in Section. 5.1) and the rule satisfies the non-staying condition ($\pi_b(x) \neq x$ for any $b$ and any $x \neq 4$).

We examine the aforementioned three conditions. First, $\pi_4 = (01)(23)$, which confirms the first condition; $\pi_4(4) = 4$. Second, $\pi_2\pi_1 = (13)$, $\pi_0\pi_1 = \text{id}$ (identity), $\pi_2\pi_3 = (13)(04)$, and $\pi_0\pi_3 = (04)$, all of which commute with each other, and hence the second condition is satisfied. Finally, by setting the state after $\lfloor 4 \rfloor^1_{\text{odd}}$ as the initial state, the permutation in a period denoted by $\Sigma'$ is calculated as

$$\Sigma' = \pi_4(\pi_2\pi_1)^p(\pi_0\pi_1)^q(\pi_2\pi_3)^q(\pi_0\pi_3)^p = \pi_4(13)^{5^{n-1}}(04)^{5^{n-1}} = (01)(23)(13)(04) = (04123), \tag{45}$$

which is a 5-cycle regardless of the value of $p$ and $q$.

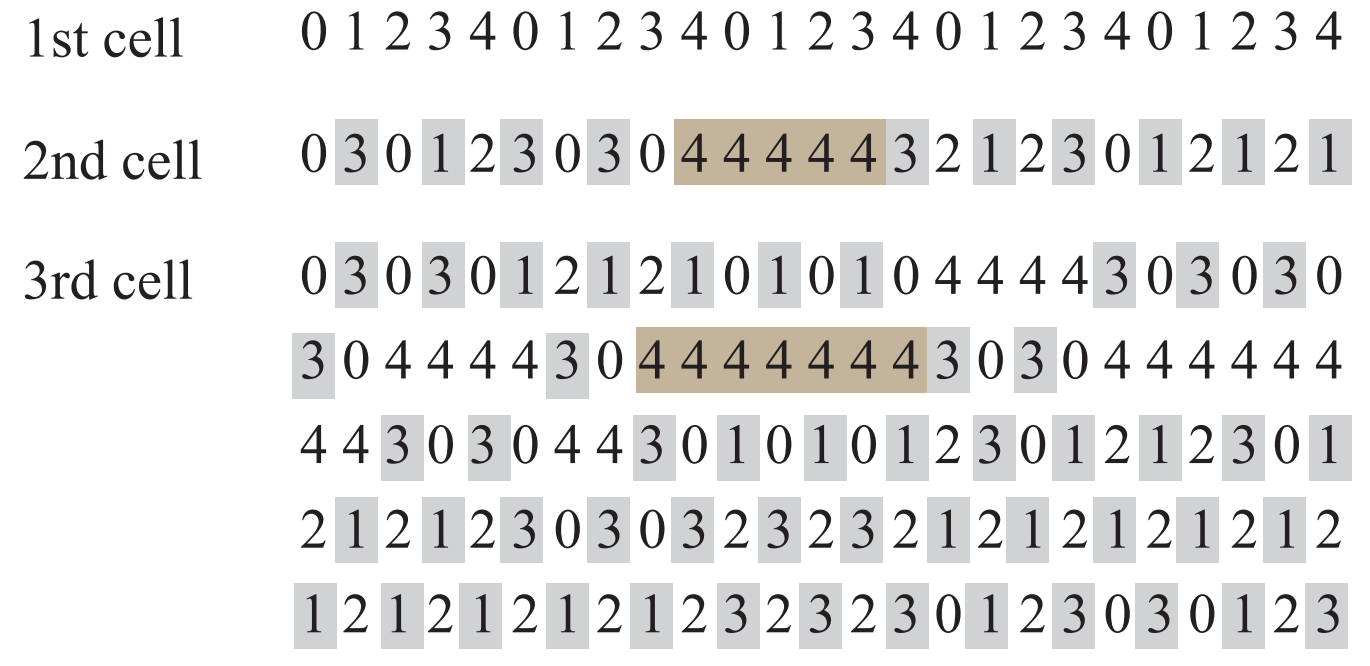

**Fig. 29** The sequence of the first, second, and third cells in Type65-2 with $(a, b, c, d) = (2, 0, 1, 3)$. The sequence in the third cell is divided into 5 lines that go from the top to the bottom. We draw 1 and 3 in light gray and $\lfloor 4 \rfloor^1_{\text{odd}}$ in brown.

## 6.5 Pattern D: Combination of patterns A and B

The structure of rules in pattern D takes the form of a combination of those in pattern A and pattern B. Namely, the structure consists of several islands, and some islands have units. In the case of pattern A, states inside an island can appear in any order. In contrast, in the case of pattern D, an island has some units and these states appear only in this order as for the case of pattern B.

### *6.5.1 D-1: Islands share a single state*

The rule in pattern D-1 has two islands which share a single state.

Type53-1; $(a, b) = (1, 2)$: $(\pi_0 = \pi_3 = (03)(12),\ \pi_1 = (12),\ \pi_2 = (012),\ \pi_4 = (034)(12))$
The structure of this rule is $(\lfloor 0,3\rfloor_{\text{even}}, [12])(\lfloor 0,3\rfloor_{\text{even}}, 4)$.

First, all of $\pi_{\lfloor 0,3\rfloor_{\text{even}}}$, $\pi_{\lfloor 0,3\rfloor_{\text{even}}}$ and $\pi_4$ induce transitions with structure $(\lfloor 0,3\rfloor_{\text{even}}, [12])$ if the present state is 0, 1 or 2, and $\pi_{\lfloor 0,3\rfloor_{\text{even}}}$ and $\pi_{[12]}$ induce transitions with structure $(\lfloor 0,3\rfloor_{\text{even}}, 4)$ if the present state is 3 or 4. In addition, $\pi_{[12]}$ induces transitions with structure $(\lfloor 0,3\rfloor_{\text{even}}, [12])$ if the present state is 0, 1, or 2, and $\pi_{[12]}$ induces transitions with structure $(\lfloor 0,3\rfloor_{\text{even}}, 4)$ if the present state is 3 or 4. Noting $\pi_{\lfloor 0,3\rfloor_{\text{even}}} = \text{id}$, we find that $\pi_{[12]}$ in $\pi_{(\lfloor 0,3\rfloor_{\text{even}}, [12])}$ can act on state 0 only when the initial state for $\pi_{(\lfloor 0,3\rfloor_{\text{even}}, [12])}$ is state 0. The above argument combined with the ergodicity shown below confirms that the structure is inherited from site $n$ to site $n+1$.

Since $\pi_{\lfloor 0,3\rfloor_{\text{even}}} = \text{id}$, by following a similar argument to pattern A, the total permutation in a single period with length $5^n$ reads

$$\Sigma = (\pi_4)^{5^{n-1}} (\pi_2\pi_1)^{5^{n-1}} = ((034)(12))^{5^{n-1}} (10)^{5^{n-1}} = \begin{cases} (034)(12)(10) & = (02134), \quad n \text{ is odd}, \\ (043)(12)(10) & = (02143), \quad n \text{ is even}, \end{cases} \tag{46}$$

both of which are 5-cycles.

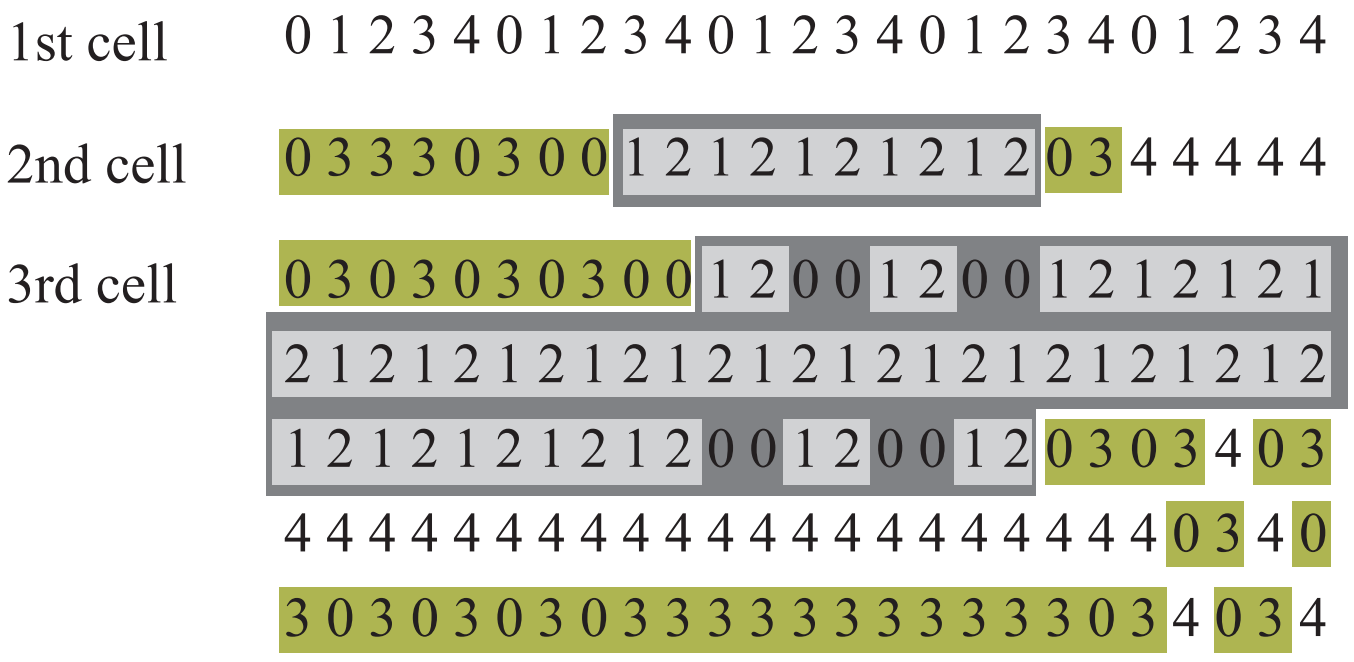

**Fig. 30** The sequence of the first, second, and third cells in Type53-1 with $(a, b) = (1, 2)$. The sequence in the third cell is divided into 5 lines that go from the top to the bottom. We draw island $(\lfloor 0,3\rfloor_{\text{even}}, [12])$ in dark gray, and [12] inside this island in light gray. In addition, we draw $\lfloor 0,3\rfloor_{\text{even}}$ in island $(\lfloor 0,3\rfloor_{\text{even}}, 4)$ in dark yellow.

Type44-1; $(a, b, c, d) = (1, 3, 2, 4)$: $(\pi_0 = \pi_2 = (13),\ \pi_1 = (132),\ \pi_3 = (013),\ \pi_4 = (04)(13))$
The structure of this rule is $([13], [00], [22], \lfloor 2\rfloor^1_{\text{odd}})(0, 4)$.

We first observe that state 4 appears only when the state after $\pi_{([13],[00],[22],\lfloor 2\rfloor^1_{\text{odd}})}$ is 0. Otherwise, the cell remains in $\{0, 1, 2, 3\}$ (the first island). In $\{0, 1, 2, 3\}$, permutations $\pi_{[13]}$, $\pi_{[00]}$, $\pi_{[22]}$, and single $\pi_2$ keep the structure $([13], [00], [22], \lfloor 2\rfloor^1_{\text{odd}})$, and $\pi_{(0,4)}$ also keeps the structure $([13], [00], [22], \lfloor 2\rfloor^1_{\text{odd}})$. In addition, in $\{0, 4\}$, $\pi_{(0,4)}$ induces a transition inside $\{0, 4\}$, which keeps the structure $(0, 4)$. Moreover, if the initial state is 4, all of the permutations $\pi_{[13]}$, $\pi_{[00]}$, $\pi_{[22]}$, and single $\pi_2$ does not change the state and the state remains at 4, which keeps the structure $(0, 4)$. The above observations, together with the ergodicity shown below, confirm the inheritance of the structure from site $n$ to site $n+1$.

We next consider the ergodicity. Since $(\pi_0)^2 = (\pi_2)^2 = \text{id}$, $\pi_{(0,4)} = (04)$, and $\pi_2 = (13)$, the total permutation in a single period with length $5^n$ reads

$$\Sigma = \pi_{(0,4)} \pi^q_{[13]} \pi_2 \pi^p_{[13]} = (04)((23)(01))^q (13)((23)(01))^p \tag{47}$$

with $p + q = 5^{n-1}$. Noticing that one of $((23)(01))^q$ or $((23)(01))^p$ is identity and the other is $(23)(01)$, we compute $\Sigma$ as

$$\Sigma = \begin{cases} (04)((23)(01))(13) = (01234), & q \text{ is odd}, \\ (04)(13)((23)(01)) = (03214), & q \text{ is even}, \end{cases} \tag{48}$$

both of which are 5-cycles.

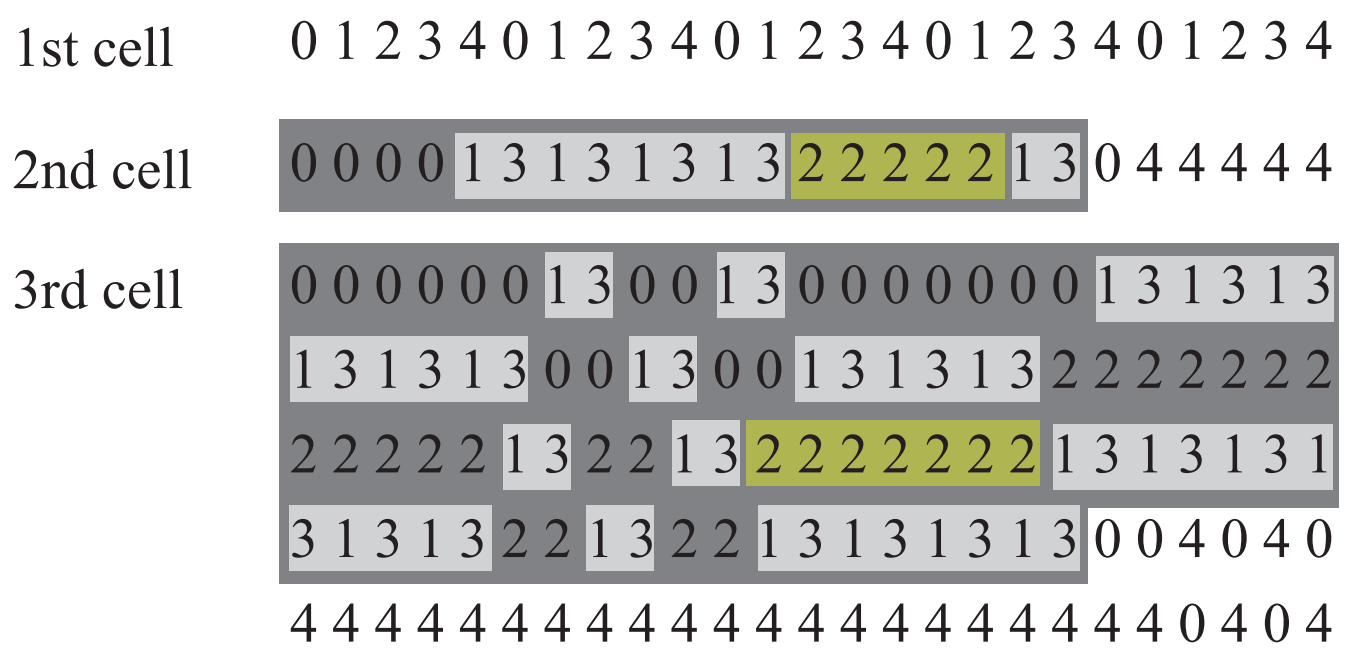

**Fig. 31** The sequence of first, second, and third cells in Type44-1 with $(a,b,c,d) = (1,3,2,4)$. The sequence in the third cell is divided into 5 lines that go from the top to the bottom. We draw island $([13],[00],[22],\lfloor 2\rfloor^1_{\text{odd}})$ in dark gray, and [13] and $\lfloor 2\rfloor^1_{\text{odd}}$ inside this island in light gray and yellow, respectively.

### *6.5.2 D-2: Islands share multiple states*

The rule in pattern D-2 has two islands which share two or more states. Since islands have multiple overlaps, the analyses become much more complicated than those for pattern D-1.

<u>Type20-1; $(a,b,c,d) = (1,2,4,3)$: $(\pi_0 = \pi_2 = \pi_4 = (04),\ \pi_1 = (012)(34),\ \pi_3 = \text{id})$</u>

The structure of this rule is $(1, \lfloor 0,2,4\rfloor_{\text{even}})(0,3,4)$. The first island consists of 1 and $\lfloor 0,2,4\rfloor_{\text{even}}$.

Noticing $\pi^2_{\lfloor 0,2,4\rfloor_{\text{even}}} = \text{id}$, we find $\pi_{(1,\lfloor 0,2,4\rfloor_{\text{even}})} = \pi_1^{5^{n-1}} = (012)^{5^{n-1}}(34)$. In addition, if the present state is 0, 1, or 2, the transition induced by $\pi_{(1,\lfloor 0,2,4\rfloor_{\text{even}})}$ satisfies the structure of $(1,\lfloor 0,2,4\rfloor_{\text{even}})$, and if the present state is 3 or 4, the transition induced by $\pi_{(1,\lfloor 0,2,4\rfloor_{\text{even}})}$ satisfies the structure of $(0,3,4)$. Moreover, since $\pi_3$ is trivial, and $\pi_0$ and $\pi_4$ act nontrivially only on 0 and 4, $\pi_{(0,3,4)}$ induces no transition if the present state is 1, 2, or 3. The overall observation with combining ergodicity confirms that the structure is inherited from site $n$ to site $n+1$.

Since the total number of state 2 appearing in the island $(1,\lfloor 0,2,4\rfloor_{\text{even}})$ is odd $(= 5^{n-1})$, the total number of states 0 and 4 appearing in the island $(1,\lfloor 0,2,4\rfloor_{\text{even}})$ is also odd, which implies the total number of states 0 and 4 in the second island $(0,3,4)$ is odd, leading to $\pi_{(0,3,4)} = (04)$. Thus, the total permutation in a single period with length $5^n$ reads

$$\Sigma = \pi_{(0,3,4)}(\pi_1)^{5^{n-1}} = \begin{cases} (04)(012)(34) = (01243) & n \text{ is odd,} \\ (04)(021)(34) = (02143) & n \text{ is even,} \end{cases} \tag{49}$$

both of which are 5-cycles.

<u>Type43-1: $(\pi_0 = \pi_2 = \pi_3 = (23),\ \pi_1 = (012)(34),\ \pi_4 = (04)(23))$</u>

The structure of this rule is $(1,\lfloor 0,2,3\rfloor_{\text{even}})(0,2,3,4)$.

Since $\pi_0 = \pi_2 = \pi_3 = (23)$, we find $\pi_{\lfloor 0,2,3\rfloor_{\text{even}}} = \text{id}$. In addition, $\pi_1 = (012)(34)$ keeps the structure $(1,\lfloor 0,2,3\rfloor_{\text{even}})$ for initial states 0, 1, or 2, and keeps the structure $(0,2,3,4)$ for initial states 3 or 4. Moreover, $\pi_{(0,2,3,4)}$ (i.e., $\pi_4 = (04)(23)$ and $\pi_0 = \pi_2 = \pi_3 = (23)$) also keeps the structure $(1,\lfloor 0,2,3\rfloor_{\text{even}})$ for initial states 1, 2, or 3, and keeps the structure $(0,2,3,4)$ for initial states 0 or 4. The above observations, together with ergodicity, confirm the inheritance of the structure from site $n$ to site $n+1$.

We note that the total number of entries 0, 2, and 3 in the island $(0,2,3,4)$ is odd, which follows from the fact that the total number of 0, 2, 3 is odd $(= 3\cdot 5^{n-1})$ and the first island $(1,\lfloor 0,2,3\rfloor_{\text{even}})$ has even number entries. With keeping in mind the fact that $\pi_4$ commutes with $\pi_0$, $\pi_2$, and $\pi_3$, we find that the total permutation in a single period with length $5^n$ reads

$$(23)(\pi_4)^{5^{n-1}}(\pi_1)^{5^{n-1}} = \begin{cases} (23)(04)(23)(012)(34) = (01243) & n \text{ is odd}, \\ (23)(04)(23)(021)(34) = (02143) & n \text{ is even}, \end{cases} \tag{50}$$

both of which are 5-cycles.

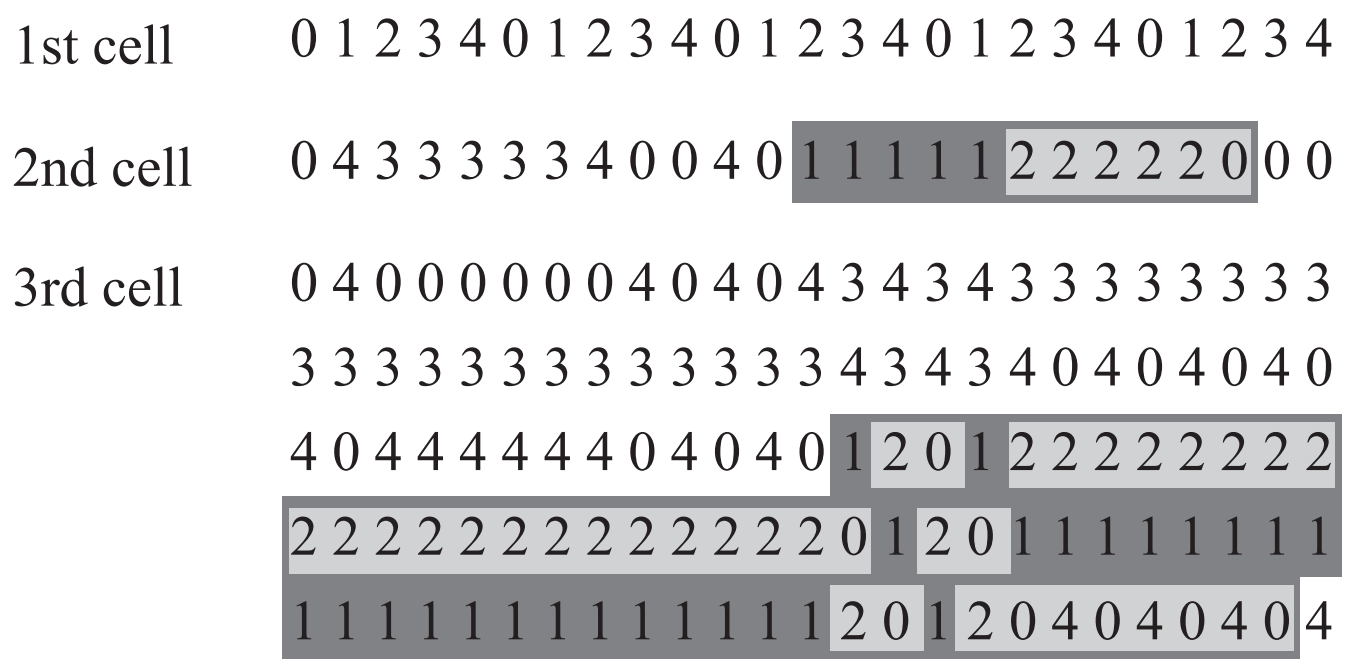

**Fig. 32** The sequence of first, second, and third cells in Type20-1 with $(a, b, c) = (1, 2, 4)$. The sequence in the third cell is divided into 5 lines that go from the top to the bottom. We draw island $(1, \lfloor 0, 2, 4 \rfloor_{\text{even}})$ in dark gray, and $\lfloor 0, 2, 4 \rfloor_{\text{even}}$ inside this island in light gray.

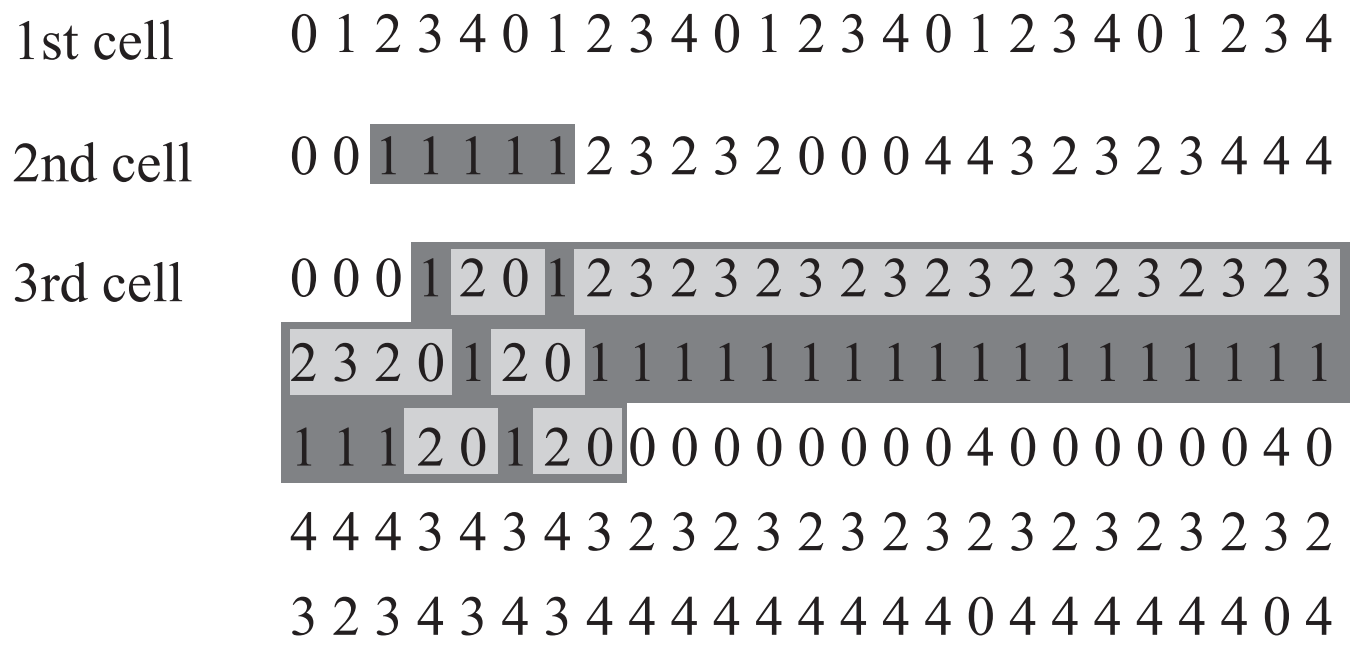

**Fig. 33** The sequence of first, second, and third cells in Type43-1. The sequence in the third cell is divided into 5 lines that go from the top to the bottom. We draw island $(1, \lfloor 0, 2, 3 \rfloor_{\text{even}})$ in dark gray, and $\lfloor 0, 2, 3 \rfloor_{\text{even}}$ inside this island in light gray.

Type27-1; $(a, b, c, d, e) = (2, 0, 1, 4, 3)$: ($\pi_0 = \pi_1 = (01)(23)$, $\pi_2 = \text{id}$, $\pi_3 = (04)$, $\pi_4 = (0123)$)

The structure of this rule is $(\lfloor 0, 1 \rfloor_{\text{even}}, \lfloor 2, 3 \rfloor_{\text{even}})(0, 1, 4)$.

We first observe that $\lfloor 0, 1 \rfloor_{\text{even}}$ yields transitions $2 \to 3 \to 2$, $0 \to 1 \to 0$, and $4 \to 4 \to 4$, which keep the structure and implies $\pi_{\lfloor 0,1 \rfloor_{\text{even}}} = \text{id}$. In addition, $\pi_2$ and $\pi_3$ act nontrivially only on states 0 and 4, which implies that if the present state is 0 or 4, this keeps the structure of $(0, 1, 4)$, and if the present state is 1, 2, or 3, this keeps the structure of $(\lfloor 0, 1 \rfloor_{\text{even}}, \lfloor 2, 3 \rfloor_{\text{even}})$. We also find $\pi_{(\lfloor 0,1 \rfloor_{\text{even}}, \lfloor 2,3 \rfloor_{\text{even}})} = (04)^{5^{n-1}} = (04)$. Further, $\pi_{(0,1,4)}$ induces a transition consisting of $\lfloor 0, 1 \rfloor_{\text{even}}$ and $\lfloor 2, 3 \rfloor_{\text{even}}$, which keeps the structure $(\lfloor 0, 1 \rfloor_{\text{even}}, \lfloor 2, 3 \rfloor_{\text{even}})$. In addition, since $\pi_{(0,1,4)}$ consists of odd number of permutations, $\pi_{(0,1,4)}$ maps $\{0, 2\}$ to $\{1, 3\}$, and vice versa.

We can compute $\pi_{(0,1,4)}$ in a similar manner to the treatment for pattern A-2 (See the case of Type27-1 with $(a, b, c, d, e) = (0, 1, 2, 3, 4)$). For brevity, we rename $\pi_0 = \pi_1$ as $\pi_f$ and regard the sequence in $(0, 1, 4)$ in site $n$ as a sequence of $f$ and 4 by identifying 0 and 1 to $f$. The island $(0, 1, 4)$ is also denoted by $(f, 4)$. Here, $|(f, 4)| = |(0, 1, 4)|$ is an odd number because the number of 4 in $(f, 4)$ is odd $(= 5^{n-1})$ and the number of $f$ in $(f, 4)$ is even. The latter follows from the fact that the total number of $f$ in $(\lfloor 0, 1 \rfloor_{\text{even}}, \lfloor 2, 3 \rfloor_{\text{even}})(0, 1, 4)$ is even $(= 2 \cdot 5^{n-1})$ and $(\lfloor 0, 1 \rfloor_{\text{even}}, \lfloor 2, 3 \rfloor_{\text{even}})$ contains an even number of $f$.

Suppose that the present state is 0, 1, 2, or 3. Setting $X := \{0,2\}$ and $Y := \{1,3\}$, we find that both $\pi_f$ and $\pi_4$ map states in $X$ to states in $Y$, and vice versa. Hence, $\pi_{(0,1,4)}$ provides a sequence where states in $X$ and states in $Y$ appear alternately. Let us consider transitions by $\pi_{(f,4)} = \pi_{(0,1,4)}$ in two periods with length $2 \cdot |(f,4)|$. With noting that $|(f,4)|$ is odd, if a $\pi_f$ in the first period acts on a state in $X$, then the same $\pi_f$ in the second period acts on a state in $Y$, and vice versa. Since $\pi_f$ acts as $+1 \mod 4$ on $X$ and as $-1 \mod 4$ on $Y$, the total action of $\pi_f$ through two periods is zero. From this observation, $(\pi_{(0,1,4)})^2 = (\pi_4)^{2\cdot 5^{n-1}} = (0123)^{2\cdot 5^{n-1}} = (02)(13)$ is obtained, which leads to $\pi_{(0,1,4)} = (0123)$ or $(0321)$. The above series of observations confirm that the structure is inherited from site $n$ to site $n+1$.

Recalling $\pi_{(\lfloor 0,1\rfloor_{\mathrm{even}},\lfloor 2,3\rfloor_{\mathrm{even}})} = (04)$, we compute the total permutation in a single period with length $5^n$ as

$$\Sigma = \begin{cases} (0123)(\pi_2)^{5^{n-1}}(\pi_3)^{5^{n-1}} = (0123)(04) = (04123), & \text{or} \\ (0321)(\pi_2)^{5^{n-1}}(\pi_3)^{5^{n-1}} = (0321)(04) = (04321), \end{cases} \tag{51}$$

both of which are 5-cycles.

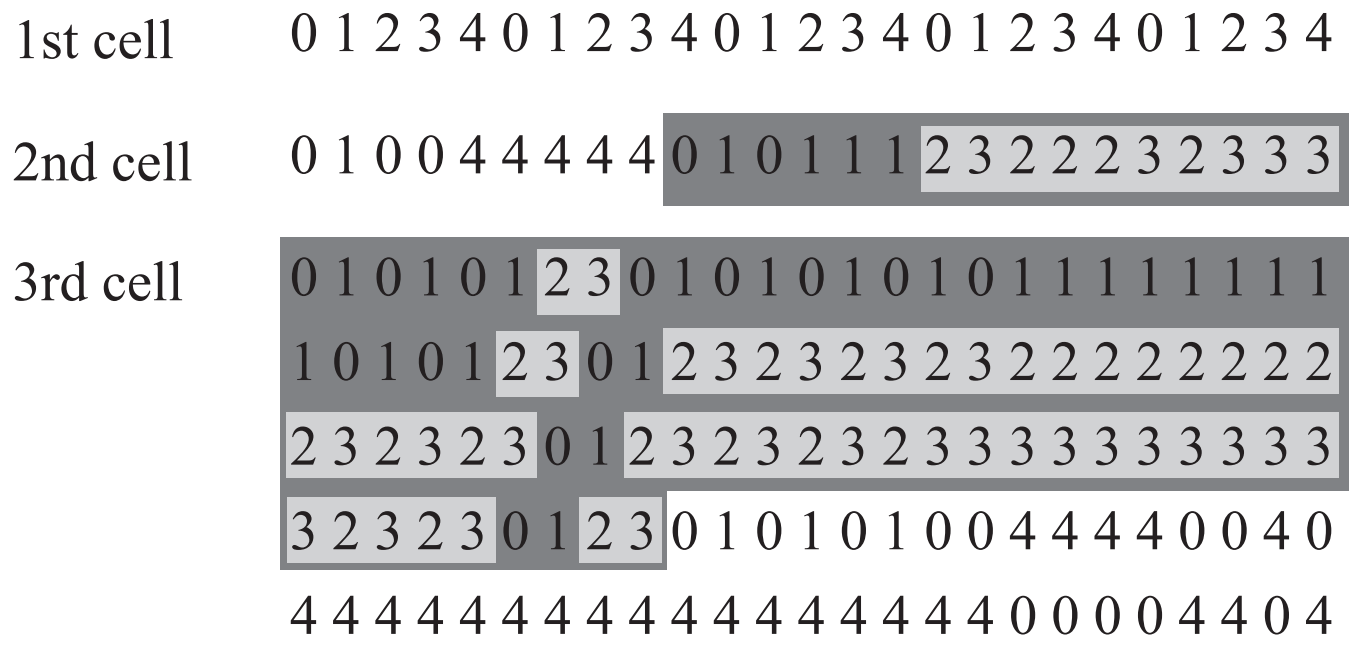

**Fig. 34** The sequence of first, second, and third cells in Type27-1 with $(a,b,c,d,e) = (2,0,1,4,3)$. The sequence in the third cell is divided into 5 lines that go from the top to the bottom. We draw island $(\lfloor 0,1\rfloor_{\mathrm{even}}, \lfloor 2,3\rfloor_{\mathrm{even}})$ in dark gray, and $\lfloor 2,3\rfloor_{\mathrm{even}}$ inside this island in light gray.

Remark that we need additional care in the case with $c = 4$ (e.g., $(a,b,c,d,e) = (2,1,4,0,3)$). In this case, the structure is the same, $(\lfloor 0,1\rfloor_{\mathrm{even}}, \lfloor 2,3\rfloor_{\mathrm{even}})(0,1,4)$, while to demonstrate the ergodicity we cannot analyze the action of $\pi_0$ and $\pi_1$ in these two island not separately but together.

<u>Type54-3; $(a,b,c,d) = (4,0,2,3)$: $(\pi_0 = (0123),\ \pi_1 = \pi_4 = (01)(23),\ \pi_2 = (23),\ \pi_3 = (04)(23))$</u>

The structure of this rule is $([23],[01],[11])([00],[01],[44],\lfloor 4\rfloor^1_{\mathrm{odd}})$.

We first examine the action of the first island ([23], [01], [11]). If the present state is 1, 2, or 3, the unit [23] yields transitions $2 \to 3 \to 2$ and $1 \to 1 \to 1$, which keep the structure ([23], [01], [11]). In addition, units [01] and [11] yield transitions $(2 \to)3 \to 2$ and $3 \to 0 \to 1 \to 2 \to 3$, which also keep the structure ([23], [01], [11]). If the present state is 0 or 4, both [01] and [11] yields transitions $(0 \to)1 \to 0$ and $(4 \to)4 \to 4$, while [23] yields transition $(0 \to)0 \to 4$ and $(4 \to)4 \to 0$, which keep the structure $([00],[10],[44],\lfloor 4\rfloor^1_{\mathrm{odd}})$.

We next examine the action of the second island $([00],[01],[44],\lfloor 4\rfloor^1_{\mathrm{odd}})$. If the state is 4, these units provide no transition and the cell remains at 4. If the state is 0, 1, 2, or 3, the unit [00] yields transition $0 \to 1 \to 2 \to 3 \to 0$, and [44] yields transitions $0 \to 1 \to 0$ and $2 \to 3 \to 2$. The unit [01] yields transition $0 \to 1 \to 2 \to 3 \to 0$ if the state is 1 or 3, and yields transitions $0 \to 1 \to 0$ and $2 \to 3 \to 2$ if the state is 0 or 2. All of these transitions keep the structure.

Notice that $\pi_{[23]} = (04)$, $\pi_{[01]} = (13)$, $\pi_{[00]} = (02)(13)$, and $\pi_{[11]} = \pi_{[44]} = \mathrm{id}$, which imply commutativity between $\pi_{[01]}$ and $\pi_{[00]}$ and that between $\pi_{[23]}$ and $\pi_{[01]}$. By setting the initial state of the period at just after applying $\lfloor 4\rfloor^1_{\mathrm{odd}}$, the total permutation $\Sigma'$ for a single period with length $5^n$ reads

$$\Sigma' = \pi_4(\pi^a_{[01]}\pi^b_{[00]})(\pi^{5^{n-1}}_{[23]}\pi^e_{[01]})(\pi^c_{[01]}\pi^d_{[00]}) = (01)(23)((02)^b(13)^{a+b})((04)^{5^{n-1}}(13)^e)((02)^d(13)^{c+d}) \tag{52}$$

with constraint $a+c+e+2b+2d=5^{n-1}$. Here, the last constraint comes from the total number of 0 in one period. Due to this constraint, we further compute the right-hand side as

$$\Sigma' = (01)(23)(13)^{5^{n-1}-b-d}(02)^b(04)(02)^d = \begin{cases} (04123) & b,d \text{ is even}, \\ (01243) & b,d \text{ is odd}, \\ (04321) & b \text{ is odd}, \ d \text{ is even}, \\ (03241) & b \text{ is even}, \ d \text{ is odd}. \end{cases} \tag{53}$$

All of the above permutations are 5-cycles.

| | |
|---|---|
| 1st cell | 0 1 2 3 4 0 1 2 3 4 0 1 2 3 4 0 1 2 3 4 0 1 2 3 4 |
| 2nd cell | 0 1 0 0 4 4 4 4 4 0 1 2 3 2 3 2 3 2 3 2 3 0 1 1 1 |
| 3rd cell | 0 1 0 1 2 3 2 3 2 3 0 1 1 1 1 1 1 1 1 1 1 1 2 3 2 |
| | 3 0 1 2 3 2 3 2 3 2 3 2 3 2 3 2 3 2 3 2 3 2 3 2 3 |
| | 2 3 2 3 0 1 0 1 0 1 2 3 2 3 2 3 2 3 2 3 2 3 0 1 0 |
| | 1 2 3 0 1 0 1 0 1 0 1 0 0 4 4 0 0 4 4 0 0 4 4 4 4 |
| | 4 4 4 4 4 4 4 4 4 4 4 4 4 0 0 4 4 0 0 4 4 0 1 0 1 |

**Fig. 35** The sequence of first, second, and third cells in Type54-3 with $(a,b,c,d)=(4,0,2,3)$. The sequence in the third cell is divided into 5 lines that go from the top to the bottom. We draw island ([23], [01], [11]) in dark gray, and units [01], [00], and $\lfloor 4 \rfloor^1_{\mathrm{odd}}$ in light gray, yellow, and brown, respectively.

### *6.5.3 D-2-2: D-2 with switching*

Rules in Pattern D-2-2 are essentially the same as those in Pattern D-2, while units at the edge of an island do not satisfy the structure owing to a smooth connection between two islands. We express units suffering this violation by the star symbol $*$.

Type48-4; $(a,b)=(1,4)$: $(\pi_0=\pi_1=(14)(23),\ \pi_2=\pi_3=(14),\ \pi_4=(01234))$

The structure of this rule is $([41]^*, \lfloor 2,3 \rfloor_{\mathrm{even}})([14]^*, [00], \lfloor 0 \rfloor^1_{\mathrm{odd}})$, where $[41]^*$ and $[14]^*$ means that at the edge of the islands these two units merge as 414 or 141 (see two green regions in Fig. 36). Another expression of this rule is $([41], \lfloor 2,3 \rfloor_{\mathrm{even}})4([14], [00], \lfloor 0 \rfloor^1_{\mathrm{odd}})1$, where 4 and 1 represent a single state 4 and 1 which do not belong to islands. Except for this point, we can treat rules in Pattern D-2-2 similar to those in Pattern D-2.

We first see that $\pi_{[41]}$ induces transitions $0\to1\to4$ and $4\to0\to0$, keeping the structure $([14]^*, [00], \lfloor 0 \rfloor^1_{\mathrm{odd}})$, and $1\to2\to3$, $2\to3\to2$, $3\to4\to1$, keeping the structure $([41]^*, \lfloor 2,3 \rfloor_{\mathrm{even}})$. Similarly, $\pi_{[14]}$ induces transitions $0\to0\to1$ and $1\to4\to0$, keeping the structure $([14]^*, [00], \lfloor 0 \rfloor^1_{\mathrm{odd}})$, and $2\to3\to4$, $3\to2\to3$, and $4\to1\to2$, keeping the structure $([41]^*, \lfloor 2,3 \rfloor_{\mathrm{even}})$. In addition, $\pi_{\lfloor 2,3 \rfloor_{\mathrm{even}}}$ induces transitions $1\to4\to1\to4\cdots$ and other states are kept unchanged, and $\pi_{[00]}$ induces transitions $1\to4\to1\to4\cdots$, $2\to3\to2\to3\cdots$, and $0\to0\to0\cdots$, both of which keep the structure. It is worth clarifying what happens with the odd length of $[41]^*$ and $[14]^*$. To see this, we consider the action of $\pi_4\pi_{([41],\lfloor 2,3 \rfloor_{\mathrm{even}})}\pi_1$. If the present state is 0, $\pi_4\pi_{([41],\lfloor 2,3 \rfloor_{\mathrm{even}})}\pi_1$ generates ([14], [00]). If the present state is 2 or 4, $\pi_4\pi_{([41],\lfloor 2,3 \rfloor_{\mathrm{even}})}\pi_1$ generates $([41]^*, \lfloor 2,3 \rfloor_{\mathrm{even}})$. If the present state is 3, $\pi_4\pi_{([41],\lfloor 2,3 \rfloor_{\mathrm{even}})}\pi_1$ generates $\lfloor 2,3 \rfloor_{\mathrm{even}}$. Care is needed if the present state is 1, where $\pi_4\pi_{([41],\lfloor 2,3 \rfloor_{\mathrm{even}})}\pi_1$ generates 4([14], [00])1.

With noting $\pi_{\lfloor 2,3 \rfloor_{\mathrm{even}}}=\mathrm{id}$ and $\pi_{[00]}=\mathrm{id}$, the total permutation $\Sigma$ for a single period with length $5^n$ is computed as

$$\Sigma = \pi_1(\pi_{[14]})^p\pi_0(\pi_{[14]})^q\pi_4(\pi_{[41]})^r, \tag{54}$$

where $p+q+r=5^{n-1}-1$. The $-1$ in the right-hand side compensates single $\pi_1$ and $\pi_4$ at the edge of the islands, which are neither included in [14] nor [41]. By setting the initial state of the period at just after $\lfloor 0\rfloor^1_{\text{odd}}$, the total permutation $\Sigma'$ reads

$$\Sigma' = \pi_0(\pi_{[14]})^q\pi_4(\pi_{[41]})^r\pi_1(\pi_{[14]})^p = \pi_0(\pi_{[14]})^{5^{n-1}} = (14)(23)((01)(24))^{5^{n-1}} = (14)(23)(01)(24) = (04321), \tag{55}$$

which is a 5-cycle.

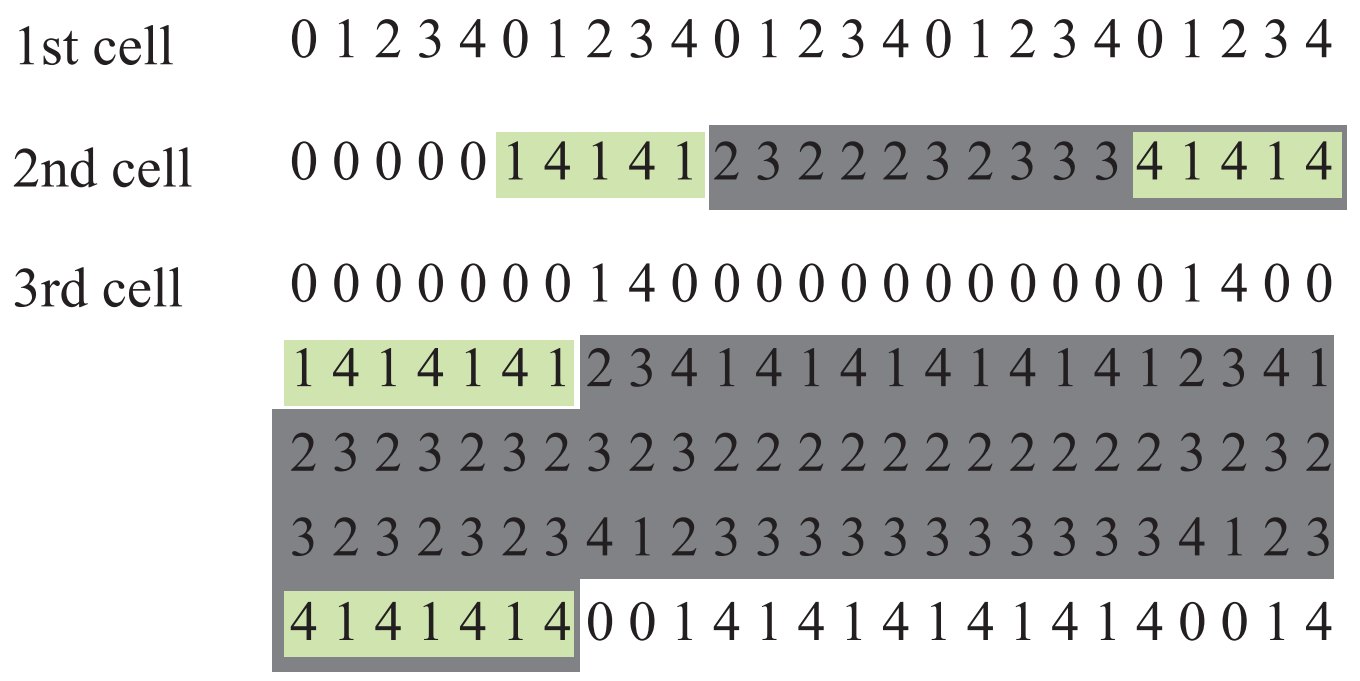

**Fig. 36** The sequence of first, second, and third cells in Type48-4 with $(a,b)=(1,4)$. The sequence in the third cell is divided into 5 lines that go from the top to the bottom. We draw island $([41]^*, \lfloor 2,3\rfloor_{\text{even}})$ in dark gray, and two joint parts of $[41]^*$ and $[14]^*$ at the edges of islands in light green.

Type50-1; $(a,b,c)=(4,2,1)$: $(\pi_0=\pi_3=(12),\ \pi_1=(043)(12),\ \pi_2=(03)(12),\ \pi_4=(0123))$

The structure of this rule is $([12], \lfloor 0,3\rfloor^*_{\text{even}})(4, \lfloor 0,3\rfloor^*_{\text{even}})$, where $\lfloor 0,3\rfloor^*_{\text{even}}$ means that at the edge of the island the unit $\lfloor 0,3\rfloor_{\text{even}}$ takes odd length (see the green regions in Fig. 37).

We first see that $\pi_{[12]}$ and $\pi_{\lfloor 0,3\rfloor_{\text{even}}}$ induce transitions $1\to 2\to 1\to 2\cdots$ for initial states 1 and 2, and $3\to 0\to 3$ (former) or $3\to 3\to 3$ (latter) for initial state 3, which keep the structure $([12], \lfloor 0,3\rfloor^*_{\text{even}})$, and $4\to 3\to 0$ and $0\to 4\to 4$ (former) or $4\to 4\to 4$ and $0\to 0\to 0$ (latter) for initial states 4 and 0, which keep the structure $(4, \lfloor 0,3\rfloor^*_{\text{even}})$. In addition, $\pi_4=(0123)$ induces transition $0\to 1\to 2\to 3\to 0$ and $4\to 4\to\cdots$, which keep the structure $([12], \lfloor 0,3\rfloor^*_{\text{even}})$ for initial states 0, 1, 2, and 3, and keep the structure $(4, \lfloor 0,3\rfloor^*_{\text{even}})$ for initial state 4. A single $\pi_0$ or $\pi_3$ plays the role of adding a single 0 or 3 if the present state is 0 or 3, which results in an odd length of the sequence of 0 and 3 in site $n+1$. We note that a single $\pi_0$ or $\pi_3$ keeps the structure [12]. The reason is as follows. Consider an extra $\pi_0$ or $\pi_3$ in $\pi_{(4,\lfloor 0,3\rfloor^*_{\text{even}})}$ at the edge of the island. If the present state is 1 or 2, the extra $\pi_0$ or $\pi_3$ adds a single 2 or 1. Here, we notice that if the state is 1 or 2, $\pi_{([12],\lfloor 0,3\rfloor^*_{\text{even}})}$ induces the transition only with alternating 1 and 2. Then, this sequence faces another extra $\pi_0$ or $\pi_3$ in $\pi_{([12],\lfloor 0,3\rfloor^*_{\text{even}})}$ at the edge of the island, and a single 2 or 1 is added again. In summary, a single 1 or 2 is added twice, and now the sequence of [12] (alternating 1 and 2) does not have any inconsistency.

With noting that $\pi_{[12]}=(04)$, $\pi_{\lfloor 0,3\rfloor_{\text{even}}}=\text{id}$, and a single 0 or 3 induces $\pi_0=\pi_3=(12)$, the total permutation $\Sigma$ for a single period with length $5^n$ is computed as

$$\Sigma = \pi_{\{0,3\}}(\pi_4)^{5^{n-1}}\pi_{\{0,3\}}(\pi_{[12]})^{5^{n-1}} = (12)((0123))^{5^{n-1}}(12)(04)^{5^{n-1}} = (12)(0123)(12)(04) = (04213), \tag{56}$$

which is a 5-cycle.

*6.5.4 D-2-3: D-2 with an island close to C*

The rules in Pattern D-2-3 are similar to those in Pattern D-2. The difference lies in the fact that the structure in an island is not close to Pattern B but to Pattern C. In fact, the shape of the transition map in Pattern D-2-3 is the same as that in Pattern C. The difference between Pattern C and Pattern D-2-3 lies in the fact that in the structure of Pattern C five elements $[\{1,3\}+\{2,4\}]$ and [00] can appear at any place, while in the structure of Pattern D-2-3 these five elements follow some order and cannot appear freely.

| | |
|---|---|
| 1st cell | 0 1 2 3 4 0 1 2 3 4 0 1 2 3 4 0 1 2 3 4 0 1 2 3 4 |
| 2nd cell | 0 0 4 4 4 4 4 3 0 0 1 2 1 2 1 2 1 2 1 2 3 3 0 3 3 |
| 3rd cell | 0 0 0 1 2 3 0 1 2 1 2 1 2 1 2 1 2 1 2 1 2 1 2 1 2 |
| | 1 2 1 2 3 0 1 2 1 2 1 2 1 2 1 2 1 2 1 2 1 2 1 2 1 |
| | 2 1 2 3 0 1 2 3 3 3 3 0 3 0 3 0 3 0 3 0 3 3 3 3 3 |
| | 3 3 3 0 1 2 3 0 0 0 0 4 4 3 0 4 4 3 0 4 4 4 4 4 4 |
| | 4 4 4 4 4 4 4 4 4 4 4 3 0 4 4 3 0 4 4 3 0 0 0 0 0 |

**Fig. 37** The sequence of first, second, and third cells in Type50-1 with $(a, b, c) = (4, 2, 1)$. The sequence in the third cell is divided into 5 lines that go from the top to the bottom. We draw island $([12], \lfloor 0, 3\rfloor^*_{\text{even}})$ in dark gray, and two joint parts of $\lfloor 0, 3\rfloor^*_{\text{even}}$ at the edges of islands in light green.

Type68-5; $(a, b, c, d, e) = (0, 1, 3, 4, 2)$: $(\pi_0 = \pi_1 = (1432),\ \pi_2 = (01234),\ \pi_3 = \pi_4 = (14)(23))$

The structure of this rule is $(\{1, 3\} + \{2, 4\})([14], [00]\lfloor 0\rfloor^1_{\text{odd}})$.

Due to the shape of the transition map, it is easy to see that all possible units appearing in the sequence are $[\{1, 3\} + \{2, 4\}]$ and $[00]$ (and a single $\lfloor 0\rfloor^1_{\text{odd}}$), which keeps the structure on units. In addition, $\pi_{([14],[00]\lfloor 0\rfloor^1_{\text{odd}})}$ yields nontrivial transitions only inside (1,2,3,4) satisfying the structure $(\{1, 3\}+\{2, 4\})$, and state 0 is kept unchanged, which satisfies the structure $([14], [00]\lfloor 0\rfloor^1_{\text{odd}})$. Moreover, direct computation ensures that $\pi_{(\{1,3\}+\{2,4\})}$ yields transitions inside $(0, 1, 4)$ if the initial state is 0 or 1, which satisfies the structure $([14], [00]\lfloor 0\rfloor^1_{\text{odd}})$, and it yields transitions inside (1,2,3,4) if the initial state is 2, 3, or 4, which keeps the structure $(\{1, 3\} + \{2, 4\})$.

Observe that permutations $\pi_{(\{1,3\}+\{2,4\})}$ and $\pi_{[00]}$ are products of permutations in $\{0, 1, 3\}$ and those in $\{2, 4\}$. Possible permutations $\pi_{(\{1,3\}+\{2,4\})}$ and $\pi_{[00]}$ restricted to $\{0, 1, 3\}$ read $\pi_{\{1,3\}+4} \to \text{id}$, $\pi_{\{1,3\}+2} \to (01)$, and $\pi_{[00]} \to (13)$. Noting that all permutations in $\{2, 4\}$ commute with each other, we conclude that four permutations $\pi_{[12]}$, $\pi_{[14]}$, $\pi_{[32]}$, and $\pi_{[34]}$ yielded by $(\{1, 3\} + \{2, 4\})$ commute with each other, and $\pi_{[14]}$ and $\pi_{[00]}$ yielded by $([14], [00]\lfloor 0\rfloor^1_{\text{odd}})$ also commute with each other. Thus, we compute the total permutation $\Sigma$ for a single period with length $5^n$ as

$$\Sigma = \pi^k_{[14]}\pi^p_{[00]}\pi_0\pi^q_{[00]}\pi^l_{[14]}\pi_{(\{1,3\}+\{2,4\})}. \tag{57}$$

where $p + q = (5^{n-1} - 1)/2$ is an even number. By setting the initial state of the period at just after $\lfloor 0\rfloor^1_{\text{odd}}$, the total permutation $\Sigma'$ reads

$$\Sigma' = (1432)(13)^q(01)^{5^{n-1}}(13)^p = \begin{cases} (1432)(13)(01)(13) & = (02143) \quad p, q \text{ are odd,} \\ (1432)(01) & = (04321) \quad p, q \text{ are even,} \end{cases} \tag{58}$$

both of which are 5-cycles.

Type70-2; $(a, b, c, d, e) = (1, 0, 2, 3, 4)$: $(\pi_0 = \pi_2 = (0123),\ \pi_1 = (01)(23),\ \pi_3 = (0321),\ \pi_4 = (043)(12))$

The structure of this rule is $(\{1, 3\}+\{0, 2\})(4+\lfloor 3+\{0, 2\}\rfloor_{\text{even}})$, where $4+\lfloor 3+\{0, 2\}\rfloor_{\text{even}}$ represents a sequence alternating 4 and "even length of a sequence alternating 3 and "0 or 2"". Examples are 444303244 and 430432, while 430004 is not an example because 3000 is not $\lfloor 3 + \{0, 2\}\rfloor_{\text{even}}$.

Due to the shape of the transition map, it is easy to see that except for state 4 all possible units appearing in the sequence take the form of $[\{1, 3\} + \{0, 2\}]$. Since state 4 newly appears only when $\pi_4$ applies to 0, we find that if the present state is 0, 1, 2, or 3, $\pi_{(\{1,3\}+\{0,2\})}$ induces a transition with structure $(\{1, 3\} + \{0, 2\})$. In addition, if the present state is 3, $\pi_{(\{1,3\}+\{0,2\})}$ induces transitions $(3 \to)0 \to 3$ or $(3 \to)2 \to 3$, which also satisfy the structure $\lfloor 3 + \{0, 2\}\rfloor_{\text{even}}$. With noting $\pi_{3+\{0,2\}} = \text{id}$ and it induces transitions $0 \to 3 \to 0$ and $3 \to 2 \to 3$, we further find that $\pi_{(4+\lfloor 3+\{0,2\}\rfloor_{\text{even}})}$ induces a transition with structure $(\{1, 3\} + \{0, 2\})$ if the present state is 1 or 2, and that with structure $(4 + \lfloor 3 + \{0, 2\}\rfloor_{\text{even}})$ if the present state is 0, 3, or 4. These observations imply the inheritance of the structure from site $n$ to site $n + 1$.

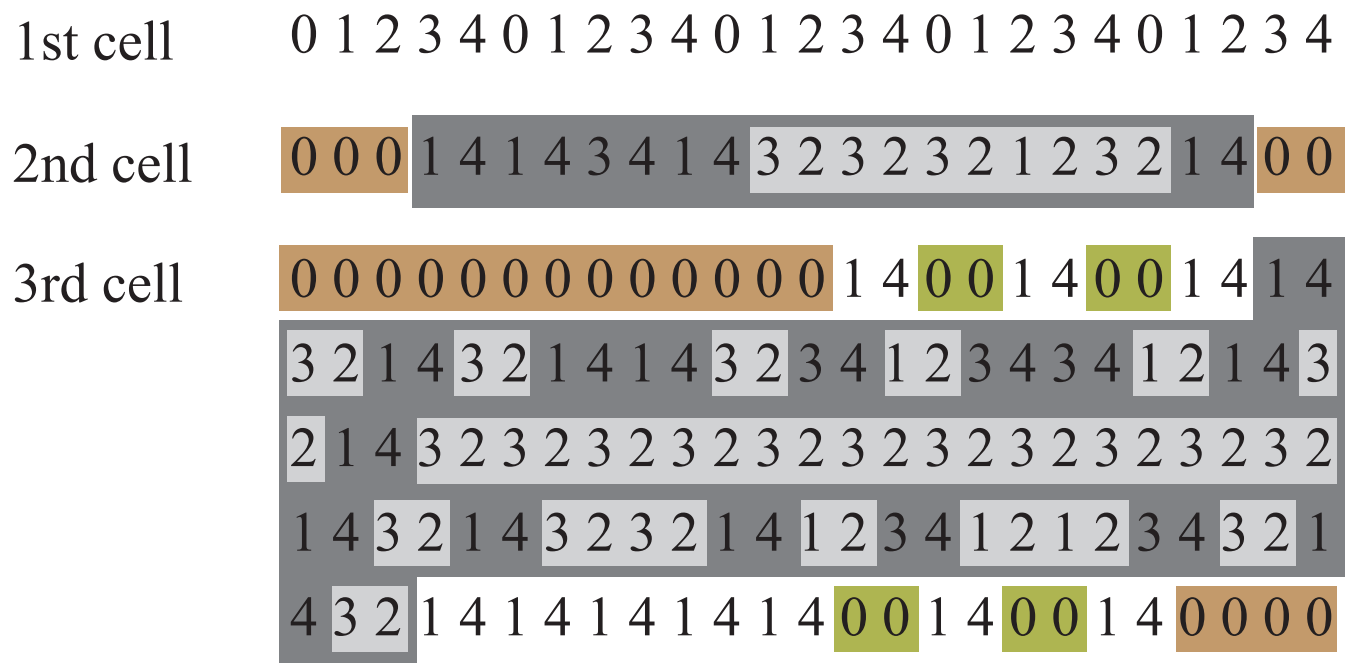

**Fig. 38** The sequence of first, second, and third cells in Type68-5 with $(a, b, c, d, e) = (0, 1, 3, 4, 2)$. The sequence in the third cell is divided into 5 lines that go from the top to the bottom. We draw island $(\{1, 3\} + \{2, 4\})$ in dark gray, and units $\{1, 3\} + 2$, $[00]$, and $\lfloor 0 \rfloor^1_{\text{odd}}$ in light gray, yellow, and brown, respectively.

With noting $\pi_{1+\{0,2\}} = (02)$ and $\pi_4 = (043)(12)$, the total permutation $\Sigma$ for a single period with length $5^n$ is computed as

$$\Sigma = (\pi_4)^{5^{n-1}} (\pi_{1+\{0,2\}})^{5^{n-1}} = ((043)(12))^{5^{n-1}} (02)^{5^{n-1}} = \begin{cases} (043)(12)(02) = (01243) & n \text{ is odd}, \\ (034)(12)(02) = (01234) & n \text{ is even}, \end{cases} \tag{59}$$

both of which are 5-cycles.

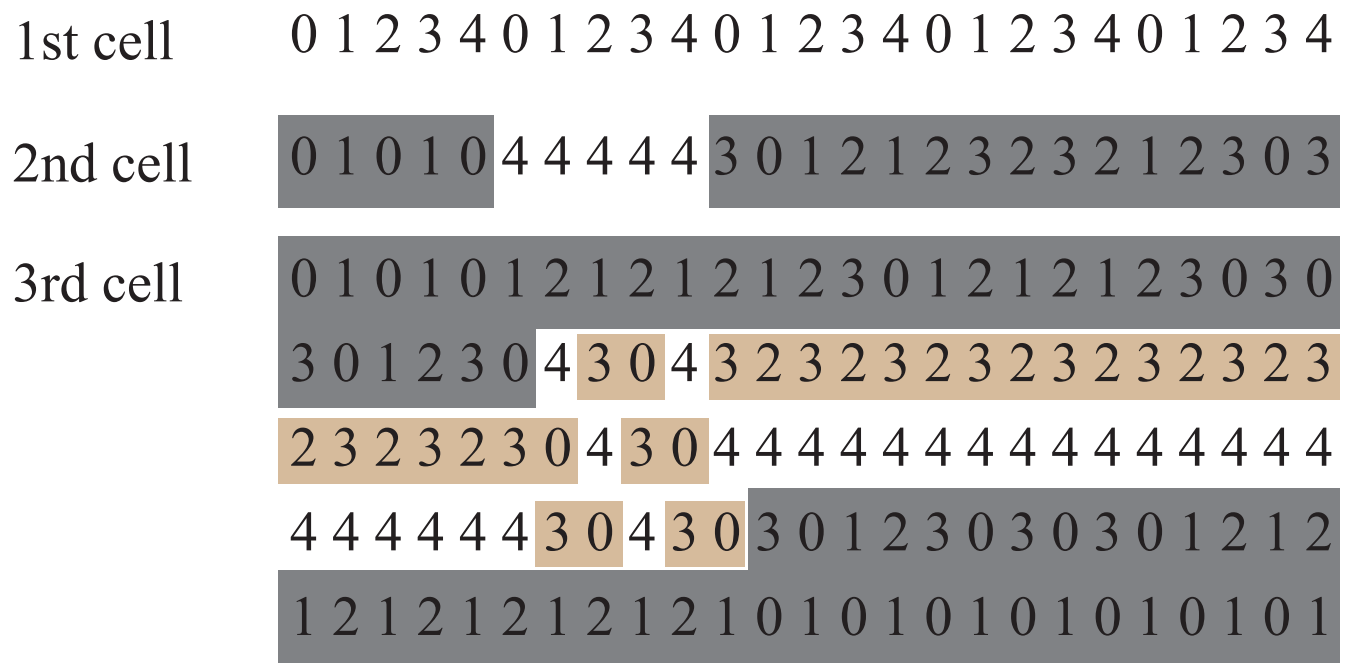

**Fig. 39** The sequence of first, second, and third cells in Type70-2 with $(a, b, c, d, e) = (1, 0, 2, 3, 4)$. The sequence in the third cell is divided into 5 lines that go from the top to the bottom. We draw island $(\{1, 3\} + \{2, 4\})$ in dark gray, and units $\lfloor 3 + \{0, 2\} \rfloor_{\text{even}}$ in the island $(4 + \lfloor 3 + \{0, 2\} \rfloor_{\text{even}})$ in light brown, respectively.

### *6.5.5 D-3: Complicated rules in islands*

Type50-5; $(a, b) = (1, 3)$: $(\pi_0 = \pi_2 = (13)$, $\pi_1 = (02)(13)$, $\pi_3 = (042)(13)$, $\pi_4 = (0123))$

The structure of this rule is $(0, 2, 4)(1 + \langle 0, 2 \rangle + 3 + \langle 0, 2 \rangle)$, where $\langle 0, 2 \rangle$ represents a queue with 0 and 2 of arbitrary length. An example of $1 + \langle 0, 2 \rangle + 3 + \langle 0, 2 \rangle$ is 12023013, which we can read as $1 + 202 + 3 + 0 + 1 + \text{blank} + 3$. Without loss of generality, we assign the island $(0, 2, 4)$ and $(1 + \langle 0, 2 \rangle + 3 + \langle 0, 2 \rangle)$ such that the total number of 0 and 2 in each island is even.

We observe that $\pi_{(0,2,4)}$ yields transitions inside $\{0, 1, 2, 3\}$, and $\pi_{(1+\langle 0,2 \rangle+3+\langle 0,2 \rangle)}$ yields transitions inside $\{0, 2, 4\}$ and $\{1, 3\}$. We first consider the action of $\pi_{(1+\langle 0,2 \rangle+3+\langle 0,2 \rangle)}$ on $\{0, 2, 4\}$. In $\{0, 2, 4\}$, both $\pi_0$ and $\pi_2$ act as the identity permutation, and thus $\pi_{(1+\langle 0,2 \rangle+3+\langle 0,2 \rangle)}$ behaves as the alternation of $\pi_3$ and $\pi_1$, $\pi_3 \pi_1 \pi_3 \pi_1 \cdots$, as long as our interest is in $\{0, 2, 4\}$. We notice that if this alternation starts from $\pi_1$ (resp. $\pi_3$), state 0 (resp. state 2) shuttles between 0 and 2, and the state never reaches 4. Since all of $\pi_0$, $\pi_1$, $\pi_2$, and $\pi_3$ act on $\{1, 3\}$ as the transposition (13), $\pi_{(1+\langle 0,2 \rangle+3+\langle 0,2 \rangle)}$ on $\{1, 3\}$ reads the identity permutation. In summary, $\pi_{(1+\langle 0,2 \rangle+3+\langle 0,2 \rangle)}$ is equal to (24) (if the alternation starts from state 1) or (04) (if the alternation starts from state 3), where we used the fact that the total number of 0 and 2 in this island is even.

We next consider the action of the permutation $\pi_{(0,2,4)}$ on $\{0,1,2,3\}$, which is a similar action to that in Pattern A-2. Setting $X = \{1,3\}$ and $Y := \{0,2\}$, we find that $\pi_4$ plays as addition of $+1 \mod 4$, and $\pi_2$ and $\pi_4$ plays as addition of $+2 \mod 4$ if and only if the state is in $X$ (They act trivially if the state is in $Y$). Following a similar argument to that for Pattern A-2, we find that $(\pi_{(0,2,4)})^2 = (02)(13)$ and hence $\pi_{(0,2,4)} = (0123)$ or $(0321)$.

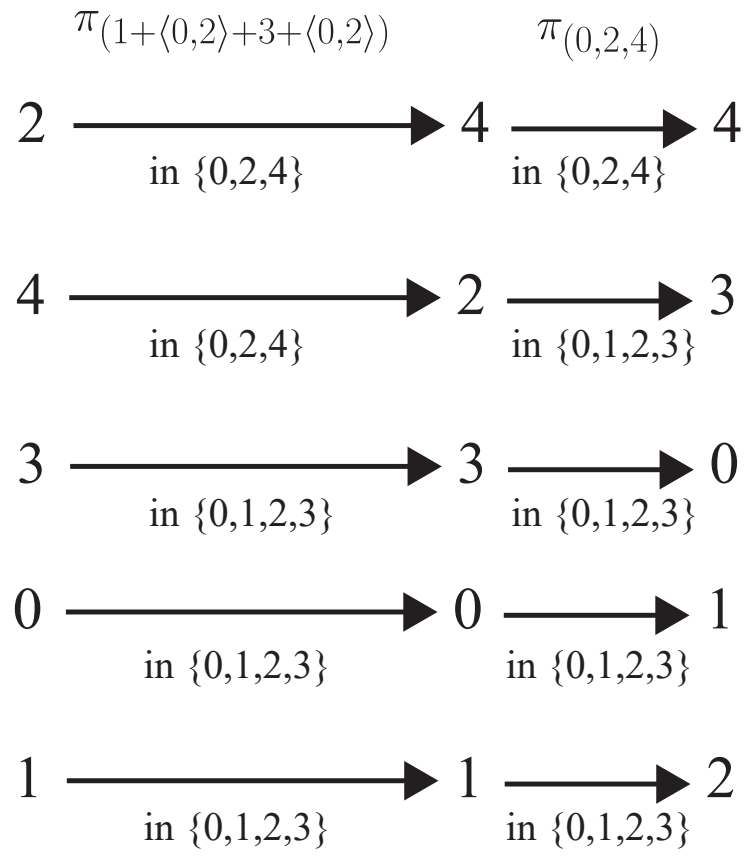

**Fig. 40** Schematic of transitions in Type50-5 with $(a, b) = (1, 3)$. We take the case that $(1 + \langle 0, 2\rangle + 3 + \langle 0, 2\rangle)$ starts from 1, and $\pi_{(0,2,4)} = (0123)$ as an example. As seen in this figure, if the structure $(0, 2, 4)(1 + \langle 0, 2\rangle + 3 + \langle 0, 2\rangle)$ holds in site $n$, it is kept in site $n + 1$. The ergodicity is also ensured.

We summarize our findings in Fig. 40, taking the case that the alternation starts from 1 and $\pi_{(0,2,4)} = (0123)$ as an example. The conservation of the structure is directly seen in this figure. The permutation in a single period with length $5^n$ reads

$$\Sigma = \pi_{(1+\langle 0,2\rangle+3+\langle 0,2\rangle)}\pi_{(0,2,4)} = \begin{cases} (24)(0123) = (01423), & \text{or} \\ (24)(0321) = (03421), & \text{or} \\ (04)(0123) = (01234), & \text{or} \\ (04)(0321) = (03214), & \end{cases} \tag{60}$$

all of which are 5-cycles.

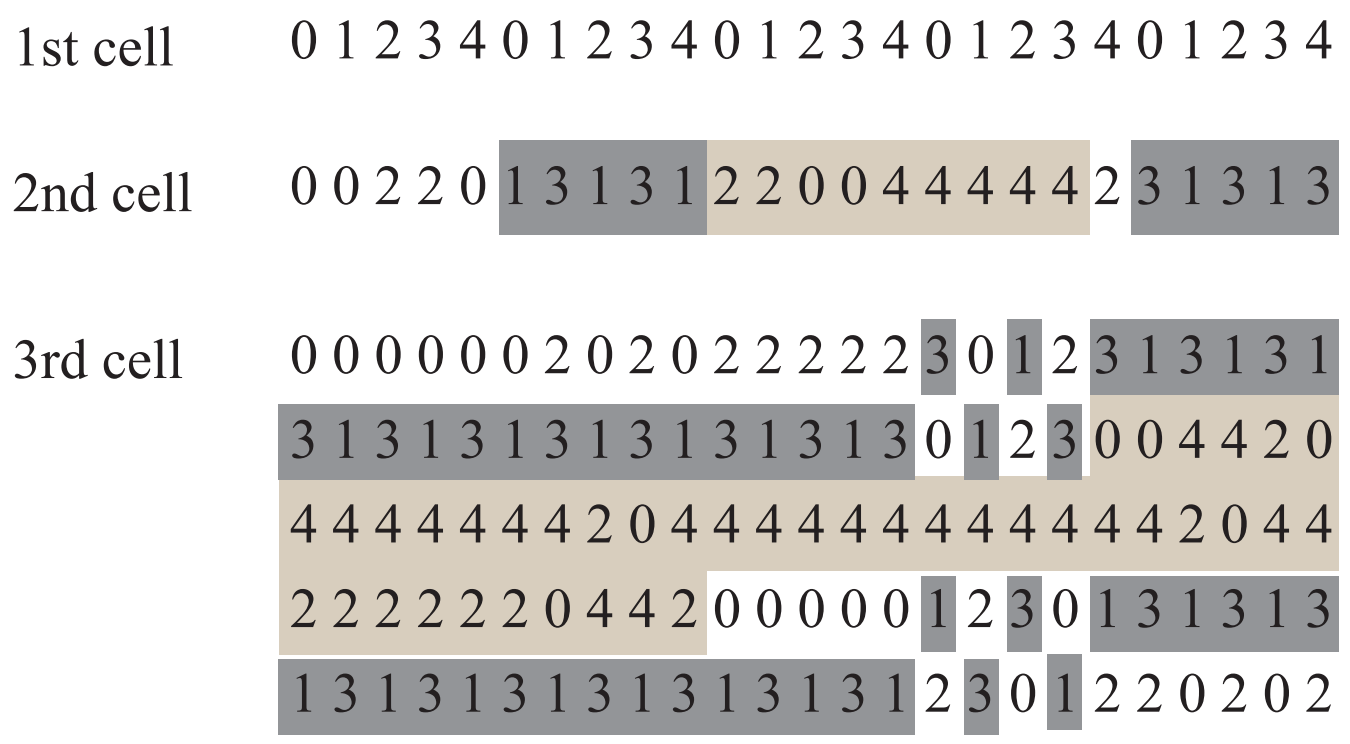

**Fig. 41** The sequence of first, second, and third cells in Type50-5 with $(a, b) = (1, 3)$. The sequence in the third cell is divided into 5 lines that go from the top to the bottom. We draw (0,2,4) in light brown, and 1 and 3 in dark gray. We can see that by removing 0 and 2 in $(1 + \langle 0, 2\rangle + 3 + \langle 0, 2\rangle)$, 1 and 3 appear alternately.

Type19-4; $(a, b, c, d) = (4, 0, 2, 3)$: ($\pi_0 = \pi_2 = (02)$, $\pi_1 = \text{id}$, $\pi_3 = (04)$, $\pi_4 = (0123)$)

The structure of this rule is $(3 + \langle 1, 0, 2|\{0, 2\}_{\text{even}}\rangle)(0, 2, 4)$, where $\langle 1, 0, 2|\{0, 2\}_{\text{even}}\rangle$ represents a queue of 1, 0, and 2, where the total number of 0 and 2 is even. We remark that the restriction on 0 and 2 is imposed only on the total number in this queue, and thus $\langle 1, 0, 2|\{0, 2\}_{\text{even}}\rangle$ includes 10121 and 11100121112 as possible examples. The island $(3 + \langle 1, 0, 2|\{0, 2\}_{\text{even}}\rangle)$ represents the alternation of 3 and $\langle 1, 0, 2|\{0, 2\}_{\text{even}}\rangle$. An example is $31113101231003\cdots$.

We can see that a nontrivial transition yielded by $(3+\langle 1, 0, 2|\{0, 2\}_{\text{even}}\rangle)$ is restricted inside $\{0, 2, 4\}$, and a nontrivial transition yielded by $(0, 2, 4)$ is restricted inside $\{0, 1, 2, 3\}$. In addition, since the state 3 always appears after an even number of 0 and 2 in $(3+\langle 1, 0, 2|\{0, 2\}_{\text{even}}\rangle)$, if the initial state is 2, the transition yielded by $(3 + \langle 1, 0, 2|\{0, 2\}_{\text{even}}\rangle)$ is restricted inside $\{0, 2\}$ and the cell never reach state 4, which satisfies the structure $(3 + \langle 1, 0, 2|\{0, 2\}_{\text{even}}\rangle)$. Hence, the structure is inherited from site $n$ to site $n + 1$.

To analyze the action of $\pi_{(0,2,4)}$ on $\{0, 1, 2, 3\}$, we employ a similar argument to Pattern A-2. Setting $X = \{1, 3\}$ and $Y := \{0, 2\}$, we find that $\pi_4$ plays as addition of $+1 \mod 4$, and $\pi_2$ and $\pi_4$ plays as addition of $+2 \mod 4$ if and only if the state is in $Y$ (They act trivially if the state is in $X$). Since $(0, 2, 4)$ contains even number of 0 and 2 and odd number ($= 5^{n-1}$) of 4, we find that $\pi_{(0,2,4)}$ is (0123) or (0321).

In the following, we consider the case of $\pi_{(0,2,4)} = (0123)$ for the sake of explanation. (The case of $\pi_{(0,2,4)} = (0321)$ is treated similarly). With noticing $\pi_{\langle 1,0,2|\{0,2\}_{\text{even}}\rangle} = \text{id}$, the permutation in a single period with length $5^n$ reads

$$\Sigma = \pi_{(0,2,4)}\pi_{(3+\langle 1,0,2|\{0,2\}_{\text{even}}\rangle)} = \begin{cases} (0123)(\pi_3)^{5^{n-1}} = (0123)(04) = (04123), & \text{or} \\ (0321)(\pi_3)^{5^{n-1}} = (0321)(04) = (04321), \end{cases} \tag{61}$$

both of which are 5-cycles.

| | |
|---|---|
| 1st cell | 0 1 2 3 4 0 1 2 3 4 0 1 2 3 4 0 1 2 3 4 0 1 2 3 4 |
| 2nd cell | 0 2 2 0 4 4 4 4 4 0 1 1 1 1 1 2 0 0 2 2 3 3 3 3 3 |
| 3rd cell | 0 2 0 2 0 1 2 3 0 1 1 1 1 1 1 1 1 1 1 1 1 1 1 1 1 |
| | 1 1 1 1 1 2 3 0 1 2 0 0 0 0 0 0 2 0 2 0 2 0 2 0 2 |
| | 2 0 2 0 2 3 0 1 2 3 3 3 3 3 3 3 3 3 3 3 3 3 3 3 3 |
| | 3 3 3 3 3 0 1 2 3 0 2 2 2 2 2 2 0 2 0 2 0 4 0 4 0 |
| | 4 4 4 4 4 4 4 4 4 4 4 4 4 4 4 4 4 4 4 4 4 0 4 0 4 |

**Fig. 42** The sequence of first, second, and third cells in Type19-4 with $(a, b, c, d) = (4, 0, 2, 3)$. The sequence in the third cell is divided into 5 lines that go from the top to the bottom. We draw 3 in $(3 + \langle 1, \{0, 2\}_{\text{even}}\rangle)$ in light gray and $(0, 2, 4)$ in dark gray. We can see that the number of 0 and 2 between each two light gray cells (state 3) is even.

## 6.6 Pattern E: Hidden alternating sequence

Rules in Pattern E employ hidden alternating sequences. Pattern E is highly exceptional so that only two rules belong to it.

Type67-3; $(a, b) = (2, 3)$: ($\pi_0 = \pi_1 = \pi_4 = (12)(34)$, $\pi_2 = (01234)$, $\pi_3 = (02143)$)

The structure of this rule is $\{2, *\} + \{3, *\}$ ($* = \{\lfloor 0, 1, 4\rfloor_{\text{odd}}, \lfloor 0, 1, 4\rfloor^1_{\text{even}}\}$). The basic form of this structure is $\{2, 0, 1, 4\} + \{3, 0, 1, 4\}$, indicating that by replacing 0, 1, and 4 by 2 or 3 properly, we can obtain an alternating sequence of 2 and 3 (see Fig. 44, where the gray numbers are replaced ones). However, the structure of Type67-3 contains a single $\{0, 1, 4\}$ in a single period, which violates the above alternating sequence. For example, in the second cell, the 24th state (the second from the right) is 3, and the subsequent 6 states drawn in brown is 000014 (due to the periodic boundary condition) whose next state is 3. If the basic form $\{2,0,1,4\}+\{3,0,1,4\}$ is satisfied, 000014 is replaced by 232323, while it is inconsistent with the fact that the state next to 000014 is 3. This means that a single $\{0,1,4\}$ is inserted besides the basic form $\{2,0,1,4\}+\{3,0,1,4\}$, which is expressed as $\lfloor 0,1,4\rceil^1_{\text{even}}$ in the structure mentioned at the beginning.

We next clarify how this rule is inherited from site $n$ to site $n+1$. To this end, we introduce a slightly modified transition map on 6 states, $0_A$, $0_B$, 1,2,3, and 4, drawn in Fig. 43. Here, the subscripts A and B are put for distinguishing two 0's in the following argument. As seen in Fig. 43, alternating $\pi_2$ and $\pi_3$ induce a sequence $0_A140_B230_A140_B23\cdots$, whose period is 6. An important observation is that all permutations (except $\lfloor 0,1,4\rceil^1_{\text{even}}$) induce transitions between states in $S_A := \{0_A,4,2\}$ and those in $S_B := \{0_B,1,3\}$ (i.e., the structure $\{2,*\}+\{3,*\}$). Namely, if the sequence starts from $S_A$ (resp. $S_B$) and the initial state is 0, 2, or 4 (resp. 0, 1, or 3), both $\pi_2$ and $\pi_3$ play the role of $+1$ shift to the right in this sequence (see Fig. 43). In addition, for any state $\pi_0$, $\pi_1$, and $\pi_4$ plays the role of $+3$ shift to the right in this sequence. In both cases, states in $S_A$ and those in $S_B$ appear alternately. We call this phase as *global phase*.

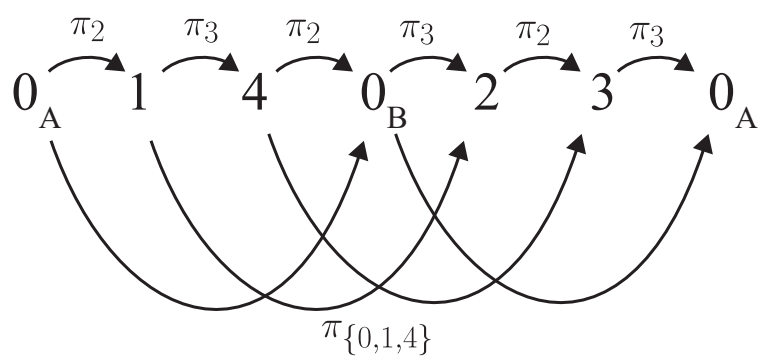

**Fig. 43** Transition map of Type67-3 with $(a,b)=(2,3)$ in the global phase. We have a base sequence with period 6.

0 0 0 0 0 0 2 1 4 3 0 0 2 1 2 3 4 0 0 1 2 3 4 0 2
2 3 2 3 2 3 3 2 2 3 3 2 3 2 3 2 3

2 1 2 1 2 1 4 3 0 0 2 1 4 3 4 0 0 1 2 3 4 0 0 1 4
3 3 3 2 2 3 3 2 2 3 2 3 2 3 2 3 2

**Fig. 44** We write the first and the third laws of the 3rd cell with supplementing 2 or 3 if the cell takes 0, 1, or 4 (numbers in gray). As demonstrated, 2 and 3 appear alternately after supplementation.

On the other hand, if the sequence starts from $S_A$ (resp. $S_B$) and the initial state is 1 or 3 (resp. 2 or 4), the state shuttles as $1\to2\to1$ or $4\to3\to4$. We call this phase as a *local phase*. These two phases are switched by $*=\lfloor 0,1,4\rceil^1_{\text{even}}$ unless the state is 0. When the state is 0, $*=\lfloor 0,1,4\rceil^1_{\text{even}}$ keeps the state in the global phase and generates $\lfloor 0,1,4\rceil^1_{\text{even}}$.

In the global phase, the permutation induced by $\{2,*\}+\{3,*\}$ in a single period is computed as follows. Since $\pi_2$ and $\pi_3$ induce $+1$ shift and the other three induce $+3$ shift, the total number of shifts in a single period is $2\cdot5^{n-1}+3\cdot3\cdot5^{n-1}=11\cdot5^{n-1}=(-1)^n \mod 6$. In the local phase, all permutations behave as (12)(34), and thus the permutation in a single period is (12)(34).

For explanation, we consider the case that the state appearing after $\lfloor 0,1,4\rceil^1_{\text{even}}$ is 3 and $n$ is even. In this case, the total permutation reads

$$\Sigma=(02143), \tag{62}$$

which is a 5-cycle. Here, if the initial state is in $\{0,1,3\}$, the system is in the global phase and $\Sigma$ maps $0\to2$, $1\to4$, and $3\to0$. In contrast, if the initial state is $\{2,4\}$, the system is in the local phase and $\Sigma$ maps $2\to1$ and $4\to3$. This permutation can be seen in the sixth column in the third cell in Fig. 45, which reads 02143 consistent with the derived $\Sigma$.

In the case that the state appearing after $\lfloor 0,1,4 \rfloor^1_{\text{even}}$ is 2 and $n$ is odd, the total permutation reads

$$\Sigma = (03412), \tag{63}$$

which is a 5-cycle. Here, if the initial state is in $\{0,2,4\}$, the system is in the global phase and $\Sigma$ maps $0 \to 1$, $2 \to 3$, and $4 \to 0$. In contrast, if the initial state is $\{1,3\}$, the system is in the local phase and $\Sigma$ maps $1 \to 2$ and $3 \to 4$. The remaining two cases are treated in a similar manner.

| | |
|---|---|
| 1st cell | 0 1 2 3 4 0 1 2 3 4 0 1 2 3 4 0 1 2 3 4 0 1 2 3 4 |
| 2nd cell | 0 0 0 1 4 3 4 3 4 3 4 3 4 0 2 1 2 1 2 1 2 1 2 3 0 |
| 3rd cell | 0 0 0 0 0 0 2 1 4 3 0 0 2 1 2 3 4 0 0 1 2 3 4 0 2 |
| | 1 2 1 2 1 2 1 2 1 2 1 2 1 2 1 2 1 2 1 2 1 2 1 2 1 |
| | 2 1 2 1 2 1 4 3 0 0 2 1 4 3 4 0 0 1 2 3 4 0 0 1 4 |
| | 3 4 3 4 3 4 3 4 3 4 3 4 3 4 3 4 3 4 3 4 3 4 3 4 3 |
| | 4 3 4 3 4 3 0 0 2 1 4 3 0 0 0 1 2 3 4 0 0 1 2 3 0 |

**Fig. 45** The sequence of first, second, and third cells in Type67-3 with $(a,b) = (2,3)$. The sequence in the third cell is divided into 5 lines that go from the top to the bottom. We draw the inconsistent unit of $\{0,1,4\}$ in brown. The 2nd cell has a single 3-"even length of $\{0,1,4\}$"-3, and the 3rd cell has a single 3-"odd length of $\{0,1,4\}$"-2, which correspond to $\lfloor 0,1,4 \rfloor^1_{\text{even}}$.

Type66-3; $(a,b) = (1,2)$: $(\pi_0 = \pi_3 = \pi_4 = (12)(34)$, $\pi_1 = (014)(23)$, $\pi_2 = (023)(14))$

The structure of this rule is $\{1,*\} + \{2,*\}$ $(* = \{\lfloor 0,3,4 \rfloor_{\text{odd}}, \lfloor 0,3,4 \rfloor^1_{\text{even}}\})$.

The proof technique is essentially the same as that for Type67-3 with $(a,b) = (2,3)$. The base sequence in the global phase is $0_A \to 1 \to 4 \to 0_B \to 2 \to 3 \to 0_A$.

One difference between this case and Type67-3 with $(a,b) = (2,3)$ is that in this case, the state can move among $\{1,2,3,4\}$ in the local phase. In the local phase, $\pi_1$ and $\pi_2$ provide transitions $1 \leftrightarrow 4$ and $2 \leftrightarrow 3$, and $\pi_0$, $\pi_3$, and $\pi_4$ provide transitions $1 \leftrightarrow 2$ and $3 \leftrightarrow 4$. Hence, by labeling $1 = (0,0)$, $2 = (1,0)$, $3 = (1,1)$, and $4 = (0,1)$, $\pi_1$ and $\pi_2$ play the role of flipping the first entry of $(x,y)$, and $\pi_0$, $\pi_3$, and $\pi_4$ the second entry of $(x,y)$. With this description, the total permutation in a single period in the local phase is just flipping the second entry ($= 3 \cdot 5^{n-1}$ flips) with keeping the first entry ($= 2 \cdot 5^{n-1}$ flips), which means the permutation (12)(34). The remaining arguments are essentially the same as those for Type67-3 with $(a,b) = (2,3)$.

| | |
|---|---|
| 1st cell | 0 1 2 3 4 0 1 2 3 4 0 1 2 3 4 0 1 2 3 4 0 1 2 3 4 |
| 2nd cell | 0 0 1 4 3 4 3 2 3 4 3 4 0 2 1 2 1 4 1 2 1 2 3 0 0 |
| 3rd cell | 0 0 0 1 2 1 2 1 4 3 4 3 4 3 0 1 4 0 0 1 4 0 2 1 2 |
| | 1 2 1 4 3 4 3 4 1 2 1 2 1 2 3 2 3 2 1 4 1 4 1 2 1 |
| | 2 1 2 3 4 3 4 3 0 0 0 0 0 0 2 3 0 1 2 3 0 1 4 3 4 |
| | 3 4 3 2 1 2 1 2 3 4 3 4 3 4 1 4 1 4 3 2 3 2 3 4 3 |
| | 4 3 4 0 0 0 0 0 2 1 2 1 2 1 4 0 2 3 4 0 2 3 0 0 0 |

**Fig. 46** The sequence of first, second, and third cells in Type66-3 with $(a,b) = (1,2)$. The sequence in the third cell is divided into 5 lines that go from the top to the bottom. We draw the inconsistent unit of $\{0,3,4\}$ in brown. The 2nd cell has a single 2-"odd length of $\{0,3,4\}$"-1, and the 3rd cell has a single 2-"odd length of $\{0,3,4\}$"-1, which correspond to $\lfloor 0,3,4 \rfloor^1_{\text{even}}$.

## 7 Generalization to CA with more than 5 states

In this section, we briefly discuss CA with states $k > 5$. We here only sketch some generalizations of patterns shown in the previous section. Remark that this section does not serve as a comprehensive description of generalization, and there will exist other generalizations of presented patterns and some unknown patterns.

### 7.1 Generalization of pattern A-1

We consider a CA with the number of states $k = 1 + \sum_{i=1}^{m} p_i$. We label these $k$ states as 0 and $(i.j)$ with $1 \le i \le m$ and $1 \le j \le p_i$, and denote the $i$-th set by $P_i := \{0, (i,1), (i,2), \dots, (i,p_i)\}$, which serves as the $i$-th island.

We suppose the following:

- There exists a permutation $\tau$ on the label of islands $\{1, 2, \dots, m\}$ such that for any state in the $i$-th island $x \in P_i$, the permutation $\pi_x$ acts nontrivially only on island $\tau(i)$. In particular, $\pi_0$ is an identity permutation.
- For any island $i \in \{1, 2, \dots, m\}$, the permutations induced by states in $P_i$ commute with each other (i.e., $\pi_x\pi_y = \pi_y\pi_x$ with $x, y \in P_i$ for any $i$).
- For any $n$ and $i \in \{1, 2, \dots, m\}$,

$$\prod_{j=1}^{p_i} (\pi_{(i,j)})^{k^{n-1}} \tag{64}$$

  is a cyclic permutation with a single cycle on $P_{\tau(i)}$.

A necessary (but not sufficient) condition for the third requirement is that $p_i + 1$ and $k$ are relatively prime for all $i$. One sufficient condition for the third requirement is the following:

- There exists an integer $n_j$ and a single-cycle permutation $\mu$ such that $\pi_{(i,j)}$ can be written as $\pi_{(i,j)} = \mu^{n_j}$.
- $k^{n-1} \sum_{j=1}^{p_i} n_j$ is relatively prime to $p_{\tau(i)} + 1$.

Then, if the leftmost cell evolves as $012 \cdots k$, this CA is ergodic. The structure of this CA is $P_1 P_2 \cdots P_m$ (while the order of islands can be permuted).

### 7.2 Generalization of pattern A-1-2

Suppose that the set of states $\{0, 1, 2, \dots, k-1\}$ have some islands $P_1, P_2, \dots, P_m$, where any state belongs to one or two islands. Corresponding to these islands, we introduce a graph with $m$ vertices such that two vertices $i$ and $j$ are connected by an edge if and only if $P_i \cap P_j \neq \emptyset$ (see Fig. 47 as an example). Let $R_i \subset \{1, 2, \dots, m\}$ be a set of integers such that permutations induced by states only in $R_i$ act nontrivially on $P_i$ (i.e., for any state $x \in P_k$ with $k \in \{1, 2, \dots, m\} \backslash R_i$, $\pi_x$ restricted to $P_i$ serves as the identity permutation). We put the following requirements:

- The graph is a tree.
- For any $i \in \{1, 2, \dots, m\}$ and any $x, y \in P_i$, $\pi_x$ and $\pi_y$ commute with each other.
- For any two islands $i$ and $j$ connected by an edge in the graph, there exists an edge $e$ in the graph satisfying the following property: Removal of $e$ shapes two connected graphs, which naturally leads to a decomposition of islands $P_1, P_2, \dots, P_m$ into two sets. We require that $R_i$ and $R_j$ belong to different sets.
- For each island $i$, there is a single-cycle permutation $\mu_i$ such that any permutation $\pi_x$ restricted to $P_i$ takes the form of $\mu_i^{n_{x,i}}$ (If $\pi_x$ acts trivially on $P_i$, we set $n_{x,i} = 0$).
- For any $i$ and $n$, $k^{n-1} \sum_x n_{x.i}$ is relatively prime to $|P_i|$.

The third condition ensures that in one period sequence the permutation on $P_i$ and that on $P_j$ act completely separately (i.e., all permutations induced by $R_i$ first act, and then those induced by $R_j$ act, or vice versa). The structure of this rule is the shortest closed path on this graph covering all islands (vertices). Taking Fig. 47 as an example to explain the third condition and the structure. Since islands C and E are connected, $R_C$ and $R_E$ should be completely separated. Thus, the choice of $R_C = \{A, B, F, G\}$ and $R_E = \{C, E\}$ is allowed (by removing edge BC, these two are separated), while the choice of $R_C = \{A, H\}$ and $R_E = \{E, F\}$ is prohibited. Examples of sequences satisfying the structure are ABFBCDCECBGHGBA and ABCDCECBFBGHGBA, which are the shortest closed paths covering all islands. The first sequence with removing islands not in $R_C$ and $R_D$ reads $AFEHA$, where islands in $R_C$ and those in $R_D$ can be decomposed contiguously. Hence, the structure in site $n$ ensures that with setting the initial state as a proper one all the permutations in $R_C$ appear first, and then all the permutations in $R_D$ appear. This ensures the ergodic visit of states in island $D$ and those in island $C$ (though the latter has breaks when the state is in island $D$ and island $E$).

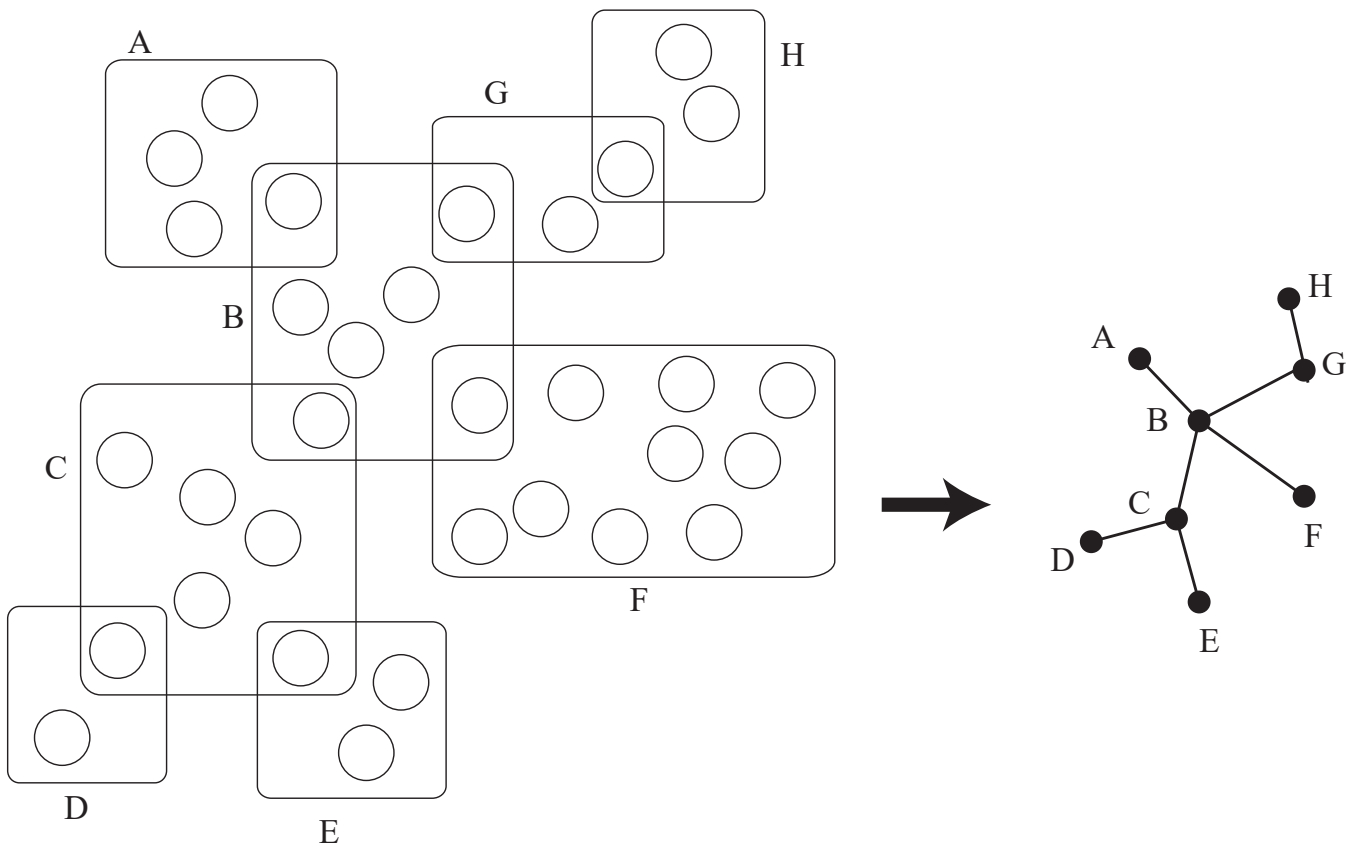

**Fig. 47** An example of state space with eight islands A, B, . . . ,H and its graph representation.

## 7.3 Generalization of pattern A-2

Consider a CA with the number of states $k = a + m^2$ with odd $k$. Here, $m^2$ states are decomposed into $m$ subsets $X_1, \dots, X_m$ with $m$ states: $X_i = \{(i, 1), \dots, (i, m)\}$. The remaining $a$ states are denoted by $\{1, \dots, a\} = A$.

We employ a coarse-grained description of $m^2$ states where we regard the $m$ subsets $X_1, \dots, X_m$ as coarse-grained states. We assign integers $1, 2, \dots, m^2$ to states $(i, j)$ as $(i, j) \leftrightarrow i + m(j-1)$. We suppose the following:

- All permutations restricted to the $m^2$ states can be divided into two classes, (i) and (ii).
- Permutations in class (i) act among subsets $X_1, \dots, X_m$ in the following manner: $\pi_y$ maps state $x$ $(1 \le x \le m^2)$ onto $x + n_y \mod m^2$, where $n_y$ and $m$ are relatively prime.
- Permutations in class (ii) permutes states only inside each subset $X_k$. $\pi_y$ maps state $(l, j)$ onto $(l, j + n_{y,l} \mod m)$, where $n_{y,l}$ and $m$ are relatively prime.
- For any $n$, $\sum_y k^{n-1} n_y \mod m$ is relatively prime to $m$.
- For any $n$, $\sum_y k^{n-1} n_y + \sum_{y',l} k^{n-1} n_{y',l} \mod m$ is relatively prime to $m$.

The fourth condition confirms that in the coarse-grained description with $X_1, \dots, X_m$, the permutation in a single period is a $k$-cycle. The fifth condition confirms that the permutation in $m$ periods, which reads $x \to x + m \sum_y k^{n-1} k^{n-1} n_y + \sum_{y',l} k^{n-1} m n_{y',l} \mod m^2$, has period $m$.

We put further assumptions on the remaining states in $A$:

- State $i$ in $A$ has its pair in the $m^2$ states denoted by $f(i)$. State $i$ and $f(i)$ are swapped by some permutations induced by states in $A$.
- Suppose that $\pi_j$ swaps $i$ and $f(i)$. Then, we require that $\pi_{f(j)}$ acts trivially on $f(i)$.

We put the second condition for the following reason: The structure of this CA has island $(j, f(j))$, which we expect to induce transitions in $(i, f(i))$. To realize it, we assume that only $j$ touches $i$ and $f(i)$, and $\pi_{f(j)}$ does not touch these two states. Since $k$ is odd, $\pi_{(j,f(j))} = (i, f(i))^{k^{n-1}} = (i, f(i))$ is a swap on $i$ and $f(i)$, not an identity permutation.

### 7.4 Generalization of pattern B

Since there are extremely many types of possible generalizations of pattern B, we here explain just one example of it for pattern B-1.

Consider a CA with the number of states $k = pm + 1$. Its driving unit is $[12\cdots p]$, and non-driving unit is $\lfloor p+1, \ldots, k-1 \rfloor_p$ and $[00\cdots 0]_p$. Here, $\lfloor \cdots \rfloor_p$ means that the length of this sequence is a multiple of $p$. In addition to these units, there also exists a single sequence of 0 $(00\cdots 0)$ whose length is not divisible by $p$.

We choose one of $1, 2, \ldots, p$ and fix it as $x$. We set permutations for $i \neq x$ as

$$\pi_i = (12\cdots p)(p+1, p+2, \cdots, 2p)\cdots(p(m-1)+1, p(m-1)+2, \cdots, pm), \tag{65}$$

and that for $x$ as

$$\pi_x = (12\cdots p, p+1, \cdots, 2p, 2p+1, \cdots, pm, 0). \tag{66}$$

Supposing $k^{n-1}$ is relatively prime to $m+1$, we conclude that this CA is ergodic.

Further generalizations to the case with multiple driving units are straightforward.

### 7.5 Generalization of pattern C

We consider a CA with state 0 and $x_{i,j}$ ($1 \le i \le p$ and $1 \le j \le m$). The number of states is $k = mp+1$. We define a set of states $X_i := \{x_{i,1}, \ldots, x_{i,m}\}$ for $1 \le i \le p$.

We assume that possible permutations induced by state $y$ is classified into two classes, (i) and (ii). A permutation in class (i) satisfies $\pi_y(0) = 0$, and $\pi(u) \in X_{i+1}$ for $u \in X_i$. Here, we identify $X_{p+1}$ as $X_1$. A permutation in class (ii) satisfies $\pi_y(0) = x_{1,s}$ and $\pi_y(x_{p,t}) = 0$.

We fix $1 \le h \le p$ and $1 \le s, t \le m$. We assume the following:

- All permutations $\pi_y$ with $y \notin X_h$ (including $y = 0$) belongs to class (i). Note that some $\pi_y$ with $y \in X_h$ may also belong to class (i).
- $\pi_0^m(y) = y$ for any $y$.
- For any $i_1, i_2, \ldots, i_p \in \{1, \ldots, m\}$, the permutations $\pi_{x_{p,i_p}} \cdots \pi_{x_{2,i_2}} \pi_{x_{1,i_1}}$ commute with each other.
- For $1 \le l \le p$, we set $\{i_{l,t}\}_{t=1}^{m\cdot k^{n-1}}$ such that $1, 2, \ldots, m$ respectively appear $k^{n-1}$ times. We assume that for any choice of $\{i_{l,t}\}_{t=1}^{m\cdot k^{n-1}}$,

$$\prod_{t=1}^{m\cdot k^{n-1}} \left(\pi_{x_{p,i_{p,t}}} \cdots \pi_{x_{2,i_{2,t}}} \pi_{x_{1,i_{1,t}}}\right) \tag{67}$$

  provides the same permutation. We denote this permutation $\Pi$, which is a product of permutations inside each $X_i$ with $i \neq p+1-h$ and inside $X_{p+1-h} \cup \{0\}$.
- $\pi_0 \Pi$ is a $k$-cycle.

If these assumptions are satisfied, then the structure of this rule is $[X_1 + X_2 + \cdots X_p]$, $[00\cdots 0]$, $[0]^1$.

## 8 Open problems

We have found various ergodic semi-infinite CA. In particular, we succeed in discovering all ergodic CA with states 3, 4, and 5. At the same time, many questions will naturally arise. Below we shall list some of them.

– As we have seen, no rules in 4-state CA are ergodic, and at present, we have discovered no ergodic rules in CA with an even number of states. It is an interesting question whether CA with 6 states or more states with an even number have ergodic rules, or all rules in CA with an even number of states are non-ergodic.
We note that the number of states of ergodic CA is not necessarily a prime. An example is the generalization of Pattern B presented in Sec. 7.4 with setting $p = 8$ and $m = 1$. In this case, the proof presented in Sec. 7.4 applies, while the number of states is 9, a composite number.

– It is practically important to grasp how many sites we need to examine to confirm the ergodicity. As encountered in Table 5.1 in Sec. 5.1, the number of candidates for ergodic rules may fool us by apparent convergence. In 5-state CA, the number of candidates is 118560 from $N = 11$ to $N = 14$, while it in fact converges to 118320 for $N \geq 15$. Since exhaustive proof as in the case of 5-state CA is hopeless for CA with many states, we need criteria when the number of candidates converges.

– Our exhaustive proof of ergodicity in 5-state CA is highly complicated, and the classification is ad hoc, and not systematic at all. A much simpler and more transparent proof which covers all CA in one fell swoop is desired, particularly for CA with many states.

– The ergodic orbits discussed in this paper are not random at all but highly ordered orbits. The order of orbits is, in some sense, similar to the counting of numbers in decimal code. In the counting in decimal code, the digit in the one's place has period 10, and that in the $10^n$'s place has period $10^{n+1}$. However, there also exists some difference between the ergodic orbits and counting in decimal code: In the ergodic orbits in our CA, the dynamics in site $n + 1$ depends only on that in site $n$, which implies local interaction. In contrast, in the counting in decimal code, the dynamics in $10^{n+1}$'s place depends on all of $10^m$'s place with $1 \leq m \leq n$, which implies global interaction. To clarify the order of ergodic orbits is an important problem.

– We have not discussed the relation between the ergodicity with one boundary condition and all boundary conditions. These two coincide in CA with a small number of states, while they may differ in CA with many states. An example is the generalization of Pattern A-1 presented in Sec. 7.1, with $m = 2$, $p_1 = p_2 = 3$ (i.e., $k = 7$) and $\pi_{(1,1)} = \pi_{(1,2)} = (0123)$, $\pi_{(1,3)} = (0321)$, $\pi_{(2,1)} = \pi_{(2,2)} = (0456)$, and $\pi_{(2,3)} = (0654)$. Then, the boundary permutation $\pi_{\mathrm{B}} = (0123456)$ induces ergodic dynamics, while the boundary permutation $\pi_{\mathrm{B}} = (0153426)$ accompanies non-ergodic orbit in site $n = 2$.
In such CA, it is an interesting question what relationship holds between the ergodicity with one boundary condition and with other boundary conditions.

– The boundary condition also affects the speed of ergodicity breaking. In the case of $k = 5$, rule $(11, 11, 11, 32, 32)$ behaves ergodic in sites less than 11, and half of the boundary conditions accompany ergodicity breaking in site $n = 11$. The remaining half of boundary conditions show ergodicity in sites less than 17, and all the boundary conditions accompany ergodicity breaking in site $n = 17$. The connection between the boundary condition and the speed of ergodicity breaking has not yet been studied even in the case of $k = 5$.

## 9 Discussion

In this paper, we introduced semi-infinite CA and thoroughly examined its ergodicity in CA with 3, 4, and 5 states by both numerical simulation and analytical calculation. We determine all ergodic rules in these CA: All ergodic CA are proven to be ergodic, and all non-ergodic CA are numerically confirmed to be non-ergodic. Although our proof of ergodicity succeeds, the proof is highly ad hock and non-systematic, and further clarification is desired.

Most of previous studies on CA treat infinite CA or CA with periodic boundary conditions. In these CA, no rule can be completely ergodic since spatial periodic patterns are conserved. On the other hand, the semi-infinite CA does not face this problem, and many rules are proven to be ergodic. One big advantage of the semi-infinite CA is that the ergodicity can be examined not only numerically but also analytically in a rigorous manner. For few-state CA, even exhaustive determination of ergodic CA is possible. The semi-infinite CA serves as a good stage to investigate properties which conventional CA cannot contain well.

**Acknowledgements** NS thanks Kunihiko Kaneko for the helpful discussion. This work is supported by JSPS KAKENHI Grants-in-Aid for Early-Career Scientists Grant Number JP19K14615 and JST ERATO Grant Number JPMJER2302, Japan.

## Appendix. A List of structures of ergodic CA

In this Appendix, we list the structures of all rules of ergodic CA. With the knowledge of this structure and the classification, one will write down the proof of ergodicity if needed.

In this table, the superscript § in pattern A-1 means that the order of the islands might be changed.

| type | parameter | class | structure |
|---|---|---|---|
| 01 | any | A-1 | (0,1,2,3)(0,4) |
| 02 | $(a,b)=\{0,3\}$ | A-1 | (0,1,2,4)(4,3) |
| | $(a,b)=\{1,4\},\{2,4\}$ (and otherwise) | A-1 | (0,1,2)(0,3,4) |
| 03 | any | A-1 | (0,1,2)(0,3,4) |
| 04 | $2\in\{a,b\}$ | A-1 | (0,1,4,3)(1,2) |
| | $3\in\{a,b\}$ (and otherwise) | A-1 | (0,1,2,4)(4,3) |
| 05 | $(a,b)=\{0,3\}$ | A-1 | (0,1,2,4)(4,3) |
| | $(a,b)=\{0,2\})$ | A-1 | (1,2)(0,1,3,4) |
| | otherwise | A-1 | (0,1,2)(0,3,4) |
| 06-1 | $a=2$ | A-1 | (0,1,4,3)(1,2) |
| | $a=3$ (and otherwise) | A-1 | (0,1,2,4)(4,3) |
| 06-2 | $0\notin\{a,b,c\}$ | A-1 | (0,1,2)(0,3,4) |
| | $0\in\{a,b,c\}$ | A-1-2 | (0,1,2)(0,4)(4,3)(0,4) |
| 06-3 | any | A-1 | (0,1,2)(0,3)(0,4) § |
| 07-1 | any | A-1-2 | (12)(01)(04)(43)(04)(01) |
| 07-2 | any | A-1-2 | (3,4)(0,1,4)(1,2)(0,1,4) |
| 07-3 | $(a,b,c)=(\{1,2\},4)$ | A-2 | $(0,1,2,3)(0,4)$ |
| | otherwise | A-1 | (0,1,2)(0,3)(0,4) § |
| 08-1 | $0\in\{a,b\}$ | A-1 | (0,1,2,4)(4,3) |
| | $0\notin\{a,b\}$ | A-1 | (0,1,2)(0,3,4) |
| 08-2 | any | A-1 | (0,1,2)(0,3,4) |
| 09-1 | $b=4$ | A-1 | (0,1,2,3)(0,4) |
| | $c=4$ | A-2 | (0,1,2,3)(0,4) |
| 09-2 | any | A-1 | (0,1,2,3)(0,4) |
| 09-3 | any | A-1 | (0,1,2,3)(0,4) |
| 09-4 | any | A-2 | (0,1,2,3)(0,4) |
| 09-5 | any | A-2 | (0,1,2,3)(0,4) |
| 10 | $0\notin\{b,c\}$ | A-1 | (0,1,2)(0,3,4) |
| | $0\in\{b,c\}$ | A-1 | (0,1,2,4)(3,4) |
| 11 | any | A-1 | (0,1,2)(0,3,4) |
| 12 | $a=2$ | A-1 | (0,1,3,4)(1,2) |
| | $a=3$ | A-1 | (0,1,2,4)(4,3) |
| 13 | any | A-1 | (0,1,2)(0,3)(0,4) § |
| 14-1 | any | A-1* | (0,1,2,3)(0,4) |
| 14-2 | any | A-1* | (0,1,2,3)(0,4) |
| 15 | any | A-1 | (0,1,2)(0,3,4) |
| 16 | any | A-2 | (0,1,2,3)(0,4) |
| 17-1 | any | A-1 | (0,1,2,3)(0,4) |
| 17-2 | any | A-1 | (0,1,2,3)(0,4) |
| 18-1 | any | A-1-2 | (1,2)(0,1)(0,4)(4,3)(0,4)(0,1) |
| 18-2 | $a\neq 0$ | A-1 | (0,1,2)(0,3)(0,4) § |
| | $a=0$ | A-1-2* | (1,2)(0,1)((0,3),(0,4))(0,1) |
| 18-3 | any | A-1 | (0,1)(0,2)(0,3)(0,4) § |
| 19-1 | $d=4$ | A-2 | (0,1,2,3)(0,4) |
| | $c=4$ | A-1 | (0,1,2,3)(0,4) |
| 19-2 | | A-1 | (0,1,2,3)(0,4) |
| 19-3 | any | A-2 | (0,1,2,3)(0,4) |
| 19-4 | $d=4$ | A-2 | (0,1,2,3)(0,4) |
| | $a=4$ | D-3 | $(3+\langle 1,\{0,2\}_{\mathrm{ev}}\rangle)(0,2,4)$ |
| 19-5 | any | A-1 | (0,1,2,3)(0,4) |
| 19-6 | any | A-1 | (0,1,2,3)(0,4) |
| 20-1 | $(a,b,c)=(3,1,2)$ | A-1-2 | (0,1,2)(0,4)(4,3)(0,4) |
| | $(a,b,c)=(1,2,4)$ | D-2 | $(1,\lfloor 0,2,4\rfloor_{\mathrm{even}})(0,4,3)$ |
| | $(a,b,c)=(2,1,4)$ | D-2 | $(2,\lfloor 0,1,4\rfloor_{\mathrm{even}})(0,4,3)$ |
| 20-2 | any | A-1-2 | (0,1,2)(0,4)(4,3)(0,4) |
| 20-3 | any | A-1-2 | (0,1,2)(0,4)(4,3)(0,4) |
| 20-4 | | D-1 | $(1,\lfloor 0,2\rfloor_{\mathrm{even}})(0,4)(4,3)(0,4)$ |
| 20-5 | | D-1 | $(2,\lfloor 0,1\rfloor_{\mathrm{even}})(0,4)(4,3)(0,4)$ |
| 21-1 | any | A-1-2 | (0,1,2)(0,4)(4,3)(0,4) |
| 21-2 | any | A-1 | (0,1,2)(0,3)(0,4) § |
| 22-1 | any | A-1-2 | (0,1,2)(0,4)(4,3)(0,4) |
| 22-2 | any | D-1 | $(2,\lfloor 0,1\rfloor_{\mathrm{even}})(0,4)(4,3)(0,4)$ |
| 22-3 | any | D-1 | $(1,\lfloor 0,2\rfloor_{\mathrm{even}})(0,4)(4,3)(0,4)$ |
| 22-4 | any | A-1 | (0,1,2)(0,3)(0,4) § |
| 22-5 | $b=2$ | A-1 | (0,1,3,4)(1,2) |
| | $b=3$ | A-1 | (0,1,2,4)(3,4) |
| 22-6 | $b=2$ | A-1 | (0,1,3,4)(1,2) |
| | $b=3$ | A-1 | (0,1,2,4)(3,4) |
| 23-1 | $e=3$ | A-2 | (0,1,2,4)(3,4) |
| | $d=3$ | A-1 | (0,1,2,4)(3,4) |
| | $e=2$ | A-1-2 | (1,2)(0,1)(0,4)(3,4)(0,4)(0,1)(1,3) |

| type | parameter | class | structure |
|---|---|---|---|
| 23-2 | any | A-1-2 | (1,2)(0,1)(0,4)(3,4)(0,4)(0,1) |
| 23-3 | any | A-1-2 | (1,2)(1,0)(0,4)(4,3)(0,4)(0,1) |
| 23-4 | any | A-2 | (0,1,2,3)(0,4) |
| 23-5 | any | A-2 | (0,1,2,3)(0,4) |
| 23-6 | any | A-2 | (0,1,2,3)(0,4) |
| 23-7 | any | A-2 | (0,1,2,3)(0,4) |
| 23-8 | any | A-1 | (0,1,2)(0,3)(0,4) § |
| 24-1 | any | A-1 | (0,1,2)(0,3,4) |
| 24-2 | any | A-1 | (0,1,2)(0,3,4) |
| 25 | any | A-1 | (0,1,2,3)(0,4) |
| 26-1 | any | A-1 | (0,1,2)(0,3,4) |
| 26-2 | any | A-1 | (0,1,2)(0,3,4) |
| 27-1 | $e = 4$ | A-2 | (0,1,2,3)(0,4) |
| | $d = 4$ | D-2 | $(\lfloor 0,1\rfloor_{\text{even}}, \lfloor 2,3\rfloor_{\text{even}})(0,1,4)$ |
| | $c = 4$ | D-2 | $(\lfloor 0,1\rfloor_{\text{even}}, \lfloor 2,3\rfloor_{\text{even}})(0,1,4)$ |
| 27-2 | any | A-2 | (0,1,2,3)(0,4) |
| 27-3 | $e = 4$ | A-2 | (0,1,2,3)(0,4) |
| | $e \in \{1,2\}$ | D-2 | $(\lfloor 1,2\rfloor_{\text{even}}, \lfloor 0,3\rfloor_{\text{even}})(0,3,4)$ |
| 27-4 | any | A-1 | (0,1,2,3)(0,4) |
| 27-5 | any | A-1 | (0,1,2,3)(0,4) |
| 27-6 | any | A-2 | (0,1,2,3)(0,4) |
| 27-7 | any | A-1 | (0,1,2,3)(0,4) |
| 27-8 | any | A-2 | (0,1,2,3)(0,4) |
| 27-9 | any | A-2 | (0,1,2,3)(0,4) |
| 28-1 | any | A-1-2 | (0,1,2)(0,4)(4,3)(0,4) |
| 28-2 | | D-1 | $(\lfloor 0,1\rfloor_{\text{even}}, 2)(0,4)(4,3)(0,4)$ |
| 28-3 | | D-1 | $(\lfloor 0,2\rfloor_{\text{even}}, 1)(0,4)(4,3)(0,4)$ |
| 29-1 | | A-1 | (0,1,2,3)(0,4) |
| 29-2 | any | A-1 | (0,1,2,3)(0,4) |
| 30 | any | A-1 | (0,1,2,3)(0,4) |
| 31 | any | A-1-2 | (0,1,2)(0,4)(4,3)(0,4) |
| 32 | any | A-1 | (0,1,2)(0,3,4) |
| 33 | $b = 2$ | A-1 | (0,1,3,4)(1,2) |
| | $b = 3$ | A-1 | (0,1,2,4)(3,4) |
| 34-1 | any | A-1 | (0,1,2,3)(0,4) |
| 34-2 | any | A-1 | (0,1,2,3)(0,4) |
| 35 | any | A-1-2 | (1,2)(0,1,4)(4,3)(0,1,4) |
| 36-1 | | A-1-2 | (0,1,2)(0,4)(4,3)(0,4) |
| 36-2 | any | A-2 | (0,1,2,3)(0,4) |
| 36-3 | any | A-2 | (0,1,2,3)(0,4) |
| 37-1 | any | A-2 | (0,1,2,3)(0,4) |
| 37-2 | any | A-2 | (0,1,2,3)(0,4) |
| 37-3 | any | A-2 | (0,1,2,3)(0,4) |
| 37-4 | any | A-2 | (0,1,2,3)(0,4) |
| 38-1 | | A-1 | (0,1,2,3)(0,4) |
| 38-2 | | A-1 | (0,1,2,3)(0,4) |
| 39 | any | A-1-2 | (0,1,2)(0,4)(4,3)(0,4) |
| 40-1 | | A-2 | (0,1,2,4)(4,3) |
| 40-2 | any | A-2 | (0,1,2,4)(4,3) |
| 40-3 | $(a,b,c,d) = (1,2,\{3,4\})$ | A-1-2 | (1,2)(0,1)(0,4,3)(0,1) |
| | $(a,b,c,d) = (4,3,\{1,2\})$ | A-1-2 | (0,1,2)(0,4)(4,3)(0,4) |
| 40-4 | any | A-2 | (0,1,2,3)(0,4) |
| 40-5 | any | A-2 | (0,1,2,3)(0,4) |
| 41 | any | D-1 | $([12], \lfloor 0\rfloor_{\text{even}})(0,3,4)$ |
| 42-1 | $a = 4$ | A-1 | (0,1,2,3)(0,4) |
| | $a \in \{2,3\}$ | B-1 | $[23], [11], \lfloor 0,4\rfloor_{\text{even}}, \lfloor 1\rfloor^1_{\text{odd}}$ |
| 42-2 | $a = 4$ | A-1 | (0,1,2,3)(0,4) |
| | $a \in \{1,2\}$ | B-1 | $[12], [33], \lfloor 0,4\rfloor_{\text{even}}, \lfloor 3\rfloor^1_{\text{odd}}$ |
| 42-3 | $b = 4$ | A-2 | (0,1,2,3)(0,4) |
| | $a = 4$ | A-1 | (0,1,2,3)(0,4) |
| | $b = 2,\ a \in \{1,3\}$ | B-2 | $1 + \{\lfloor 2\rfloor_{\text{even}}, \lfloor 2\rfloor^1_{\text{odd}}\} + 3 + \lfloor 0,4\rfloor_{\text{even}}$ |
| 42-4 | any | A-1 | (0,1,2,3)(0,4) |
| 42-5 | any | A-2 | (0,1,2,3)(0,4) |
| 42-6 | any | A-2 | (0,1,2,3)(0,4) |
| 42-7 | any | A-2 | (0,1,2,3)(0,4) |
| 42-8 | any | A-2 | (0,1,2,3)(0,4) |
| 43-1 | | D-2 | $(1, \lfloor 0,2,3\rfloor_{\text{even}})(0,2,3,4)$ |
| 43-2 | any | D-1 | ([00],[12])(0,4,3) |
| 43-3 | any | A-1-2 | (0,1,2)(0,4)(4,3)(0,4) |
| 43-4 | | D-2 | $(1, \lfloor 0,2,4\rfloor_{\text{even}})(0,2,3,4)$ |

| type | parameter | class | structure |
|---|---|---|---|
| 44-1 | $d = 4$ | D-3 | $([13], [00], [22], \lfloor 2\rceil^1_{\mathrm{odd}})(0,4)$ |
| | $d = 2$ | B-2 | $1 + 3 + \{\lfloor 2\rceil_{\mathrm{even}}, \lfloor 2\rceil^1_{\mathrm{odd}}, \lfloor 0,4\rceil_{\mathrm{even}}\}$ |
| 44-2 | any | B-2 | $1 + \{\lfloor 2\rceil_{\mathrm{even}}, \lfloor 2\rceil^1_{\mathrm{odd}}\} + 3 + \lfloor 0,4\rceil_{\mathrm{even}}$ |
| 45-1 | $b \in \{1,2\}, c \notin \{1,2\}$ | D-1 | $([12], [00], \lfloor 0\rceil^1_{\mathrm{odd}})(0,4)(4,3)(0,4)$ |
| | $(a,b,c) = (0,1,2)$ | B-3 | $[12]^k + 0 + \lfloor 4,3\rceil_{\mathrm{even}} + \{\lfloor 0\rceil_{\mathrm{odd}}, \lfloor 0\rceil^1_{\mathrm{even}}\}$ |
| | $(a,b,c) = (0,2,1)$ | B-3 | $[12]^k + \{\lfloor 0\rceil_{\mathrm{odd}}, \lfloor 0\rceil^1_{\mathrm{even}}\} + \lfloor 4,3\rceil_{\mathrm{even}} + 0$ |
| 45-2 | $a = 3, b \in \{1,2\}$ or $a = 4, c \in \{1,2\}$ | D-1 | ([12], [00])([33],[00])(0,4) |
| | $a \in \{1,2\}, b, c \in \{3,4\}$ | D-1 | $([12], [00], \lfloor 0\rceil^1_{\mathrm{odd}})(0,3)(0,4)$ § |
| | $a, c \in \{1,2\}$ | B-3 | $[12]^k + 0 + \{[44]^k, \lfloor 4\rceil^1_{\mathrm{odd}}\} + \lfloor 0,3\rceil_{\mathrm{even}}$ |
| | $a, b \in \{1,2\}$ | B-3 | $[12]^k + 0 + \{[33]^k, \lfloor 3\rceil^1_{\mathrm{odd}}\} + \lfloor 0,4\rceil_{\mathrm{even}}$ |
| 46 | any | D-1 | $([12], [00])(0,3,4)$ |
| 47-1 | | A-2 | (0,1,2,4)(4,3) |
| 47-2 | any | A-2 | (0,1,2,4)(4,3) |
| 47-3 | any | A-1-2 | (1,2)(2,0)(0,4)(4,3)(0,4)(0,1) |
| 47-4 | any | A-2 | (0,1,2,3)(0,4) |
| 47-5 | any | A-2 | (0,1,2,3)(0,4) |
| 47-6 | any | A-2 | (0,1,2,3)(0,4) |
| 47-7 | any | A-2 | (0,1,2,3)(0,4) |
| 48-1 | any | B-1 | $[34]^k + \{[00], \lfloor 0\rceil_{\mathrm{odd}}\} + \lfloor 1,2\rceil_{\mathrm{even}}$ |
| 48-2 | any | B-1 | $[12]^k + \lfloor 3,4\rceil_{\mathrm{even}} + \{[00], \lfloor 0\rceil_{\mathrm{odd}}\}$ |
| 48-3 | any | D-2 | $([23], \lfloor 1,4\rceil_{\mathrm{even}})(0,1,4)$ |
| 48-4 | any | D-2-2 | $([41]^*, \lfloor 2,3\rceil_{\mathrm{even}})([14]^*, [00], \lfloor 0\rceil^1_{\mathrm{odd}})$ |
| 49-1 | | D-1 | $([23], \lfloor 0,1\rceil_{\mathrm{even}}, \lfloor 1\rceil^1_{\mathrm{odd}})(0,4)$ |
| 49-2 | | D-1 | $([12], \lfloor 0,3\rceil_{\mathrm{even}}, \lfloor 3\rceil^1_{\mathrm{odd}})(0,4)$ |
| 49-3 | any | A-2 | (0,1,2,3)(0,4) |
| 49-4 | any | A-2 | (0,1,2,3)(0,4) |
| 49-5 | any | A-2 | (0,1,2,3)(0,4) |
| 50-1 | $c = 4$ | D-2 | $([12], \lfloor 0,3\rceil_{\mathrm{even}})(0,3,4)$ |
| | $b = 4$ | D-2 | $([12], \lfloor 0,3\rceil_{\mathrm{even}})(0,3,4)$ |
| | $a = 4$ | D-2-2 | $([12], \lfloor 0,3\rceil^*_{\mathrm{even}})(4, \lfloor 0,3\rceil^*_{\mathrm{even}})$ |
| 50-2 | any | B-1 | $[30], \lfloor 1,2\rceil_{\mathrm{even}}, [44], \lfloor 4\rceil_{\mathrm{odd}}$ |
| 50-3 | $c = 4$ | D-2 | $([12], \lfloor 0,3\rceil_{\mathrm{even}})(0,3,4)$ |
| | $b = 4$ | D-2 | $([12], \lfloor 0,3\rceil_{\mathrm{even}})(0,3,4)$ |
| | $a = 4$ | D-2-2 | $([12], \lfloor 0,3\rceil^*_{\mathrm{even}})([44], \lfloor 0,3\rceil^*_{\mathrm{even}}, \lfloor 4\rceil^1_{\mathrm{odd}})$ |
| 50-4 | any | B-2 | $3 + \{\lfloor 4\rceil_{\mathrm{even}}, \lfloor 4\rceil^1_{\mathrm{odd}}\} + 0 + \lfloor 1,2\rceil_{\mathrm{even}}$ |
| 50-5 | any | D-3 | $(0,2,4)(1 + \langle 0,2\rangle + 3 + \langle 0,2\rangle)$ |
| 51 | $a, b \in \{1,2\}$ | D-1 | ([12],[00])(0,3,4) |
| | $b = 1, c = 2$ | B-3 | $[12]^k + 0 + \lfloor 3,4\rceil_{\mathrm{even}} + \{\lfloor 0\rceil_{\mathrm{odd}}, \lfloor 0\rceil^1_{\mathrm{even}}\}$ |
| | $b = 2, c = 1$ | B-3 | $[12]^k + \{\lfloor 0\rceil_{\mathrm{odd}}, \lfloor 0\rceil^1_{\mathrm{even}}\} + \lfloor 3,4\rceil_{\mathrm{even}} + 0$ |
| 52-1 | any | D-2 | $([34], \lfloor 0,2\rceil_{\mathrm{even}})(0,1,2)$ |
| 52-2 | any | D-2 | $([34], \lfloor 0,2\rceil_{\mathrm{even}})(0,1,2)$ |
| 53-1 | any | D-2 | $([12], \lfloor 0,3\rceil_{\mathrm{even}})(\lfloor 0,3\rceil_{\mathrm{even}}, 4)$ |
| 53-2 | any | D-2 | $([12], \lfloor 0,4\rceil_{\mathrm{even}})(\lfloor 0,4\rceil_{\mathrm{even}}, 3)$ |
| 54-1 | any | B-1 | $[23], \lfloor 0,1,4\rceil_{\mathrm{even}}, \lfloor 0,1,4\rceil^1_{\mathrm{odd}}$ |
| 54-2 | any | B-1 | $[12], \lfloor 0,3,4\rceil_{\mathrm{even}}, \lfloor 0,3,4\rceil^1_{\mathrm{odd}}$ |
| 54-3 | $(a,b) = (0,4)$ | D-2 | $([23], \lfloor 0,1\rceil_{\mathrm{even}})(\lfloor 0,1\rceil_{\mathrm{even}}, [44], \lfloor 4\rceil^1_{\mathrm{odd}})$ |
| | $(a,b) = (4,0)$ | D-2 | $([23], [01], [11])([00], [01], [44], \lfloor 4\rceil^1_{\mathrm{odd}})$ |
| 54-4 | $(a,d) = (0,4)$ | D-2 | $([12], \lfloor 0,3\rceil_{\mathrm{even}})(\lfloor 0,3\rceil_{\mathrm{even}}, [44], \lfloor 4\rceil^1_{\mathrm{odd}})$ |
| | $(a,b) = (4,0)$ | D-2 | $([12], [30], [33])([00], [30], [44], \lfloor 4\rceil^1_{\mathrm{odd}})$ |
| 54-5 | $(a,b) = (0,4)$ | A-1 | (0,1,2,3)(0,4) |
| | $(a,b) = \{1,3\}$ | B-2 | $1 + \{[22], \lfloor 2\rceil^1_{\mathrm{odd}}\} + 3 + \lfloor 0,4\rceil_{\mathrm{even}}$ |
| 54-6 | any | A-2 | (0,1,2,3)(0,4) |
| 54-7 | any | A-2 | (0,1,2,3)(0,4) |
| 54-8 | any | A-2 | (0,1,2,3)(0,4) |
| 54-9 | any | A-2 | (0,1,2,3)(0,4) |
| 55-1 | any | D-2 | $([12], \lfloor 0,4\rceil_{\mathrm{even}})(0,4,3)$ |
| 55-2 | any | D-1 | $([12], [00])(0,4)(4,3)(0,4)$ |
| 56-1 | | A-2 | (0,1,2,3)(0,4) |
| 56-2 | any | A-2 | (0,1,2,3)(0,4) |
| 57 | any | B-1 | $[123], [000]\lfloor 0,4\rceil^1_{l_n}$ with $l_n \equiv 2 \cdot 5^{n-1} \mod 3$, |
| 58-1 | $(a,b,c,d) = (1,2,3,4)$ | B-1 | $3, 4, \lfloor 0,1,2\rceil_{\mathrm{even}}, \lfloor 0,1,2\rceil^1_{\mathrm{odd}}$ |
| | $(a,b,c,d) = (3,4,1,2)$ | B-1 | $1, 2, \lfloor 0,3,4\rceil_{\mathrm{even}}, \lfloor 0,3,4\rceil^1_{\mathrm{odd}}$ |
| 58-2 | any | B-2 | $1 + \{[22], \lfloor 2\rceil^1_{\mathrm{odd}}\} + 3 + \lfloor 0,4\rceil_{\mathrm{even}}$ |
| 58-3 | any | B-1 | $[13], [22], \lfloor 2\rceil^1_{\mathrm{odd}}, \lfloor 0,4\rceil_{\mathrm{even}}$ |
| 59 | $(a,b,c,d) = (1,3,2,4)$ | B-1 | $[401], [223], \lfloor 2,3\rceil_{l_n}$ with $l_n \equiv 2 \cdot 5^{n-1} \mod 3$, |
| | $(a,b,c,d) = (2,4,1,3)$ | B-1 | $[401], [233], \lfloor 2,3\rceil_{l_n}$ with $l_n \equiv 2 \cdot 5^{n-1} \mod 3$, |
| 60-1 | any | B-1 | $[12], \lfloor 0,4,3\rceil_{\mathrm{even}}, \lfloor 0,4,3\rceil^1_{\mathrm{odd}}$ |
| 60-2 | any | B-1 | $[12], \lfloor 0,4,3\rceil_{\mathrm{even}}, \lfloor 0,4,3\rceil^1_{\mathrm{odd}}$ |
| 60-3 | any | B-1 | $[12], \lfloor 0,4,3\rceil_{\mathrm{even}}, \lfloor 0,4,3\rceil^1_{\mathrm{odd}}$ |

| type | parameter | class | structure |
|---|---|---|---|
| 61 | | B-1 | [231], [040], $\lfloor 0,4 \rfloor_{l_n}$ with $l_n \equiv 2 \cdot 5^{n-1} \mod 3$, |
| 62 | any | B-2 | $1 + * + 2 + *$ $(* \in \{\lfloor 0,3,4 \rfloor_{\mathrm{even}}, \lfloor 0,3,4 \rfloor^1_{\mathrm{odd}}\})$ |
| 63-1 | any | B-1 | [12], [34], [00], $\lfloor 0 \rfloor^1_{\mathrm{odd}}$ |
| 63-2 | any | C | $[\{1,3\} + \{2,4\}]$, [00], $\lfloor 0 \rfloor^1_{\mathrm{odd}}$ |
| 63-3 | any | C | $[\{1,3\} + \{2,4\}]$, [00], $\lfloor 0 \rfloor^1_{\mathrm{odd}}$ |
| 63-4 | any | C | $[\{1,3\} + \{2,4\}]$, [00], $\lfloor 0 \rfloor^1_{\mathrm{odd}}$ |
| 64-1 | | B-1 | [12], $\lfloor 0,3,4 \rfloor_{\mathrm{even}}$, $\lfloor 0,3,4 \rfloor^1_{\mathrm{odd}}$ |
| 64-2 | any | B-2 | $1 + * + 2 + *$ $(* \in \{\lfloor 0,3,4 \rfloor_{\mathrm{even}}, \lfloor 0,3,4 \rfloor^1_{\mathrm{odd}}\})$ |
| 64-3 | | B-1 | [23], $\lfloor 0,1,4 \rfloor_{\mathrm{even}}$, $\lfloor 0,1,4 \rfloor^1_{\mathrm{odd}}$ |
| 65-1 | $b = 4$ | D-2 | ([12], [30])($\lfloor 4 \rfloor_{\mathrm{even}}$, $\lfloor 4 \rfloor^1_{\mathrm{odd}}$, [30]) |
| | $a = 4$ | D-2 | ([12], [30])([30], [44], $\lfloor 4 \rfloor^1_{\mathrm{odd}}$) |
| | otherwise | C | $[\{3,1\} + \{0,2\}]$, [44], $\lfloor 4 \rfloor^1_{\mathrm{odd}}$ |
| 65-2 | any | C | $[\{3,1\} + \{0,2\}]$, [44], $\lfloor 4 \rfloor^1_{\mathrm{odd}}$ |
| 65-3 | any | C | $[\{3,1\} + \{0,2\}]$, [44], $\lfloor 4 \rfloor^1_{\mathrm{odd}}$ |
| 65-4 | any | C | $[\{3,1\} + \{0,2\}]$, [44], $\lfloor 4 \rfloor^1_{\mathrm{odd}}$ |
| 65-5 | $a, b \in \{0,2\}$ or $a, b \in \{1,3\}$ | C | $[\{0,2\} + \{3,1\}]$, [44], $\lfloor 4 \rfloor^1_{\mathrm{odd}}$ |
| | $a, b \in \{0,3\}$ | B-2 | $0 + \lfloor 12 \rfloor_{\mathrm{even}} + 3 + \{\lfloor 4 \rfloor_{\mathrm{even}}, \lfloor 4 \rfloor^1_{\mathrm{odd}}\}$ |
| | $a, b \in \{1,2\}$ | B-1 | [12], $\lfloor 0,3,4 \rfloor_{\mathrm{even}}$, $\lfloor 0,3,4 \rfloor^1_{\mathrm{odd}}$ |
| | $a = 4$ or $b = 4$ | D-2-2 | ([12], $[30]^*$)(4, $[03]^*$) |
| 65-6 | any | C | $[\{0,2\} + \{3,1\}]$, [44], $\lfloor 4 \rfloor^1_{\mathrm{odd}}$ |
| 65-7 | any | C | $[\{0,2\} + \{3,1\}]$, [44], $\lfloor 4 \rfloor^1_{\mathrm{odd}}$ |
| 65-8 | any | C | $[\{0,2\} + \{3,1\}]$, [44], $\lfloor 4 \rfloor^1_{\mathrm{odd}}$ |
| 66-1 | $a, b \in \{3,4\}$ | B-1 | [34], $\lfloor 0,1,2 \rfloor_{\mathrm{even}}$, $\lfloor 0,1,2 \rfloor^1_{\mathrm{odd}}$ |
| | $a, b \in \{1,2\}$ | B-1 | [12], $\lfloor 0,3,4 \rfloor_{\mathrm{even}}$, $\lfloor 0,3,4 \rfloor^1_{\mathrm{odd}}$ |
| | otherwise | D-1 | ([12], [00], $\lfloor 0 \rfloor^1_{\mathrm{odd}}$)([34], [00]) |
| 66-2 | any | B-1 | $[\{1,3\} + \{2,4\}]$, [00], $\lfloor 0 \rfloor^1_{\mathrm{odd}}$ |
| 66-3 | $(a,b) = \{3,4\}$ | E | $\{3,*\} + \{4,*\}$ $(* = \{\lfloor 0,1,2 \rfloor_{\mathrm{odd}}, \lfloor 0,1,2 \rfloor^1_{\mathrm{even}}\})$ |
| | $(a,b) = \{2,3\}$ | E | $\{2,*\} + \{3,*\}$ $(* = \{\lfloor 0,1,4 \rfloor_{\mathrm{odd}}, \lfloor 0,1,4 \rfloor^1_{\mathrm{even}}\})$ |
| | $(a,b) = \{1,4\}$ | E | $\{1,*\} + \{4,*\}$ $(* = \{\lfloor 0,2,3 \rfloor_{\mathrm{odd}}, \lfloor 0,2,3 \rfloor^1_{\mathrm{even}}\})$ |
| | $(a,b) = \{1,2\}$ | E | $\{1,*\} + \{2,*\}$ $(* = \{\lfloor 0,3,4 \rfloor_{\mathrm{odd}}, \lfloor 0,3,4 \rfloor^1_{\mathrm{even}}\})$ |
| 66-4 | any | B-2 | $3 + * + 4 + *$ $(* = \{\lfloor 0,1,2 \rfloor_{\mathrm{even}}, \lfloor 0,1,2 \rfloor^1_{\mathrm{odd}}\})$ |
| 67-1 | any | B-1 | [12], [34], [00], $\lfloor 0 \rfloor^1_{\mathrm{odd}}$ |
| 67-2 | any | B-1 | [34], $\lfloor 0,1,2 \rfloor_{\mathrm{even}}$, $\lfloor 0,1,2 \rfloor^1_{\mathrm{odd}}$ |
| 67-3 | $(a,b) = (3,4)$ | B-2 | $3 + * + 4 + *$ $(* = \{\lfloor 0,1,2 \rfloor_{\mathrm{even}}, \lfloor 0,1,2 \rfloor^1_{\mathrm{odd}}\})$ |
| | $(a,b) = (2,1)$ | B-2 | $1 + * + 2 + *$ $(* = \{\lfloor 0,3,4 \rfloor_{\mathrm{even}}, \lfloor 0,3,4 \rfloor^1_{\mathrm{odd}}\})$ |
| | $(a,b) = (2,3)$ | E | $\{2,*\} + \{3,*\}$ $(* = \{\lfloor 0,1,4 \rfloor_{\mathrm{odd}}, \lfloor 0,1,4 \rfloor^1_{\mathrm{even}}\})$ |
| | $(a,b) = (1,4)$ | E | $\{1,*\} + \{4,*\}$ $(* = \{\lfloor 0,2,3 \rfloor_{\mathrm{odd}}, \lfloor 0,2,3 \rfloor^1_{\mathrm{even}}\})$ |
| 68-1 | any | B-1 | [12], [34], [00], $\lfloor 0 \rfloor^1_{\mathrm{odd}}$ |
| 68-2 | $a = 0$ | B-1 | [23], $\lfloor 4,1 \rfloor_{\mathrm{even}}$, [00], $\lfloor 0 \rfloor^1_{\mathrm{odd}}$ |
| | $c = 0$ | B-2 | $4 + \{\lfloor 0 \rfloor_{\mathrm{even}}, \lfloor 0 \rfloor^1_{\mathrm{odd}}\} + 1 + [23]^{\otimes k}$ |
| 68-3 | any | C | $[\{1,3\} + \{2,4\}]$, [00], $\lfloor 0 \rfloor_{\mathrm{odd}}$ |
| 68-4 | any | C | $[\{1,3\} + \{2,4\}]$, [00], $\lfloor 0 \rfloor_{\mathrm{odd}}$ |
| 68-5 | $c = 0$ | C | $[\{1,3\} + \{2,4\}]$, [00], $\lfloor 0 \rfloor_{\mathrm{odd}}$ |
| | $a = 0$ | D-2-3 | $(\{1,3\} + \{2,4\})$([14], [00]$\lfloor 0 \rfloor^1_{\mathrm{odd}}$) |
| 68-6 | any | C | $[\{1,3\} + \{2,4\}]$, [00], $\lfloor 0 \rfloor_{\mathrm{odd}}$ |
| 68-7 | $c = 0$ | C | $[\{1,3\} + \{2,4\}]$, [00], $\lfloor 0 \rfloor_{\mathrm{odd}}$ |
| | $a = 0$ or $b = 0$ | D-2-3 | $(\{1,3\} + \{2,4\})$([14], [00]$\lfloor 0 \rfloor^1_{\mathrm{odd}}$) |
| 68-8 | any | C | $[\{1,3\} + \{2,4\}]$, [00], $\lfloor 0 \rfloor_{\mathrm{odd}}$ |
| 69 | $(b,d) = (2,4)$ | E | $\{2,4\} + \{3, \lfloor 0,1 \rfloor_{\mathrm{odd}}, \lfloor 0,1 \rfloor^1_{\mathrm{even}}\}$ |
| | $(b,d) = (1,3)$ | E | $\{1.3\} + \{2, \lfloor 0,4 \rfloor_{\mathrm{odd}}, \lfloor 0,4 \rfloor^1_{\mathrm{even}}\}$ |
| 70-1 | any | C | $[\{0,2\} + \{1,3\}]$, [44], $\lfloor 4 \rfloor^1_{\mathrm{odd}}$ |
| 70-2 | $a = 4$ | C | $[\{0,2\} + \{1,3\}]$, [44], $\lfloor 4 \rfloor^1_{\mathrm{odd}}$ |
| | $e = 4$ | D-2-3 | $(\{1,3\} + \{0,2\})(4 + \lfloor 3 + \{0,2\} \rfloor_{\mathrm{even}})$ |
| 70-3 | any | C | $[\{0,2\} + \{1,3\}]$, [44], $\lfloor 4 \rfloor^1_{\mathrm{odd}}$ |
| 70-4 | $d = 4$ | C | $[\{0,2\} + \{1,3\}]$, [44], $\lfloor 4 \rfloor^1_{\mathrm{odd}}$ |
| | $c = 4$ | D-2-3 | $(\{1,3\} + \{0,2\})(4 + \lfloor 3 + \{0,2\} \rfloor_{\mathrm{even}})$ |
| 70-5 | $d = 4$ | C | $[\{0,2\} + \{1,3\}]$, [44], $\lfloor 4 \rfloor^1_{\mathrm{odd}}$ |
| | $a, c = 4$ | D-2-3 | $(\{1,3\} + \{0,2\})(4 + \lfloor 3 + \{0,2\} \rfloor_{\mathrm{even}})$ |
| | $e = 4$ | D-2-3 | $(\{1,3\} + \{0,2\})(\lfloor 0 + \{1,3\} \rfloor_{\mathrm{even}} + 4)$ |
| 71-1 | | B-1 | [1234], [0000], $\lfloor 1 \rfloor_{l_n}$ with $l_n \equiv 1 \mod 4$ |
| 71-2 | any | C | $[\{1,3\} + \{2,4\}]$, [00], $\lfloor 0 \rfloor_{\mathrm{odd}}$ |
| 71-3 | any | D-2-3 | $(\{1,3\} + \{2,4\})$([14], [00]$\lfloor 0 \rfloor^1_{\mathrm{odd}}$) |
| 72-1 | any | D-2 | ([23], [41])(0,1,4) |
| 72-2 | any | D-2 | ([23], [41])(0,1,4) |